%% file: HWW_PRD_submit.tex
\RequirePackage{lineno}
\documentclass[prd,twocolumn,showpacs,superscriptaddress,floatfix]{revtex4}

\usepackage{ifthen}
\usepackage{lineno}
\usepackage{epsfig}
\usepackage{amsmath}
\usepackage{comment}
\usepackage{longtable}
\usepackage{subfigure}
\usepackage{color}
\usepackage{xspace}
\usepackage{slashbox}
\usepackage{amsfonts}

\usepackage{multirow}
\usepackage{array} 
\usepackage{booktabs}
\usepackage{dcolumn}% Align table columns on decimal point
\usepackage[abs]{overpic}

\newcommand{\Met}{\ifmmode  E\kern-0.50em\raise0.10ex\hbox{/}_{T}
                 \else
                            \mbox{$E\kern-0.50em\raise0.10ex\hbox{/}_{T}$}
                 \fi
                 }
\newcommand{\VecMet}{\ifmmode  \vec{E}\kern-0.50em\raise0.10ex\hbox{/}_{T}
                 \else
                            \mbox{$\vec{E}\kern-0.50em\raise0.10ex\hbox{/}_{T}$}
                 \fi
                 }
\newcommand{\MetSpec}{\ifmmode  E\kern-0.50em\raise0.10ex\hbox{/}_{T}^{\rm spec}
                 \else
                            \mbox{$E\kern-0.50em\raise0.10ex\hbox{/}_{T}^{\rm spec}$}
                 \fi
                 }
\newcommand{\MetDeltaPhi}{\ifmmode  \Delta\phi_{\Met,{\rm nearest}}
                 \else
                            \mbox{$\Delta\phi_{\Met,{\rm nearest}}$}
                 \fi
                 }
\newcommand{\MetSig}{\ifmmode  E\kern-0.50em\raise0.10ex\hbox{/}_{T}^{\rm sig}
                 \else
                            \mbox{$E\kern-0.50em\raise0.10ex\hbox{/}_{T}^{\rm sig}$}
                 \fi
                 }

\begin{document}

%\linenumbers

\title{Searches for the Higgs boson decaying to 
{\boldmath $W^+W^-\rightarrow\ell^+\nu\ell^-{\bar{\nu}}$} with the CDF~II detector}

\input{November2012_Authors-1}

\date{\today}

\begin{abstract}
We present a search for a standard model Higgs boson decaying 
to two $W$ bosons that decay to leptons using the full data 
set collected with the CDF~II detector in $\sqrt{s}=1.96$~TeV 
$p\bar{p}$ collisions at the Fermilab Tevatron, corresponding 
to an integrated luminosity of 9.7~fb${}^{-1}$.  We obtain no 
evidence for production of a standard model Higgs boson with 
mass between 110 and 200~GeV/$c^2$, and place upper limits on 
the production cross section within this range.  We exclude 
standard model Higgs boson production at the 95\% confidence 
level in the mass range between 149 and 172~GeV/$c^2$, while 
expecting to exclude, in the absence of signal, the range 
between 155 and 175~GeV/$c^2$.  We also interpret the search 
in terms of standard model Higgs boson production in the 
presence of a fourth generation of fermions and within the 
context of a fermiophobic Higgs boson model.  For the specific 
case of a standard model-like Higgs boson in the presence of 
fourth-generation fermions, we exclude at the 95\% confidence 
level Higgs boson production in the mass range between 124 
and 200~GeV/$c^2$, while expecting to exclude, in the absence 
of signal, the range between 124 and 221~GeV/$c^2$. 
\end{abstract}


\maketitle

% Introduction
\section{Introduction}
\label{sec:intro}

In the standard model of particle physics (SM), the electroweak force is 
characterized by a gauge theory of the $\mathrm{(SU(2)_L \times U(1)_Y)}$ 
symmetry group ~\cite{Glashow:1961aa,Weinberg:1967aa,Salam:1969aa}.  This 
symmetry is broken, which introduces differences in the observed phenomenology 
of electromagnetic and weak interactions.  The mechanism of symmetry breaking 
in the SM is known as the Higgs mechanism~\cite{Englert:1964aa,Higgs:1964aa,Higgs:1964ab,Guralnik:1964aa},
which introduces a complex doublet of scalar fields.  The self-interaction 
of these fields introduces a potential term in the electroweak Lagrangian, 
which has a minimum at a nonzero value of the field.  At sufficiently low 
energies (the Fermi scale and below), the electroweak Lagrangian is 
approximated by an effective Lagrangian, which is no longer symmetric under 
the full gauge group but rather retains only U(1)$_{\mathrm{EM}}$ symmetry, 
leading to additional terms.  Three of these terms are identified with the 
masses of the $W^\pm$ and $Z$ vector bosons, and the fourth results in an  
associated scalar boson known as the Higgs boson.  The masses of the leptons 
and quarks also require that electroweak symmetry is broken and are generated 
in the SM through Yukawa interactions with the scalar Higgs field.

Owing to its central position in the understanding of the phenomenology of 
the electroweak force, the discovery of the Higgs boson~\cite{atlasobs,cmsobs} 
was an important milestone for particle physics.  Properties of the Higgs boson, 
including production rates and decay branching ratios, are highly sensitive to 
physics beyond the SM.  Many models, such as supersymmetry, require extended 
Higgs sectors with additional multiplets of scalar fields, resulting in 
additional Higgs bosons, some of which interact very differently from the 
SM-predicted Higgs boson.

The possible mass range for the SM Higgs boson ($m_H$) is constrained 
by theoretical and experimental results.  The $W$ boson mass $M_W$, 
the $Z$ boson mass $M_Z$, and the top-quark mass $m_t$ are modified by 
self-energy terms involving the Higgs boson as a virtual particle in 
processes with amplitudes involving one or more loops, which depend on the 
mass of the Higgs boson.  This, in turn, allows for a prediction of the 
Higgs boson mass using precision measurements of $M_W$, $M_Z$, and $m_t$.  
The most recent average of available $W$ boson mass measurements is 
$M_W=80.385\pm0.015$~GeV/$c^2$~\cite{lepewwg}, and the most recent average 
of top-quark mass measurements is $m_t=173.2\pm0.9$~GeV/$c^2$~\cite{mtwa}. 
These mass measurements are combined with other precision electroweak 
measurements to calculate an allowed range of $m_H = 94^{+29}_{-24}$~GeV/$c^2$ 
at the 68\% confidence level (C.L.) or less than $152$~GeV/$c^2$ at the 
95\% C.L.~\cite{Collaboration:2009jr}.  In addition, direct searches at the 
LEP collider excluded SM Higgs boson production for masses below $114.4$~GeV/$c^2$ 
at the 95\% C.L.~\cite{Barate:2003sz}.  A combination of the direct LEP searches 
with indirect constraints indicates that the SM Higgs boson should have a mass 
below $171$~GeV/$c^2$ at the 95\% C.L.~\cite{Collaboration:2009jr}.

A previously unknown boson with a mass of approximately 125~GeV/$c^2$, 
compatible with the SM Higgs boson, has been observed in data collected 
from $\sqrt{s}=$~7--8~TeV $pp$ collisions at the CERN Large Hadron Collider 
(LHC) by the ATLAS~\cite{atlasobs} and CMS~\cite{cmsobs} collaborations.  
The new boson was observed with high significance in the {\it ZZ} and 
$\gamma\gamma$ decay modes and at a somewhat lower level of significance 
in the {\it WW} decay mode.  Updated ATLAS~\cite{atlasww} and CMS~\cite{cmsww} 
searches focusing on $H\rightarrow W^+W^-$ decay and using additional 
data provide strengthened evidence for this decay mode.  Since the 
phenomenology of the Higgs mechanism is characterized by its interactions 
with $W$ and $Z$ bosons, observation of the Higgs boson in the {\it WW} 
decay mode and refined measurements of the corresponding branching ratio 
are of critical importance.

For higher Higgs boson masses, $m_H > 130$~GeV/$c^2$, where the decay to two $W$ 
bosons dominates~\cite{Djouadi:1997yw}, a SM Higgs boson is primarily observable
at the Tevatron via gluon-fusion production through a top-quark loop ($ggH$), 
with subsequent decay to a pair of $W^*$ bosons~\cite{hhg,Glover:1988fn,Han:1998ma,Han:1998sp}.
This decay mode provides a low-background search topology, when both $W$ bosons 
decay leptonically.  The main backgrounds to $H\rightarrow W^+W^-\rightarrow\ell^+
\nu\ell^-{\bar{\nu}}$ are Drell-Yan (DY) production of oppositely-charged leptons, 
$p{\bar{p}}\rightarrow W^+W^-$, $W^\pm Z$, {\it ZZ}, $t{\bar{t}}$, $W+$jets, 
and $W+\gamma$ processes.  Events consistent with the $\ell^+\nu\ell^-{\bar{\nu}}$
final state are selected by requiring two oppositely-charged leptons and 
a significant overall imbalance in measured transverse energies within the 
event (missing transverse energy or $\Met$).  CDF reconstructs electrons and 
muons with high efficiency and minimal contamination from jets misidentified 
as leptons (fakes).  We treat separately tau leptons decaying hadronically, 
which are harder to reconstruct and significantly contaminated with fakes.  
Missing transverse energy associated with the unobserved neutrinos provides 
discrimination against backgrounds that do not contain leptonically decaying 
$W$ bosons, such as DY production.

A potential Higgs boson signal is distinguishable from the other background 
processes with real $\Met$ generated from neutrinos based on unique kinematic 
properties associated with the Higgs boson decay.  The fact that the Higgs 
boson is a scalar particle induces a spin correlation between the $W$ bosons, 
which manifests itself as a preference for the charged leptons in the final 
state to be emitted in similar directions to one another.  The non-resonant 
$p{\bar{p}}\rightarrow W^+W^-$ background has a very different spin 
structure~\cite{Dittmar:1996sp}, resulting in a different distribution of the 
angle between the two charged leptons.

In addition to the $ggH$ production mechanism, the SM Higgs boson is expected 
to be produced in association with a $W$ or $Z$ vector boson ({\it WH}, {\it ZH}, 
or, collectively, {\it VH} production), and in vector boson fusion (VBF), where
a virtual pair of W bosons or Z bosons fuse to form a Higgs boson, usually with 
recoiling jets.  Including these additional production mechanisms expands 
acceptance by approximately 50\% for $m_H = 160$~GeV/$c^2$, compared to searching 
for only the $ggH$ production process~\cite{Aaltonen:2008ec}.  These additional 
production mechanisms were included in the most recent CDF results~\cite{Aaltonen:2010cm},
which were combined with similar results from the D0 collaboration~\cite{Abazov:2010ct}
to exclude at 95\%~C.L. a SM Higgs boson in the mass range between $162$ and 
$166$~GeV/$c^2$~\cite{Aaltonen:2010yv}.

For lower Higgs boson masses, $m_H<130$~GeV/$c^2$, the decay $H\rightarrow 
b{\bar{b}}$ dominates.  A direct search for the SM Higgs boson in the process 
$gg\rightarrow H\rightarrow b{\bar{b}}$ would be overwhelmed by nonresonant, 
multijet backgrounds.  Hence, Tevatron searches in this mass region focus 
on the $WH\rightarrow\ell\nu b{\bar{b}}$~\cite{cdfWH,d0WH}, $ZH\rightarrow
\ell^+\ell^- b{\bar{b}}$~\cite{cdfZH,d0ZH}, and $ZH\rightarrow\nu{\bar{\nu}}
b{\bar{b}}$~\cite{cdfMetbb,d0Metbb} processes.  The combination of Tevatron 
searches in these decay modes~\cite{Tevbb} resulted in first evidence for 
{\it VH} production in association with $H\rightarrow b\bar{b}$ decay. 
Despite the low SM Higgs boson branching ratio to $W$ bosons within this mass 
range, the {\it WW} decay mode still contributes significantly to combined 
Tevatron search sensitivities because it is accessible within a final state 
originating from $ggH$ production.

In this paper we present a search for the production of SM Higgs bosons 
with subsequent decay to two oppositely-charged $W^{(*)}$ bosons using a 
sample of $\sqrt{s} = 1.96$~TeV proton-antiproton ($p\bar{p}$) collision 
data corresponding to 9.7~fb$^{-1}$ of integrated luminosity collected 
with the CDF~II detector at the Fermilab Tevatron.  This result improves 
on previous CDF results~\cite{Aaltonen:2010cm,HidasPhD,GrisoPhD,CarlsPhD} 
by including more data, using improved analysis techniques, and incorporating 
additional search topologies such as dilepton pairs with invariant mass 
below 16~GeV/$c^2$ and trilepton events from {\it VH} production, where the 
third lepton results from the decay of the associated weak vector boson.  

This paper is organized as follows.  
Section~\ref{sec:theory} describes the phenomenology of Higgs boson production and decay, 
Sec.~\ref{sec:strategy} describes the analysis strategy, 
Sec.~\ref{sec:ec:cdf} describes the CDF~II detector,
Sec.~\ref{sec:evtsel} describes the event selection,
Sec.~\ref{sec:BGmodeling} describes the background modeling,
Sec.~\ref{sec:Multivaraite} describes the multivariate techniques used to
separate the expected signal events from the background events,
Sec.~\ref{sec:analysis} describes each analysis sample, 
Sec.~\ref{sec:syst} summarizes systematic uncertainties on signal and 
background predictions, and
Sec.~\ref{sec:finalresults} describes the procedures used for interpreting
the data and the final results.

\section {Phenomenology of Higgs boson production and decay}  
\label{sec:theory}

Higgs boson searches in hadron collisions rely both on accurate 
predictions of Higgs boson production and decay rates and on  
accurate kinematic modeling of the resulting events.  The theoretical
community has provided calculations of all relevant signal production 
cross sections at next-to-next-to-leading order (NNLO) accuracy in 
the strong-interaction coupling constant $\alpha_{s}$, and also 
differential cross sections for $ggH$ production at the same order.  
These calculations, in conjunction with Monte Carlo simulation tools 
for modeling the signal and background processes as well as the 
response of the CDF~II detector to the particles originating from 
these processes, are critical inputs to this search.   

The dominant Higgs boson production mechanism over the mass range of 
interest in $p\bar{p}$ collisions is $ggH$.  Because of the large Yukawa 
coupling of the top quark to the Higgs boson, the largest contribution 
to the cross section comes from the top-quark-loop amplitude.  However, 
loops involving other quark flavors are incorporated within the calculations.  
Calculations of the inclusive cross section for $ggH$ production in hadron 
collisions have progressed from leading order (LO)~\cite{Georgi:1977gs}, to 
next-to-leading order (NLO)~\cite{Dawson:1990zj,Djouadi:1991tka,Spira:1995rr}, 
to next-to-next-to-leading order (NNLO)~\cite{Harlander:2002wh,
Anastasiou:2002yz,Ravindran:2003um}, and finally to the NNLO calculations 
described in Refs.~\cite{deFlorian:2009hc} and~\cite{Anastasiou:2008tj}, 
which are used here.

The expected cross section for this process ranges from 1385~fb at $m_H = 110$~GeV/$c^2$ 
to 189.1~fb at $m_H = 200$~GeV/$c^2$~\cite{deFlorian:2009hc,Anastasiou:2008tj}, as 
summarized in Table~\ref{table:xsec}.  These cross section predictions are obtained 
from calculations at NNLO in perturbative QCD, incorporating contributions from both 
top- and bottom-quark loops, effects of finite quark masses, electroweak contributions 
from two-loop diagrams~\cite{Aglietti:2004nj}, interference effects from mixing 
of electroweak and QCD contributions~\cite{Anastasiou:2008tj}, leading logarithmic 
resummation of soft gluon contributions~\cite{Catani:2003zt,deFlorian:2009hc}, and 
MSTW2008 NNLO parton distribution functions (PDFs)~\cite{Martin:2009iq}.  Consistent 
results are obtained from calculations based on substantially different techniques 
and independent groups.

The NLO prediction for the $ggH$ production cross section at the Tevatron is
typically a factor of two larger than the LO prediction, and the NNLO prediction 
is another factor of 1.4 larger.  Uncertainties in the NNLO cross section 
calculation are evaluated by studying the effect on the result of factorization 
and renormalization scale choices.  The largest variation is obtained when the two 
scales are varied together.  We take an uncertainty on the production cross section 
corresponding to the shift observed when these scales are varied upwards and 
downwards by factors of two.  Calculations that have been performed including the 
primary amplitudes at next-to-next-to-next-to leading order (NNNLO) indicate that no 
additional large modification of the cross section is expected~\cite{Moch:2005ky}.

The NNLO generator programs {\sc fehip}~\cite{Anastasiou:2004xq,Anastasiou:2005qj}
and {\sc hnnlo}~\cite{Catani:2007vq,Grazzini:2008tf} and studies based on these 
programs~\cite{Anastasiou:2009bt} are used to tune the leading order simulation, 
which models the kinematic properties of final state particles originating from 
$ggH$ production, and to assess systematic uncertainties associated with this 
modeling.

In the search described here, events are separated into samples in which 
the leptonically decaying $W^+W^-$ system is observed to recoil against 
zero-, one-, or two-or-more parton jets.  Jet reconstruction, discussed 
in Sec.~\ref{sec:evtsel}, collects the energy depositions associated with 
particles produced in the hadronization and fragmentation of partons 
originating from the $p\bar{p}$ interaction.  We normalize the yields of
simulated $ggH$ events based on the inclusive cross section calculations 
described above, but assign differential uncertainties incorporating 
calculations of the exclusive one-or-more parton jet and two-or-more 
parton jet cross sections from Refs.~\cite{Anastasiou:2009bt} 
and~\cite{Campbell:2010cz}, respectively.  We follow the prescription 
of Refs.~\cite{Dittmaier:2012vm} and~\cite{Stewart:2011cf}, propagating 
scale uncertainties associated with the inclusive cross section, the 
one-or-more parton jet cross section, and the two-or-more parton jet 
cross section through the subtractions needed to obtain the exclusive 
zero-, one-, and two-or-more parton jet cross sections.  We follow the 
prescription of Refs.~\cite{Botje:2011sn} and~\cite{Dittmaier:2011ti} 
in evaluating the effects of PDF uncertainties on the production cross 
sections.

This search includes substantial additional acceptance for the Higgs boson 
by incorporating potential signal contributions from {\it VH} and VBF production.
The cross sections for these production processes are roughly $\mathcal{O}(0.1)$ 
of those for $ggH$ production.  In the mass range between $110$ to $200$~GeV/$c^2$, 
the {\it WH}, {\it ZH}, and VBF production cross sections vary from 204 to 
19.1~fb, 120 to 13.0~fb, and 82.8 to 21.7~fb, respectively, as summarized in 
Table~\ref{table:xsec}.

The cross sections for {\it VH} and VBF production have been calculated at NNLO in 
Refs.~\cite{Baglio:2010um,Ferrera:2011bk,Assamagan:2004mu,Brein:2003wg,Ciccolini:2003jy} 
and~\cite{Bolzoni:2010xr,Assamagan:2004mu,Berger:2004pca}, respectively.  The 
VBF cross sections are adjusted for electroweak corrections computed at NLO in 
Refs.~\cite{Ciccolini:2007ec} and~\cite{Ciccolini:2007jr}.  All calculations 
are based on MSTW2008 NNLO parton distribution functions~\cite{Martin:2009iq}.  
Uncertainties on {\it VH} and VBF production cross sections are typically much lower 
than those associated with $ggH$ cross section calculations due to the smaller 
amount of color in the quark initial states, the pure tree-level electroweak 
nature of the lowest-order amplitudes, as well as their dependence on quark PDFs, 
which are known more precisely than the gluon PDF at high Bjorken~$x$.

The {\it VH} and VBF production mechanisms result in signal events with topologies 
and kinematic distributions strikingly different than those associated with $ggH$
production.  A significant fraction of these events have partons in the final state
additional to the Higgs boson decay products.  Leptonic decays of the vector boson 
produced in association with a Higgs boson that decays to $W^+W^-$ leads to events 
with three or four charged leptons or, in other cases, in which one of the $W$ bosons 
from the Higgs boson decays hadronically, to dilepton events containing two leptons 
with the same charge.  Although the production rates associated with these types of 
events are small, the resulting event topologies are minimally contaminated by other 
SM backgrounds.  Overall, the inclusion of the additional Higgs boson production 
mechanisms increases the sensitivity of the search by roughly 30\%.

\begin{table*}[t]
  \setlength{\extrarowheight}{3pt}
\begin{ruledtabular}
\begin{center}
\caption{\label{table:xsec}
(N)NLO production cross sections and decay branching ratios to $W^+W^-$ for 
the SM Higgs boson; $ggH$ production cross sections and decay branching ratios 
to $W^+W^-$ for the SM-like Higgs boson in SM4; and the decay branching ratios 
to $W^+W^-$ for the fermiophobic Higgs boson in FHM as functions of Higgs boson 
mass.}
%\begin{small}
\begin{tabular}{l*{8}{c}}
\toprule
$m_H$ & $\sigma_{ggH}$ & $\sigma_{WH}$ & $\sigma_{ZH}$ & $\sigma_{VBF}$ & 
$\mathcal{B}(H\rightarrow W^+W^-)$ & $\sigma_{ggH}^{\rm{SM4}}$ & $\mathcal{B}^{\rm{SM4}}
(H\rightarrow W^+W^-)$ & $\mathcal{B}^{\rm{FHM}}(H\rightarrow W^+W^-)$ \\ 
(GeV/$c^2$) & (fb) & (fb) & (fb) & (fb) & (\%) & (fb) & (\%) & (\%) \\ \hline
   110 &    1385    &    204  &   120   &  82.8  & 4.82 &  12310  & 2.83  & 85.3 \\
   115 &    1216    &    175  &   104   &  76.5  & 8.67 &  10730  & 5.05  & 86.6 \\
   120 &    1072    &    150  &  90.2   &  70.7  & 14.3 &   9384  & 8.34  & 86.9 \\
   125 &   949.3    &    130  &  78.5   &  65.3  & 21.6 &   8240  & 12.9  & 86.8 \\
   130 &   842.9    &    112  &  68.5   &  60.5  & 30.5 &   7259  & 18.8  & 86.7 \\
   135 &   750.8    &   97.2  &  60.0   &  56.0  & 40.3 &   6414  & 26.0  & 86.6 \\
   140 &   670.6    &   84.6  &  52.7   &  51.9  & 50.4 &   5684  & 34.6  & 86.8 \\
   145 &   600.6    &   73.7  &  46.3   &  48.0  & 60.3 &   5050  & 44.3  & 87.4 \\
   150 &   539.1    &   64.4  &  40.8   &  44.5  & 69.9 &   4499  & 55.3  & 88.6 \\
   155 &   484.0    &   56.2  &  35.9   &  41.3  & 79.6 &   4018  & 68.1  & 90.9 \\
   160 &   432.3    &   48.5  &  31.4   &  38.2  & 90.9 &   3595  & 85.0  & 95.1 \\
   165 &   383.7    &   43.6  &  28.4   &  36.0  & 96.0 &   3221  & 94.2  & 97.5 \\
   170 &   344.0    &   38.5  &  25.3   &  33.4  & 96.5 &   2893  & 95.2  & 97.5 \\
   175 &   309.7    &   34.0  &  22.5   &  31.0  & 95.8 &   2604  & 94.8  & 96.6 \\
   180 &   279.2    &   30.1  &  20.0   &  28.7  & 93.2 &   2349  & 92.5  & 93.9 \\
   185 &   252.1    &   26.9  &  17.9   &  26.9  & 84.4 &   2122  & 83.1  & 84.8 \\
   190 &   228.0    &   24.0  &  16.1   &  25.1  & 78.6 &   1920  & 77.1  & 78.8 \\
   195 &   207.2    &   21.4  &  14.4   &  23.3  & 75.7 &   1740  & 74.5  & 75.9 \\
   200 &   189.1    &   19.1  &  13.0   &  21.7  & 74.1 &   1580  & 73.0  & 74.2 \\
\bottomrule
\end{tabular}
%\end{small}
\end{center}
\end{ruledtabular}
\end{table*}

The decay branching ratios used in this search are listed in 
Table~\ref{table:xsec}~\cite{Dittmaier:2011ti}.  The partial widths for 
all decay processes are computed with {\sc hdecay}~\cite{Djouadi:1997yw} 
with the exception of those that result in four fermion ($4f$) final 
states, $H\rightarrow W^+W^-\rightarrow 4f$ and $H \rightarrow ZZ
\rightarrow 4f$, for which the partial widths are computed with 
{\sc prophecy4f}~\cite{Bredenstein:2006rh,Bredenstein:2006ha}.  
Branching ratios are computed from the relative fractions of the total 
partial widths.  The SM branching ratio for a Higgs boson decaying to a 
pair of $W$ bosons, which is 4.82\% at $m_H = 110$~GeV/$c^2$, becomes 
dominant for $m_H > 135$~GeV/$c^2$, increasing to above 90\% near the 
threshold to produce both $W$ bosons on mass shell at $m_H = 160$~GeV/$c^2$ 
and decreasing to 74\% at $m_H = 200$~GeV/$c^2$, where decay to two $Z$ bosons 
becomes significant.  

Extensions to the SM can significantly modify the Higgs boson production cross 
sections and the $H\rightarrow W^+W^-$ branching ratio.  If the SM is extended 
to include a fourth sequential generation of heavy fermions (SM4), $ggH$ 
production of a SM-like Higgs boson is significantly enhanced and branching 
ratios are modified~\cite{Kribs:2007nz}.  Table~\ref{table:xsec} lists $ggH$ 
production cross sections for the SM4 model assuming masses of 400~GeV/$c^2$ 
and 450~GeV/$c^2$+$10\ln(m_H/115)$~GeV/$c^2$ for fourth-generation down-type 
and up-type quarks, respectively~\cite{Anastasiou:2010bt}.  Modified branching 
ratios for $H\rightarrow W^+W^-$ within the SM4 model assuming that the fourth 
generation charged lepton and neutrino are sufficiently heavy to be inaccessible 
as Higgs boson decay products are also listed in Table~\ref{table:xsec}.

In the case of a fermiophobic (FHM) Higgs boson, the $ggH$ production 
cross section is highly suppressed, but as shown in Table~\ref{table:xsec}, 
the $H \rightarrow W^+W^-$ branching ratio is significantly larger than in 
the SM, particularly in the mass range 110~$<m_H<$~150~GeV/$c^2$~\cite{Brucher:1999tx}.  
In the FHM model, the {\it WH}, {\it ZH}, and VBF production cross sections are 
assumed to be the same as those in the SM.  

\section{Analysis strategy} 
\label{sec:strategy}

The single most challenging aspect of searching for the Higgs 
boson in the $H \rightarrow W^+W^- \rightarrow \ell^+ \nu \ell^- 
\bar{\nu}$ ($\ell =$ $e$, $\mu$) decay channel is the very small 
production rate of these events.  Even when incorporating tau
lepton decays to electrons and muons, we expect, based on 
production cross sections and branching ratios, 170 signal 
events to be produced in Tevatron collisions corresponding 
to an integrated luminosity of 10~fb$^{-1}$, for a SM Higgs 
boson of mass  $m_H =$~125~GeV/$c^2$.  The search sensitivity 
depends on the fraction of these events that can be retained 
for final analysis.  We select events containing two 
reconstructed charged leptons and an overall imbalance in 
measured transverse energies originating from the multiple 
neutrinos.  After applying a loose set of kinematic criteria 
to the most inclusive two-charged-lepton candidate sample, we 
select about 25\% of the available signal.

Since the remaining background contributions are typically 
$\mathcal{O}(10^2)$ times larger than that of the expected 
signal, simple event counting is not feasible.  We
construct detailed models for the kinematic distributions
of events originating from each of the various signal and 
SM background processes.  Based on these models, potential 
signal events within the data sample are identified by 
exploiting differences between the kinematic properties of 
signal and background events.  To obtain the best possible 
signal-to-background separation, candidate events are 
classified into multiple subsamples tailored to isolate 
contributions from specific signal and background production 
processes.  Potential signal in each sample is then isolated 
using multivariate techniques, which offer increased search 
sensitivity relative to conventional approaches based on 
one-dimensional selection requirements on directly observed 
quantities.  The multivariate techniques allow for 
simultaneous analysis of multiple kinematic input variables 
and the correlations between them.   

\section{The CDF~II detector}
\label{sec:ec:cdf}

The Collider Detector at Fermilab (CDF) \cite{Blair:1996kx:cdftdr,
Abe:1988me:cdfoverview,Acosta:2004yw,Acosta:2004hw,Abulencia:2005ix} 
is a general-purpose particle detector with a cylindrical layout and 
azimuthal and forward-backward symmetry~\cite{coordsys}.

The silicon tracking system (SVX)~\cite{L00,L00a,SVX,ISL} and open-cell 
drift chamber (COT)~\cite{COT} are used to measure the momenta of 
charged particles and identify secondary vertices from the decays of 
bottom quarks, which have finite lifetimes.  The COT is segmented 
into eight concentric superlayers of wire planes with alternating axial 
and $\pm2^{\circ}$ stereo angle stringing.  The active volume covers the 
radial range from 40 to 137~cm and is located within a superconducting 
solenoid with a 1.4 T magnetic field parallel to the beam axis.  Tracking  
efficiency within the COT is nearly 100\% in the range $|\eta| \leq 1$; 
and with the addition of silicon detector information, tracks can be 
reconstructed within the wider range of $|\eta| < 1.8$.  The momentum 
resolution is $\sigma(p_T)/p_T^2$~$\approx$~0.001~GeV$^{-1}$ for tracks 
within $|\eta| \leq$~1 and degrades with increasing $|\eta|$.

Electromagnetic (EM) and hadronic (HAD) calorimeters~\cite{CEM,PEM,CHAWHA}, which 
are lead-scintillator and iron-scintillator sampling devices, respectively, surround 
the solenoid and measure the energy flow of interacting particles. They are segmented 
into projective towers, each one covering a small range in pseudorapidity and azimuth. 
The calorimeters have complete azimuthal coverage over $|\eta| < 3.6$. The central 
region $|\eta| < 1.1$ is covered by the central electromagnetic calorimeter (CEM) 
and the central and end-wall hadronic calorimeters (CHA and WHA). The forward region 
$1.1 < |\eta| < 3.6$ is covered by the end-plug electromagnetic calorimeter (PEM) and 
the end-plug hadronic calorimeter (PHA).  

Energy deposition in the electromagnetic calorimeters is used to identify and measure 
the energy of electrons and photons.  The energy resolution for an electron with 
transverse energy $E_T$ (measured in GeV) is given by $\sigma(E_T)/E_T \approx 13.5\%
/\sqrt{E_T} \oplus 1.5\%$ and $\sigma(E_T)/E_T \approx 16.0\%/\sqrt{E_T} \oplus 1\%$ 
for those identified in the CEM and PEM, respectively.  Deposits in the electromagnetic
and hadronic calorimeter towers are used to identify and measure the energies of the 
clustered groups of particles originating from parton showers (jets).  The resolution 
of calorimeter jet energy measurements is approximately $\sigma(E_T) \approx 0.1 E_T 
+ 1.0$~GeV~\cite{JES}. The CEM and PEM calorimeters also contain strip detectors with 
two-dimensional readout, which are located at the depth corresponding approximately
to the maximum shower development for an electron.  These detectors aid in the 
identification of electrons and photons by providing position information that helps 
to distinguish them from $\pi^0$ decay products.

Beyond the calorimeters are muon detectors~\cite{Artikov:2004ew:Muons}, which provide 
muon identification in the range $|\eta| < 1.5$.  Muons are detected in four separate 
subdetectors. Central muons with $p_T > 1.4$~GeV/$c$ penetrate on average the five 
absorption lengths of the calorimeter and are detected in the four layers of planar 
multiwire drift chambers of the central muon detector (CMU).  A second set of drift 
chambers, the central muon upgrade (CMP), sits behind an additional 60~cm of steel 
and detects muons with $p_T > 2.2$~GeV/$c$. The CMU and CMP chambers cover an equivalent 
range in pseudorapidity, $|\eta| < 0.6$.  Central muon extension (CMX) chambers cover 
the pseudorapidity range from $0.6 < |\eta| < 1.0$ and thus complete muon system 
coverage over the full fiducial region of the COT. Muons in the pseudorapidity range 
$1.0 < |\eta| < 1.5$ are detected in the forward barrel muon (BMU) chambers.

The Tevatron collider luminosity at the CDF interaction point is determined 
using multicell gas Cherenkov detectors~\cite{Acosta:2001zu:clc} located in 
the pseudorapidity range $3.7 < |\eta| < 4.7$, which measure the average number 
of inelastic $p\bar{p}$ collisions per bunch crossing. 

The CDF online event selection system (trigger) is designed with three sequential 
decision levels to cope with high event rates.  The first level relies on dedicated 
hardware to reduce high event rates from the effective beam-crossing frequency of 
1.7~MHz to roughly 15~kHz.  The second level uses a mixture of dedicated hardware 
and fast software algorithms to analyze more completely the available trigger 
information.  This level reduces the event rate to roughly 1~kHz, the maximum 
detector-readout rate.  The third level is an array of computers that run a fast 
version of the offline event reconstruction algorithms on the full detector readout, 
selecting events for permanent storage at a rate of up to 150~Hz.

\section{Event selection}
\label{sec:evtsel}

The search is based on events containing two or three charged 
lepton candidates with $p_T > 10$~GeV/$c$.  Events are 
recorded online if they meet the criteria of either one of two 
single-electron triggers or one of four single-muon triggers.  
The central electron trigger requires a CEM energy cluster with 
$E_T > 18$~GeV matched to a reconstructed COT track with $p_T > 
8$~GeV/$c$.  The forward electron trigger requires a PEM energy 
cluster with $E_T > 20$~GeV and an overall missing transverse 
energy of at least 15~GeV in the calorimeter.  The four muon 
triggers are based on track segments in one or more muon chambers 
(CMU + CMP, CMU, CMP, and CMX) matched to reconstructed COT 
tracks with $p_T > 18$~GeV/$c$.  For each event, the charged 
lepton consistent with having satisfied the trigger is required 
to have $p_T > 20$~GeV/$c$, to ensure uniform trigger efficiency.  
Trigger efficiencies are measured from observed $W \rightarrow 
\ell \nu$ and $Z \rightarrow \ell \ell$ decays~\cite{Acosta:2004uq}.  
To ensure that the charged lepton candidates are consistent with 
having been produced in a single interaction, the $z$ positions 
of each candidate's reconstructed track at the point of closest 
approach to the beamline are required to lie within 4~cm of one 
another.  In addition, the few events (less than 0.1\% of total) 
containing reconstructed leptons with energies in excess of 
400~GeV are attributed to mismeasurements and removed.

\subsection{Lepton identification}
\label{sec:lepsel}

Electron and muon candidates are constructed from combinations of measurements 
in various subdetectors. Because the coverage of these subdetectors varies 
over $\eta$ and $\phi$, selection criteria for individual lepton candidates 
depend on their trajectory within the detector.  The general goal is to use 
all available information to suppress contributions from jets misidentified 
as leptons, while not rejecting candidates just because they are detected 
in less instrumented portions of the detector. As a result, we use four 
categories of electron candidates, eight categories of muon candidates, and 
two final categories of candidates likely to be either an electron or muon 
but indistinguishable on the basis of available information.

\subsubsection{Electron identification}

Identification of electron candidates is based on reconstructed showers in the EM 
calorimeter with a $E_T$ of at least 10~GeV after correcting for energy leakage 
into the HAD calorimeter. For the central region ($|\eta|<1.1$), we employ both 
a cut-based and a multivariate likelihood-based method, combining information 
from the calorimeter, tracking, and shower-maximum detectors.  The cut-based method 
requires that the shower energy within the HAD calorimeter ($E_{{\rm HAD}}$) 
must be less than 5\% of that in the EM calorimeter ($E_{{\rm EM}}$) and that the 
distribution of shower energies in the calorimeter towers and shower maximum 
detector is consistent with those of an electron.  The shower orientation must be 
geometrically matched to a reconstructed track with a measured $p_T$ such that the 
ratio of the shower $E_T$ to the track $p_T$ lies between 0.5 and 2.0.  The track 
is also required to pass standard quality requirements.

If a central electron candidate fails the above selection, it can still be used
as a likelihood-based electron. The likelihood function is constructed based on 
variables used in the cut-based version such as the ratio of $E_{{\rm HAD}}$ to 
$E_{{\rm EM}}$, the ratio of $E_T$ to $p_T$, and the shapes of calorimeter and 
shower-maximum energy distributions.  Signal likelihood templates are constructed 
from the unbiased electron candidates in $Z \rightarrow ee$ events.  Background 
likelihood templates are constructed from loose electron candidates in inclusive 
dijet events.

A combination of cut-based and likelihood-based selections is used to 
identify electron candidates in the forward region of the calorimeter, 
$1.2 < |\eta| <2.0$.  A specialized track-finding algorithm that uses 
locations of the reconstructed calorimeter shower and primary vertex 
to define a search road for hits in the SVX is used to increase the 
selection efficiency.  A similar set of kinematic and shower shape 
variables to those employed in central electron selection are used as 
the basis for the cut-based selection and as inputs in the formation 
of a forward-candidate likelihood function.

\subsubsection{Muon identification}

Muon candidates are constructed from reconstructed tracks with $p_T >$ 10~GeV/$c$. 
Eight separate categories of reconstructed muon candidates are used.  In six
of these, the track can be matched with hits from one or more of the muon 
detector systems.  The separate categories are for candidates associated with 
hits in both central muon detectors, in only the inner or outer central muon 
detectors, where the track trajectory is consistent with having passed through 
an uninstrumented gap in the other, in one of two portions of the extended muon 
detector, and in the forward muon detector.  This categorization provides a 
mechanism for matching muon candidates with specific sets of event triggering 
criteria.  Muon candidate tracks are also required to point toward calorimeter 
energy depositions consistent with those expected from a minimum-ionizing particle.
The last two muon categories apply to tracks matched only to energy depositions 
consistent with having originated from minimum-ionizing particles in either 
the central ($|\eta| < 1.1$) or forward ($1.2 < |\eta| < 2.0$) calorimeters.  
The inclusion of these categories ensures high selection efficiencies for muons 
that pass through regions of missing muon detector coverage.

\subsubsection{Isolation requirements}

To improve the separation of charged leptons produced in the decays of $W$ 
and $Z$ bosons from those produced in the decays of heavy-flavor hadrons, 
electron and muon candidates are required to be isolated from other observed 
particle activity within the event.  In particular, we require lepton 
candidates to satisfy both calorimeter and track isolation requirements.  
The sums over measured transverse energies in individual calorimeter towers 
and the transverse momenta of reconstructed particles whose trajectories lie 
within a cone of $\Delta R \equiv \sqrt{(\Delta \eta)^2 + (\Delta \phi)^2} 
< 0.4$ around the candidate must be less than 10\% of the electron $E_{T}$ 
or muon $p_{T}$.  An exception is the case of the likelihood-based electron 
selection, for which the isolation variables are included as additional 
inputs in the construction of the likelihood function.

For the targeted $H \rightarrow W^+W^-$ decay process, the spin correlation 
between the two leptonically-decaying $W$ bosons tends to result in leptons 
with trajectories close to one another.  In roughly 10\% of cases, the 
leptons lie within each other's isolation cones, and the energy deposits 
and tracks associated with one lepton cause the other lepton to 
fail its isolation requirements.  To avoid this issue and recover lost 
signal acceptance, isolation calculations are modified to exclude from 
the search cone all calorimeter tower energies and reconstructed tracks 
associated with other lepton candidates that meet nonisolation-related 
criteria.  

\subsubsection{Isolated tracks}

Two additional lepton categories are defined for tracks that extrapolate
geometrically to noninstrumented regions of the calorimeters and have no 
matches with track segments in the muon detectors.  Such tracks, which 
meet quality and isolation requirements, comprise one further lepton category.  
Since the candidates in this category are not distinguishable as electrons 
or muons, either of the possibilities are allowed in each event.  Electrons 
that pass though nonactive regions of the calorimeter may radiate 
bremsstrahlung photons thus failing isolation requirements because of 
photon energy deposition in surrounding EM towers.  Such electrons 
are recovered into a second track-based category containing track 
candidates that fail the standard calorimeter isolation criteria but 
satisfy a modified criterion, in which EM energy depositions from 
towers adjacent to the track candidate are subtracted from the total 
measured energy within the isolation cone.

\subsection{Lepton identification efficiency determination}
\label{sec:lepsel}

Selection requirements reduce the probability for electrons and muons to 
be identified as lepton candidates.  In order to account for a potential
mismodeling of this efficiency in the simulation, the efficiency is 
measured from observed $Z\rightarrow\ell^+\ell^-$ decays.  The events are 
collected using the single central electron and muon trigger paths.  One 
of the reconstructed lepton candidates (referred to as the tag) must 
satisfy all cut-based selection criteria and be identified as consistent  
with the lepton that triggered the online selection of the event.  The 
second candidate (known as the probe) is only required to pass minimal 
requirements, for which the expected efficiency approaches 100\% and is
therefore assumed to be well modeled in the simulation.  The dilepton 
invariant mass is required to lie within $\pm15$~GeV/$c^2$ of the $Z$ boson 
mass to ensure that the event samples contain primarily real dilepton 
events from $Z\rightarrow\ell^+\ell^-$ production. 
 
Based on these samples, the measured efficiency for an additional set 
of test criteria applied on the probe lepton is simply the fraction 
of the probes that satisfy the full criteria.  A small complication 
arises when tagged leptons also satisfy probe-lepton criteria due to 
overlapping selection requirements.  For these cases, events in which 
both candidates are identified as tags need to be counted twice in 
the efficiency calculation.  Events that do not meet the test criteria 
have nonnegligible background contributions from $W$+jet and multijet 
production.  Measured efficiencies need to be corrected to account 
for the presence of background within these events.  The background 
contributions are estimated using a linear extrapolation across the 
$Z$ boson signal mass range based on events counts within sideband 
regions on both sides of the signal range.  

Separate efficiency calculations are made for each of the lepton 
categories.  The measured efficiencies are defined as an average 
over those for each of the individual probe candidates from the 
$Z\rightarrow\ell^+\ell^-$ sample.  Hence, the measured efficiencies 
are applied as corrections to the detector simulation, relying 
on its description of $p_T$, $\eta$, and $\phi$ dependence, but 
correcting the average efficiency to that measured directly from 
observed events.  Measurements based on observed events deviate 
from those obtained in simulation by up to $6\%$ with uncertainties 
of $1$-$2\%$.  The efficiencies are measured separately for several 
data-taking periods.  Observed effects of additional $p\bar{p}$ 
collisions within individual beam crossings (``pile-up'') are 
found to be well modeled in the simulation.

We validate the estimate of trigger and lepton selection efficiencies 
and their proper inclusion in the simulation by measuring the DY 
production cross section from independent, inclusive dilepton 
samples, each corresponding to one possible same-flavor combination 
of the lepton categories.  In the case of $Z\rightarrow e^+e^-$, we 
measure cross sections from 11 independent samples constructed from 
two triggerable electron categories, two nontriggerable categories, 
and two isolated track categories.  For $Z\rightarrow\mu\mu$ 
events we extract measurements from 35 independent samples based on 
five triggerable muon categories, three nontriggerable categories, 
and one isolated track category.  The 46 independent measurements 
are found to agree within $\pm$~5\%, consistent with the $\pm$~3\% 
uncertainties assigned to the trigger and lepton selection 
efficiency measurements.  The cross sections measured from samples 
containing events with one forward electron candidate are observed 
to be on average about 10\% smaller than those of the other samples.  
This effect is attributed to reduced track reconstruction efficiency 
in the forward region of the detector ($|\eta|>1.2$), where COT 
coverage is reduced.  Since track reconstruction, which is used 
to define probe leptons in this region, is not fully efficient, 
an additional correction is required.  This factor is obtained 
directly from the extracted DY cross sections as the ratio of 
averaged measurements from event samples with and without forward 
electron candidates. 

\subsection{Tau lepton identification}

Decays of tau leptons to electrons and muons (roughly 35\% of total 
branching ratio) are identified within the lepton categories, and 
the additional acceptance from leptonic $\tau$ decays is included 
within all background and signal estimates.  In the remaining 65\% 
of cases, tau leptons undergo a hadronic decay $\tau\to X_{h}\nu_{\tau}$, 
where $X_{h}$ can be a charged pion, kaon, or a short-lived intermediate 
resonance that decays to final states containing neutral or charged 
pions and kaons.  Additional signal acceptance is obtained by identifying 
tau lepton candidates produced via these decay modes.

The pions and kaons produced in tau lepton decays are expected to deposit 
significant energy in neighboring calorimeter towers. The reconstruction of 
hadronically-decaying tau lepton candidates is therefore based on a narrow 
calorimeter cluster with a maximum of three matched tracks.  The sum of 
measured transverse energies from calorimeter towers contained within the 
cluster is labeled as $E^{\tau}_{\rm{clus}}$, and the matching track with 
the highest $p_T$ is referred to as the tau lepton {\it seed track}.  Signal 
and isolation cones are defined around the seed track direction where the 
opening angle of the signal cone depends on the calorimeter cluster energy, 
$\theta_{\rm{sig}}=\min(0.17,~5/E^{\tau}_{\rm{clus}}~[{\rm GeV}])$~radians, 
and the opening angle of the isolation cone is fixed at 0.52~radians.  
Neutral pions within the signal cone are reconstructed by combining position 
information from the shower-maximum detector with energy depositions 
measured in the EM calorimeter.  Tracks and reconstructed $\pi^0$ 
candidates matched to the calorimeter cluster are combined to reconstruct 
the {\it visible momentum} of the tau lepton candidate.  A detailed 
description of the techniques used for reconstructing hadronically-decaying 
tau leptons is provided in Ref.~\cite{Abulencia:2007iy}.

Additional requirements are imposed to improve the purity of 
hadronically-decaying tau lepton candidates.  Candidates are required 
to have one track (1-prong) or three tracks (3-prong), where the absolute 
value of the sum of the charges of the reconstructed particles is one.  
The visible transverse momentum of the candidate is required to exceed 
15~GeV/$c$ or 20~GeV/$c$ for 1-prong and 3-prong tau lepton candidates, 
respectively.  The mass reconstructed from the visible momentum must 
also be consistent with the tau lepton mass.  To reduce background 
contamination from parton jets, which are expected to produce wider 
energy clusters than those of hadronically-decaying tau leptons, low 
activity in both the calorimeter and tracking systems is required in 
the region between the outer edges of the signal and isolation cones.  
Contamination from electrons is reduced by limiting the relative 
fractions of EM and HAD energy within the reconstructed 
calorimeter cluster.
 
\subsection{Jet identification}
\label{sec:ec:jets}

Calorimeter jets are reconstructed using a fixed cone algorithm~\cite{JES}
with a radius of $\Delta R=\sqrt{\Delta\phi^2+\Delta\eta^2} = 0.4$. 
Corrections are applied to measured jet energies to compensate for 
nonlinearities and nonuniformities in the response of the calorimeter, 
excess energy deposited within the jet cone from sources other than 
the assumed parent parton, and missing energy from the parent parton 
deposited outside the jet cone~\cite{JES}.  In this search we only 
consider jets with corrected $E_{T}>15$~GeV and within the pseudorapidity 
region $\left|\eta\right|<2.5$. Jets are also required to be separated 
($\Delta R > 0.4$) from identified leptons.

To reduce backgrounds originating from $t\bar{t}$ production, 
events with exactly two oppositely-charged leptons and two or more 
reconstructed jets are vetoed if any of the jets can be identified as 
likely to have originated from a bottom quark.  This identification 
is made by reconstructing within a jet secondary track vertices 
consistent with the decay of longer-lived hadrons produced in the 
hadronization of heavy quarks~\cite{secvtx}.

\subsection{Missing transverse energy}
\label{sec:ec:met}

Neutrinos escape detection and their energies cannot be directly measured.
Their presence is inferred from an imbalance of observed transverse 
energies within a event, \Met, which is defined as the magnitude of 
$-\Sigma_i E_{T}^{i} \hat{n}_{i}$, where $\hat{n}_{i}$ is the unit vector 
in the azimuthal plane that points from the beamline to the $i$th calorimeter 
tower.  The \Met is corrected by subtracting the energy deposited in the 
calorimeter by minimum-ionizing muons and adding back their measured $p_{T}$.  
Energy corrections applied to calorimeter jets are also accounted for in the 
\Met determination through the subtraction of raw jet energies and addition 
of corrected jet energies.

The primary purpose of \Met requirements is to significantly reduce 
backgrounds from DY processes, which have large production cross 
sections but result in final states containing charged leptons but 
no neutrinos.  Since any remaining DY background after the 
application of \Met requirements necessarily results from detector 
energy mismeasurements, we also use a modified $\MetSpec$ variable 
defined as 
\begin{equation}
   \MetSpec \equiv \left\{ 
   \begin{array}{ll} 
   \Met                       & \mbox{ if } \MetDeltaPhi > \frac{\pi}{2} \\
   \Met\sin({\MetDeltaPhi})   & \mbox{ if } \MetDeltaPhi < \frac{\pi}{2}, \\
   \end{array} \right.
\end{equation}
where $\MetDeltaPhi$ is the angle between the $\Met$ and the closest 
lepton or jet transverse momentum vector.  An undermeasurement of 
the lepton or jet momentum leads the $\Met$ to be aligned with the
direction of the corresponding candidate, and for these cases the 
$\sin({\MetDeltaPhi})$ term significantly reduces the value of 
$\MetSpec$ with respect to the nominal $\Met$.

\subsection{Data sample selections}
\label{sec:Selection}

\begin{table*}[t]
  \setlength{\extrarowheight}{3pt}
\begin{ruledtabular}
\begin{center}
\caption{\label{tbl:samples}
Summary of names assigned to the Higgs boson search samples and their 
associated control samples along with the background processes targeted 
by each control sample.
} 
\begin{tabular}{llc}
\toprule
Search sample(s)                       & Associated control sample(s)                                            & Background    \\
                                       &                                                                         & targeted      \\ 
\hline 
OS Base (0 Jet, high $s/b$ Leptons)      & SS Base                                                                 & $W$+jets      \\
OS Base (0 Jet, low $s/b$ Leptons)       & OS Base (Intermediate $\MetSpec$)                                       & DY            \\
OS Base (1 Jet, high $s/b$ Leptons)      &                                                                         &               \\
OS Base (1 Jet, low $s/b$ Leptons)       &                                                                         &               \\ \hline
OS Base ($\ge$2 Jets)                      & SS Base ($\ge$2 Jets)                                                       & $W$+jets      \\
                                       & OS Base ($\ge$2 Jets, Intermediate $\MetSpec$)                              & DY            \\ 
                                       & OS Base ($\ge$2 Jets, $b$-tagged)                                           & $t\bar{t}$    \\ \hline
OS Inverse $M_{\ell\ell}$              & SS Inverse $M_{\ell\ell}$                                               & $W\gamma$     \\
                                       & OS Inverse $M_{\ell\ell}$ (Intermediate $\MetSpec$)                     & DY            \\ \hline
OS Hadronic Tau ($e$ + $\tau_{\rm{had}}$)   & OS Hadronic Tau ($e + \tau_{\rm{had}}$, high $\Delta\varphi(\vec{p}_T(\tau),\vec{p}_T(\ell))$) & $W$+jets \\
OS Hadronic Tau ($\mu$ + $\tau_{\rm{had}}$) & OS Hadronic Tau ($e + \tau_{\rm{had}}$, low $\Met$)                          & Multijet      \\
             & OS Hadronic Tau ($\mu + \tau_{\rm{had}}$, low $\Met$, low $\Delta\varphi(\vec{p}_T(\ell),\VecMet)$) & $Z/\gamma^\ast\rightarrow\tau\tau$ \\ \hline
SS ($\ge$1 Jets)                           & SS ($\ge$1 Jets, low $\Met$)                                                & DY            \\
                                       & SS (0 Jet)                                                              & $W$+jets      \\ \hline
Trilepton {\it WH}                           & Trilepton {\it WH} (Intermediate $\Met$)                                      & $Z\gamma$     \\ 
Trilepton {\it WH} ($\ell + \ell + \tau_{\rm{had}}$) & Trilepton {\it WH} ($\ell + \ell + \tau_{\rm{had}}$, Intermediate $\Met$)       & $Z$+jets      \\ \hline
Trilepton {\it ZH} (1 Jet)                   & Trilepton {\it ZH} (0 Jet)                                                    & {\it WZ}          \\
Trilepton {\it ZH} ($\ge$2 Jets)                 &                                                                         &               \\ 
\bottomrule
\end{tabular}
\end{center}
\end{ruledtabular}
\end{table*}

We define multiple independent data samples based on various kinematic 
selection requirements such as the number of reconstructed jets and 
leptons and the measured $\Met$ or $\MetSpec$.  The construction of 
multiple samples enhances the ability to separate potential signal 
and background contributions.  Statistical independence of the samples 
allows convenient combination of results based on distinct subsamples 
to preserve maximum sensitivity.  Additional control samples are 
constructed to tune or test modeling of specific background processes.  
Typically, these control samples are based on the kinematic selections
used for defining one of the search subsamples, where one or more  
criteria has been modified to further enhance the dominant background 
contribution.  Tuning parameters used to improve the agreement between
data and simulation are obtained from specific control regions and 
incorporated, where applicable, into background modeling across all 
data samples used in the search.

Table~\ref{tbl:samples} summarizes the 13 data samples used in this 
search as well as the 15 associated control samples.  The specific 
kinematic criteria associated with each grouping of search samples 
and its associated control sample(s) are described in the following 
subsections.

\subsubsection{Opposite-sign base selection (0 or 1 jet)}
\label{sec:osSelection}

Events with exactly two opposite-sign (OS) electron or muon candidates 
and one or zero reconstructed jets are included in the base selection.  
The main background contributions to this event sample are from the 
DY process, where the observed $\Met$ originates from mismeasurements 
of lepton or jet energies; $W\gamma$ and $W$+jets, where a photon or 
jet is misidentified as a lepton; and direct $W^+W^ \rightarrow \ell^+
\nu \ell^- \bar{\nu}$ production, which has an equivalent final state 
as the signal.  To suppress DY background, we require $\MetSpec > 25$~GeV. 
This criterion is released to $\MetSpec > 15$~GeV for electron-muon 
events, for which the DY background contribution is significantly 
reduced.  We also require the candidates to have $M_{\ell\ell} > 
16$~GeV/$c^2$ to suppress $W\gamma$ background contributions. 

We separate the selected events into four further samples based on 
whether they contain a reconstructed jet and the qualities of the 
two lepton candidate types.  Events with central lepton candidates 
are considered as having high signal-to-background (high $s/b$), while 
events with one or more forward lepton candidates are considered as 
having low signal-to-background (low $s/b$).  The additional subdivision 
of events allows further isolation of specific background contributions.  
Contributions from $W\gamma$ and $W$+jets are more significant in the 
low $s/b$ samples, while the relative mix of {\it WW} and DY contributions is 
significantly different for events with and without a reconstructed jet.

We construct two additional control samples based on the generic selection 
criteria associated with these search samples.  Events containing same-sign 
(SS) dileptons that otherwise satisfy the signal sample criteria form the 
SS Base control region, which is used to test $W$+jets background modeling.  
The OS Base (Intermediate $\MetSpec$) control sample contains events with 
same-flavor ($e^{+}e^{-}$ or $\mu^{+}\mu^{-}$) dileptons and $\MetSpec$ 
between 15 and 25~GeV that otherwise satisfy search sample criteria.  This 
control sample is used to tune the DY modeling applied to the associated 
search samples.  
      
\subsubsection{Opposite-sign base selection ($\ge$2 jets)}
\label{sec:os2Selection}

Events that satisfy the criteria for the OS Base selection but contain two 
or more reconstructed jets are classified separately.  The largest background 
contribution to this sample is from the $t\bar{t}\rightarrow b\ell^+\nu 
\bar{b}\ell^-\bar{\nu}$ process.  To help reduce this background, events are 
rejected from the search sample if any of the reconstructed jets are tagged 
as consistent with having originated from a bottom-quark decay by the 
{\sc secvtx} algorithm~\cite{secvtx}, which identifies displaced track 
vertices within jets.  Even after application of this veto, $t\bar{t}$ 
production is still the single largest source of background events to this 
search sample.    

To test background modeling, three additional control samples are defined.  
Same-sign dilepton events, which otherwise satisfy the signal sample criteria, 
form the SS Base ($\ge$2 Jets) control sample, which is again used to test 
$W$+jets background modeling.  Similarly, the DY modeling for this search 
sample is tested using the OS Base ($\ge$2 Jets,Intermediate $\MetSpec$) 
control sample, which contains same-flavor dilepton events with $\MetSpec$\ 
between 15 and 25~GeV that satisfy remaining search sample criteria.  Events 
that are rejected from the search sample exclusively due to the identification 
of one or more jets as being consistent with bottom-quark decays form the 
OS Base ($\ge$2 Jets, $b$-tagged) control sample used to test $t\bar{t}$ 
modeling.       
    
\subsubsection{Opposite-sign inverse $M_{\ell\ell}$ selection}
\label{sec:mllSelection}

Events that fail the $M_{\ell\ell}> 16$~GeV/$c^2$ requirement but otherwise 
satisfy OS Base (0 or 1 jet) selection criteria are collected into another 
independent search sample.  The primary source of background events in this 
search sample is $W\gamma$ production, where the photon is misidentified as 
an electron.  Dilepton events originating from the decays of heavy-flavor 
hadrons are mostly removed by tighter $\Met$ requirements on events with 
reconstructed dilepton mass ($M_{\ell\ell}$) consistent with $J/\psi$ and 
$\Upsilon$ meson decays.  We define $\Met$ significance as the ratio of the 
measured $\Met$ to the scalar sum of measured transverse energies for all 
reconstructed jets and leptons.  For events with $M_{\ell\ell}< 6$~GeV/$c^2$ 
and $8.5 < M_{\ell\ell}< 10.5$~GeV/$c^2$, the $\Met$ significance is required 
to be greater than four.       

Same-sign dilepton events that pass the other selection requirements of 
this search sample form the SS Inverse $M_{\ell\ell}$ control sample, which 
is used to tune the $W\gamma$ background modeling.  Validation of the DY 
modeling used in association with this search sample is based on the OS 
Inverse $M_{\ell\ell}$ (Intermediate $\MetSpec$) control sample made up 
of same-flavor events with $\MetSpec$ between 15 and 25~GeV that otherwise 
satisfy sample selection criteria.

\subsubsection{Opposite-sign Hadronic Tau selection}
\label{sec:tauDilepton}

While tau lepton decays to electrons and muons are incorporated within the search 
samples, signal acceptance is enhanced by including events containing one electron 
or muon candidate and one hadronically-decaying tau lepton candidate in separate 
search samples.  Because events in these samples are collected by the same trigger 
selections, the single electron or muon is necessarily responsible for having 
triggered the event and is therefore required to have $p_T > 20$~GeV/$c$.  
 
Additional selection criteria are applied to reduce background 
contributions, which are significantly larger in this sample.  To minimize 
contributions from processes with final states without neutrinos such as 
DY $Z/\gamma^\ast\rightarrow\ell\ell$ ($\ell = e$ or $\mu$), multijet, 
and  $\gamma$+jet production, the observed $\Met$ is required to exceed 
20~GeV.  Dilepton invariant mass, $M(\tau\ell)$, is also required to be 
above 20~GeV/$c^2$ to reduce backgrounds from the decays of heavy-flavor 
hadrons.  The DY $Z/\gamma^\ast\rightarrow\tau\tau$ background contribution 
is removed by requiring a minimum angle of 1.5~radians between the dilepton 
transverse momentum and the missing transverse energy, $\Delta\varphi
(\vec{p}_T(\ell) + \vec{p}_T(\tau),\VecMet)$.  Similarly, the dominant 
$W$+jets background contribution is suppressed by requiring a maximum 
angle of 1.5~radians between the transverse momenta of the two leptons, 
$\Delta\varphi(\vec{p}_T(\tau),\vec{p}_T(\ell))$.  To take advantage of 
differing background compositions, events are separated into two search 
samples based on the presence of an electron or muon candidate.

Background modeling for these search samples is validated using 
three control samples.  The $W$+jets-dominated OS Hadronic Tau 
(high $\Delta\varphi(\vec{p}_T(\tau),\vec{p}_T(\ell))$) sample is 
constructed by selecting events with $\Delta\varphi(\vec{p}_T(\tau),
\vec{p}_T(\ell)) > 2.0$~radians that otherwise satisfy search sample 
criteria.  The multijet-dominated OS Hadronic Tau ($e + \tau_{\rm{had}}$, 
low $\Met$) sample is composed of events containing electron 
candidates, which fail the search sample criteria solely on the basis 
of an observed $\Met < 20$~GeV.  The OS Hadronic Tau ($\mu + \tau_{\rm{had}}$, 
low $\Met$, low $\Delta\varphi(\vec{p}_T(\ell),\VecMet)$) sample 
contains events with muon candidates, for which the observed $\Met < 
20$~GeV and $\Delta\varphi(\vec{p}_T(\ell),\VecMet) < 0.5$~radians.
This control sample is used to validate DY $Z/\gamma^\ast\rightarrow
\tau\tau$ background modeling and hadronically-decaying tau lepton 
reconstruction efficiencies. 

\subsubsection{Same-sign dilepton selection}
\label{sec:ssSelection}

Events with exactly two same-sign electron or muon candidates form an 
additional search sample.  Higgs boson production in association with 
a $W$ or $Z$ boson can result in a final state containing same-sign 
leptons when, for example, two $W^+$ bosons (one from the original 
associated production and the other from a subsequent $H\rightarrow 
W^+ W^-$ decay) decay leptonically.  The remaining $W$ boson from 
the Higgs boson decay most often decays hadronically, leading to the 
production of jets within the event.  Hence, events in this search 
sample are required to have at least one reconstructed jet.      

An important background contribution to the same-sign event sample 
is DY $Z \rightarrow \ell^+ \ell^-$ production, where one of the 
lepton charges is misreconstructed, or a bremsstrahlung photon 
converts into a $e^+ e^-$ pair within the detector, creating the 
potential for the original lepton to be reconstructed with an 
incorrect charge.  Lepton candidates of this type are referred to 
as {\it tridents}.  To help reduce DY background contamination, 
events containing forward electron candidates, which are affected 
by significant charge mismeasurement rates, are rejected.  In 
addition, since looser likelihood criteria tend to select trident 
candidates, central electrons in these events are required to pass 
tight cut-based selection.  Backgrounds from DY processes are 
further reduced by requiring events to have $\Met > 10$~GeV.  The 
other significant sources of background events for this sample 
are $W$+jets and $W\gamma$ production, where a jet or photon is 
misidentified as a lepton.  To reduce backgrounds from these 
sources, events are required to have $M_{\ell\ell} > 16$~GeV/$c^2$ 
and the minimum $p_T$ criterion on the nontriggered lepton in 
these events is increased from 10 to 20~GeV/$c$.

Two associated control samples are formed to validate background 
modeling for this search sample.  Events that satisfy the search 
sample criteria apart from containing no reconstructed jets 
form the SS (0 Jet) sample, which is dominated by background 
contributions from $W$+jets production.  The SS ($\ge$1 Jets, low 
$\Met$) control sample is composed of events with $\Met < 10$~GeV 
that otherwise satisfy the search sample criteria.  This sample 
is used to test DY background modeling and the modeling for 
trident events.  

\subsubsection{Trilepton {\it WH} selection}
\label{sec:whTrilepSelection}

We also incorporate separate search samples for events containing 
exactly three charged lepton candidates.  Such final states 
are contributed by Higgs boson production in association with 
a $W(\rightarrow\ell\nu)$ boson and decaying as $H\rightarrow   
W(\rightarrow\ell\nu) W(\rightarrow\ell\nu)$.  Events containing 
three leptons of the same charge are not consistent with the 
corresponding final state and are rejected.  To increase signal 
acceptance, events with a single tau lepton candidate serving as 
one of the three lepton candidates are included in this search 
sample.  

Because of differing background contributions, events are classified 
into two separate search samples based on whether they contain a 
tau lepton candidate.  In addition, events containing a same-flavor, 
opposite-sign pair of lepton candidates with an invariant mass 
within $\pm15$~GeV/$c^2$ of the $Z$ boson mass are removed and 
assigned to Trilepton {\it ZH} search samples described in the following 
section.  The dominant backgrounds to the Trilepton {\it WH} search 
samples are $Z\gamma$ and $Z$+jets production, where the $Z$ is 
produced off-shell and a jet or photon is misidentified as a lepton.  
Because these processes lead to final states without neutrinos, 
we require events in these samples to have $\Met > 20$~GeV. 

To validate modeling of the primary backgrounds, we construct 
two associated control samples from events with $\Met$ between 
10 and 20~GeV that otherwise satisfy the search sample criteria.  
The Trilepton {\it WH} (intermediate $\Met$) control sample contains 
events with no tau lepton candidates and is used to validate 
$Z\gamma$ background modeling.  Events containing a tau lepton 
candidate form the Trilepton {\it WH} ($\ell + \ell + \tau_{\rm{had}}$, 
intermediate $\Met$) control sample used for testing $Z$+jets 
background modeling.    

\subsubsection{Trilepton {\it ZH} selection}
\label{sec:zhTrilepSelection}

A similar production mode for signal events with exactly three leptons 
is associated Higgs boson production with a $Z$ boson and subsequent 
$H \rightarrow W^+ W^-$ decay.  A same-flavor, opposite-sign lepton 
pair is produced in a leptonic decay of the $Z$ boson.  A third lepton 
can originate from the leptonic decay of either $W$ boson produced in 
the Higgs boson decay.  Events containing three leptons with the same 
charge are inconsistent with the signal final state and rejected from 
the search sample.   The remaining $W$ boson from the Higgs boson 
decay must decay hadronically, leading to the production of jets.  
Hence, events in this search sample are required to have at least one 
reconstructed jet.     

Events with exactly one and two or more jets are separated into two 
search samples.  Determination of a transverse Higgs boson mass is 
possible in events containing at least two jets due to the availability 
of all decay products in the assumed final state (the transverse energy 
of the single neutrino is inferred from the $\Met$).  The statistical 
independence of these search samples with respect to the Trilepton {\it WH} 
samples is maintained by selecting only the events that contain a 
same-flavor, opposite-sign lepton pair within $\pm15$~GeV/$c^2$ of 
the $Z$ boson mass.  Because of large background contributions from 
on-shell $Z$+jets production, events containing tau lepton candidates are 
not included within the Trilepton {\it ZH} search samples.  Events in these 
samples are also required to have observed $\Met > 10$~GeV to further 
reduce $Z\gamma$ and $Z$+jets background contributions. 

A single associated control sample is formed to test the background
modeling used for these search samples.  The Trilepton {\it ZH} (0 Jet) 
control sample consists of events with no reconstructed jets that 
otherwise satisfy search sample criteria.  Contributions from {\it WZ} 
production are the single largest source of events to this control 
sample.

\section {Background modeling}  
\label{sec:BGmodeling}

We exploit differences between the kinematic features of signal and 
background events to enhance search sensitivity.  Hence, accurate 
modeling of all contributing processes is essential.  We model 
contributions from all signal and most background processes using 
Monte Carlo event generators interfaced to a {\sc geant}-based 
simulation of the CDF~II detector~\cite{ref::GEANT}.  Events that 
contain a falsely identified ({\it fake}) lepton candidate produced 
within the shower of a parton jet are more difficult to model using 
simulation.  Therefore, data-driven methods are generally employed 
for modeling these backgrounds.

Many of the relevant signal and background processes are modeled with 
{\sc pythia}~\cite{Sjostrand:2006za}, which is a leading order (LO) 
event generator that incorporates higher-order corrections through 
parton-shower algorithms.  Events are generated with {\sc pythia}
version 6.216 using the CTEQ5L~\cite{Lai:1999wy,cteq2} parton 
distribution functions (PDFs) and the set of input parameters that 
best match underlying event distributions in CDF data~\cite{ref::tuneA}.  
For background processes more sensitive to higher-order contributions,  
next-to-leading order (NLO) generators are used and interfaced with 
{\sc pythia} to model the showering and fragmentation of generated 
initial and final state particles.  We incorporate simulated event 
samples generated with both {\sc mc@nlo}~\cite{Frixione:2002ik} and 
{\sc madgraph}~\cite{ref::madgraph}.  Because NLO event generators 
include first-order radiative effects, the scale of radiative 
corrections applied in subsequent {\sc pythia} shower modeling is 
cut off at the lower bound of that applied within the original 
event generation.  In other cases, contributions from orders above 
NLO play an important role and {\sc alpgen}~\cite{ref::alpgen} 
interfaced with {\sc pythia} is used for generating samples.  Here, 
independent samples for the LO process plus $n = $ 0, 1, 2, and 3
or more additional partons are generated, and a matching algorithm 
is used to remove overlapping contributions.  These contributions 
originate from, for example, an {\sc alpgen} LO plus 0 parton event 
which gains an additional hard radiation through the {\sc pythia} 
showering and becomes a LO plus 1 parton event.  Modeling of the 
$W\gamma$ and $Z\gamma$ production processes is generally achieved
with a dedicated LO generator~\cite{Baur:1992cd} interfaced with 
{\sc pythia} to incorporate initial state radiative effects.  
Normalizations of predicted event rates are based on theoretical 
cross section calculations performed at the highest available order. 

Nonresonant {\it WW} production in conjunction with subsequent leptonic 
decays of both $W$ bosons results in a final state similar to that 
of the primary signal.  Because of the relevance of {\it WW} backgrounds,
NLO generators are generally used to model them.  In particular, {\sc 
mc@nlo} is used to simulate events originating from {\it WW} production 
in the OS Base (0 Jet and 1 Jet), OS Inverse $M_{\ell\ell}$, and SS 
($\ge$1 Jets) search samples.  The {\sc mc@nlo} generator does not 
simulate the small but potentially signal-like contributions to {\it WW} 
production originating from gluon fusion~\cite{ref::ggWW}.  To account 
for this contribution, events are reweighted as a function of the angular 
separation in the transverse plane between the two generator-level 
leptons, $\Delta\phi_{\ell\ell}$, to incorporate the extra 
contribution predicted in Ref.~\cite{ref::ggWW}.  Uncertainties 
on the correction are obtained from alternate re-weightings that 
correspond to halving or doubling the predicted contribution of 
the unmodeled production modes.  In the OS Base ($\ge$2 Jets) search 
sample, the presence of multiple reconstructed jets requires 
inclusion of NNLO contributions in the {\it WW} background model.  
Therefore, events generated with {\sc alpgen} are used for modeling 
the {\it WW} contribution.  For the OS Hadronic Tau ($e + \tau_{\rm{had}}$ 
and $\mu + \tau_{\rm{had}}$) search samples, {\it WW} background contributions 
have a reduced significance with respect to those from other sources 
and are therefore modeled using {\sc pythia}.  Because events 
originating from direct {\it WW} production share the same final states 
with potential signal events, it is not possible to define independent 
{\it WW} background-rich data control regions for testing the modeling.  
Instead, the primary validation of this modeling comes from using it 
to extract a measurement of the {\it WW} production cross section 
directly from the search samples (see Sec.~\ref{sec:finalresults}).     
    
\begin{figure*}[t]
\begin{center}
\subfigure{
\includegraphics[width=0.45\textwidth]{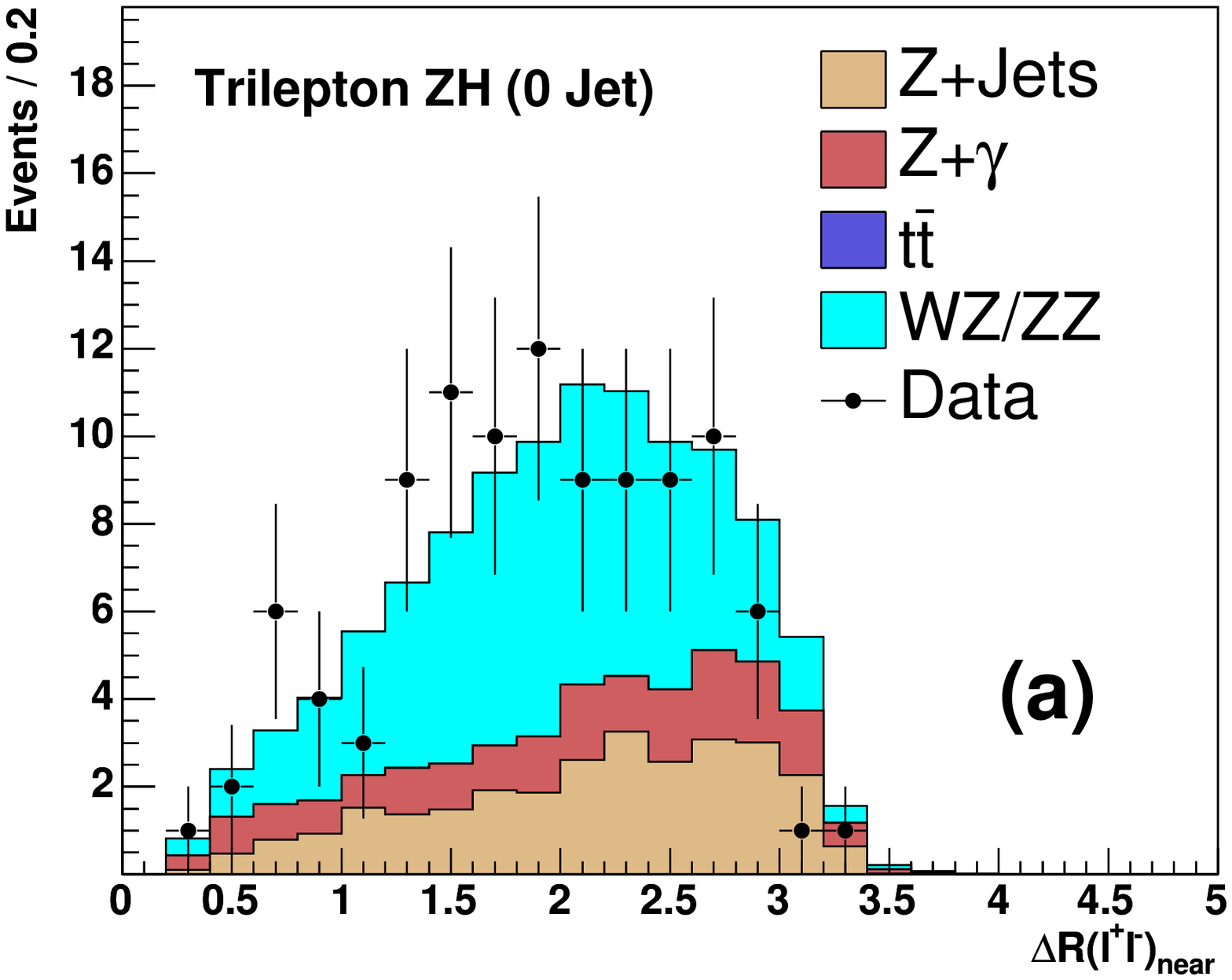}
}
\subfigure{
\includegraphics[width=0.45\textwidth]{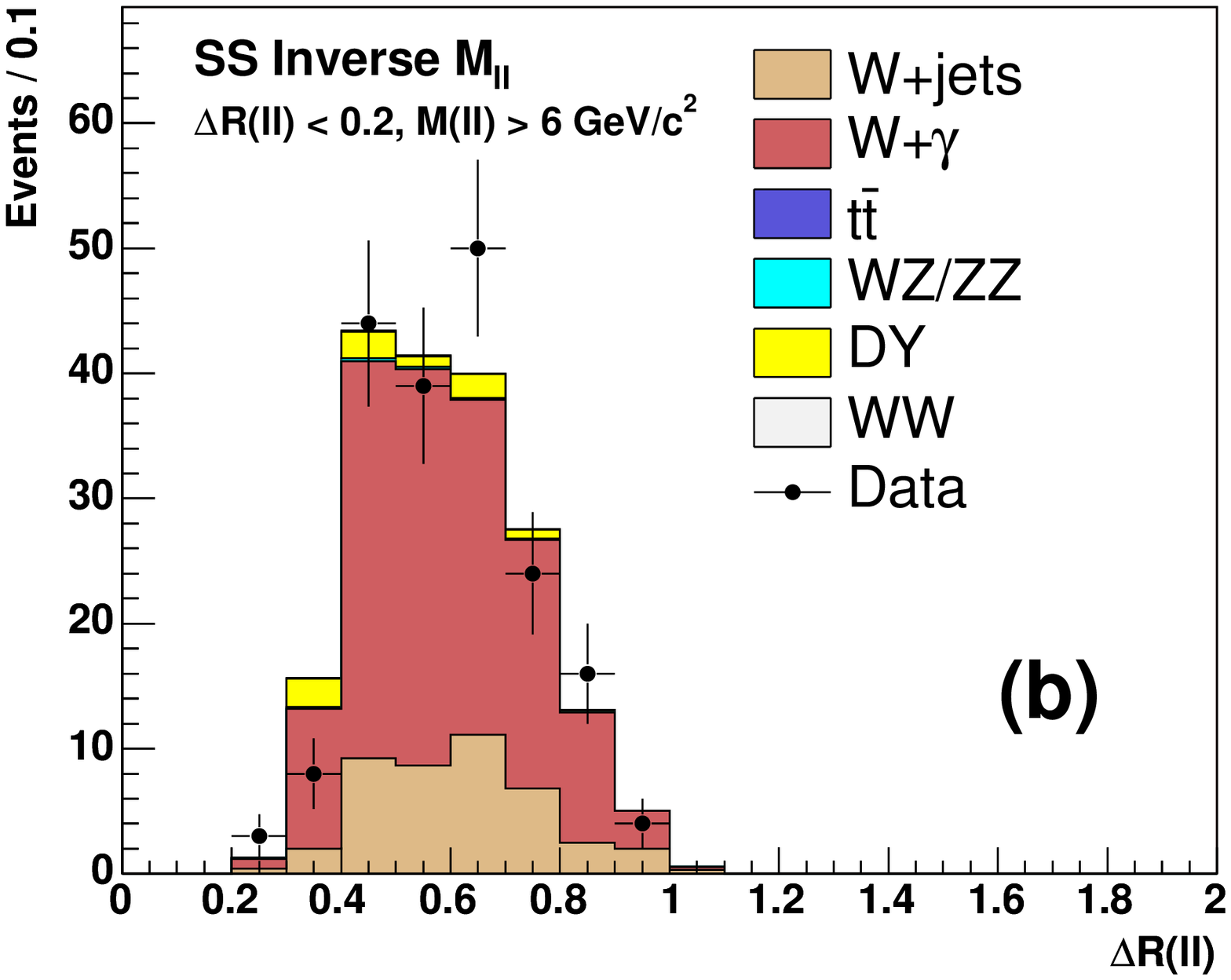}
}
\subfigure{
\includegraphics[width=0.45\textwidth]{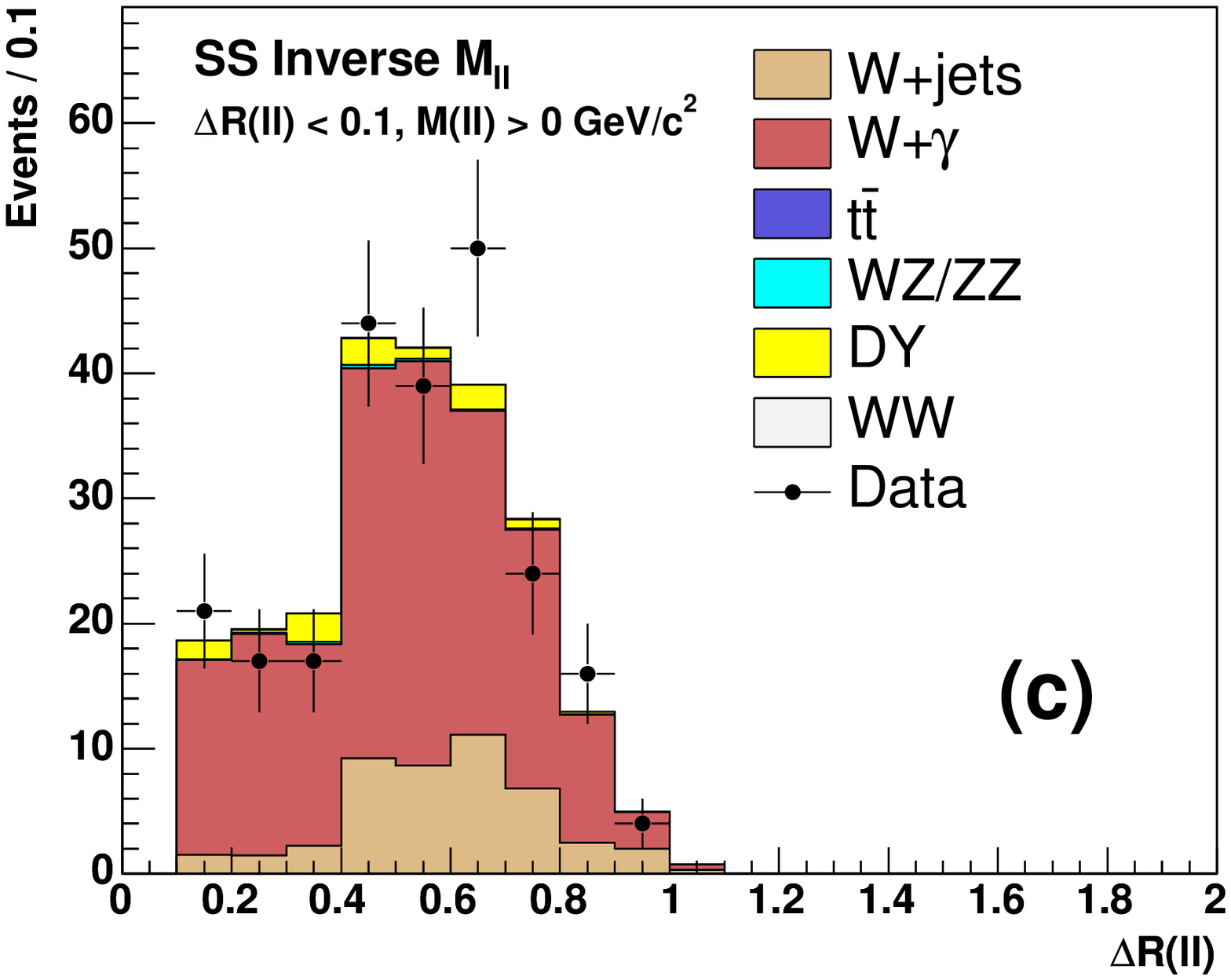}
}
\subfigure{
\includegraphics[width=0.45\textwidth]{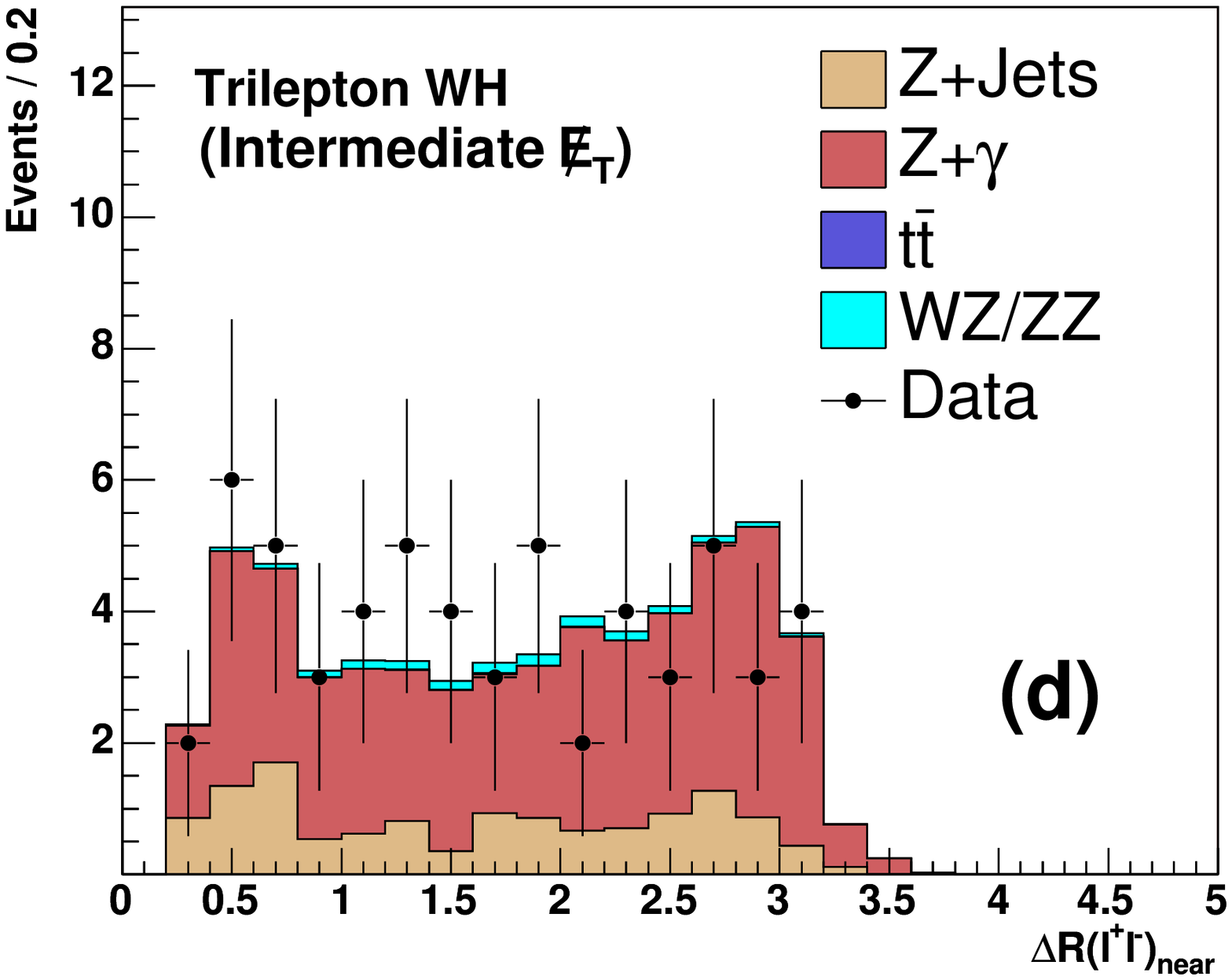}
}
\end{center}
\caption{Comparisons of observed and predicted kinematic distributions 
from independent data control samples used for validating the modeling 
of individual background processes contributing to search samples.  
(a) Dilepton angular separation, $\Delta R(\ell^+\ell^-)_{{\rm near}}$, from 
Trilepton {\it ZH} (0 Jet) control sample testing {\sc pythia} {\it WZ} 
event model.  (b) $\Delta R(\ell\ell)$ from SS Inverse $M_{\ell\ell}$ 
control sample testing the Baur $W\gamma$ event model.  (c) $\Delta 
R(\ell\ell)$ from SS Inverse $M_{\ell\ell}$ control sample testing 
{\sc madgraph} $W\gamma$ event model.  (d) $\Delta R(\ell^+\ell^-)_{{\rm near}}$ 
from Trilepton {\it WH} (Intermediate $\Met$) control sample testing the 
Baur $Z\gamma$ event model.  Normalizations for background event yields 
are taken directly from the modeling of the individual processes.}
\label{fig::CR_WZ_Wg_Wg_Zg}
\end{figure*}

Backgrounds from {\it WZ} and {\it ZZ} production in the dilepton sample 
are significantly smaller than those from {\it WW} production.  In 
addition, when two leptons are produced in the decay of one $Z$ 
boson, the most probable hadronic decay of the extra $Z$ or 
$W$ boson leads to events containing multiple jets at LO.  We 
therefore mostly rely on events generated with {\sc pythia} to 
model event contributions from these processes.  The {\sc pythia} 
{\it WZ} and {\it ZZ} event samples include $\gamma^*$ contributions 
based on lower $m_{Z/\gamma^*}$ thresholds of 2 and 15~GeV/$c^2$, 
respectively.  Event contributions from {\it WZ} and {\it ZZ} production
to the trilepton search samples are more significant and 
higher-order contributions are more relevant in the modeling of 
events containing multiple jets.  Hence, independent {\sc alpgen} 
{\it WZ} and {\it ZZ} event samples are used for modeling event 
contributions from these processes in the Trilepton {\it ZH} ($\ge$2 Jets) 
search sample.  The {\sc pythia} {\it WZ} background model is tested 
in the Trilepton {\it ZH} (0 Jet) control sample.  An example of the 
agreement between observed and predicted kinematic distributions 
for this control sample is shown in Fig.~\ref{fig::CR_WZ_Wg_Wg_Zg}(a).  
The background model is further validated by determining the {\it WZ} 
production cross section directly from the Trilepton {\it WH} search 
sample (see Sec.~\ref{sec:finalresults}).  The modeling of {\it ZZ} 
background contributions is similarly tested by measuring the 
{\it ZZ} cross section in $ZZ \rightarrow \ell\ell\nu\nu$ within the 
OS Base search samples (see Sec.~\ref{sec:finalresults}).

Samples produced using the Baur LO event generator~\cite{Baur:1992cd} 
are used in most cases for modeling $W\gamma$ and $Z\gamma$ 
contributions to the search samples.  Generated events are required 
to have a minimum angular separation of 0.2 radians between the 
photon and charged lepton(s) produced in the boson decay.  The 
photon is also required to have a minimum $p_T$ of at least 4~GeV/$c$.  
Modeling of the $W\gamma$ background is tested using the SS 
Inverse $M_{\ell\ell}$ control region, for which the $W\gamma$ 
event contribution is expected to be greater than 75\%.  Based 
on this control sample, a scale factor of 0.71 on the overall 
normalization of the $W\gamma$ sample is obtained.  An example 
of the agreement between observed and predicted kinematic 
distributions in this sample (for the Baur model after scaling) 
is shown in Fig.~\ref{fig::CR_WZ_Wg_Wg_Zg}(b).  The $W\gamma$ 
background contributions are of particular importance in the 
OS Inverse $M_{\ell\ell}$ search sample.  For this sample 
only, {\sc madgraph} is used to model $W\gamma$ background 
contributions.  The minimum threshold on the angular separation 
between the photon and charged lepton(s) is reduced to 0.1 
radians, which expands the search reach in the low $M_{\ell\ell}$ 
region.  The {\sc madgraph} model is also validated with the 
SS Inverse $M_{\ell\ell}$ control sample.  In this case, the 
normalization of the model agrees with data, and no scaling is 
needed.  An example of the agreement between observed and predicted 
kinematic distributions for this control sample (for the {\sc 
madgraph} model) is shown in Fig.~\ref{fig::CR_WZ_Wg_Wg_Zg}(c).  
Validation of the Baur modeling of the $Z\gamma$ process is 
obtained from the Trilepton {\it WH} (intermediate $\Met$) control 
sample, and an example of the agreement between observed and 
predicted kinematic distributions for this control sample is 
shown in Fig.~\ref{fig::CR_WZ_Wg_Wg_Zg}(d). 
    
\begin{figure*}[t]
\begin{center}
\subfigure{
\includegraphics[width=0.45\textwidth]{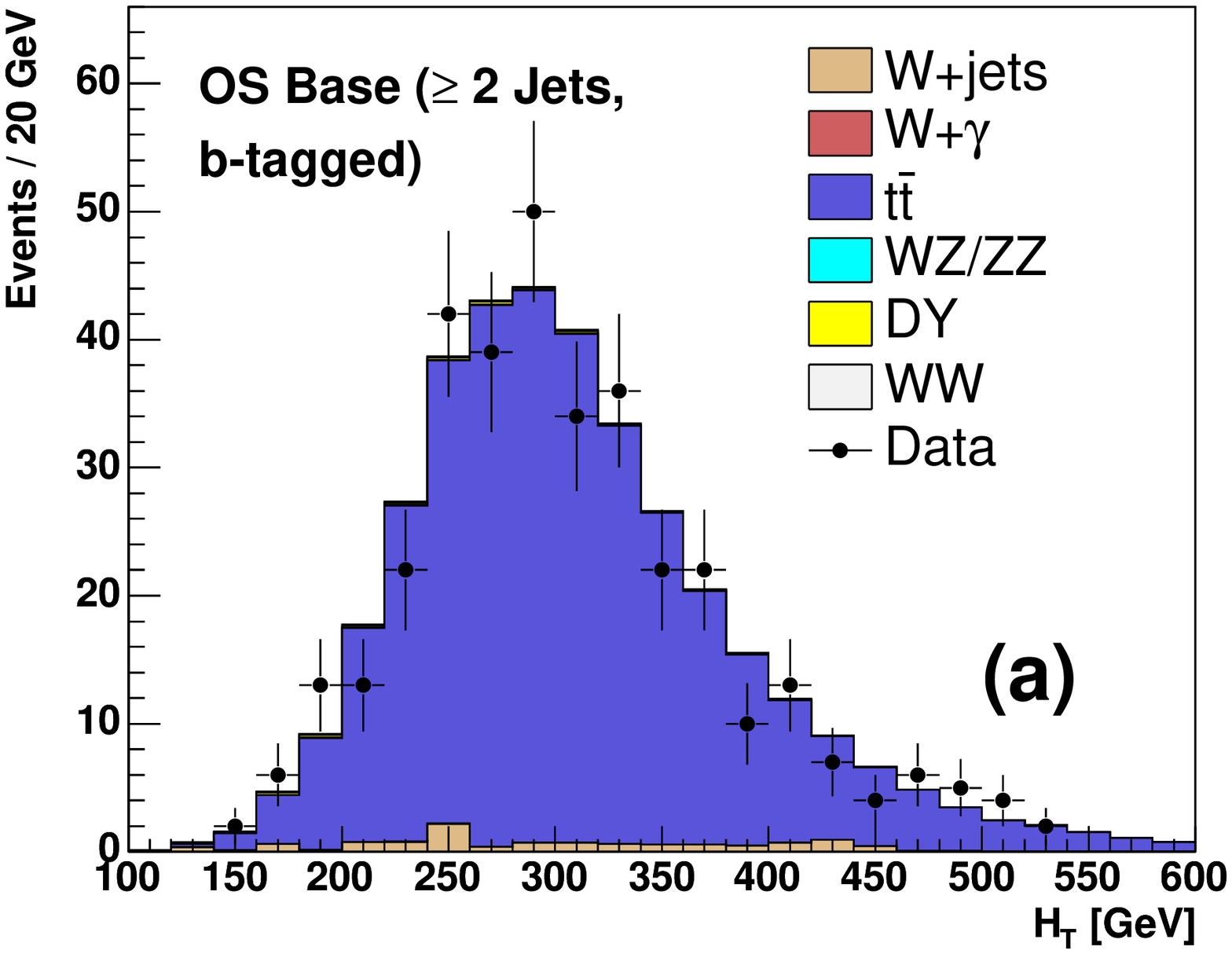}
}
\subfigure{
\includegraphics[width=0.45\textwidth]{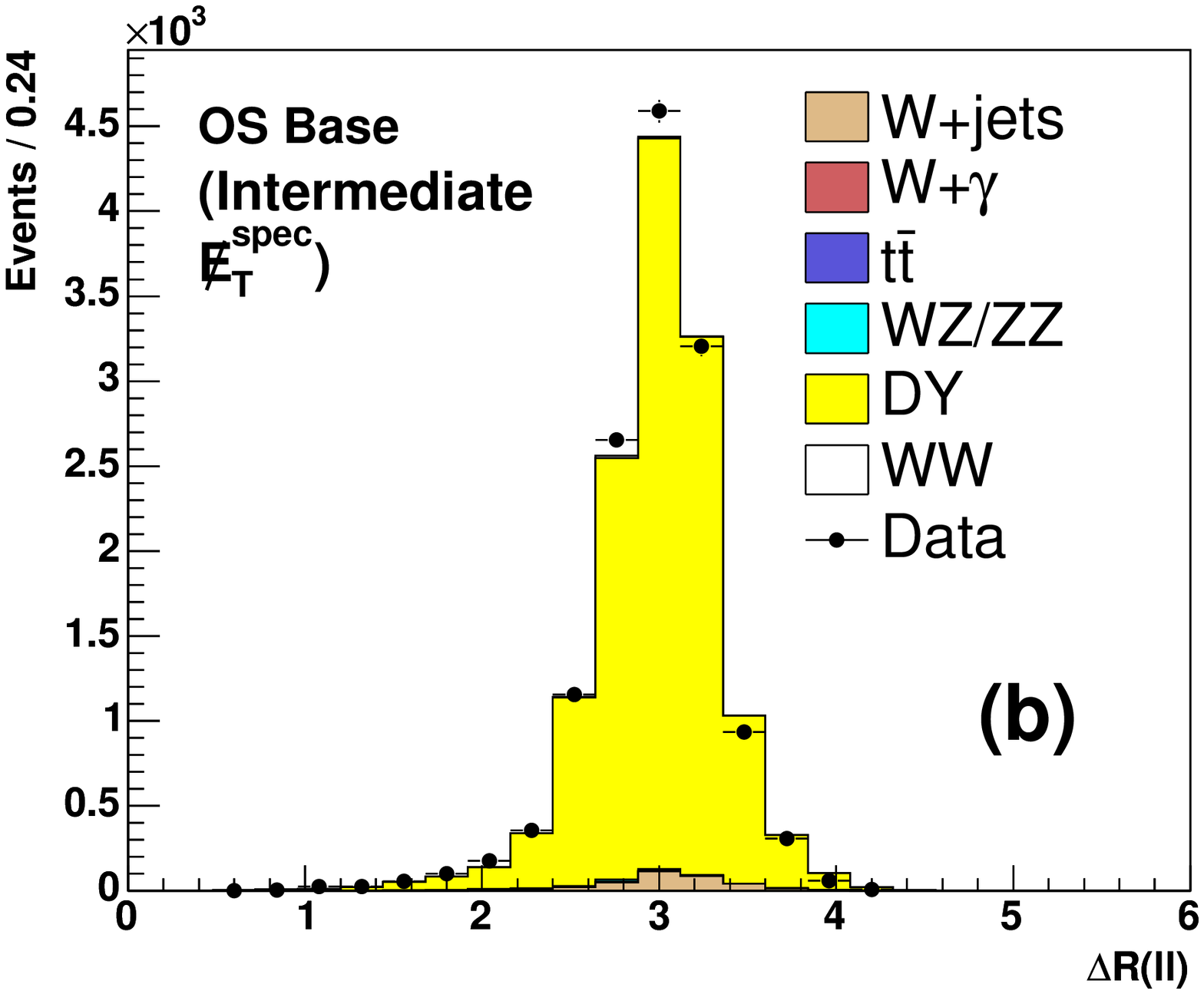}
}
\subfigure{
\includegraphics[width=0.45\textwidth]{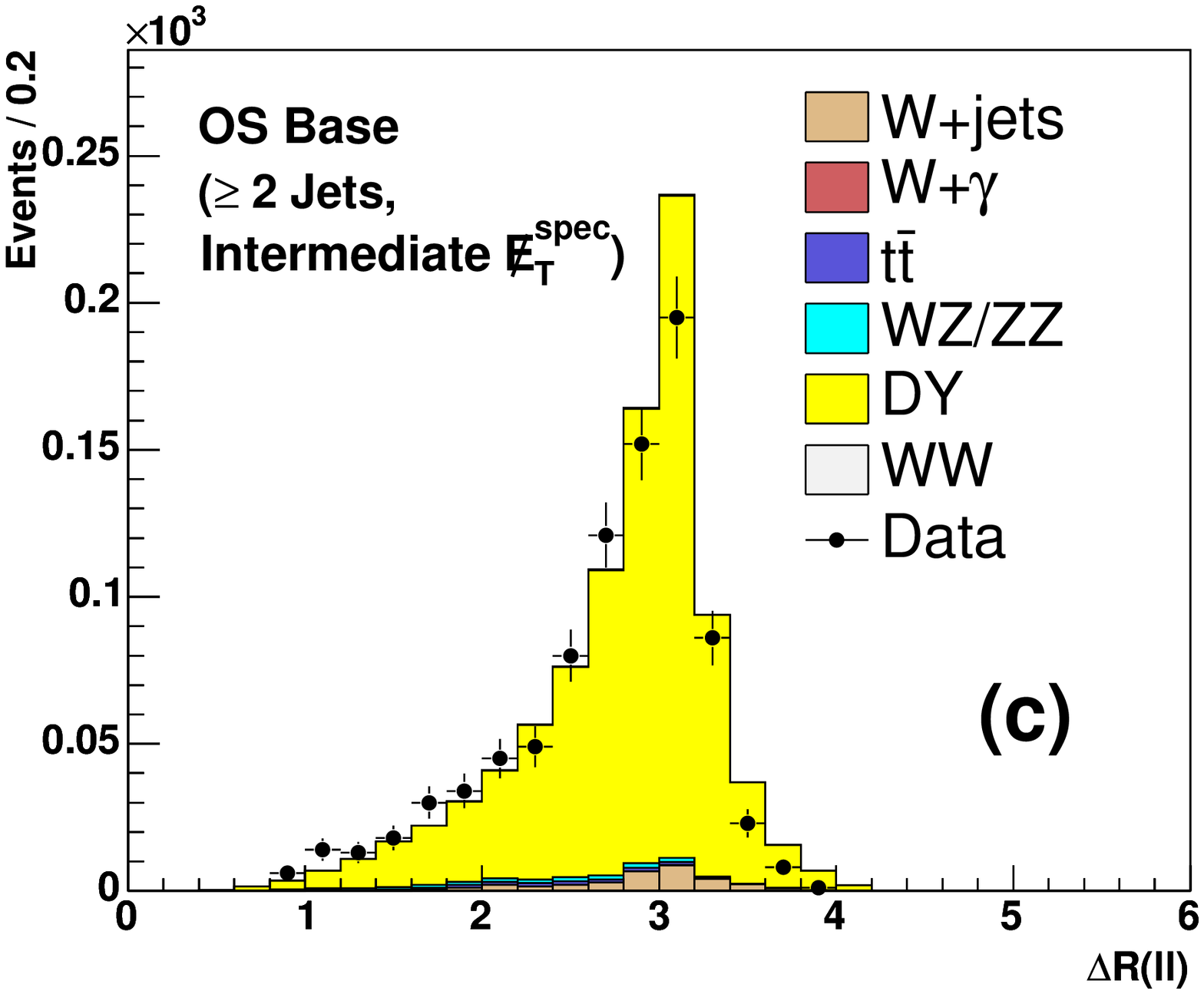}
}
\subfigure{
\includegraphics[width=0.45\textwidth]{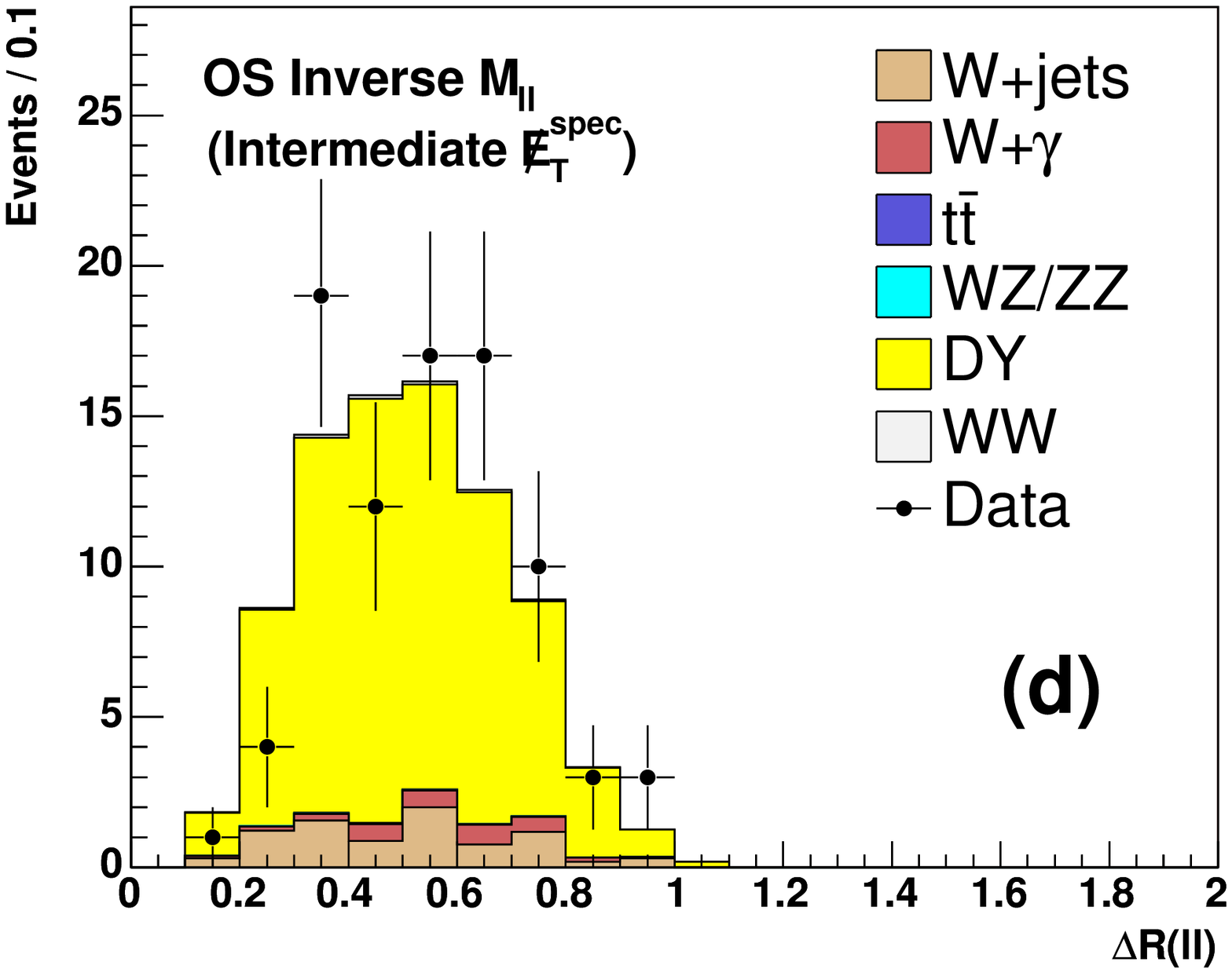}
}
\end{center}
\caption{Comparisons of observed and predicted kinematic distributions 
from independent data control samples used for validating the modeling 
of individual background processes contributing to search samples.  
(a) Sum of measured lepton and jet transverse energies and missing 
transverse energy, $H_T$, from OS Base ($\ge$2 Jets, b-tagged) control 
sample testing {\sc pythia} $t\bar{t}$ event model.  (b) $\Delta 
R(\ell\ell)$ from OS Base (Intermediate $\MetSpec$) control sample 
testing tuned {\sc pythia} DY event model.  (c) $\Delta R(\ell
\ell)$ from OS Base ($\ge$2 Jets, Intermediate $\MetSpec$) control 
sample testing {\sc alpgen} DY event model.  (d) $\Delta R(\ell\ell)$ 
from OS Inverse $M_{\ell\ell}$ (Intermediate $\MetSpec$) control 
sample testing {\sc madgraph} DY event model.  Normalizations for 
background event yields are taken directly from the modeling of the 
individual processes.}
\label{fig::CR_tt_DY_DY_DY}
\end{figure*}

Dilepton events originate from the process $t\bar{t}\rightarrow 
W^+bW^-\bar{b} \rightarrow \ell^+\nu b \ell^-\bar{\nu} \bar{b}$.  
The presence of two bottom quarks in the final state implies LO 
contributions to all search samples including those that contain 
events with multiple reconstructed jets.  Event samples obtained 
from {\sc pythia} are therefore used for modeling $t\bar{t}$ 
background contributions across all search samples.  Events 
containing jets tagged as $b$-quark candidates are removed from 
the OS ($\ge$2 Jets) search sample.  For the special case of 
modeling the $t\bar{t}$ background contribution within this 
sample, a standard CDF scale factor~\cite{Acosta:2004hw} 
(1.04~$\pm$~0.05) that corrects Monte Carlo $b$-tagging 
inefficiency to match that observed in data is applied.  A 
second scale factor (1.02~$\pm$~0.02) is used to account for 
the small fraction of events in data in which silicon-tracker 
information required for tagging $b$-quark jets is missing.  
The OS Base ($\ge$2 Jets, $b$-tagged) control sample, which is 
expected to have a $t\bar{t}$ contribution greater than 95\%, 
is used to validate the modeling.  Since events in this control 
sample are required to have at least one $b$-tagged jet, a 
reciprocal set of scale factors are applied to the modeled 
$t\bar{t}$ contribution.  An example of the agreement between 
observed and predicted kinematic distributions for this control 
sample is shown in Fig.~\ref{fig::CR_tt_DY_DY_DY}(a). 

Modeling of background contributions in the search samples associated 
with DY ($Z/\gamma^*$) production is particularly complicated.
Inclusive production is generally very well-modeled with {\sc pythia}.  
However, because of minimum missing transverse energy requirements, the 
search samples contain DY background contributions originating 
from only a small subset of this inclusive production.  In particular, 
since dilepton events originating from DY production do not 
involve neutrinos, missing transverse energy is necessarily generated 
from the mismeasurement of lepton and jet energies.  For the OS Base 
(0 Jet and 1 Jet) search samples, DY modeling is based on 
{\sc pythia}-generated event samples.  The OS Base (Intermediate 
$\MetSpec$) control sample is used to validate and tune these samples.
Initially, we observe poor modeling of the observed maximum measured 
missing transverse energies associated with specific values of 
$Z/\gamma^*$ transverse momenta.  In addition to mismeasurememnts of 
particles that recoil against the $Z/\gamma^*$ in the hard interaction, 
soft scattering processes that are not necessarily well modeled in the 
simulation can be mismeasured, thus producing events with larger 
missing transverse energies than expected from simulation.

To mimic these unmodeled effects, a constant offset is added to the 
missing transverse energy within each simulated event.  The value of 
this offset is such that the best match is achieved in the relevant 
kinematic distributions between data and simulation within the OS 
Base (Intermediate $\MetSpec$) control sample.  The resulting offset 
is +4~$\pm$~2~GeV.  The tuned simulated events are reweighted to
reproduce observed event counts correctly in the OS Base (Intermediate 
$\MetSpec$) control sample.  Independent reweightings are obtained for 
simulated events within the dilepton invariant mass ranges of 16--36, 
36--56, 56--76, 76--106, and greater than 106~GeV/$c^2$.  An example 
of the agreement between observed and predicted kinematic distributions 
for the control sample after applying this tuning procedure is shown 
in Fig.~\ref{fig::CR_tt_DY_DY_DY}(b).   
 
In the OS Base ($\ge$2 Jets) search sample, DY contributions 
from NNLO are significant and {\sc alpgen}-generated events are 
used for modeling the background.  A similarly defined OS Base 
($\ge$2 Jets, Intermediate $\MetSpec$) data control region is used 
to validate the {\sc alpgen} event modeling.  Owing to the 
presence of two high-$E_T$ jets within each event, effects from 
unmodeled energies associated with soft scattering processes are 
reduced, and the untuned event model is found to be sufficient.  
An example of the agreement between observed and predicted 
kinematic distributions for this control sample is shown in 
Fig.~\ref{fig::CR_tt_DY_DY_DY}(c).

\begin{figure*}[t]
\begin{center}
\subfigure{
\includegraphics[width=0.45\textwidth]{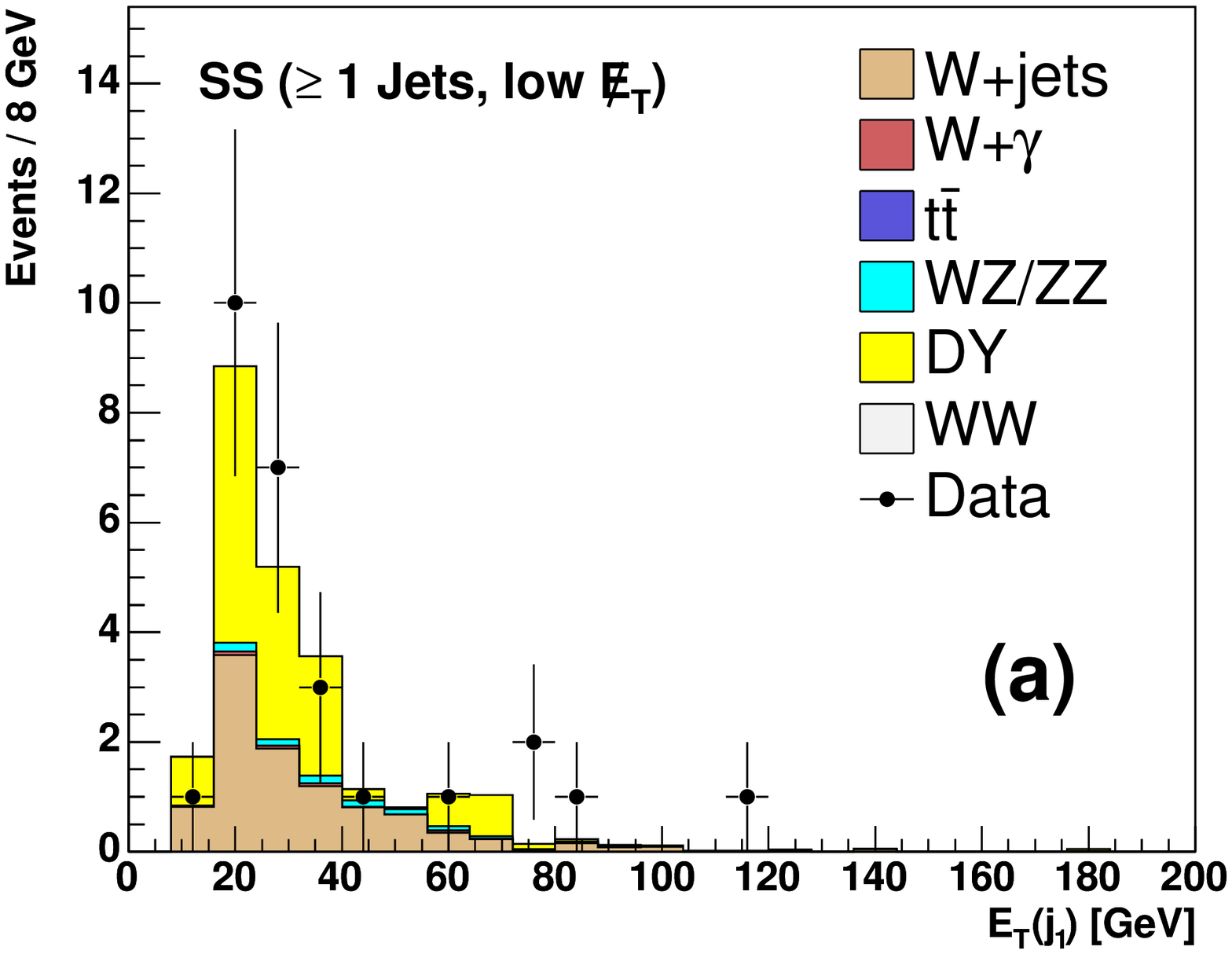}
}
\subfigure{
\includegraphics[width=0.45\textwidth]{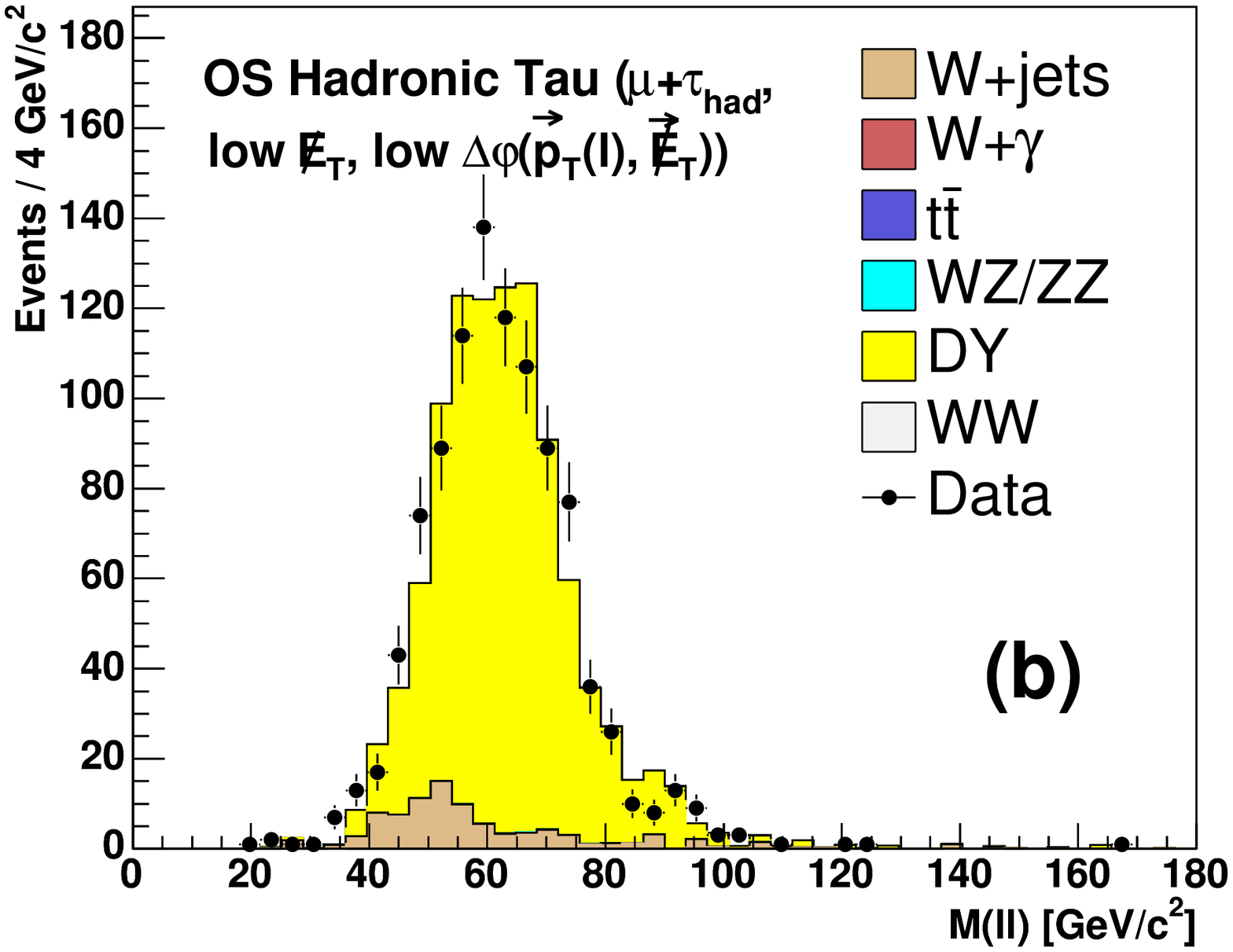}
}
\subfigure{
\includegraphics[width=0.45\textwidth]{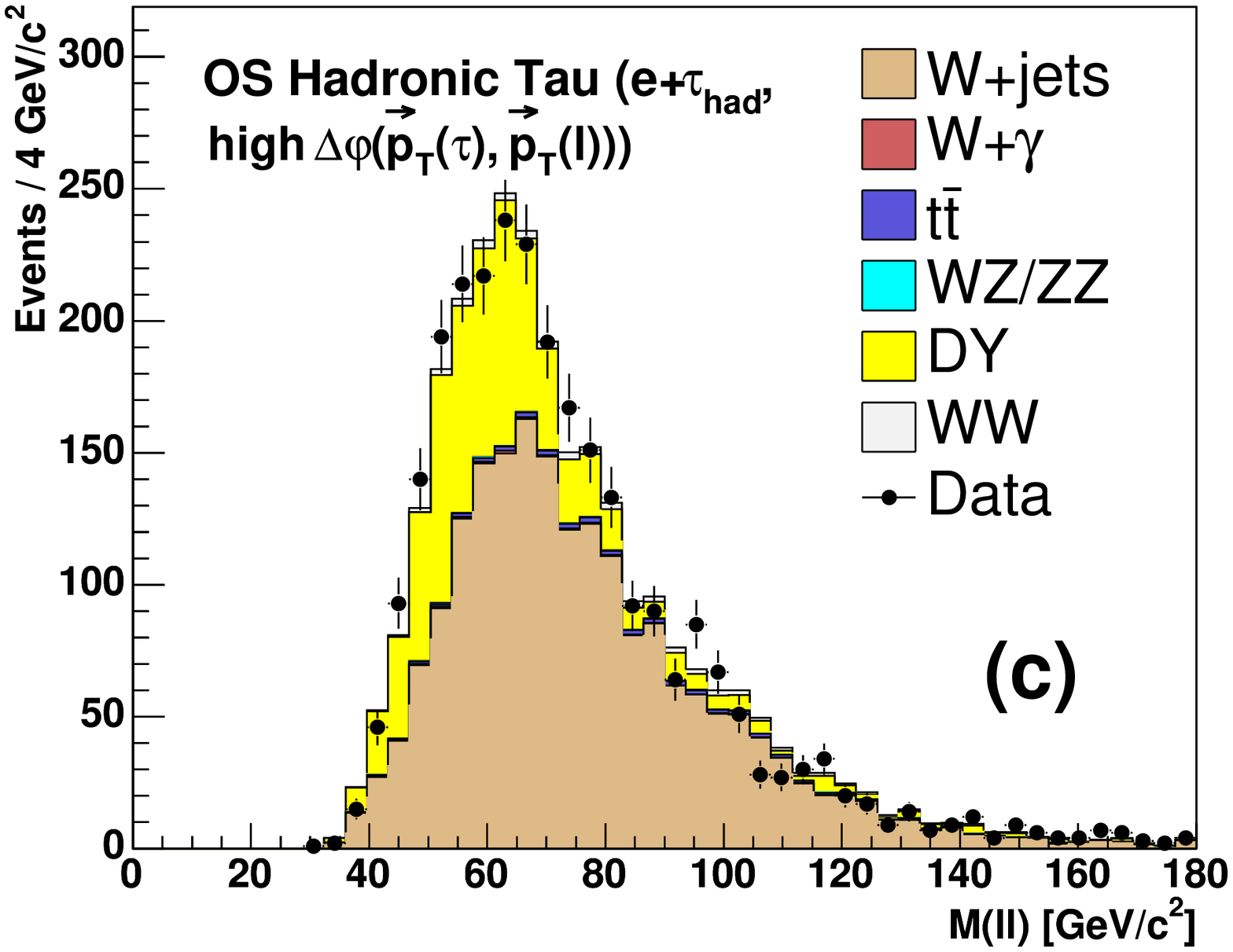}
}
\subfigure{
\includegraphics[width=0.45\textwidth]{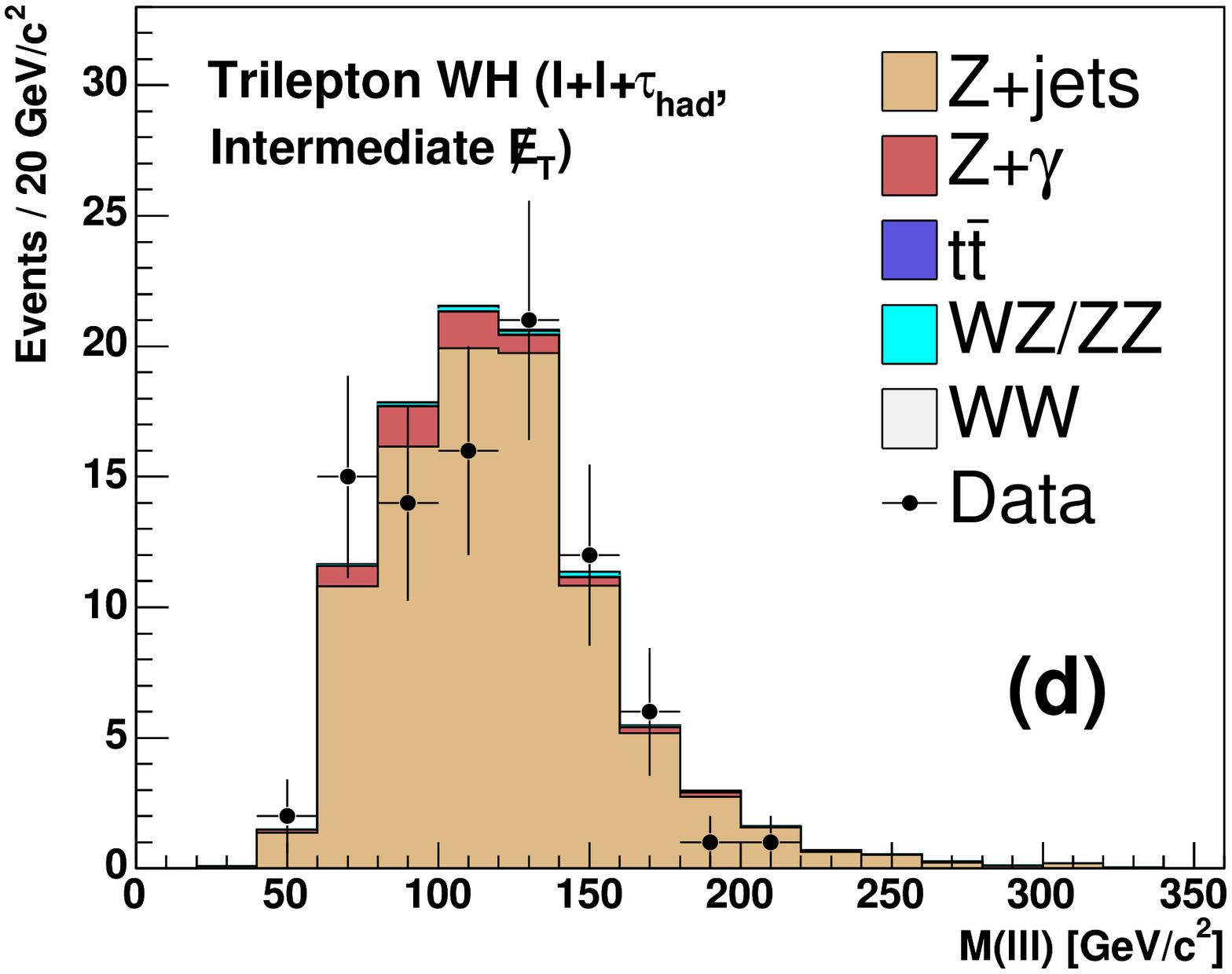}
}
\end{center}
\caption{Comparisons of observed and predicted kinematic distributions 
from independent data control samples used for validating the modeling 
of individual background production processes contributing to the 
search samples.  (a) Transverse energy of leading jet, $E_T(j_1)$, from 
SS ($\ge$1 Jets, low $\Met$) control sample testing {\sc pythia} DY
event model for tridents.  (b) Invariant mass of muon and tau lepton pair,
$M_{\mu,\tau}$ from OS Hadronic tau ($\mu + \tau_{\rm{had}}$, low $\Met$, 
low $\Delta\phi(\vec{p}_{T}(\ell),\VecMet)$) control sample testing 
{\sc pythia} DY $Z/\gamma^* \rightarrow \tau^+ \tau^-$ event model.  
(c) Invariant mass of electron and tau lepton pair, $M_{e,\tau}$ from 
OS Hadronic tau ($e + \tau_{\rm{had}}$, high $\Delta\phi(\vec{p}_T(e),
\vec{p}_T(\tau))$) control sample testing {\sc alpgen} $W$+Jets event 
model.  (d) Trilepton invariant mass, $M_{\ell \ell \tau_{\rm{had}}}$, 
from Trilepton {\it WH} ($\ell + \ell + \tau_{\rm{had}}$, Intermediate 
$\Met$) control sample testing {\sc alpgen} $Z$+Jets event model.  
Normalizations for background event yields are taken directly from 
the modeling of the individual processes.}
\label{fig::CR_DY_DY_Wj_Zj}
\end{figure*}
  
A unique set of production processes is associated with DY
contributions to the OS Inverse $M_{\ell\ell}$ search sample.  In 
this sample, events originate primarily from simple $2 \rightarrow 
2$ scattering processes, in which the $Z/\gamma^*$ is radiated from 
a final state quark.  This mechanism allows for the production of 
events with low mass $Z/\gamma^*$ bosons of sufficient $p_T$ such 
that significant missing transverse energy can result from the 
mismeasurement of associated recoil particle energies.  Since 
this process is not modeled by {\sc pythia}, {\sc madgraph} is used  
to model DY contributions in this sample.  This modeling is 
validated using the OS Inverse $M_{\ell\ell}$ (Intermediate $\MetSpec$) 
control sample.  The lack of $e$--$\mu$ dilepton events within this 
sample indicates that cascade decays of bottom quarks are not an 
appreciable background.  Dileptons from charmonium and bottomonium 
decays can be observed within this control sample but are vetoed as 
described in Sec.~\ref{sec:evtsel}.   An example of the agreement 
between observed and predicted kinematic distributions for this 
control sample is shown in Fig.~\ref{fig::CR_tt_DY_DY_DY}(d).
   
Finally, DY background contributions to the SS ($\ge$1 Jets) search 
sample come primarily from $Z \rightarrow e^+e^-$ production, in which 
a photon radiated from one of the electrons subsequently converts into
an additional $e^+e^-$ pair within the detector material.  Resulting 
trident electron candidates with two neighboring charged particles 
often have misreconstructed charges due to issues associated with 
the sharing of hits between tracks.  Since background contributions 
from tridents can be significant in this search sample, electron 
candidates in these events are required to satisfy tight selection 
criteria.  With this requirement, trident event contributions are 
substantially reduced and {\sc pythia}-generated samples are found 
to provide a good model for the remaining background.  The model is 
validated with the SS ($\ge$1 Jets, low $\Met$) control sample, and 
an example of the agreement between observed and predicted kinematic 
distributions for this sample is shown in Fig.~\ref{fig::CR_DY_DY_Wj_Zj}(a).

Background contributions from DY $Z/\gamma^* \rightarrow \tau^+ 
\tau^-$ decays represent another special case.  This process results 
in nonnegligible event contributions to both OS Hadronic Tau search 
samples and the $e$--$\mu$ components of other dilepton search samples.
The neutrinos produced in subsequent decays of the $\tau$ leptons into 
electrons and muons introduce missing transverse energy in these 
events, increasing their probability to be accepted in the search 
samples even without significant energy mismeasurements.  We use 
{\sc pythia} interfaced with {\sc tauola}~\cite{tauola} to model 
this process and validate the modeling using the OS Hadronic Tau 
($\mu + \tau_{\rm{had}}$, low $\Met$, low $\Delta\phi(\vec{p}_T(\ell),
\VecMet)$) control sample.  Figure~\ref{fig::CR_DY_DY_Wj_Zj}(b) shows 
an example of the agreement between observed and predicted kinematic 
distributions for this sample.  The overall agreement within this 
control region is also used for assigning uncertainties on the 
efficiency for reconstructing and identifying hadronically-decaying 
tau lepton candidates, which is obtained directly from simulation.  

\begin{figure*}
\begin{center}
\subfigure{
\includegraphics[width=0.45\textwidth]{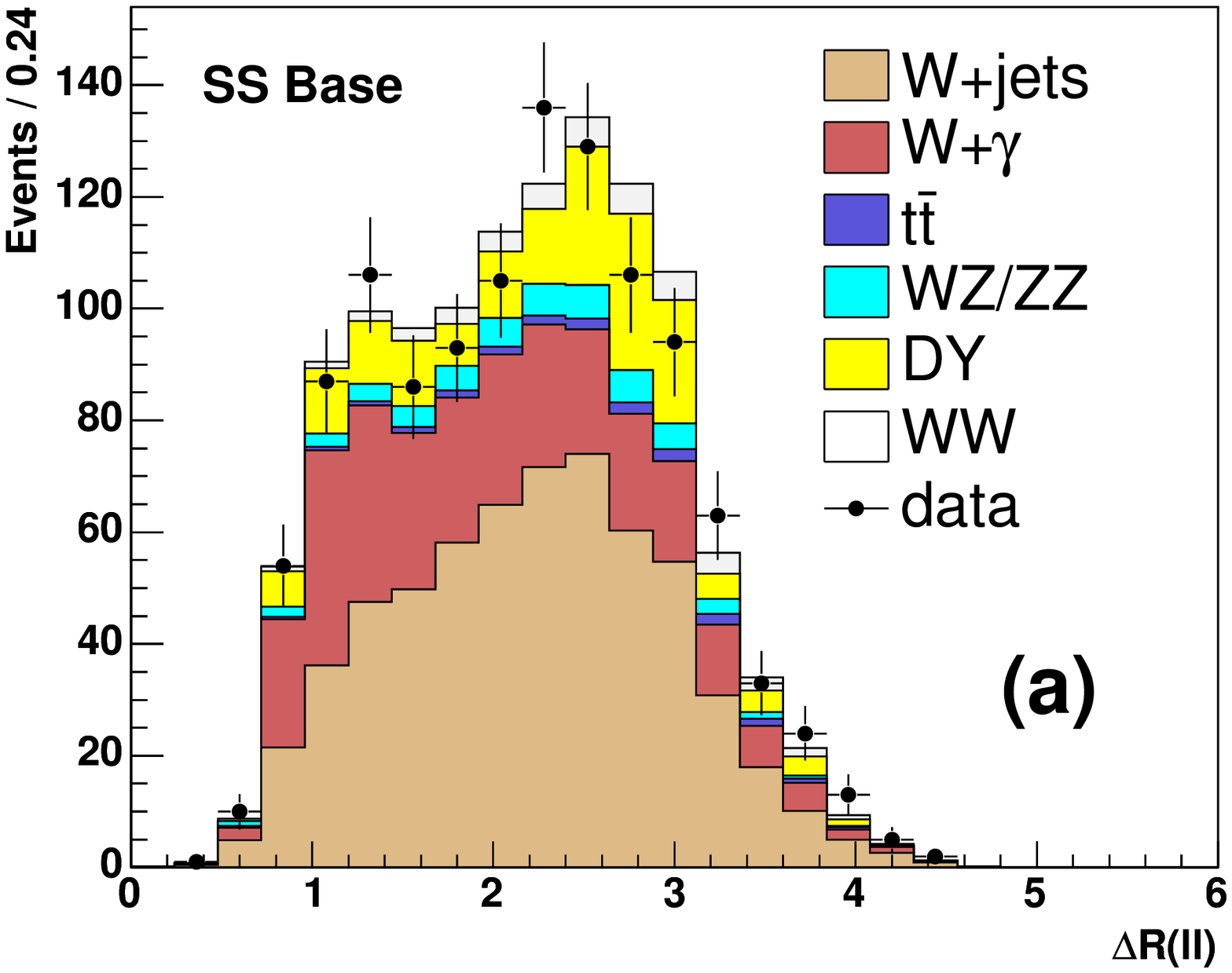}
}
\subfigure{
\includegraphics[width=0.45\textwidth]{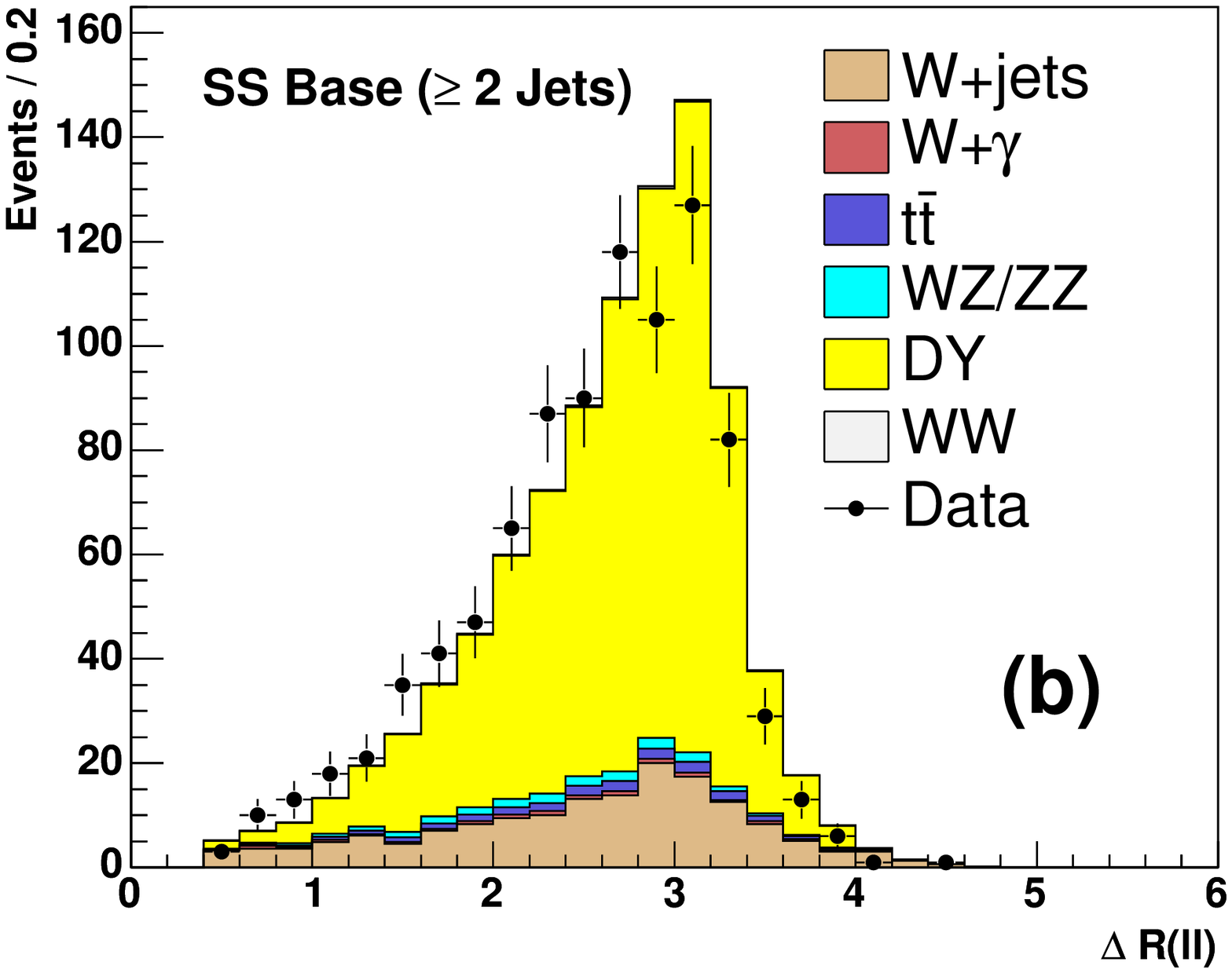}
}
\subfigure{
\includegraphics[width=0.45\textwidth]{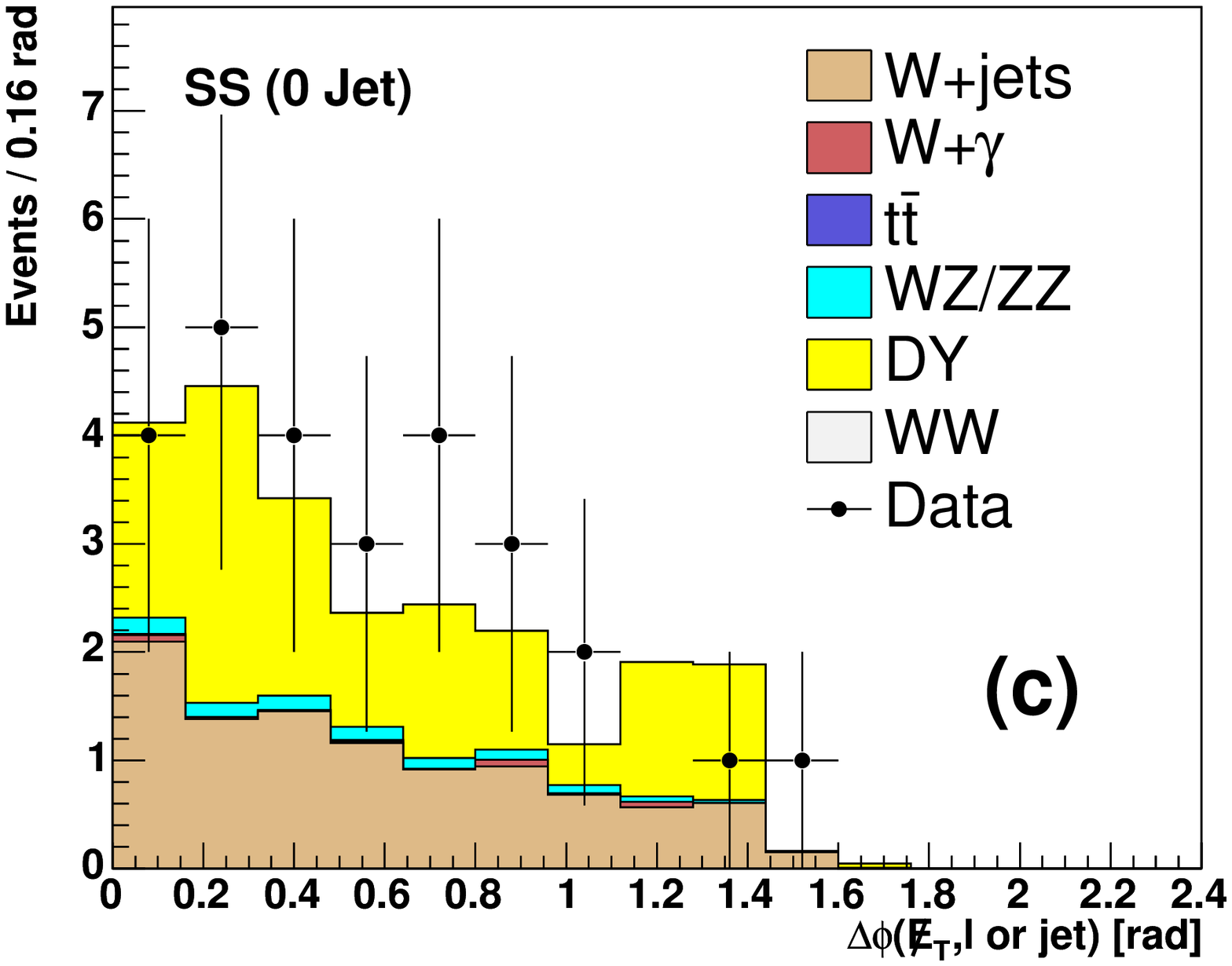}
}
\subfigure{
\includegraphics[width=0.45\textwidth]{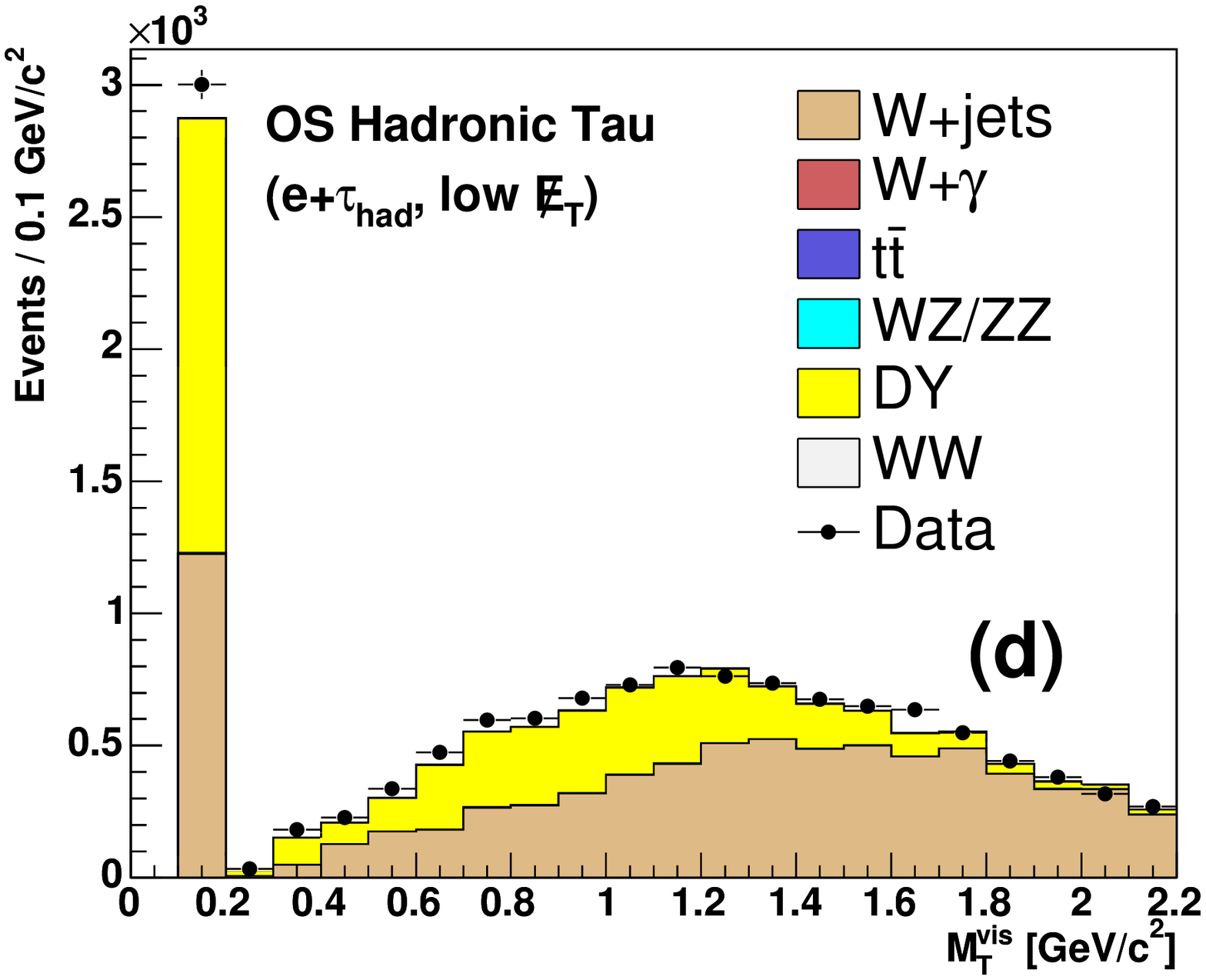}
}
\end{center}
\caption{Comparisons of observed and predicted kinematic distributions 
from independent data control samples used for validating the modeling 
of individual background production processes contributing to the 
search samples.  (a) $\Delta R(\ell\ell)$ from SS Base control sample 
testing data-driven $W$+jets event model.  (b)  $\Delta R(\ell\ell)$ 
from SS Base ($\ge$2 Jets) control sample testing data-driven $W$+jets 
event model. (c) Azimuthal opening angle between missing transverse 
energy and nearest lepton or jet, $\Delta\phi$($\Met$,$\ell$ or jet), 
from SS (0 Jet) control sample testing data-driven $W$+jets event 
model.  (d) Visble transverse mass of hadronically-decaying tau lepton 
candidate, $M_{T}^{{\rm vis}}$, from OS Hadronic Tau ($e + \tau_{\rm{had}}$, 
low $\Met$) control sample testing data-driven multijet event model.
Normalizations for background event yields are taken directly from 
the modeling of the individual processes.}
\label{fig::CR_Wj_Wj_Wj_mj}
\end{figure*}

Dilepton background contributions from events containing one 
real lepton and a jet misidentified as a lepton originate from 
a high-production cross section process, $W$ bosons produced 
in association with additional partons, in combination with 
low-probability and hard-to-simulate detector-level effects 
that allow a jet to be reconstructed as a lepton.  Similarly,
trilepton background contributions from events with two real 
leptons and a third jet misidentified as a lepton originate 
from $Z$ boson production in association with jets.  Because 
of the difficulties associated with simulating these processes, 
we rely mostly on data-driven background modeling.  However, 
the probability for a jet to mimic the signature of a 
hadronically-decaying tau lepton candidate is significantly 
larger than that of an electron or muon candidate.  We therefore 
rely on {\sc alpgen}-generated events for modeling $W$+jets 
contributions in the OS Hadronic Tau search samples and the 
$Z$+jets contribution in the Trilepton {\it WH} ($\ell + \ell 
+ \tau_{\rm{had}}$) sample.  We rely on the OS Hadronic Tau (high 
$\Delta\phi(\vec{p}_T(\tau), \vec{p}_T(\ell))$) and Trilepton 
{\it WH} ($\ell + \ell + \tau_{\rm{had}}$, Intermediate $\Met$) 
control samples, respectively, for validating the two 
{\sc alpgen} background models.  Examples of the agreement 
between observed and predicted kinematic distributions for the 
two samples are shown in Figs.~\ref{fig::CR_DY_DY_Wj_Zj}(c) 
and~\ref{fig::CR_DY_DY_Wj_Zj}(d).  

For the remaining search samples, in which a parton jet is misidentified 
as an electron or muon candidate, a data-driven technique is used for 
modeling contributions from $W$+jets and $Z$+jets production.  The 
technique relies on parametrization of the probability for a jet to 
be misidentified as a lepton. This parametrization is obtained from 
data using events collected by single-jet triggers with varying 
energy thresholds.  For each electron and muon category used in the 
searches, an associated {\it fakeable} lepton candidate is defined 
based on relaxed identification requirements.  To avoid trigger biases,
we ignore the highest $E_T$ jet reconstructed within each single-jet 
triggered event.  The total number of remaining jets that satisfy the 
fakeable-lepton selection criteria forms the denominator of the jet 
{\it fake rate} for the associated lepton type.  The number of these 
jets that additionally satisfy the full charged-lepton identification 
selection forms the fake rate numerator, which is corrected for the 
expected contribution of real high-$p_T$ leptons in these samples from 
simulated $W$ and $Z$ boson events. Fake rates are parametrized as a 
function of lepton $p_T$ and are typically of the order of a few percent.
Modeling of dilepton $W$+jets background contributions is obtained by 
applying the measured fake rates as weights to events collected using 
standard high-$p_T$ single-lepton triggers that are found to contain 
exactly one fully-selected lepton candidate and one or more fakeable 
lepton candidates.  Similarly, modeling of trilepton $Z$+jets 
background contributions is obtained from a sample of events with 
exactly two fully-selected lepton candidates and one or more fakeable 
candidates.  A correction is applied to the weights of individual events 
for which the fakeable candidate is associated with a lepton category 
that cannot be responsible for triggering collection of the event.  This 
correction accounts for the missing contribution of events containing 
leptons from the same categories, in which the triggered lepton is the 
fake lepton.

Several control samples are used to validate the data-driven background
modeling for $W$+jets production.  We use the SS Base and SS Base ($\ge$2 
Jets) control samples to validate $W$+jets modeling in the OS dilepton 
search samples.  Examples of the agreement between observed and 
predicted kinematic distributions for the two samples are shown in 
Figs.~\ref{fig::CR_Wj_Wj_Wj_mj}(a) and~\ref{fig::CR_Wj_Wj_Wj_mj}(b).  
Although we use the same data-driven technique to model $W$+jets 
backgrounds in the SS ($\ge$1 Jets) search sample, several of the looser 
lepton categories, which are a dominant source of fake backgrounds in 
the OS dilepton search samples, are not used for selecting events for 
the SS dilepton sample.  Therefore, we independently validate $W$+jets 
modeling for this sample using the SS (0 Jet) control sample.  An example 
of the agreement between observed and predicted kinematic distributions 
for this sample is shown in Fig.~\ref{fig::CR_Wj_Wj_Wj_mj}(c).

Finally, background contributions from dijet and photon-jet production
to the OS Hadronic Tau search samples are also modeled directly from data.  
Events containing an electron or muon candidate and a hadronically-decaying 
tau lepton candidate with the same charge that otherwise satisfy search 
sample criteria are used to model these background sources, which contribute
events containing two fake leptons.  Electroweak contributions to the 
same-sign sample are subtracted based on estimates obtained from simulated 
event samples.  This background model is tested using the OS Hadronic 
Tau ($e + \tau_{\rm{had}}$, low $\Met$) control sample.   An example of the 
agreement between observed and predicted kinematic distributions for 
this sample is shown in Fig.~\ref{fig::CR_Wj_Wj_Wj_mj}(d).

\section{Multivariate techniques}
\label{sec:Multivaraite}

Three multivariate techniques are used to obtain the best possible 
separation of event contributions from a potential signal from those 
originating from background processes.  These are the matrix-element 
method, artificial neural networks, and boosted decision trees.  One 
or a combination of these techniques is applied to the analysis of 
each search sample.  

\subsection{Matrix-element method}
\label{sec:ME}

The matrix-element method (ME) uses an event-by-event calculation of the 
probability density for each contributing process to produce the observed 
event. This method is based on simulation of the relevant processes and 
has been applied to a number of other measurements~\cite{Aaltonen:2011tmass,
Aaltonen:2010tmass,Aaltonen:2009tmass,Aaltonen:2009tmasslj,Abulencia:2006mi,
Aaltonen:2008sy,Aaltonen:2010sy,Aaltonen:2010wwwz,Aaltonen:2012wh,
Aaltonen:2009ZllH,Hsu:2008zz}.  If all details of the collision properties 
and the detector response are modeled in the ME calculation, this method 
provides the optimal sensitivity to the signal. However, there are several 
approximations used in the calculations: theoretical differential cross 
sections are implemented only at leading order, a simple parametrization 
of the detector response is used, and for some small ({\it WZ} and 
$t\bar{t}$) or difficult-to-model (DY) backgrounds, a probability 
density is not calculated.  

The event probability density for a given process is calculated as
\begin{equation}
\label{eqn:master_prob}
P(\vec{x}_{\rm{obs}}) =
  {1 \over \langle\sigma\rangle }
  \int \frac {d \sigma_{{\rm{LO}}} (\vec{y}) }{ d\vec{y} }
  \epsilon (\vec{y}) G(\vec{x}_{\rm{obs}},\vec{y}) d\vec{y},
\end{equation}
where the elements of $\vec{y}$ ($\vec{x}_{{\rm{obs}}}$) are the true (observed) 
values of the lepton momenta and $\Met$, $d \sigma_{{\rm{LO}}} / d\vec{y}$ is the
parton-level differential cross section  from {\sc mcfm} v3.4.5~\cite{Campbell:1999ah},
$\epsilon(\vec{y})$ is a parametrization of the detector acceptance and selection 
efficiencies, and $G(\vec{x}_{\rm{obs}},\vec{y})$ is the transfer function representing 
the detector resolution and a {\sc pythia}-based estimate of transverse momentum 
of the $\ell\ell\Met$ system due to the initial state radiation. The constant 
$\langle\sigma\rangle$ normalizes the total event probability to unity. This 
calculation integrates the theoretical differential cross section over the 
missing information due to two unobserved neutrinos in the final state. We form 
a likelihood ratio ($LR$) discriminant, which is the signal probability density 
divided by the sum of signal and background probability densities,
\begin{equation}
\label{eqn:lrhww}
LR_{ggH}(\vec{x}_{\rm{obs}})\equiv\frac{P_{H}(\vec{x}_{\rm{obs}})}{P_{H}(\vec{x}_{\rm{obs}}) + 
\sum_i k_i P_{i}(\vec{x}_{\rm{obs}})},
\end{equation}
where $k_i$ are the expected fractions of {\it WW}, {\it ZZ}, $W\gamma$, and 
$W$+jets background events.  An analogous likelihood ratio, $LR_{WW}$, is similarly 
formed by treating direct {\it WW} production as the signal.  The ME method is used 
in conjunction with an artificial neural network and only in the OS Base (0 Jet) 
search samples as defined in Sec.~\ref{sec:evtsel}.

\subsection{Neural networks}
\label{sec:NN}

Artificial neural networks~\cite{Feindt:2006pm} are used to discriminate 
potential signal events from background events.  A three-layer feed-forward 
network is constructed with $N_I$ input nodes in the first layer, $N_I +1$ 
nodes in the second layer, and one output node in the third and final 
layer for each search sample relying on this approach.  The single output
parameter of the network, referred to as the discriminant, is used to enhance 
the separation between signal and background.  The number of variables 
being considered, $N_I$, varies depending on the search sample.  Events in 
the simulated or data-driven background samples are weighted such that the 
sum of the weights is equal to the number of generated and simulated signal 
events.  Only input variables with accurately modeled distributions are used.
A separate neural network is trained for each Higgs boson mass considered.  
Variables with less discriminating power are determined for each value of 
$m_H$ and removed, resulting in differing sets of network inputs for each 
value of $m_H$.  The selection of kinematic input variables for the neural 
network is based on kinematic properties of the production and decay of the 
Higgs boson.  Correlated variables are discarded, resulting in the minimal 
set of discriminant variables.  For the OS Base (0 Jet) search samples, 
matrix-element likelihood ratios were included as inputs to the neural 
network and resulted in only 5\% improvements in overall search sensitivity, 
demonstrating that the neural network is able to determine the input 
variables needed to describe the full kinematic properties of the events 
and efficiently separate signal and background.  Comparable results are 
obtained using an alternative neural network algorithm~\cite{Hocker:2007ht}.

\subsection{Boosted decision trees}
\label{sec:BDT}

To discriminate signal from backgrounds in the OS Hadronic Tau and Trilepton 
WH ($\ell + \ell + \tau_{\rm{had}}$) search samples, a boosted decision tree 
(BDT) algorithm~\cite{Breiman:1984lb,Hocker:2007ht} is used.  The use 
of BDTs for these samples provides a simple mechanism for incorporating 
hadronically-decaying tau lepton identification variables, which have a 
significant role in separating potential signal from dominant $W$+jets 
background contributions.  A set of criteria is applied sequentially to 
the variables provided as input to the tree.  A boosting procedure is 
applied to enhance the separation performance and make the decision robust 
against statistical fluctuations in the training samples.  New trees are 
derived from the same training sample by reweighting the events that are 
misclassified.  In this way, each tree is extended to a forest of trees 
and the final decision is based on a weighted majority vote of all trees 
within the forest~\cite{foot:bdt}.

\begin{table*}[t]
  \setlength{\extrarowheight}{3pt}
\begin{ruledtabular} 
\begin{center} 
\caption{\label{tbl:YieldsDilepton}
Summary of predicted and observed event yields for seven dilepton 
search samples formed from electron and muon candidates.  Expected 
signal yields are shown for potential SM Higgs boson masses of 125 
and 165~GeV/$c^2$.  Normalizations for background event yields are 
taken directly from the modeling of the individual processes.}
%{\scriptsize
\begin{tabular}{l*{7}{c}}
\toprule                                                                                                                     
Process &   OS 0 Jet          &    OS 0 Jet          &   OS 1 Jet        &  OS 1 Jet        &  OS $\ge$2 Jets    &  OS Inverse $M_{\ell\ell}$ &  SS $\ge$1 Jets \\
        &   High $s/b$ Lep.     &    Low $s/b$ Lep.      &   High $s/b$ Lep.   &  Low $s/b$ Lep.    &                &                            &             \\
\hline                                                                                                                     
$t\bar t$ & 2.93 $\pm$ 0.93   &    0.99 $\pm$ 0.26   &   75  $\pm$ 15    &  24.5 $\pm$ 4.6  &  287 $\pm$ 42  &  1.82$\pm$0.35    &  0.58$\pm$0.08 \\
DY      & 230  $\pm$ 63     &    230  $\pm$ 63     &   239 $\pm$ 55    &  176  $\pm$ 41   &  155 $\pm$ 66  &  23.9$\pm$4.9     &  16.4$\pm$4.6  \\  
{\it WW}      & 661  $\pm$ 66     &    308  $\pm$ 31     &   183 $\pm$ 22    &  78.0 $\pm$ 9.6  &  53  $\pm$ 12  &  37.5$\pm$3.6     &  0.07$\pm$0.02 \\
{\it WZ}      & 29.1 $\pm$ 4.4    &    15.5 $\pm$ 2.4    &   26.4$\pm$ 3.6   &  16.1 $\pm$ 2.2  &  11.7$\pm$ 2.2 &  0.96$\pm$0.13    &  14.6$\pm$2.0  \\
{\it ZZ}      & 42.1 $\pm$ 6.0    &    21.4 $\pm$ 3.0    &   11.5$\pm$ 1.7   &  5.71 $\pm$ 0.82 &  5.3 $\pm$ 1.0 &  0.29$\pm$0.04    &  2.43$\pm$0.33 \\
$W+$jets  & 137  $\pm$ 33     &    443  $\pm$ 67     &   54  $\pm$ 15    &  163  $\pm$ 26   &  80  $\pm$ 15  &  56.3$\pm$7.8     &    45$\pm$17   \\
$W\gamma$ & 68.3 $\pm$ 8.6    &    181  $\pm$ 23     &   9.9 $\pm$ 1.5   &  31.6 $\pm$ 4.9  &  7.7 $\pm$ 1.9 &  171 $\pm$ 14     &  5.59$\pm$0.85 \\
\hline                                                                                                                     
Total background & 1170$\pm$120  &  1200$\pm$110  &  599$\pm$78    &  495  $\pm$ 56   &  600 $\pm$ 98  &  291 $\pm$ 19     &    85$\pm$18   \\
\hline                                                                                                                     
\multicolumn{8}{c}{$M_H = 125$~GeV/$c^2$}\\                                                                         
\hline                                             
$ggH$             & 6.9  $\pm$ 2.1   &  2.4   $\pm$ 0.7   &  2.8 $\pm$ 1.2    &  0.91$\pm$ 0.39 &  1.07$\pm$0.53 &  1.81$\pm$0.30  &   --       \\
{\it WH}          & 0.41 $\pm$ 0.07  &  0.16  $\pm$ 0.03  &  0.87 $\pm$ 0.14  &  0.30$\pm$ 0.05 &  1.59$\pm$0.22 &  0.10$\pm$0.02  &  1.25$\pm$0.17\\
{\it ZH}          & 0.25 $\pm$ 0.04  &  0.08  $\pm$ 0.01  &  0.27 $\pm$ 0.04  &  0.10$\pm$ 0.02 &  0.76$\pm$0.10 &  0.06$\pm$0.01  &  0.18$\pm$0.02\\
VBF         & 0.04 $\pm$ 0.01  &  0.013 $\pm$ 0.003 &  0.23 $\pm$ 0.04  &  0.07$\pm$ 0.01 &  0.55$\pm$0.09 &  0.05$\pm$0.01  &  --        \\
\hline                                                                                                                   
Total signal & 7.6 $\pm$ 2.1    &  2.6  $\pm$ 0.7    &   4.2 $\pm$ 1.2   &   1.4$\pm$ 0.4  &  3.98$\pm$0.71  &  2.02$\pm$0.30 &  1.43$\pm$0.17\\   
\hline         
\multicolumn{8}{c}{$M_H = 165$~GeV/$c^2$}\\                                                                         
\hline                                             
$ggH$              & 21.6 $\pm$ 6.4   &  7.3  $\pm$ 2.2  &  10.9 $\pm$ 4.6   &  3.5 $\pm$ 1.5  &   5.0$\pm$2.5  &  4.02$\pm$0.66  &    --       \\
{\it WH}           & 0.53 $\pm$ 0.09  &  0.19 $\pm$ 0.03 &  1.47 $\pm$ 0.23  &  0.47$\pm$ 0.08 &  4.35$\pm$0.61 &  0.14$\pm$0.02  &  2.69$\pm$0.36\\
{\it ZH}           & 0.55 $\pm$ 0.08  &  0.15 $\pm$ 0.02 &  0.57 $\pm$ 0.09  &  0.18$\pm$ 0.03 &  2.16$\pm$0.29 &  0.11$\pm$0.02  &  0.39$\pm$0.05\\
VBF          & 0.19 $\pm$ 0.04  &  0.06 $\pm$ 0.01 &  1.05 $\pm$ 0.18  &  0.30$\pm$ 0.05 &  2.51$\pm$0.41 &  0.15$\pm$0.03  &    --       \\
\hline                                                                                                                    
Total signal & 22.9 $\pm$ 6.5   &  7.7  $\pm$ 2.2  &  14.0 $\pm$ 4.7   &  4.4 $\pm$ 1.5  &  14.0$\pm$2.9  &  4.41$\pm$0.68  &  3.08$\pm$0.41\\
\hline                                                                                                                     
Data         & 1136             &  1402              &  545              &  488            &  596           &  319            &  87\\   
\bottomrule
\end{tabular}
%}
\end{center}
\end{ruledtabular}
\end{table*}

\section{Analysis outcomes}
\label{sec:analysis}

Higgs boson search results from the 13 search samples defined in 
Sec.~\ref{sec:evtsel} using the multivariate techniques described in 
Sec.~\ref{sec:Multivaraite} are presented here.  Kinematic event variables 
used as inputs to the multivariate algorithms are chosen to achieve the 
best possible separation of potential signal within each search sample 
from background contributions.  Relative contributions of different signal 
and background production processes vary significantly across samples.
Therefore, the multivariate outputs used to classify events within each 
search sample are based on unique sets of input variables, designed to 
take advantage of the distinct kinematic properties of potential signal 
and background events within each sample.  Each multivariate output is 
trained to distinguish potential signal from backgrounds based on the 
modeling described in Sec.~\ref{sec:BGmodeling}.

\subsection{Dilepton search samples}
\label{sec:DileptonEvents}

The numbers of expected events from each contributing signal and background 
process are compared in Table~\ref{tbl:YieldsDilepton} with the total number 
of observed events in each of the seven dilepton search samples formed from 
electron and muon candidates.  Background and signal predictions, which 
are shown for potential Higgs boson masses of 125 and 165~GeV/$c^2$, are 
taken from the models described in Sec.~\ref{sec:BGmodeling}.

A summary of the kinematic variables used as inputs to the multivariate algorithms 
for separating potential signal from background contributions in these seven search 
samples is shown in Table~\ref{tbl:Dilepton Variables}.  Important input variables 
for the diboson search samples include the charged-lepton transverse momenta, the 
angular separation of the lepton trajectories, and angles between the lepton and 
jet momenta in the events. The scalar sums of transverse momenta, including or 
excluding $\Met$, are also considered.  

\begin{table*}[t]
  \setlength{\extrarowheight}{3pt}
\begin{ruledtabular} 
\begin{center} 
\caption{\label{tbl:Dilepton Variables}
Summary of kinematic variables used as inputs to the multivariate algorithms for 
separating signal and background contributions in the dilepton search samples.}
{\scriptsize
%\begin{tabular}{|p{2.0cm}|p{7.5cm}|p{0.8cm}|p{0.8cm}|p{0.8cm}|p{0.8cm}|p{0.8cm}|p{0.8cm}|}         
\begin{tabular}{llcccccc}         
\toprule                                                                                                                     
Variable                           & Definition            & OS  & OS  & OS   & OS             & SS   & OS       \\
%Name                              &                       & 0   & 1   & $\ge$2   & Inverse        & $\ge$1   & Hadronic \\
%                                  &                       & Jet & Jet & Jets & $M_{\ell\ell}$ & Jets & Tau      \\
                                   &                       & 0 Jet   & 1 Jet   & $\ge$2 Jets   & Inverse $M_{\ell\ell}$ & $\ge$1 Jets & Hadronic Tau \\
\hline                                                                 
$E(\ell_1)$                        & Energy of the leading lepton                                                           &     &  \checkmark  &     &  \checkmark  &     &     \\     
$E(\ell_2)$                        & Energy of the subleading lepton                                                       &     &     &     &  \checkmark  &     &     \\     
$p_{T}(\ell_1)$                    & Transverse momentum of the leading lepton                                              &  \checkmark  &  \checkmark  &  \checkmark  &  \checkmark  &  \checkmark  &  \checkmark  \\       
$p_{T}(\ell_2)$                    & Transverse momentum of the subleading lepton                                          &  \checkmark  &  \checkmark  &  \checkmark  &  \checkmark  &  \checkmark  &  \checkmark  \\    
$\Delta\phi(\ell\ell)$             & Azimuthal angle between the leptons                                                    &  \checkmark  &     &  \checkmark  &  \checkmark  &     &  \checkmark  \\ 
$\Delta\eta(\ell\ell)$             & Difference in pseudorapidities of the leptons                                          &     &     &     &     &     &  \checkmark  \\ 
$\Delta R(\ell\ell)$               & $((\Delta\eta(\ell\ell))^2 + (\Delta\phi(\ell\ell))^2)^{1/2}$                          &  \checkmark  &  \checkmark  &  \checkmark  &  \checkmark  &     &  \checkmark  \\       
$M(\ell\ell)$                      & Invariant mass of dilepton pair                                                        &  \checkmark  &  \checkmark  &  \checkmark  &     &     &  \checkmark  \\      
$E_{T}(j_1)$                       & Transverse energy of the leading jet                                                   &     &     &  \checkmark  &     &  \checkmark  &     \\    
$E_{T}(j_2)$                       & Transverse energy of the subleading jet                                               &     &     &  \checkmark  &     &     &     \\    
$\eta(j_1)$                        & Pseudorapidity of the leading jet                                                      &     &     &  \checkmark  &     &     &     \\    
$\eta(j_2)$                        & Pseudorapidity of the subleading jet                                                  &     &     &  \checkmark  &     &     &     \\    
$\Delta\phi(jj)$                   & Azimuthal angle between two leading jets                                               &     &     &  \checkmark  &     &     &     \\    
$\Delta\eta(jj)$                   & Difference in pseudorapidities of two leading jets                                     &     &     &  \checkmark  &     &     &     \\    
$\Delta R(jj)$                     & $((\Delta\eta(jj))^2 + (\Delta\phi(jj))^2)^{1/2}$                                      &     &     &  \checkmark  &     &     &     \\      
$M(jj)$                            & Invariant mass of two leading jets                                                     &     &     &  \checkmark  &     &     &     \\    
$N_{\rm{jets}}$                    & Number of jets in event                                                                &     &     &     &     &  \checkmark  &     \\    
$\Sigma E_{T}$(jets)               & Scalar sum of transverse jet energies                                                  &     &     &  \checkmark  &     &  \checkmark  &     \\    
$\Sigma E_{T}$($\ell$,jets)        & Scalar sum of lepton $p_T$ and jet (if any) $E_T$                                      &     &     &  \checkmark  &  \checkmark  &     &     \\
$|\Sigma \vec{E}_T|$               & Magnitude of vector sum of lepton $p_T$ and jet (if any) $E_T$                         &     &     &     &  \checkmark  &     &     \\
$\Met$                             & Missing transverse energy                                                              &     &     &     &     &  \checkmark  &  \checkmark  \\    
$\Sigma E_{T}$($\ell$,$\Met$)      & Scalar sum of transverse lepton momenta and the \Met                                   &     &     &  \checkmark  &  \checkmark  &     &     \\     
$\Delta\phi$($\Met$,$\ell$)        & Azimuthal angle between $e$ or $\mu$ candidate and \Met                                &     &     &     &     &     &  \checkmark  \\
$\Delta\phi$($\Met$,$\tau$)        & Azimuthal angle between $\tau$ candidate and \Met                                      &     &     &     &     &     &  \checkmark  \\
$\Delta\phi$($\ell\ell$,$\Met$)    & Azimuthal angle between $\vec{p}_T(\ell_1) + \vec{p}_T(\ell_2)$ and the \Met           &     &  \checkmark  &  \checkmark  &     &     &     \\    
$\Delta\phi$($\Met$,$\ell$ or jet) & Azimuthal angle between the \Met and nearest lepton or jet                             &     &  \checkmark  &     &  \checkmark  &  \checkmark  &     \\
$\MetSpec$                         & Projection of \Met on nearest lepton or jet                                            &  \checkmark  &  \checkmark  &  \checkmark  &  \checkmark  &  \checkmark  &     \\
                                   & or \Met if $\Delta\phi$($\Met$,$\ell$ or jet) $>\pi/2$                                 &     &     &     &     &     &     \\
\MetSig                            & $\Met$/($\Sigma E_{T}$($\ell$,jets))$^{1/2}$                                           &     &  \checkmark  &  \checkmark  &  \checkmark  &  \checkmark  &  \checkmark  \\    
$M_T$($\ell$,$\Met$)               & Transverse mass of $e$ or $\mu$ candidate and \Met                                     &     &     &     &     &     &  \checkmark  \\    
$M_T$($\tau$,$\Met$)               & Transverse mass of $\tau$ candidate and \Met                                           &     &     &     &     &     &  \checkmark  \\    
$M_T$($\ell$,$\ell$,$\Met$)        & Transverse mass of the two leptons and the \Met                                        &  \checkmark  &  \checkmark  &  \checkmark  &     &     &     \\    
$M_T$($\ell$,$\ell$,$\Met$,jets)   & Transverse mass of the two leptons, all jets, and the \Met                             &     &     &  \checkmark  &     &     &     \\   
$H_T$                              & Scalar sum of lepton $p_T$, jet $E_T$, and the \Met                                    &  \checkmark  &     &  \checkmark  &  \checkmark  &     &     \\
$C$                                & Centrality based on leptons, jets and the \Met                                         &     &  \checkmark  &     &     &     &     \\    
$A$                                & Aplanarity based on leptons, jets and the \Met                                         &     &     &  \checkmark  &     &     &     \\    
$LR(HWW)$                          & ME-based likelihood for $ggH$ Higgs boson production                                   &  \checkmark  &     &     &     &     &     \\
$LR(WW)$                           & ME-based likelihood for nonresonant  $W^{+}W^{-}$ production                           &  \checkmark  &     &     &     &     &     \\
cos($\Delta\phi$($\ell\ell$))$_{\rm{CM}}$ & Cosine of the azimuthal angle between the leptons in the                             &     &  \checkmark  &  \checkmark  &     &     &     \\
                                     & Higgs boson rest frame                                                               &     &     &     &     &     &     \\    
cos($\psi$($\ell_2$))$_{\rm{CM}}$         & Cosine of angle between subleading lepton and Higgs boson                                 &     &     &  \checkmark  &     &     &     \\
                                     & in Higgs boson rest frame                                                            &     &     &     &     &     &     \\    
\bottomrule                                                                                                                     
\end{tabular}
}
\end{center}
\end{ruledtabular}
\end{table*}

\begin{figure*}[t]
\begin{center}
\subfigure{
\includegraphics[width=0.45\linewidth]{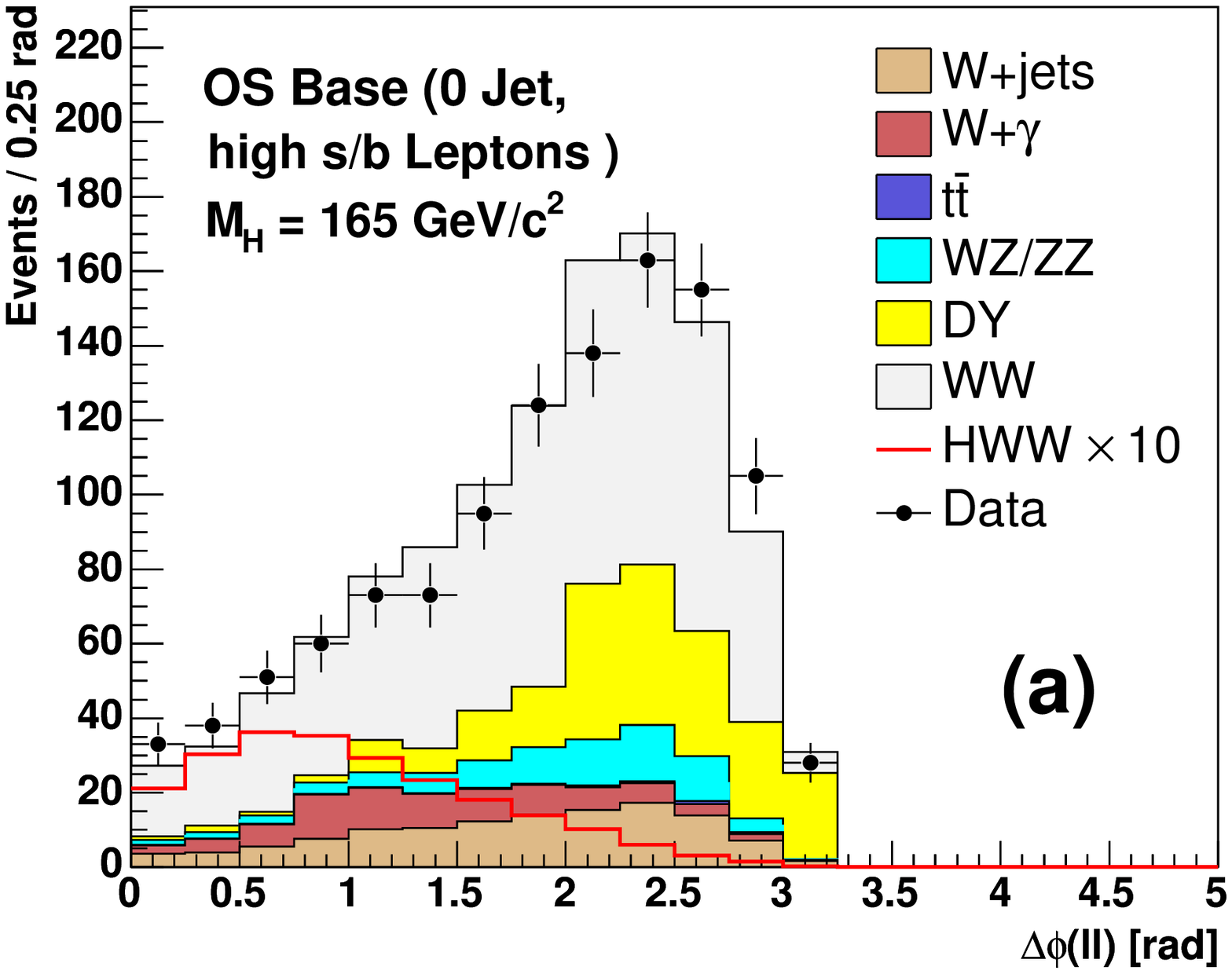}
}
\subfigure{
\includegraphics[width=0.45\linewidth]{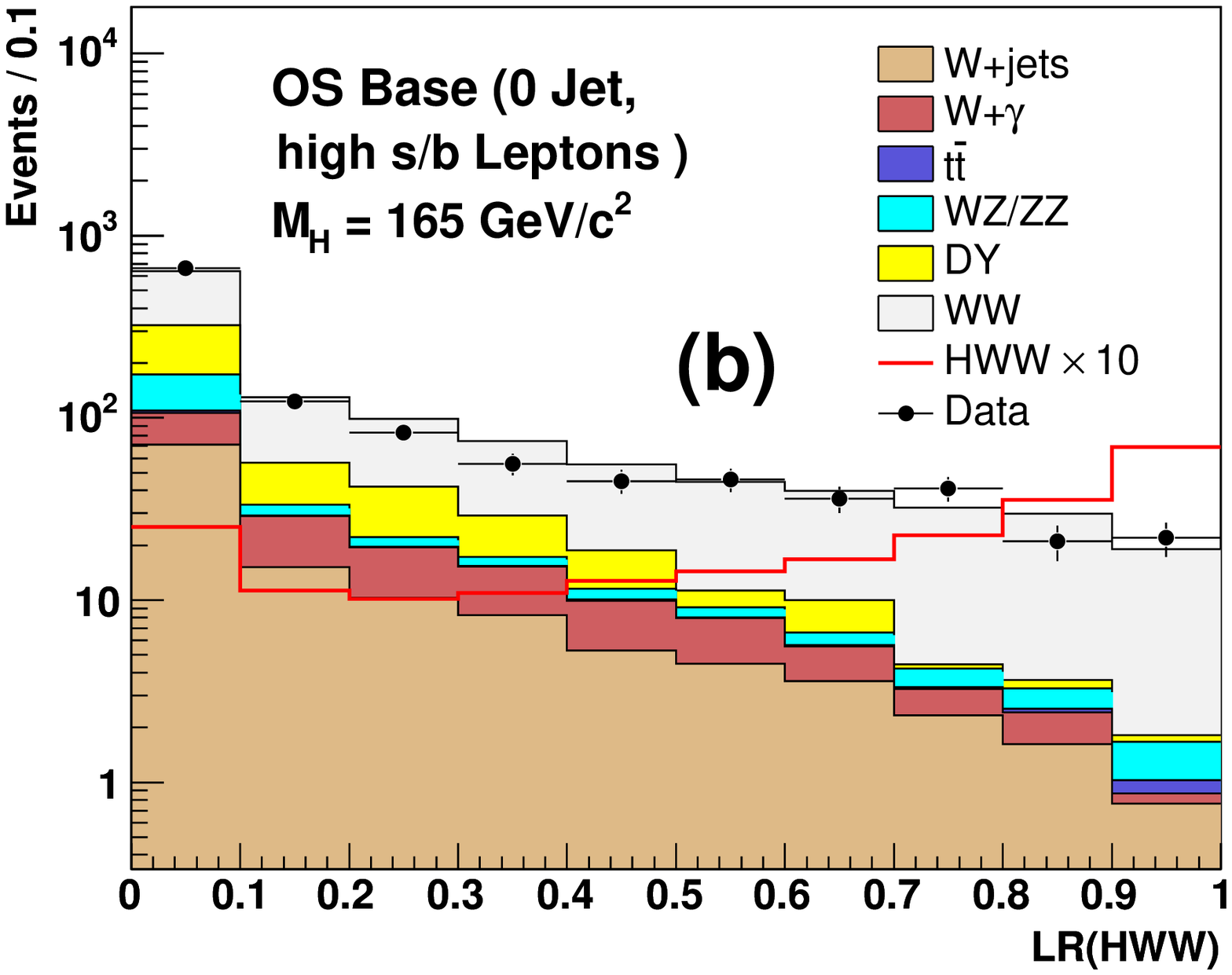}
}
\subfigure{
\includegraphics[width=0.45\linewidth]{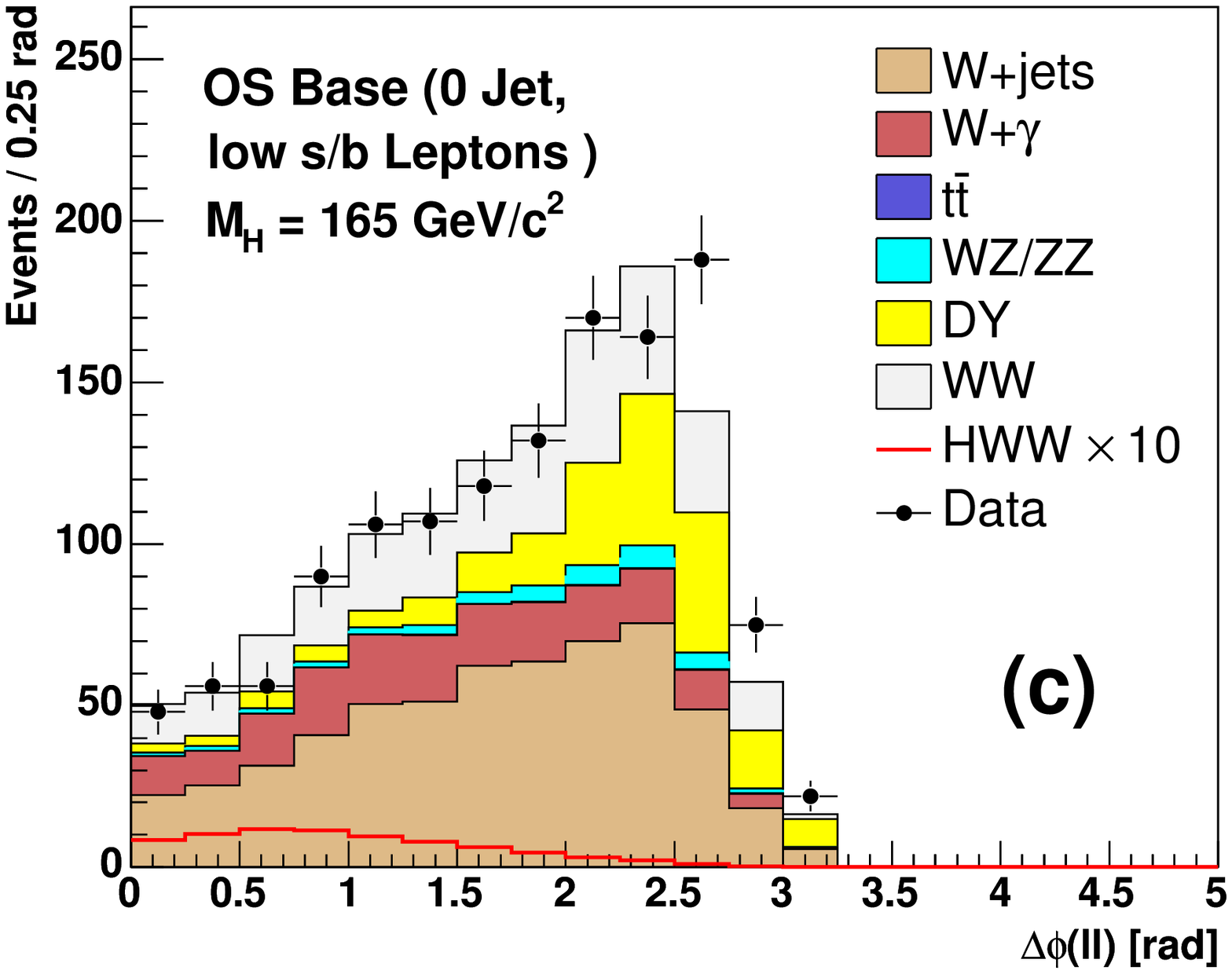}
}
\subfigure{
\includegraphics[width=0.45\linewidth]{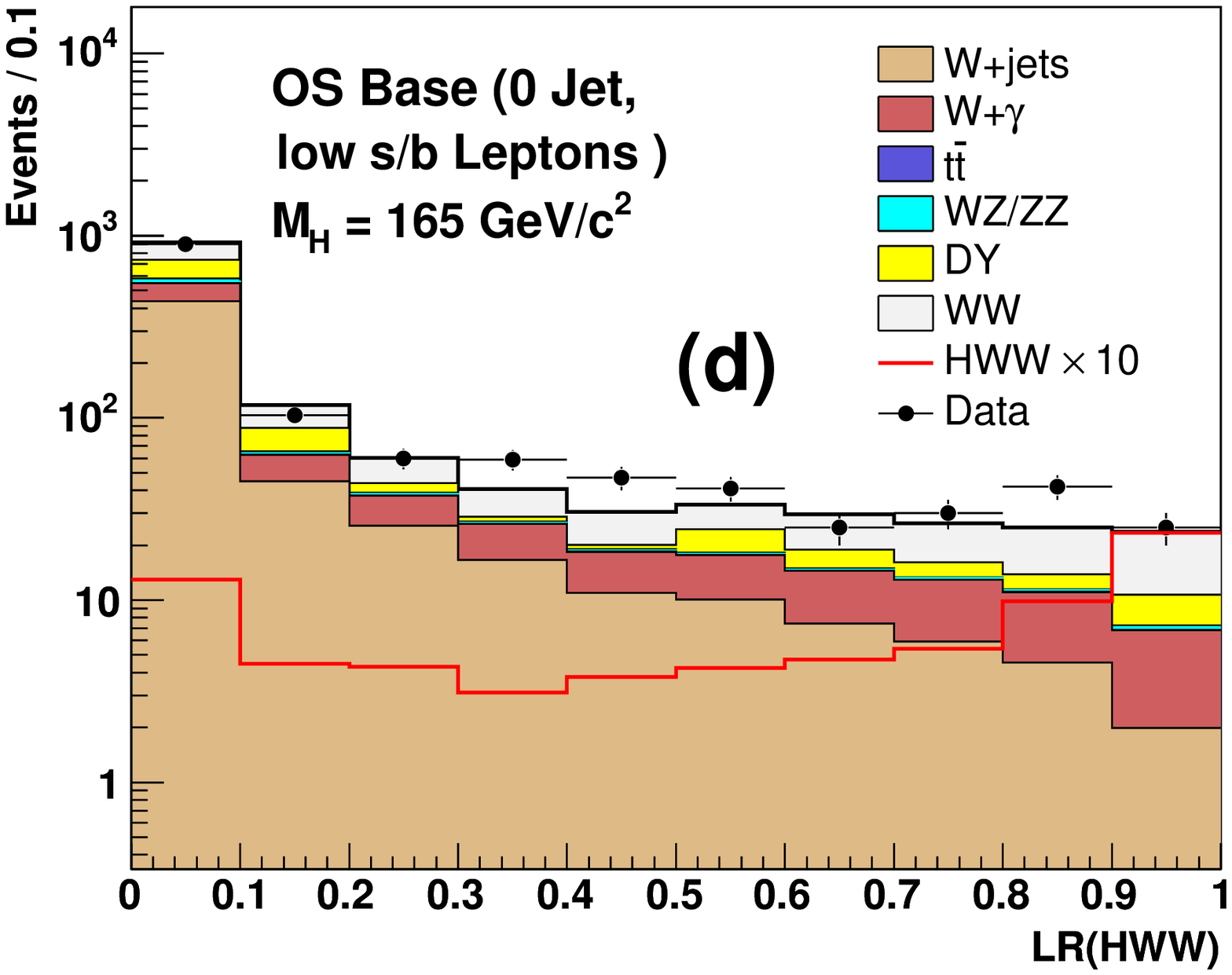}
}
\end{center}
\caption{Predicted and observed distributions of kinematic input variables providing 
the largest separation between potential signal and background contributions in 
the (a,b) OS Base (0 Jet, high $s/b$ Leptons) and (c,d) OS Base (0 Jet, low $s/b$ 
Leptons) search samples.  The overlaid signal predictions correspond to the sum of 
four production modes ($ggH$, {\it WH}, {\it ZH}, and VBF) for a Higgs boson with 
mass of 165~GeV/$c^2$ and are multiplied by a factor of 10 for visibility.
Normalizations for background event yields are those obtained from the final fit 
used to extract search limits.}
\label{fig::NNInputsDilA1}
\end{figure*}

\begin{figure*}[t]
\begin{center}
\subfigure{
\includegraphics[width=0.45\textwidth]{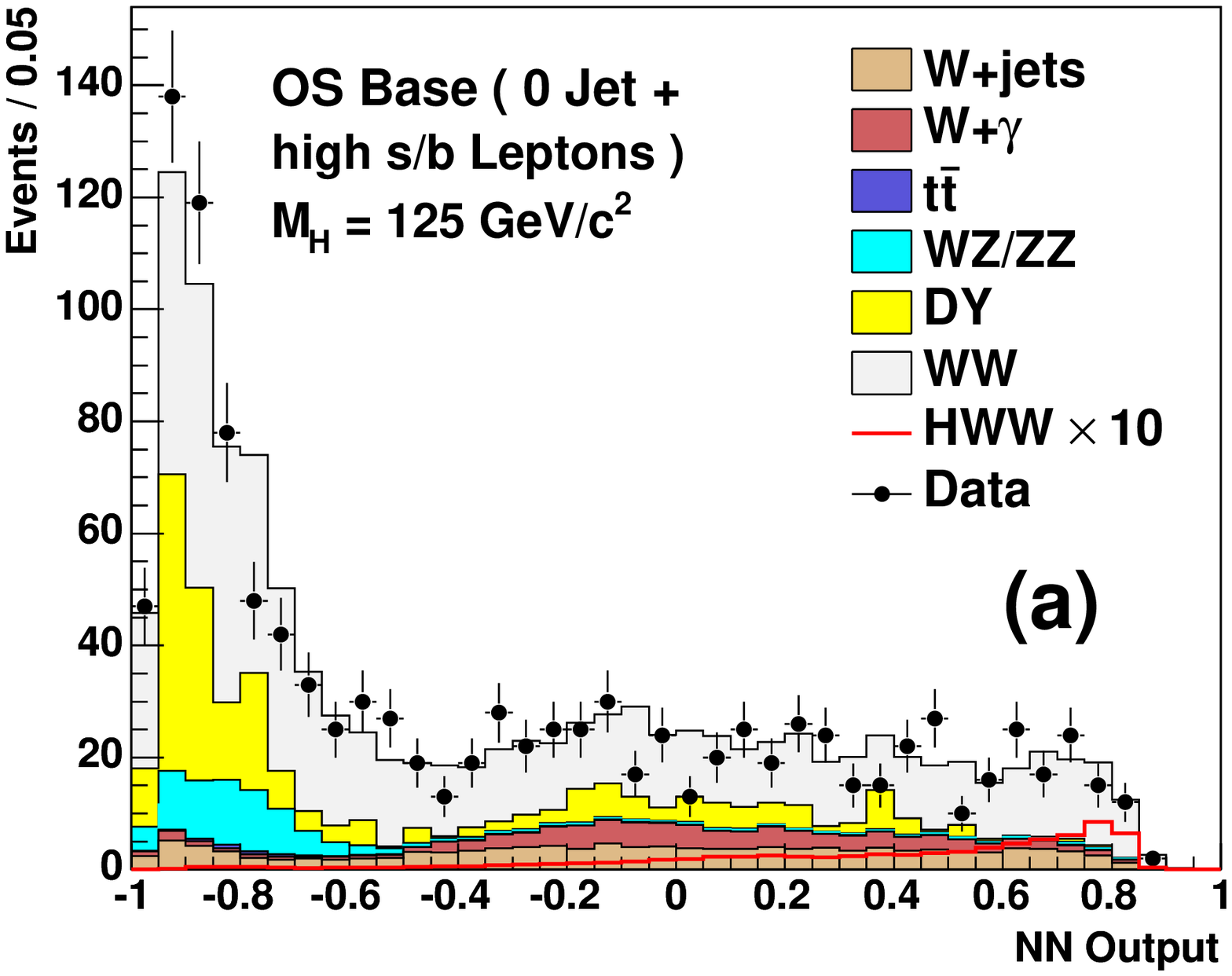}
}
\subfigure{
\includegraphics[width=0.45\textwidth]{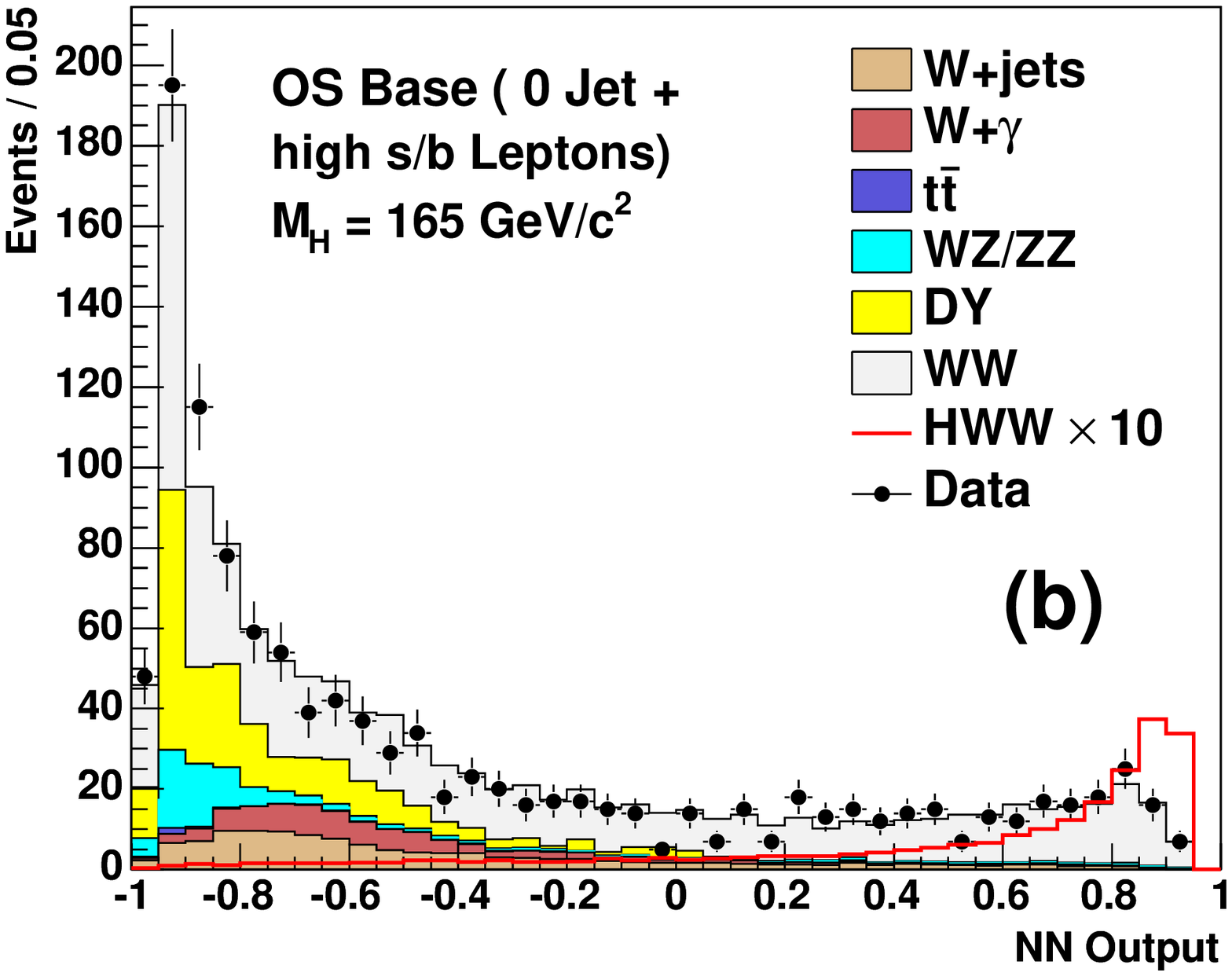}
}
\subfigure{
\includegraphics[width=0.45\textwidth]{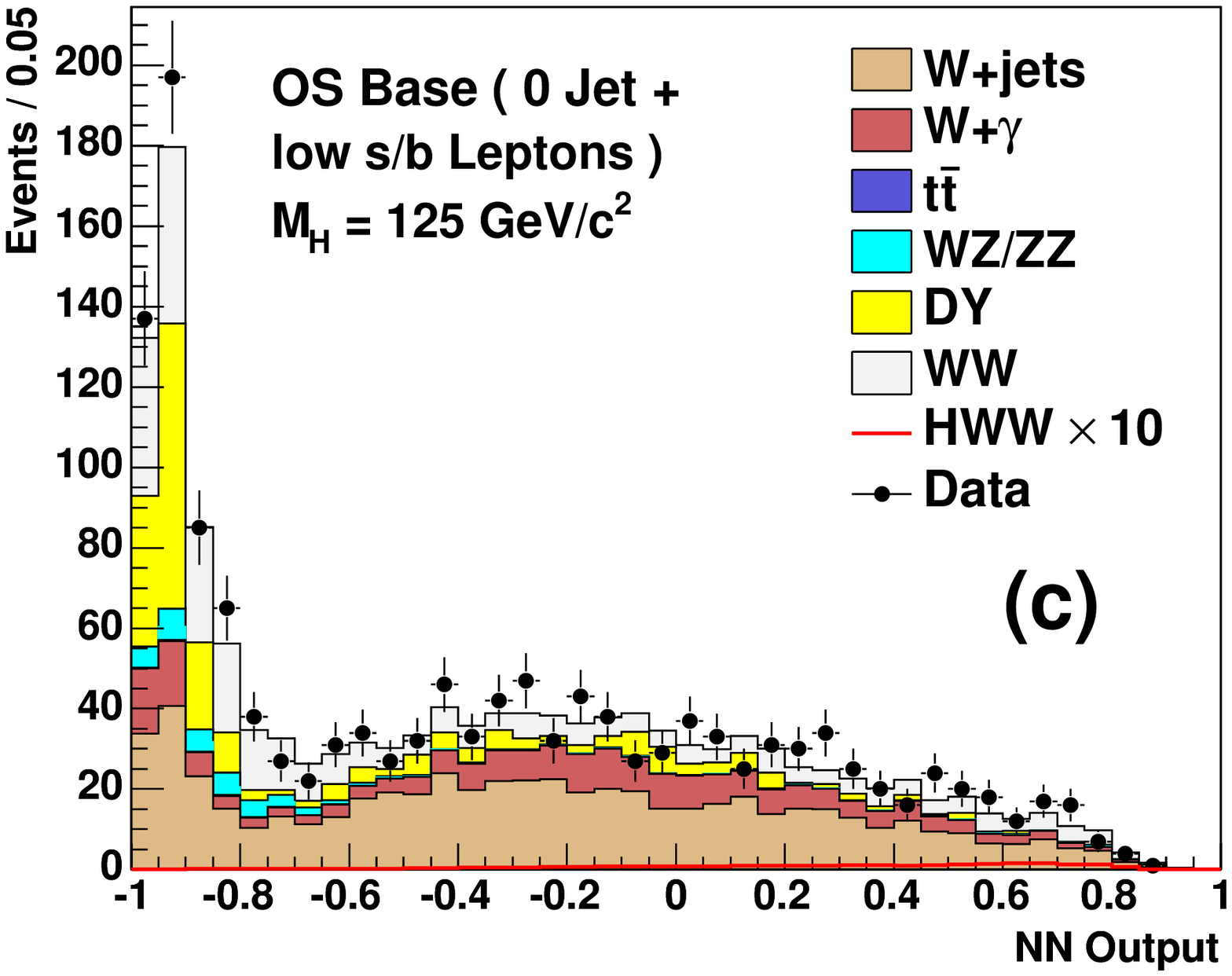}
}
\subfigure{
\includegraphics[width=0.45\textwidth]{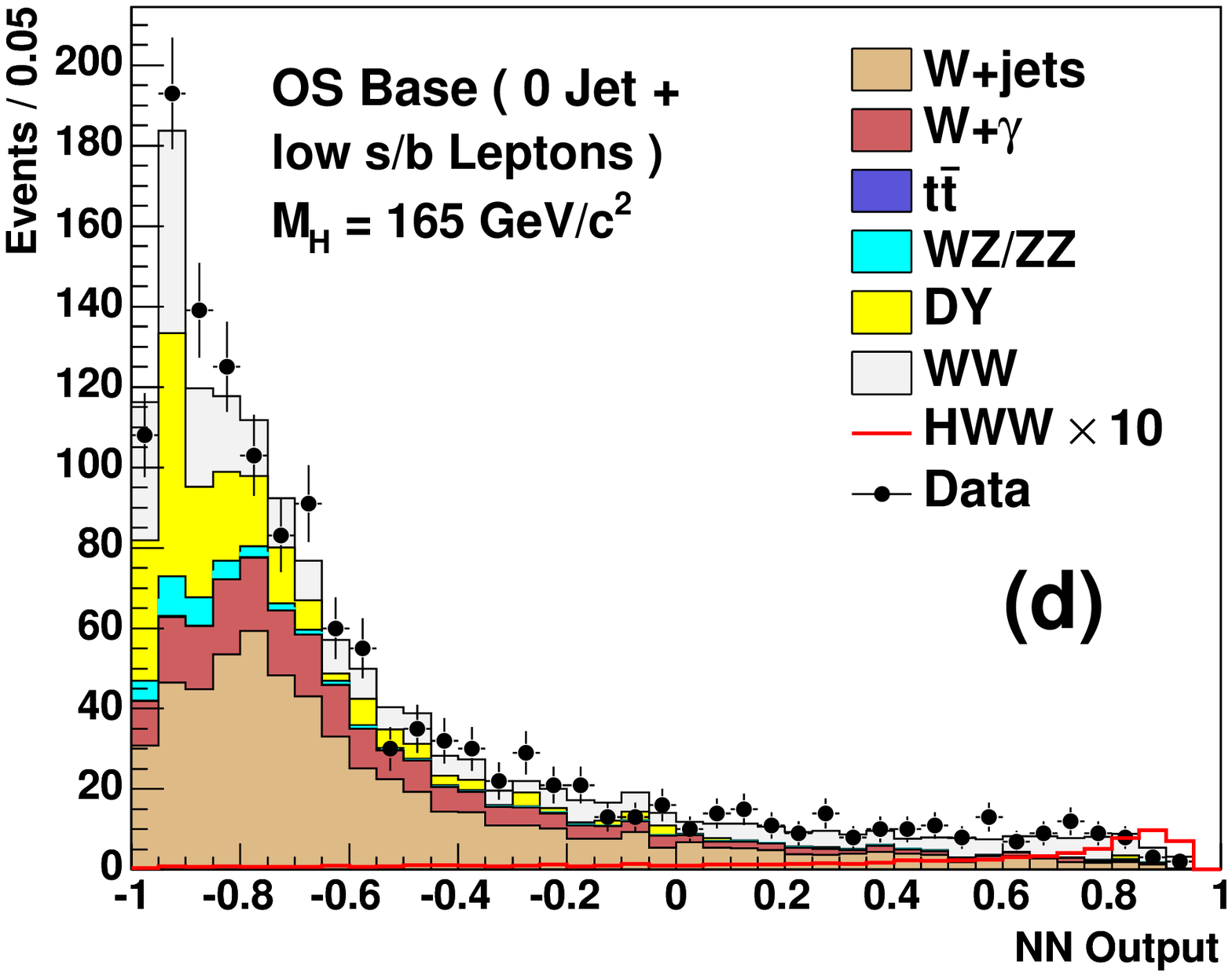}
}
\end{center}
\caption{Predicted and observed distributions of neural network output variables 
for networks trained to separate potential Higgs boson events from background 
contributions in the (a,b) OS Base (0 Jet, high $s/b$ Leptons) and (c,d) OS Base 
(0 Jet, low $s/b$ Leptons) search samples for Higgs boson mass hypotheses of 
125~and~165~GeV/$c^2$.  The overlaid signal predictions correspond to the sum of 
four production modes ($ggH$, {\it WH}, {\it ZH}, and VBF) and are multiplied by 
a factor of 10 for visibility.  Normalizations for background event yields are 
those obtained from the final fit used to extract search limits.}
\label{fig:TemplatesDilA1}
\end{figure*}

\begin{figure*}[t]
\begin{center}
\subfigure{
\includegraphics[width=0.45\linewidth]{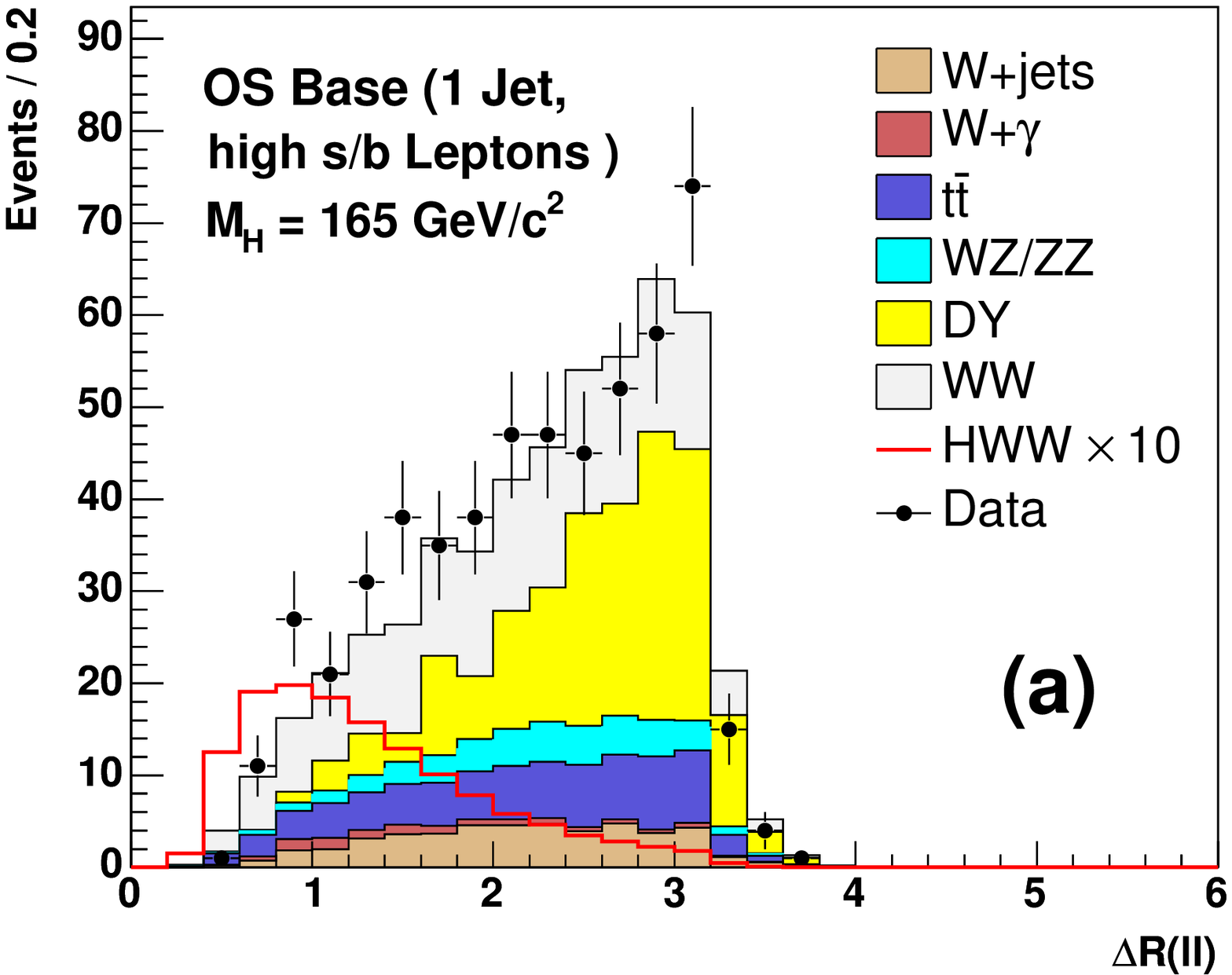}
}
\subfigure{
\includegraphics[width=0.45\linewidth]{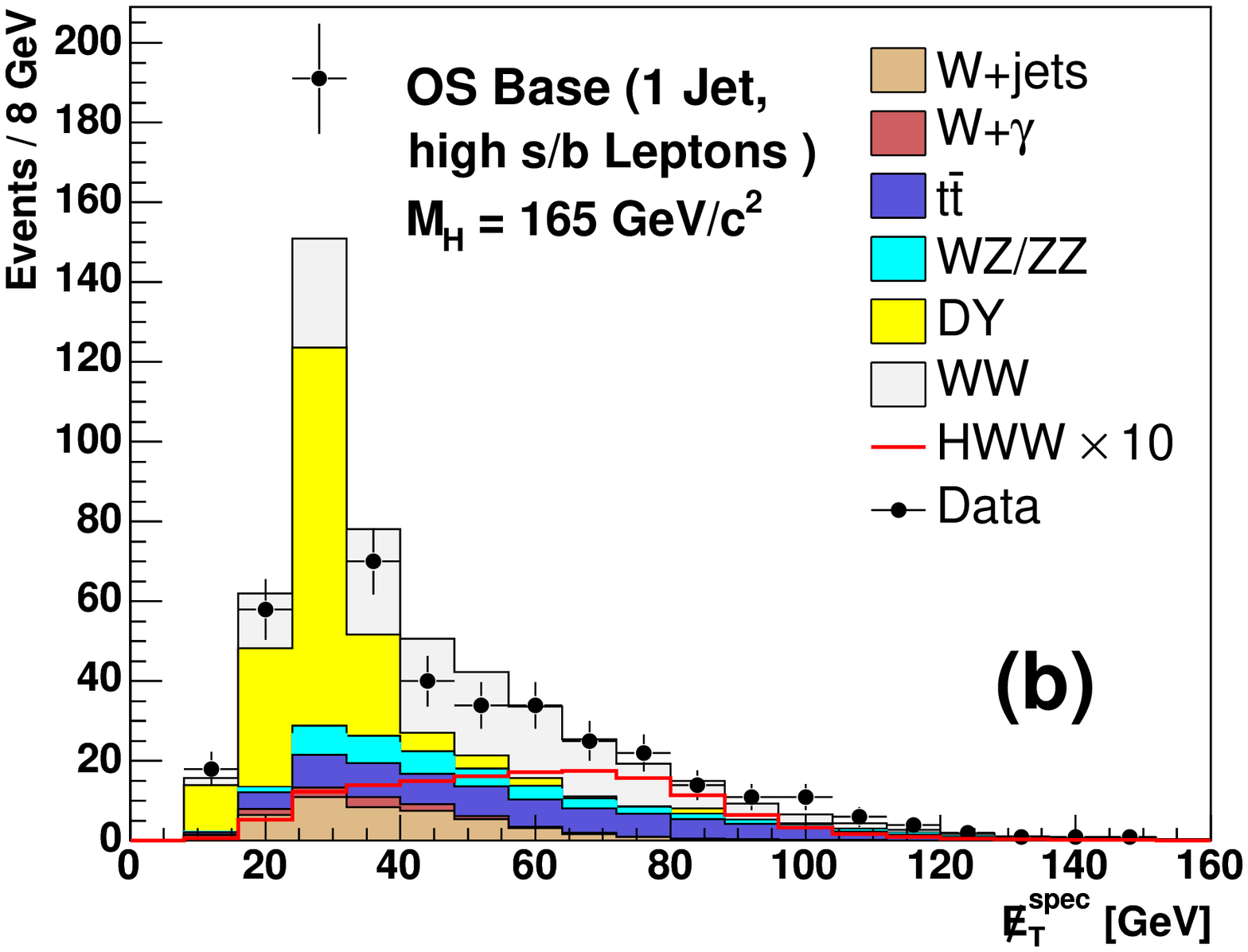}
}
\subfigure{
\includegraphics[width=0.45\linewidth]{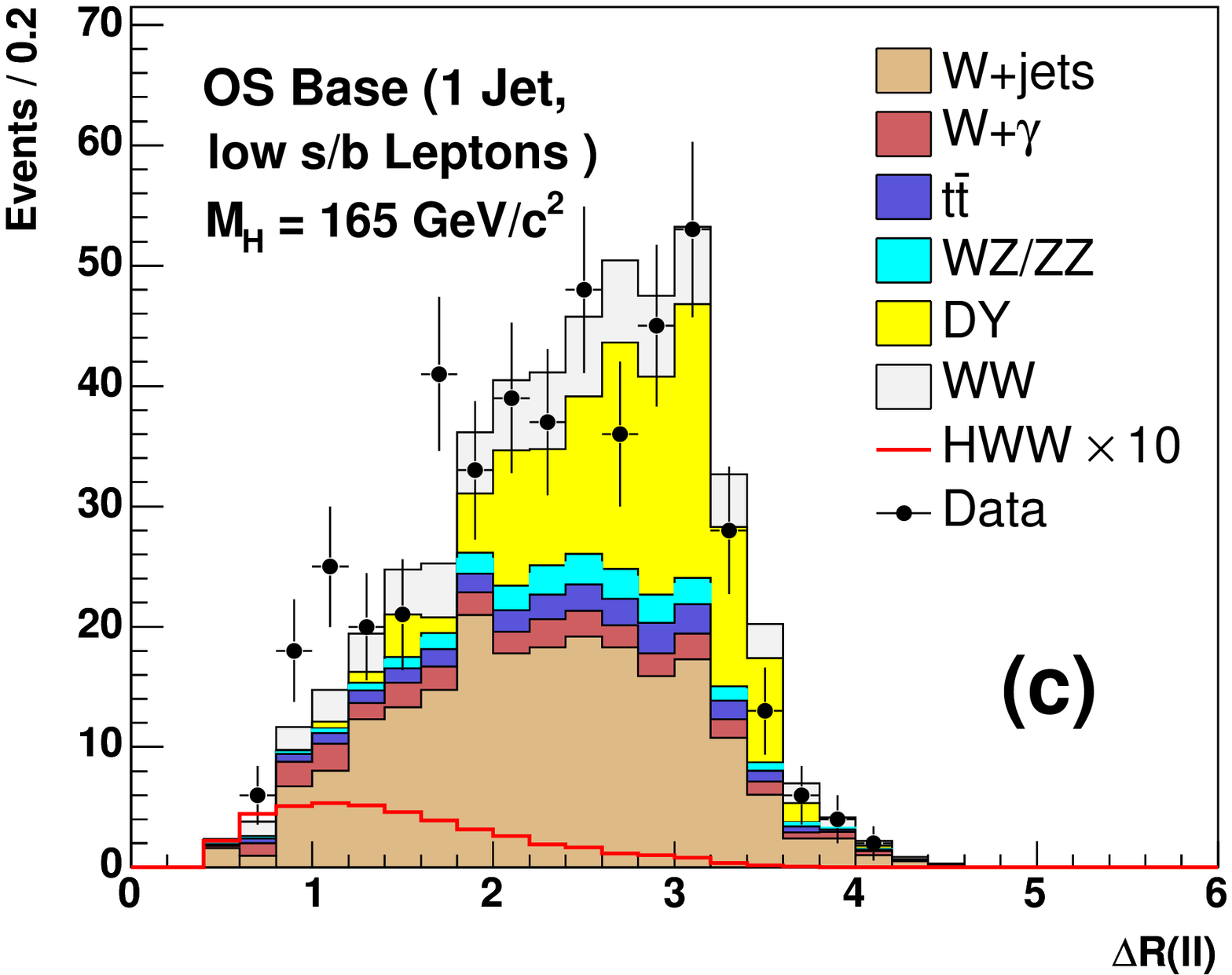}
}
\subfigure{
\includegraphics[width=0.45\linewidth]{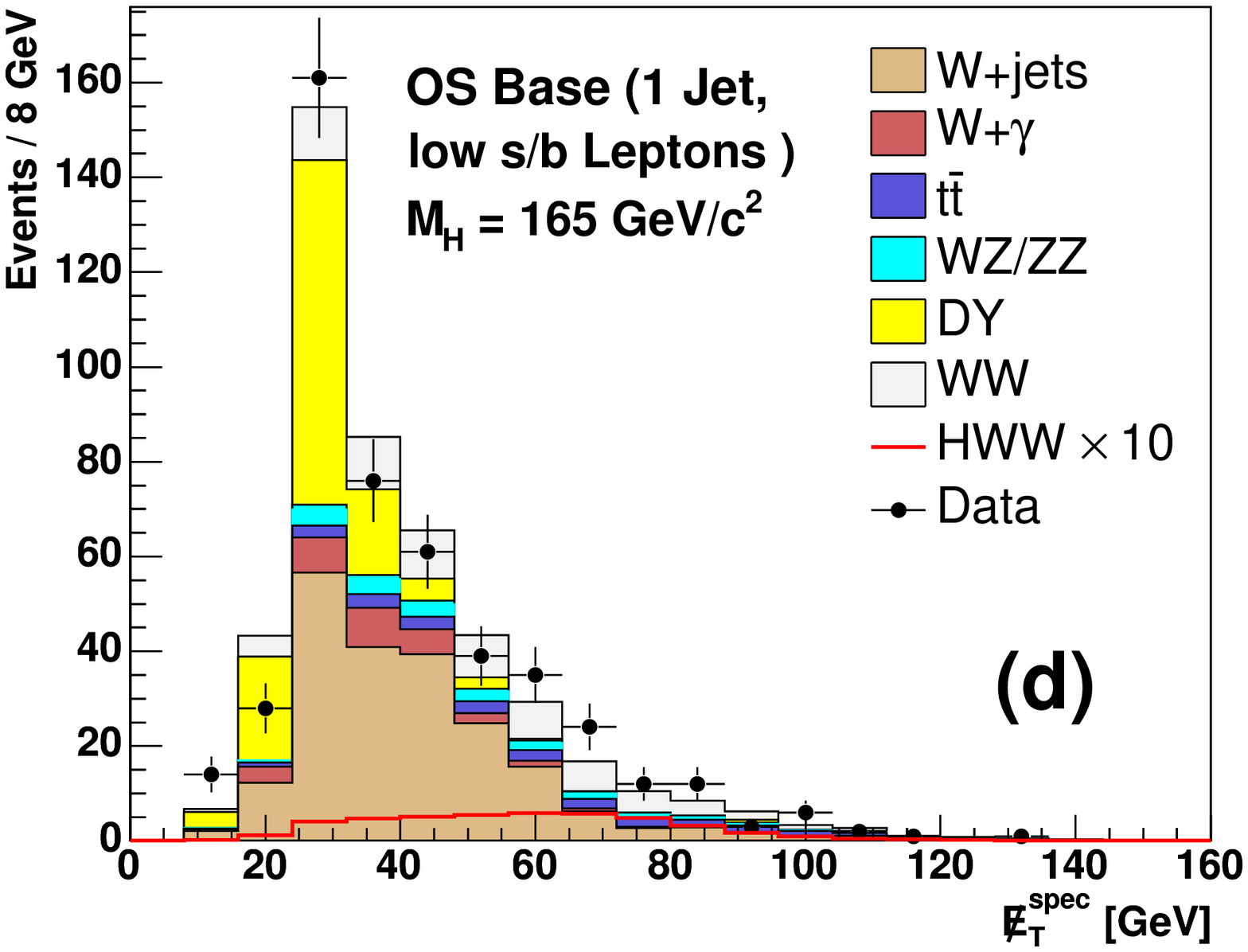}
}
\end{center}
\caption{Predicted and observed distributions of kinematic input variables providing 
the largest separation between potential signal and background contributions in 
the (a,b) OS Base (1 Jet, high $s/b$ Leptons) and (c,d) OS Base (1 Jet, low $s/b$ 
Leptons) search samples.  The overlaid signal predictions correspond to the sum of 
four production modes ($ggH$, {\it WH}, {\it ZH}, and VBF) for a Higgs boson with 
mass of 165~GeV/$c^2$ and are multiplied by a factor of 10 for visibility.
Normalizations for background event yields are those obtained from the final fit 
used to extract search limits.}
\label{fig::NNInputsDilA2}
\end{figure*}

\begin{figure*}[t]
\begin{center}
\subfigure{
\includegraphics[width=0.45\textwidth]{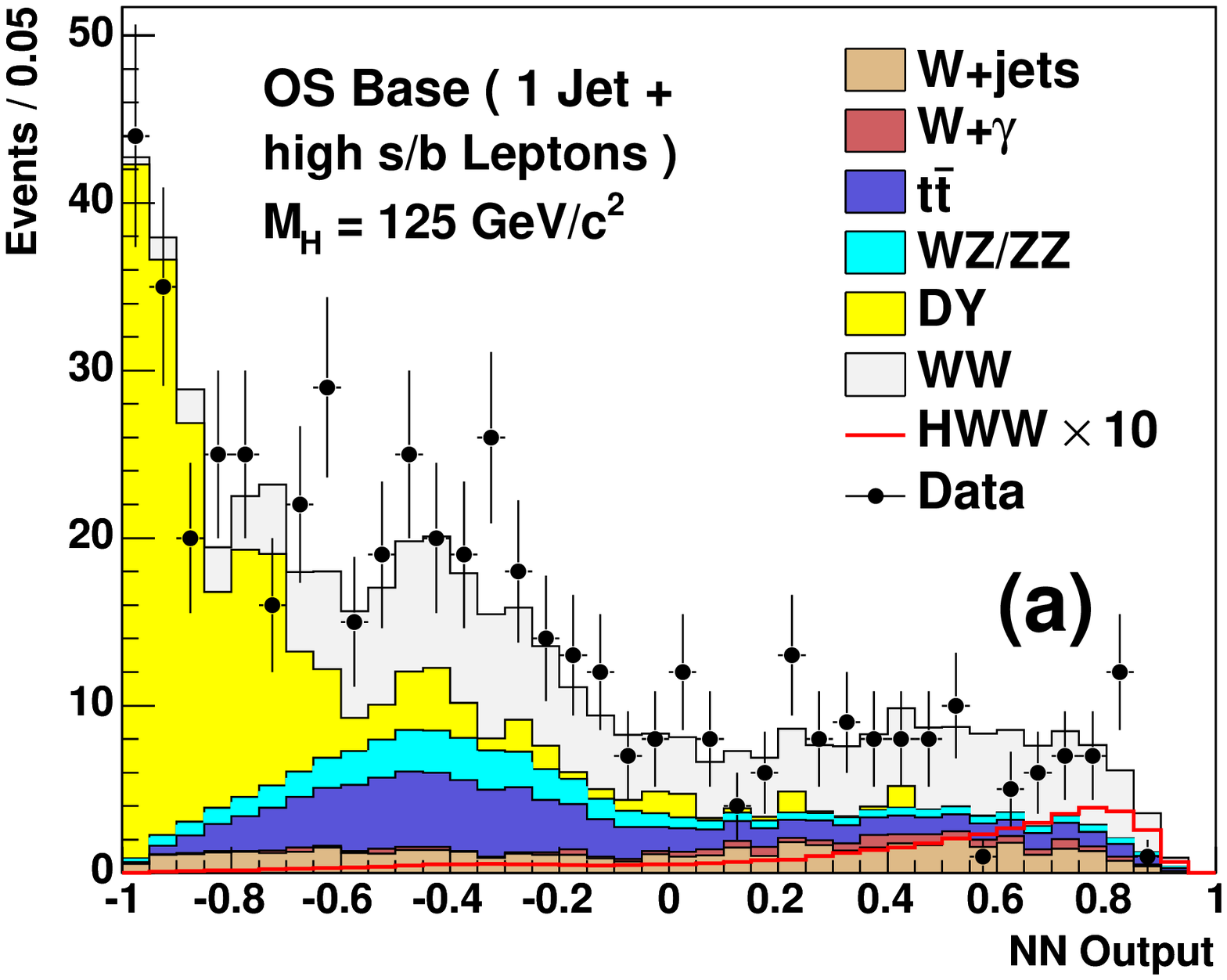}
}
\subfigure{
\includegraphics[width=0.45\textwidth]{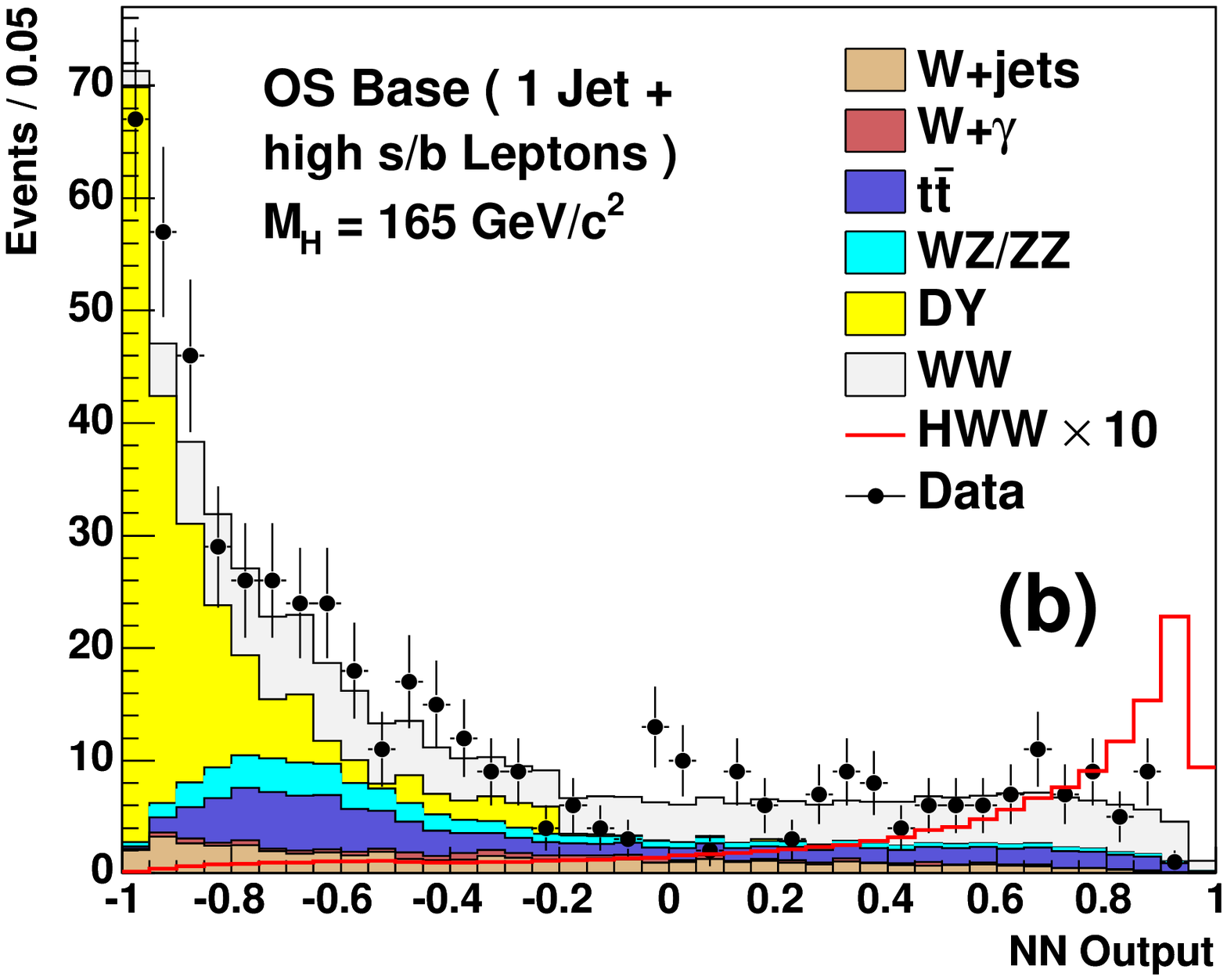}
}
\subfigure{
\includegraphics[width=0.45\textwidth]{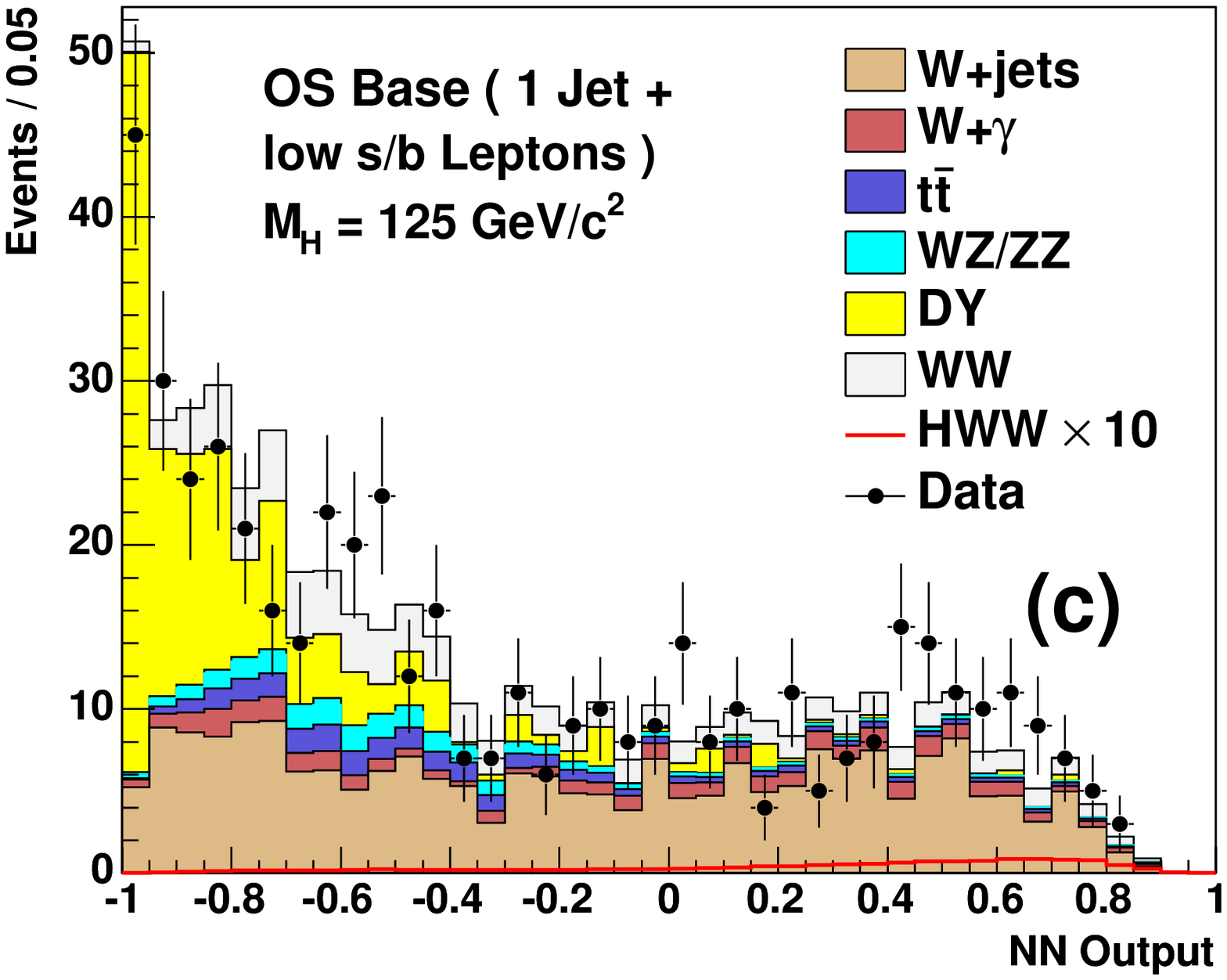}
}
\subfigure{
\includegraphics[width=0.45\textwidth]{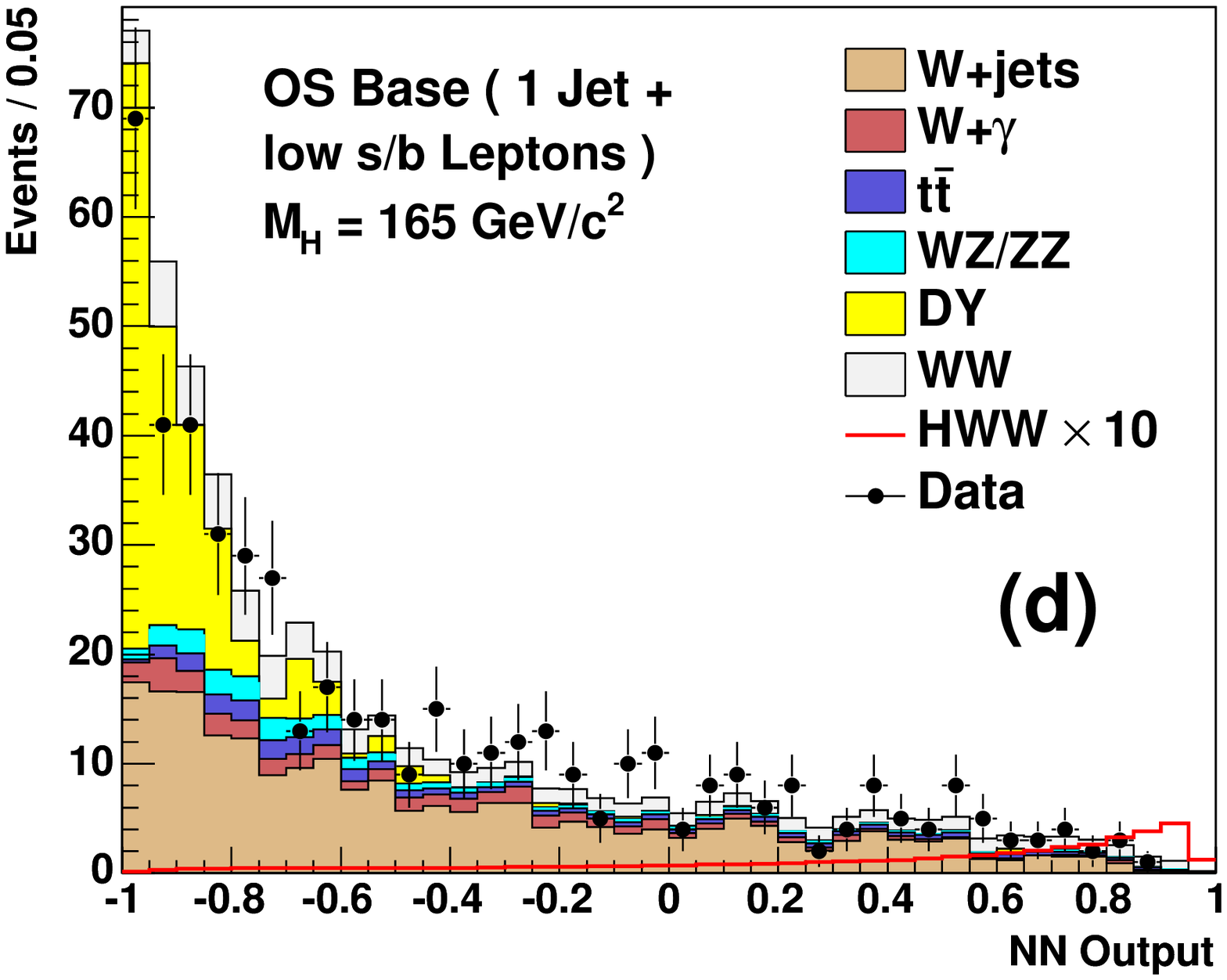}
}
\end{center}
\caption{Predicted and observed distributions of neural network output variables 
for networks trained to separate potential Higgs boson events from background 
contributions in the (a,b) OS Base (1 Jet, high $s/b$ Leptons) and (c,d) OS Base 
(1 Jet, low $s/b$ Leptons) search samples for Higgs boson mass hypotheses of 
125~and~165~GeV/$c^2$.  The overlaid signal predictions correspond to the sum of 
four production modes ($ggH$, {\it WH}, {\it ZH}, and VBF) and are multiplied by 
a factor of 10 for visibility.  Normalizations for background event yields are 
those obtained from the final fit used to extract search limits.}
\label{fig:TemplatesDilA2}
\end{figure*}

The OS 0 Jet search samples have the best individual sensitivity to a potential 
Higgs boson signal.  The dominant Higgs boson production process contributing 
to these samples is $ggH$, but small ($\approx$~5\%) contributions from other 
production mechanisms are considered.  The primary background contribution (over 
40\%) to these samples is from direct $W^+ W^-$ production and neural networks are 
trained specifically to distinguish this background from potential $ggH$-produced 
Higgs boson events.  In this case, the neural network input variables include 
matrix-element likelihood ratios, $LR(HWW)$ and $LR(WW)$, along with the following 
eight kinematic event variables: $\Delta\phi(\ell\ell)$, $\Delta R(\ell\ell)$, 
$M(\ell\ell)$, $p_{T}(\ell_1)$, $p_{T}(\ell_2)$, $H_T$, $M_T$($\ell$,$\ell$,$\Met$), 
and $\MetSpec$.  Distributions of the most discriminating among these variables,
$\Delta\phi(\ell\ell)$ and $LR(HWW)$, are shown in Figs.~\ref{fig::NNInputsDilA1}(a) 
and~\ref{fig::NNInputsDilA1}(b) for the High $s/b$ Leptons sample and in 
Figs.~\ref{fig::NNInputsDilA1}(c) and~\ref{fig::NNInputsDilA1}(d) for the Low $s/b$ 
Leptons sample.  These variables are sensitive to the spin correlations between 
the two $W$ bosons produced in the decay of the spin-0 Higgs boson, which tend to 
result in events with collinear leptons.  Separate neural networks are trained 
for each tested Higgs boson mass using combined samples of modeled signal and 
background events containing both high and low $s/b$ lepton candidates.  These 
networks are then applied independently to both the high and low $s/b$ Leptons 
search samples.  Examples of neural network output distributions for Higgs boson 
masses of 125 and 165~GeV/$c^2$ are shown in Figs.~\ref{fig:TemplatesDilA1}(a) 
and~\ref{fig:TemplatesDilA1}(b) for the high $s/b$ Leptons sample and in  
Figs.~\ref{fig:TemplatesDilA1}(c) and~\ref{fig:TemplatesDilA1}(d) for the 
low $s/b$ Leptons sample.  These distributions illustrate the ability of the 
neural network to efficiently separate potential signal events from background 
contributions with the exception of direct $W^+ W^-$ production, which is 
indistinguishable from signal in a portion of phase space.

For the OS Base 1 Jet search samples, the {\it VH} and VBF Higgs boson production 
mechanisms contribute more significantly, accounting for $\approx$~25\% of 
the potential signal.  Background contributions from DY events, which 
contain significant missing energy due to jet energy mismeasurements, are also 
relevant.  Neural networks for these search samples are based on the following 
12 kinematic input variables: $\Delta R(\ell\ell)$, $M(\ell\ell)$, 
$p_{T}(\ell_1)$, $p_{T}(\ell_2)$, $M_T$($\ell$,$\ell$,$\Met$), $\MetSpec$, 
$E(\ell_1)$, $\Delta\phi$($\Met$,$\ell$ or jet), $\Delta\phi$($\ell\ell$,$\Met$), 
cos($\Delta\phi$($\ell\ell$))$_{{\rm CM}}$, \MetSig, and $C$.  Distributions of the
most discriminating among these variables, $\Delta R(\ell\ell)$ and $\MetSpec$, 
are shown in Figs.~\ref{fig::NNInputsDilA2}(a) and~\ref{fig::NNInputsDilA2}(b) 
for the high $s/b$ Leptons sample and in Figs.~\ref{fig::NNInputsDilA2}(c) 
and~\ref{fig::NNInputsDilA2}(d) for the low $s/b$ Leptons sample.  The $\Delta 
R(\ell\ell)$ variable provides good discrimination against significant 
$W^+ W^-$ contributions, while the $\MetSpec$ variable is useful for separating 
the signal from larger DY contributions.  Training of the neural networks 
is based on combined samples containing events with both high and low $s/b$ 
lepton candidates.  The resulting networks are then applied separately to 
the two search samples containing the events with high and low $s/b$ leptons.   
Examples of neural network output distributions for Higgs boson masses 
of 125 and 165~GeV/$c^2$ are shown in Figs.~\ref{fig:TemplatesDilA2}(a) 
and~\ref{fig:TemplatesDilA2}(b) for the high $s/b$ Leptons sample and in  
Figs.~\ref{fig:TemplatesDilA2}(c) and~\ref{fig:TemplatesDilA2}(d) for the 
low $s/b$ Leptons sample.  

\begin{figure*}[t]
\begin{center}
\subfigure{
\includegraphics[width=0.45\linewidth]{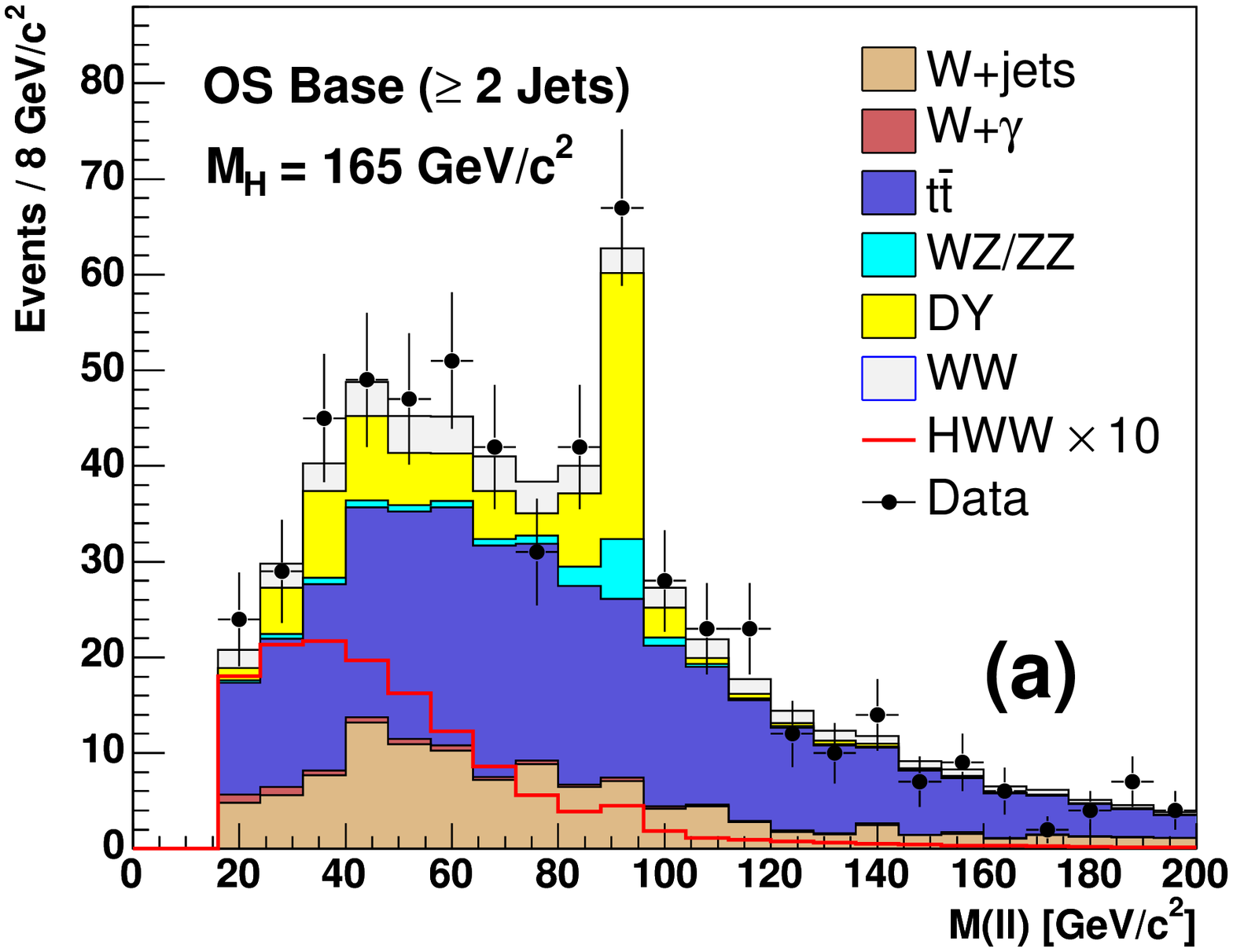}
}
\subfigure{
\includegraphics[width=0.45\linewidth]{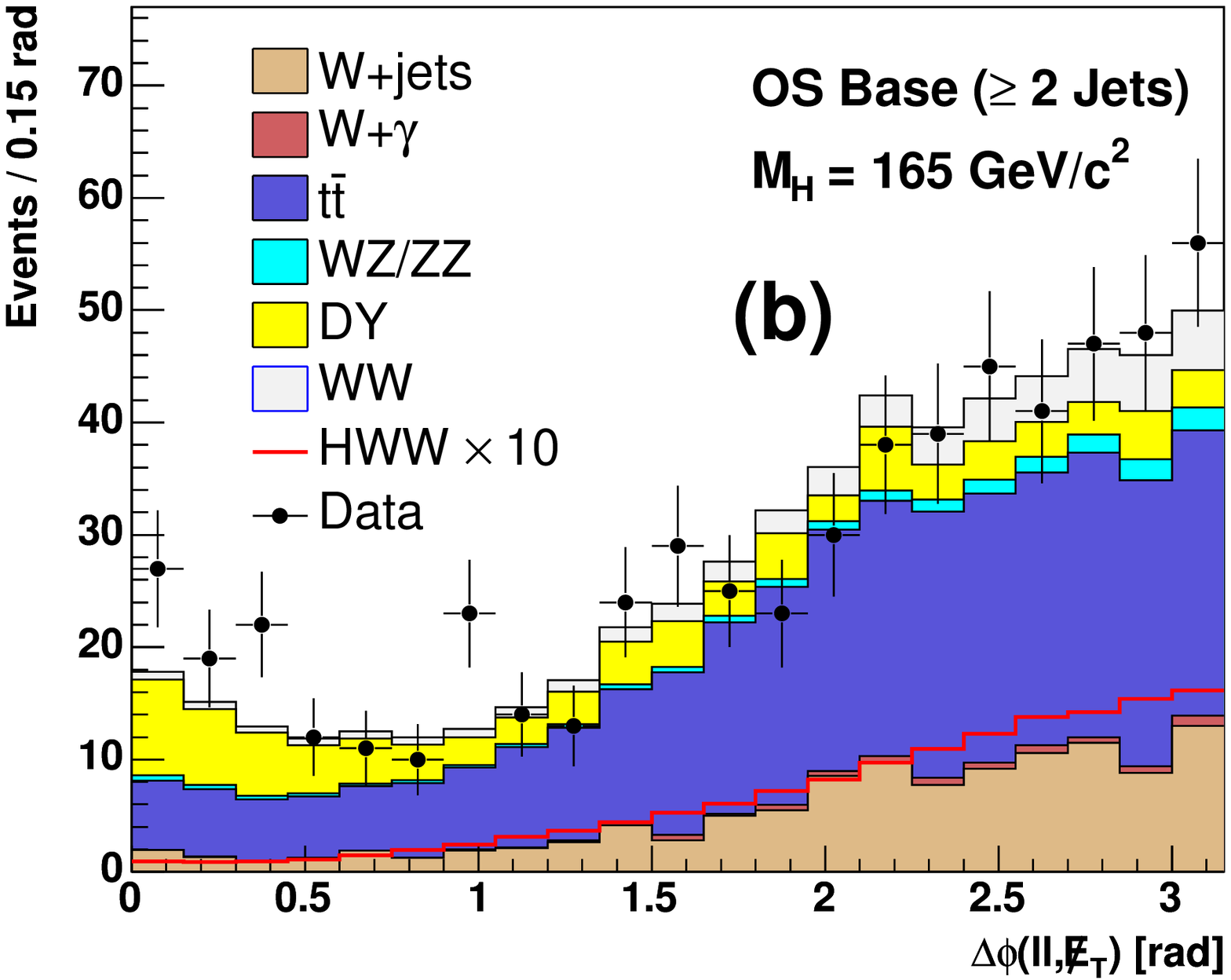}
}
\subfigure{
\includegraphics[width=0.45\linewidth]{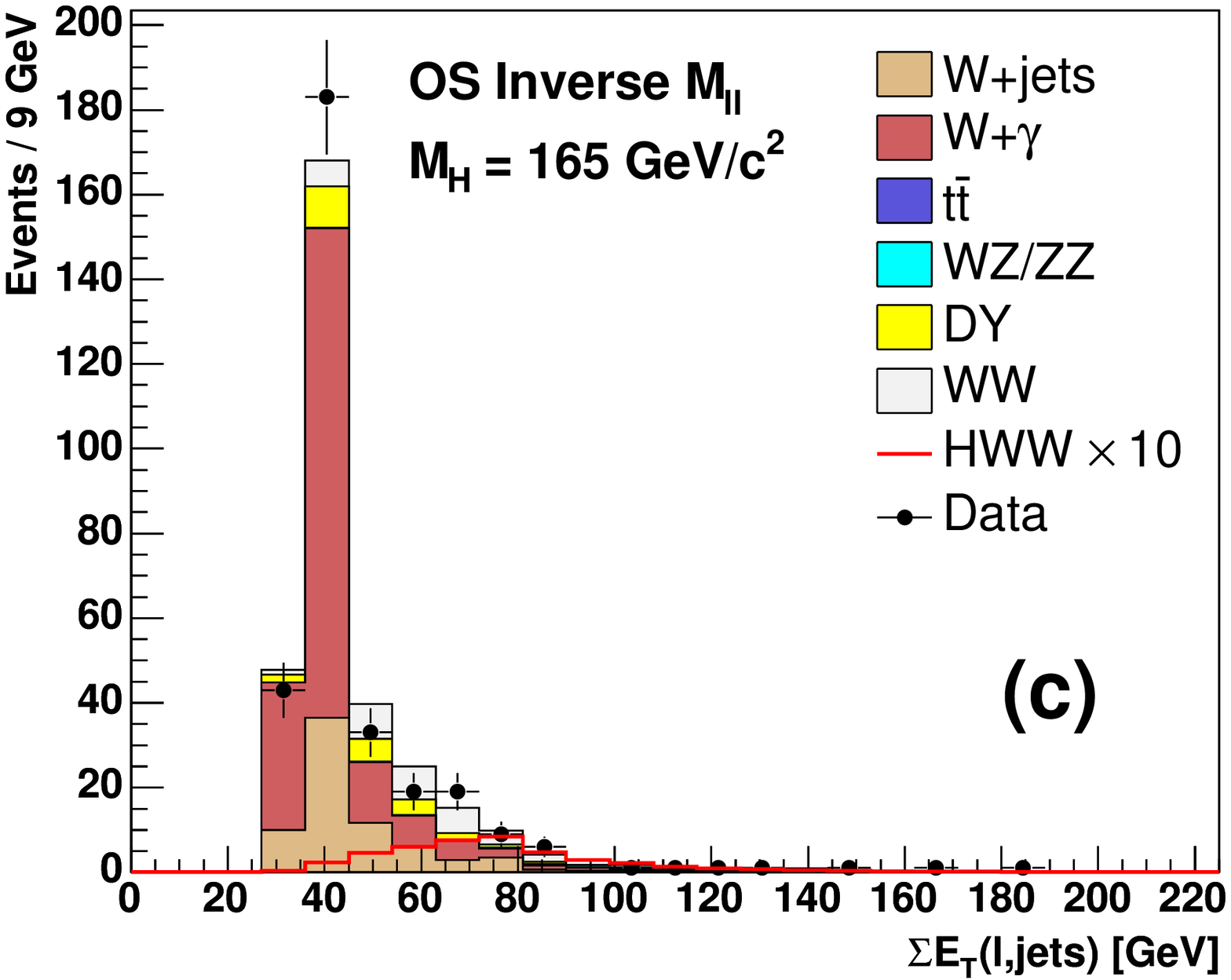}
}
\subfigure{
\includegraphics[width=0.45\linewidth]{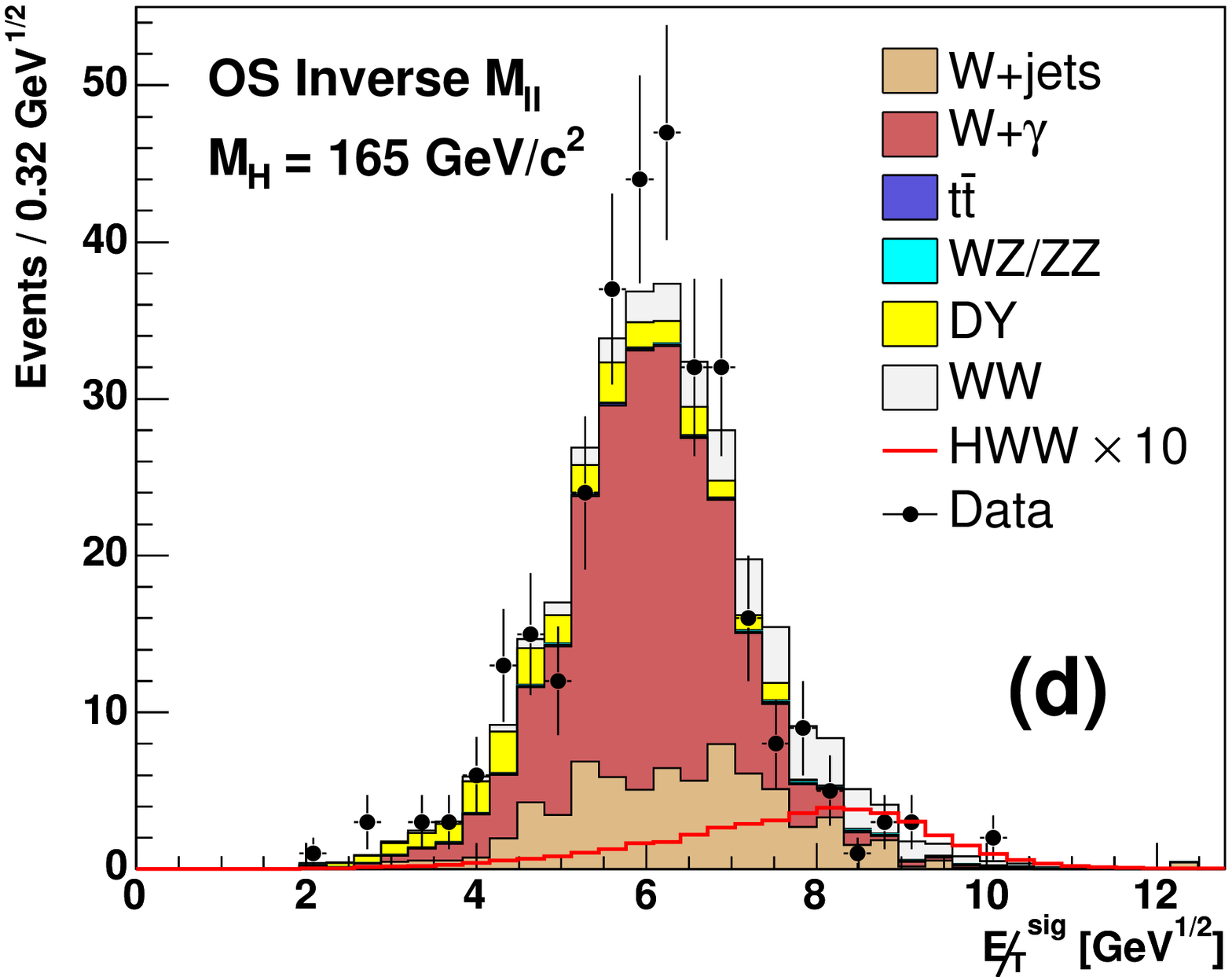}
}
\end{center}
\caption{Predicted and observed distributions of kinematic input variables providing 
the largest separation between potential signal and background contributions in the 
(a,b) OS Base ($\ge$2 Jets) and (c,d) OS Inverse $M_{\ell\ell}$ search samples.  The 
overlaid signal predictions correspond to the sum of four production modes ($ggH$, 
{\it WH}, {\it ZH}, and VBF) for a Higgs boson with mass of 165~GeV/$c^2$ and are 
multiplied by a factor of 10 for visibility.  Normalizations for background event 
yields are those obtained from the final fit used to extract search limits.}
\label{fig::NNInputsDilA3}
\end{figure*}

\begin{figure*}[t]
\begin{center}
\subfigure{
\includegraphics[width=0.45\textwidth]{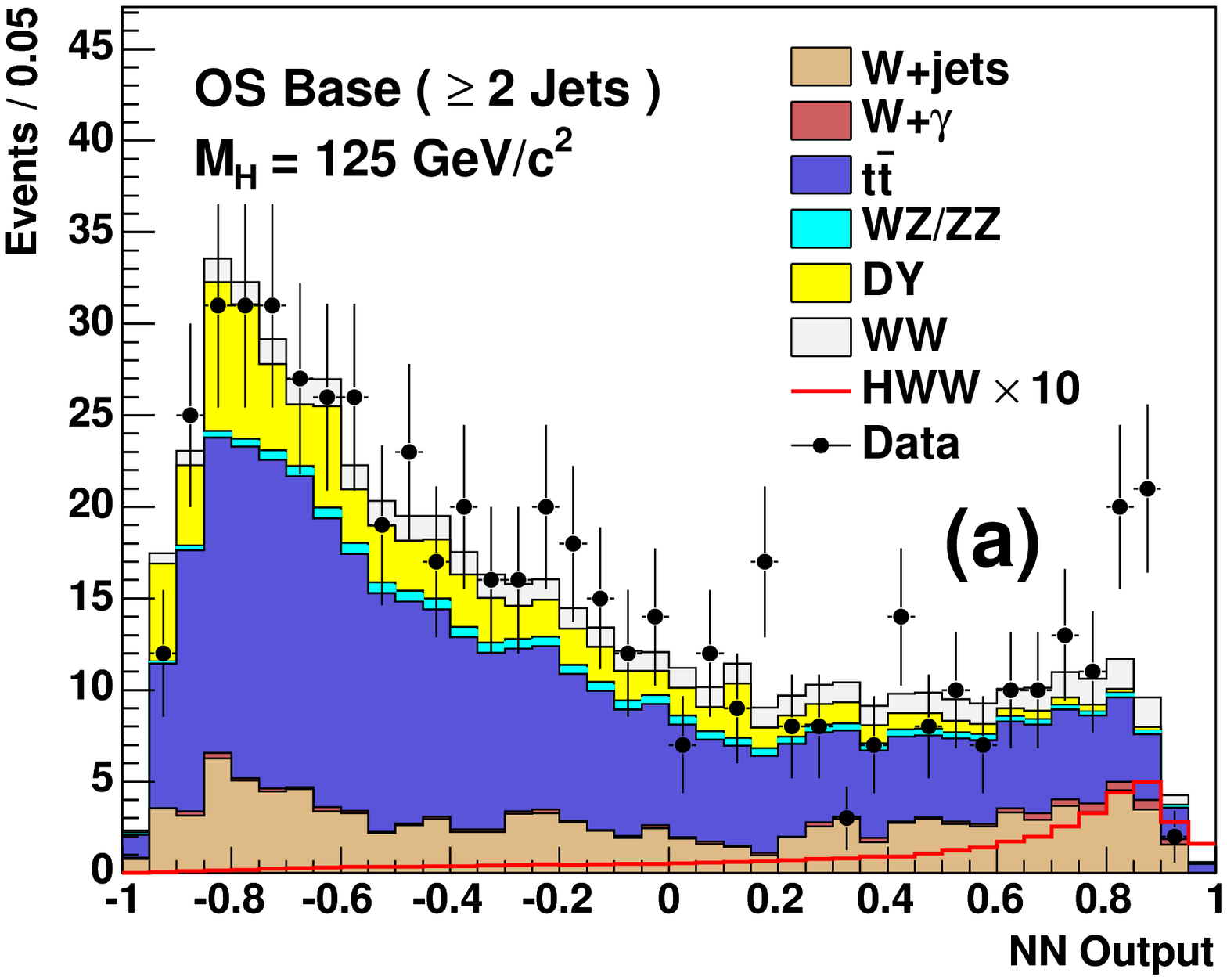}
}
\subfigure{
\includegraphics[width=0.45\textwidth]{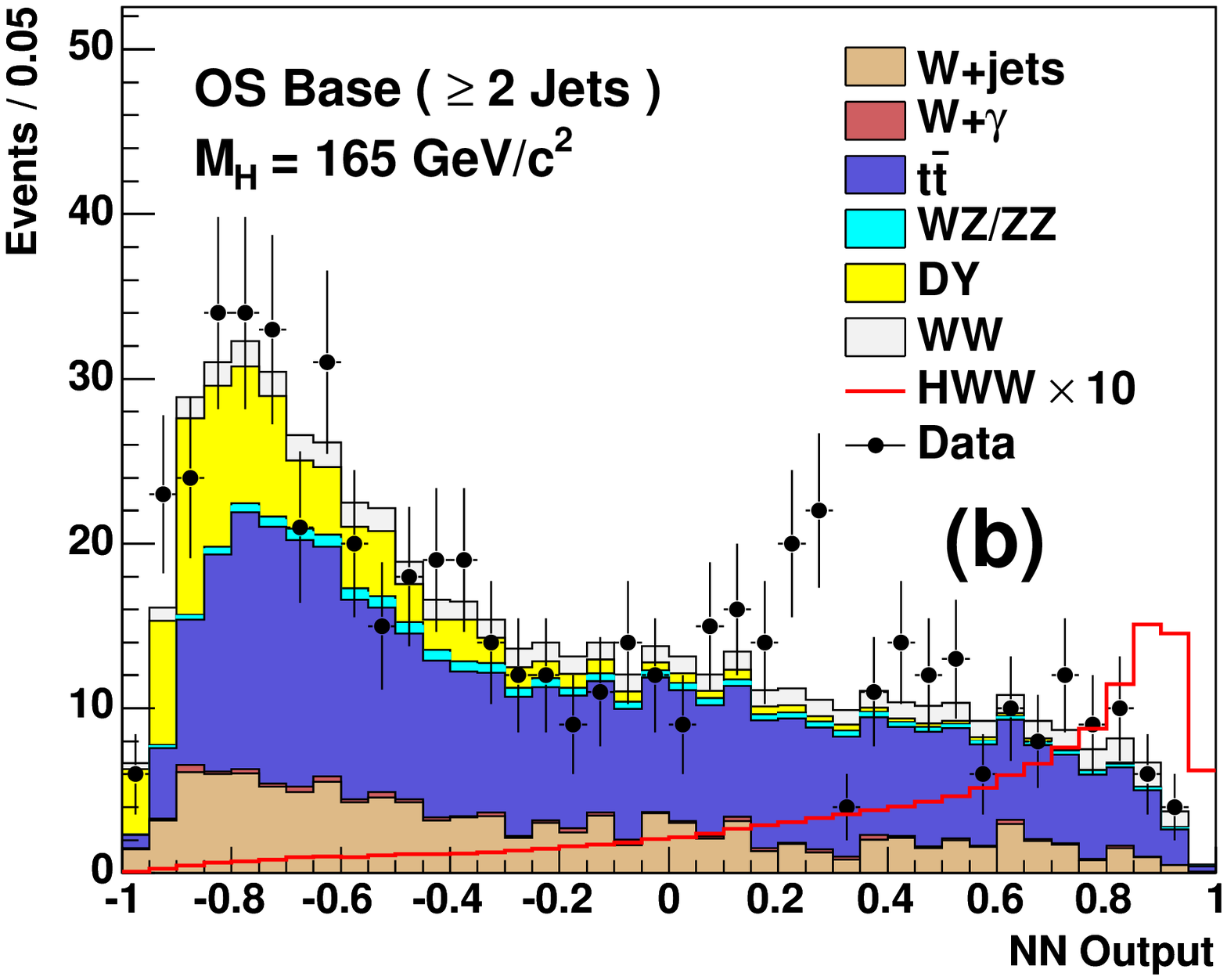}
}
\subfigure{
\includegraphics[width=0.45\textwidth]{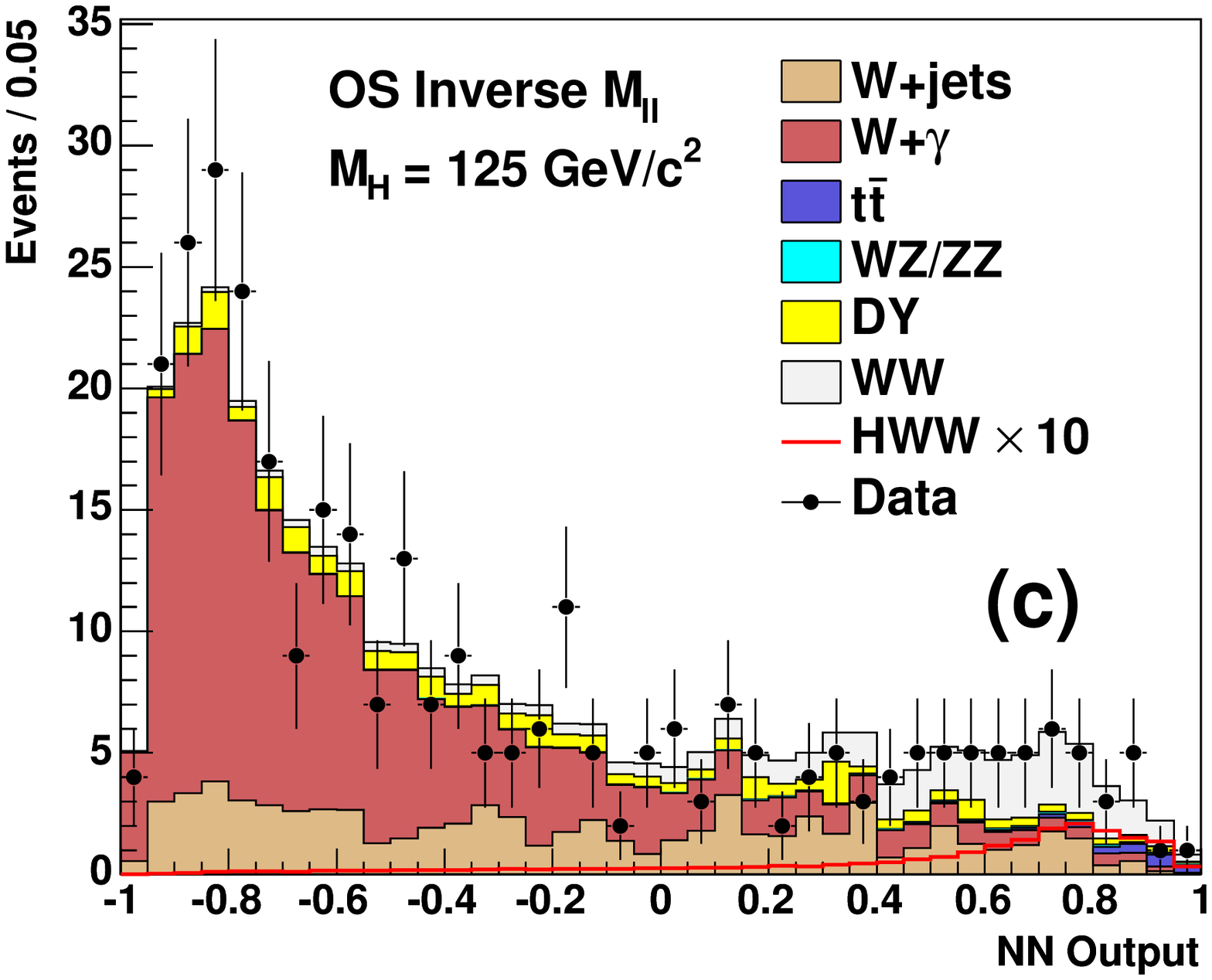}
}
\subfigure{
\includegraphics[width=0.45\textwidth]{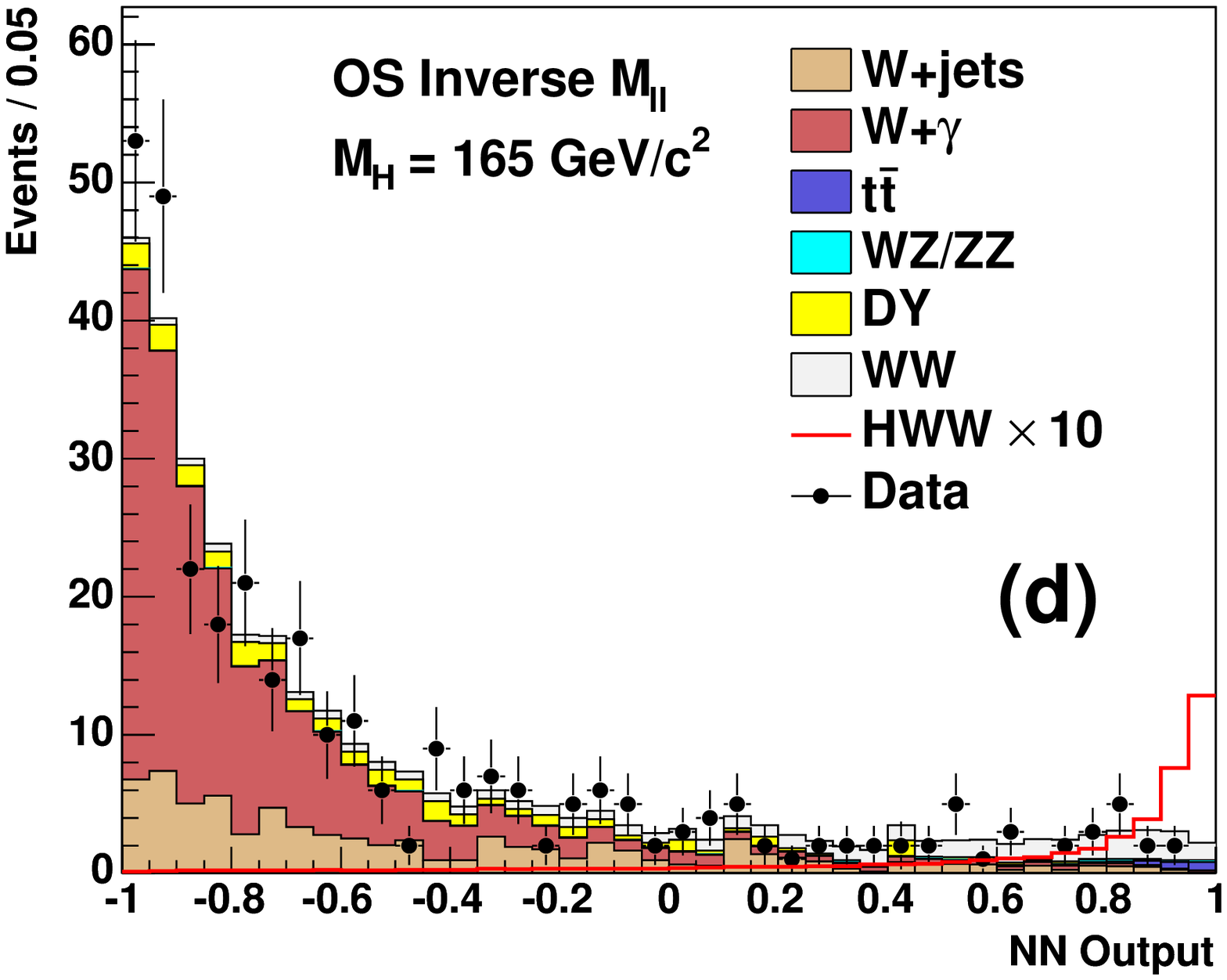}
}
\end{center}
\caption{Predicted and observed distributions of neural network output variables 
for networks trained to separate potential Higgs boson events from background 
contributions in the (a,b) OS Base ($\ge$2 Jets) and (c,d) OS Inverse $M_{\ell\ell}$
search samples for Higgs boson mass hypotheses of 125~and~165~GeV/$c^2$.  The 
overlaid signal predictions correspond to the sum of four production modes ($ggH$, 
{\it WH}, {\it ZH}, and VBF) and are multiplied by a factor of 10 for visibility.
Normalizations for background event yields are those obtained from the final fit 
used to extract search limits.}
\label{fig:TemplatesDilA3}
\end{figure*}

\begin{figure*}[t]
\begin{center}
\subfigure{
\includegraphics[width=0.45\linewidth]{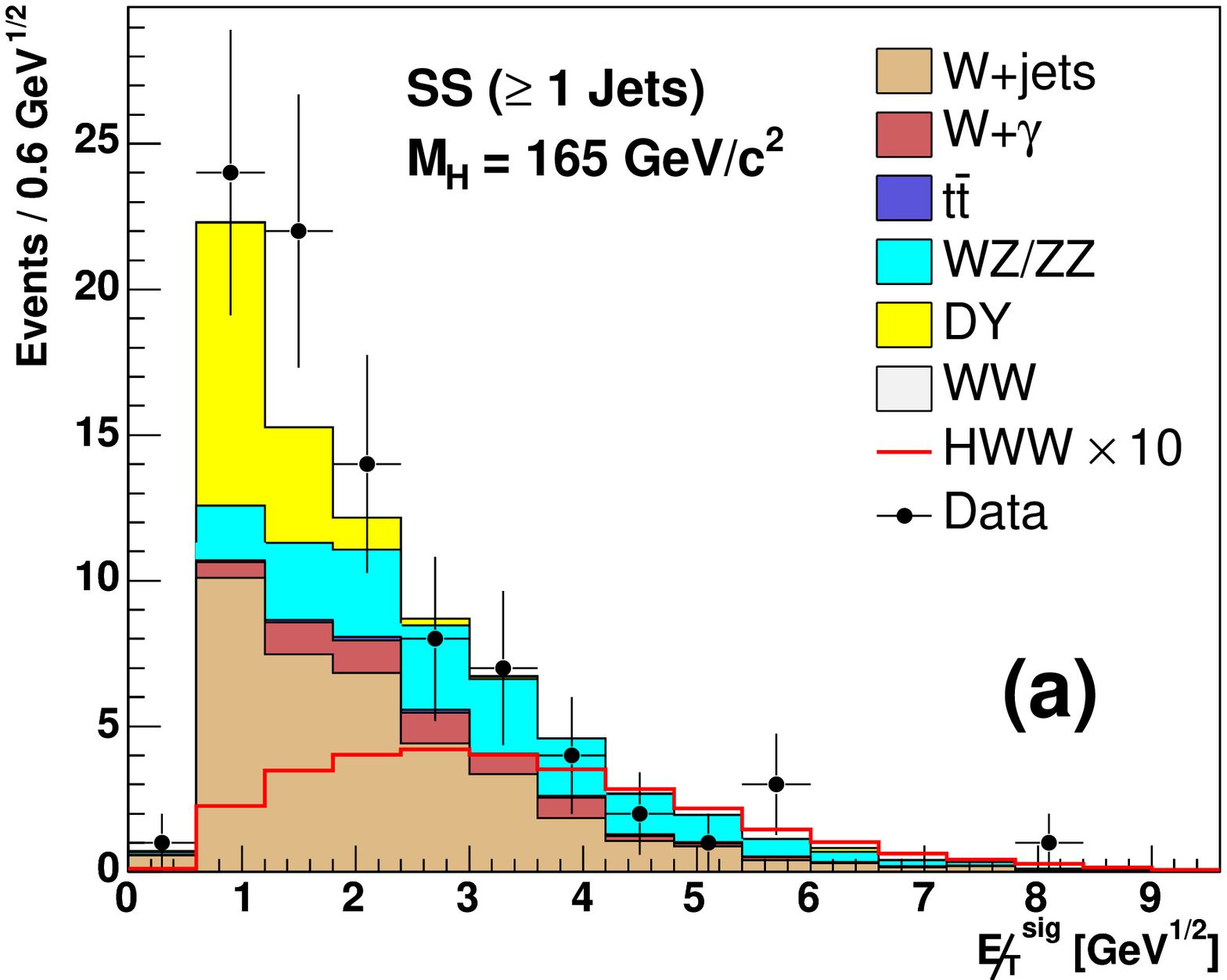}
}
\subfigure{
\includegraphics[width=0.45\linewidth]{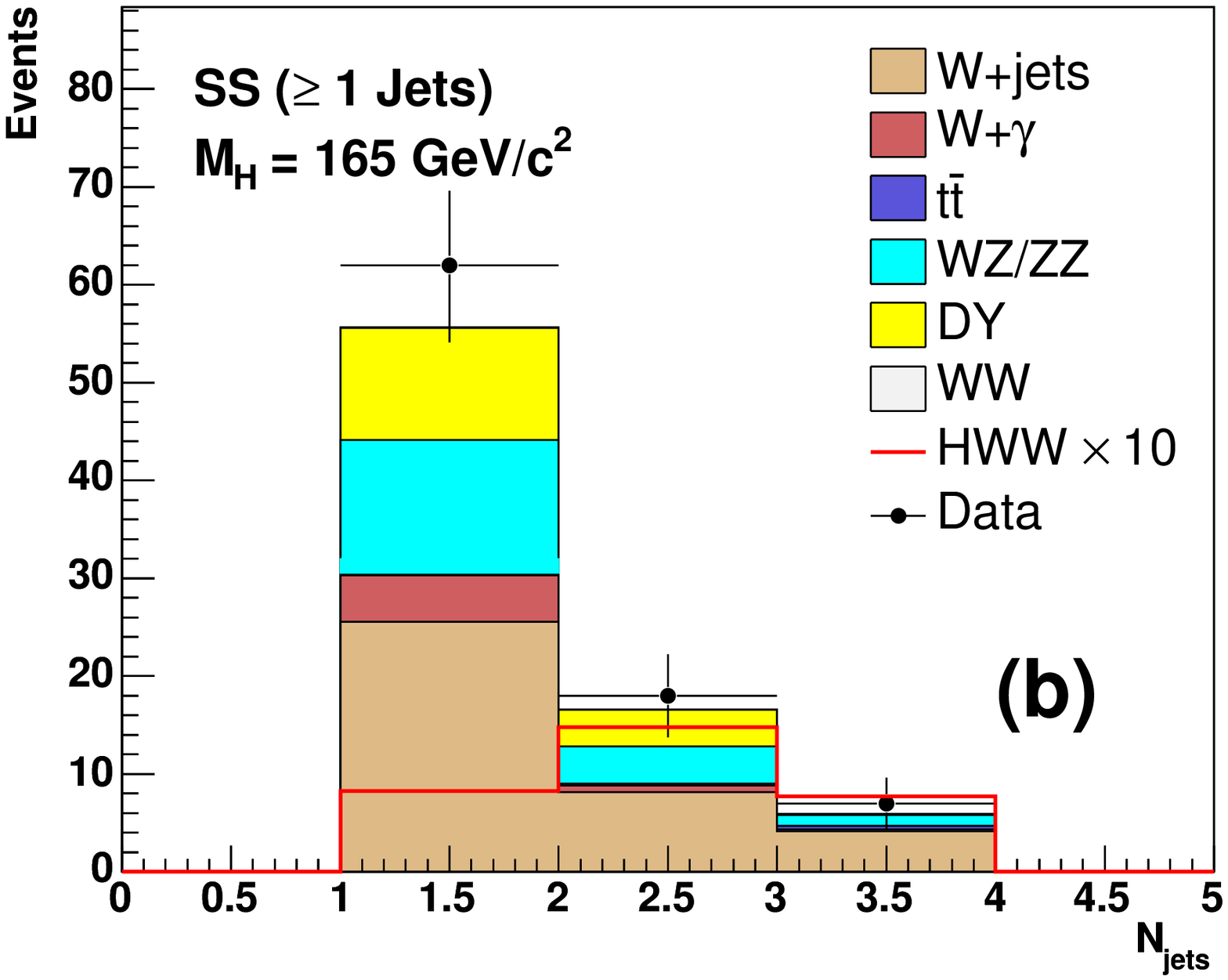}
}
\end{center}
\caption{Predicted and observed distributions of kinematic input variables providing 
the largest separation between potential signal and background contributions in the 
SS ($\ge$1 Jets) search sample.  The overlaid signal predictions correspond to the 
sum of two production modes ({\it WH} and {\it ZH}) for a Higgs boson with mass of 
165~GeV/$c^2$ and are multiplied by a factor of 10 for visibility.  Normalizations 
for background event yields are those obtained from the final fit used to extract 
search limits.}
\label{fig::NNInputsDilA4}
\end{figure*}

\begin{figure*}[t]
\begin{center}
\subfigure{
\includegraphics[width=0.45\textwidth]{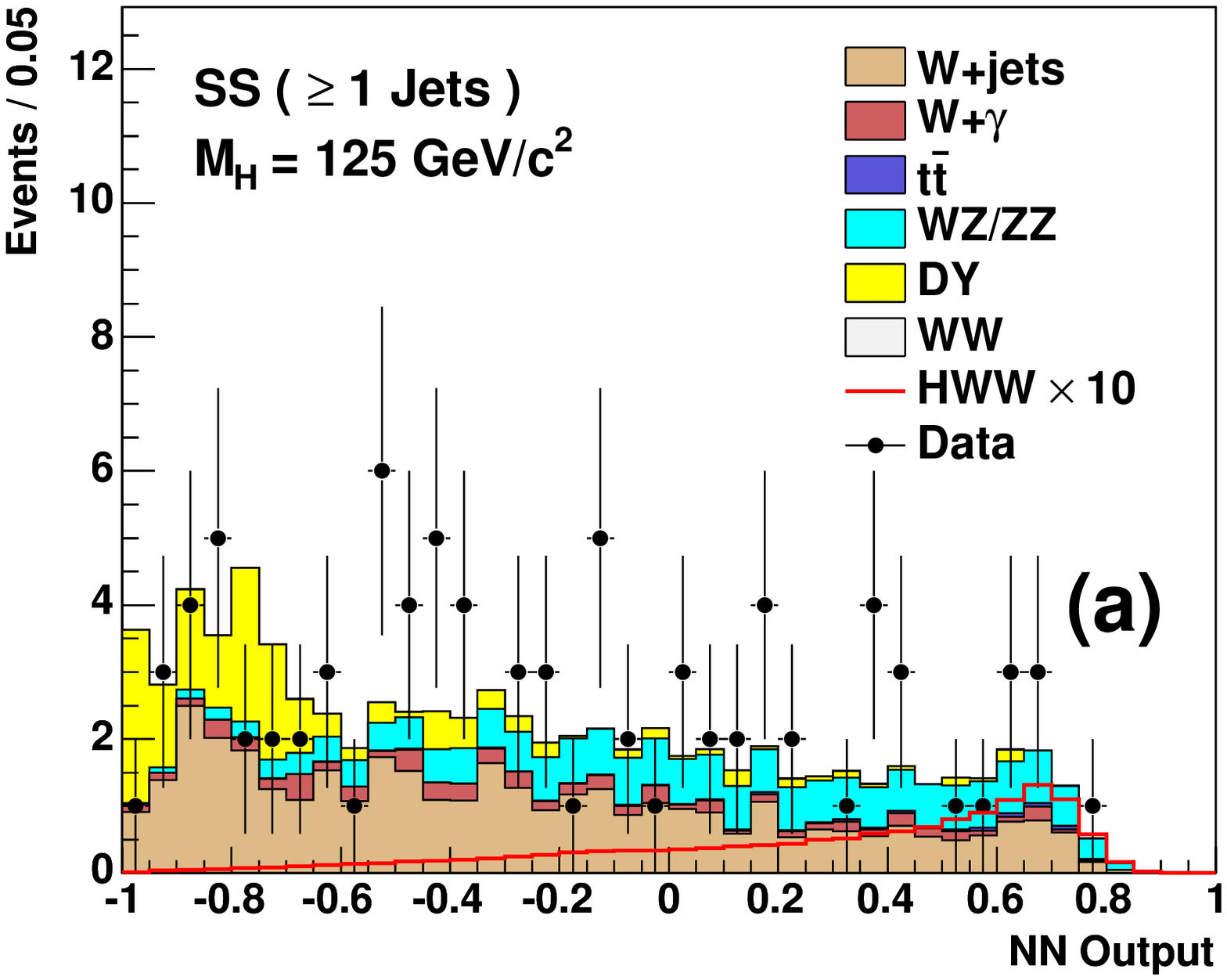}
}
\subfigure{
\includegraphics[width=0.45\textwidth]{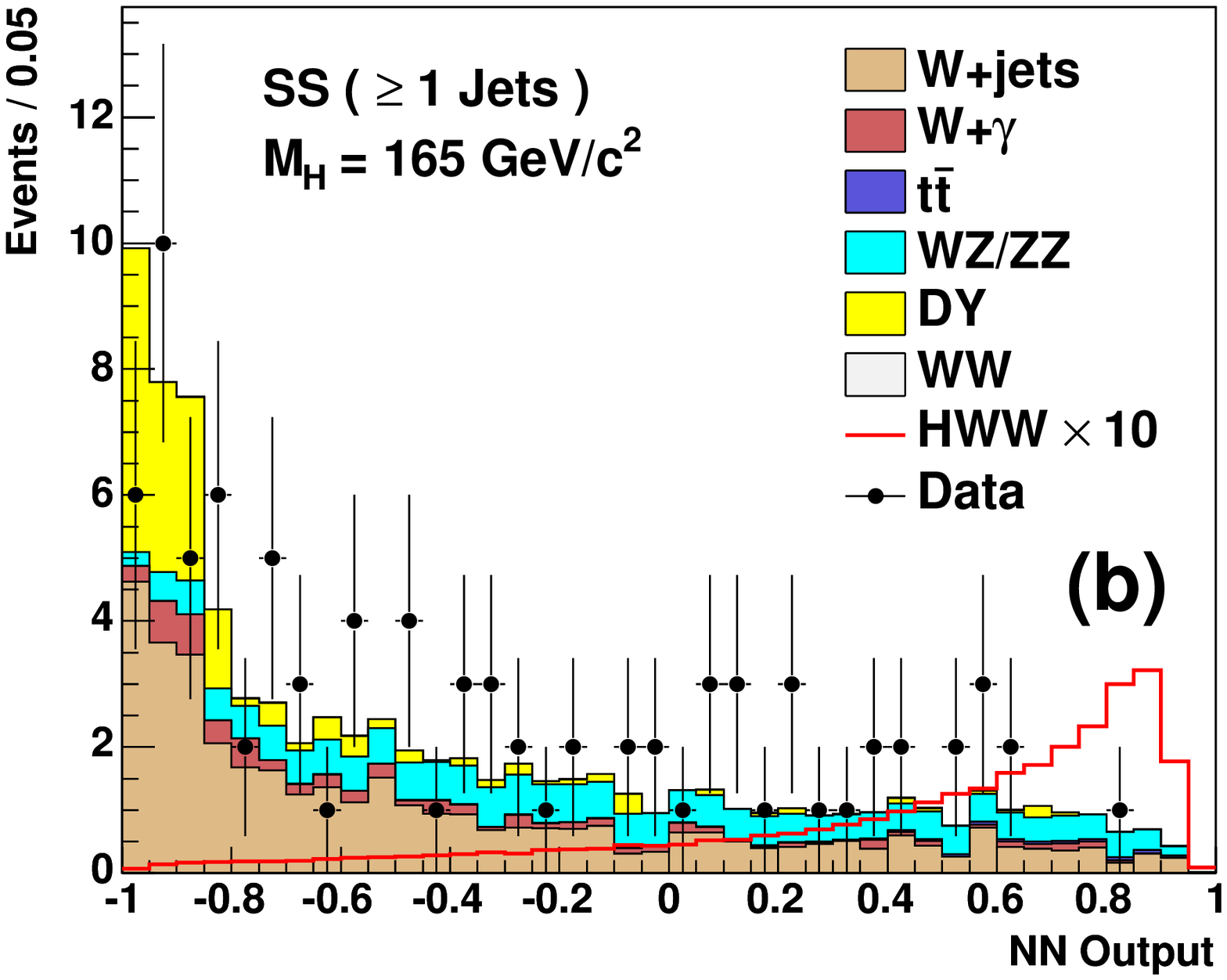}
}
\end{center}
\caption{Predicted and observed distributions of neural network output variables 
for networks trained to separate potential Higgs boson events from background 
contributions in the SS ($\ge$1 Jets) search sample for Higgs boson mass hypotheses 
of 125~and~165~GeV/$c^2$.  The overlaid signal predictions correspond to the sum of 
two production modes ({\it WH} and {\it ZH}) and are multiplied by a factor of 10 
for visibility.   Normalizations for background event yields are those obtained from 
the final fit used to extract search limits.}
\label{fig:TemplatesDilA4}
\end{figure*}

In the OS $\ge$2 Jets search sample the {\it VH} and VBF Higgs boson production mechanisms 
account for $\approx$~65\% of the total expected signal.  Even after rejecting 
events with a jet tagged as likely to have originated from a bottom quark, roughly
50\% of background events are estimated to originate from $t\bar{t}$ production.  
Two neural networks are trained to distinguish signal from background.  One network 
distinguishes $ggH$ production from background contributions without using jet 
kinematic information.  The second network incorporates jet-related variables as 
inputs and is trained to separate {\it VH} and VBF production, which result in events 
with multiple jets at LO, from background contributions.  A single, final discriminant 
is obtained by taking the higher of the two discriminant values obtained from the individual 
neural networks.  We follow this approach to avoid dependence on the {\sc pythia} modeling 
of the higher-order processes within $ggH$ production, which yield the small fraction 
of $ggH$ events containing multiple jets.  Higgs boson events from $ggH$ production 
are dominantly selected by the first network minimizing any potential mismodeling 
effects.  The 17 kinematic input variables used for both networks are $\Delta\phi(\ell\ell)$, 
$\Delta R(\ell\ell)$, $M(\ell\ell)$, $p_{T}(\ell_1)$, $p_{T}(\ell_2)$, $H_T$, $M_T$
($\ell$,$\ell$,$\Met$), $M_T$($\ell$,$\ell$,$\Met$,jets), $\MetSpec$, $\Sigma E_{T}$
($\ell$,$\Met$), $\Delta\phi$($\ell\ell$,$\Met$), $\Sigma E_{T}$($\ell$,jets), 
cos($\Delta\phi$($\ell\ell$))$_{{\rm CM}}$, cos($\psi$($\ell_2$))$_{{\rm CM}}$, $\MetSig$, 
$A$, and $\Sigma E_{T}$(jets).  The additional 8 jet-related variables used as inputs to 
the network trained for separating {\it VH} and VBF production are $M(jj)$, $\Delta\phi(jj)$, 
$\Delta\eta(jj)$, $\Delta R(jj)$, $E_{T}(j_1)$, $E_{T}(j_2)$, $\eta(j_1)$, and $\eta
(j_2)$.  In the case of this second network, the 4 combinations of the total 23 
variables most discriminating for Higgs boson mass of 125, 140, 160, and 185~GeV/$c^2$, 
are reused as inputs to networks trained for neighboring mass values.   Distributions 
of the variables found to contain the largest discriminating power, $M(\ell\ell)$ 
and $\Delta\phi$($\ell\ell$,$\Met$), are shown in Figs.~\ref{fig::NNInputsDilA3}(a) 
and~\ref{fig::NNInputsDilA3}(b).  Using a large number of network input variables 
makes it possible to separate the large number of signal and background processes 
that contribute to this sample.  Specific variables are targeted, for example, at 
identifying the $W$ boson spin correlation associated with the decay of the spin-0
Higgs boson, the hadronic decay of a third vector boson associated with {\it VH} production,
the large rapidity gap present between the additional jets originating from VBF 
production, the high overall energy in events from top-quark pair production, and 
the $Z$ boson associated with either DY or direct {\it WZ} and {\it ZZ} production.   
Examples of neural network output distributions for this search sample are shown 
in Figs.~\ref{fig:TemplatesDilA3}(a) and~\ref{fig:TemplatesDilA3}(b) for Higgs 
boson masses of 125 and 165~GeV/$c^2$, respectively.

Including the OS Inverse $M_{\ell\ell}$ search sample leads to an overall gain 
in signal acceptance of approximately 35\%, with respect to that of the combined 
OS Base search samples, for a Higgs boson with $m_H=$~125~GeV/$c^2$.  In this 
sample the dominant signal contribution is from $ggH$ production, although smaller 
contributions from {\it VH} and VBF production are considered.  The largest background 
contribution is associated with $W\gamma$ production.  The 13 kinematic variables 
used as inputs to the neural network trained for separating signal and background 
are $\Delta\phi(\ell\ell)$, $\Delta R(\ell\ell)$, $p_{T}(\ell_1)$, $p_{T}(\ell_2)$, 
$H_T$, $\MetSpec$, $E(\ell_1)$, $E(\ell_2)$, $\Sigma E_{T}$, $|\Sigma \vec{E}_T|$, 
$\MetSig$, $\Delta\phi$($\Met$,$\ell$ or jet), and $\Sigma E_{T}$($\ell$,jets).   
Distributions of the most discriminating among these variables, $\Sigma 
E_{T}$($\ell$,jets) and $\MetSig$, are shown in Figs.~\ref{fig::NNInputsDilA3}(c) 
and~\ref{fig::NNInputsDilA3}(d).  The two variables exploit the higher total event 
energy expected from a high-mass Higgs boson decay and the absence of neutrinos 
in events originating from $W\gamma$ production.   Examples of neural network 
output distributions for this search sample are shown in 
Figs.~\ref{fig:TemplatesDilA3}(c) and~\ref{fig:TemplatesDilA3}(d) for Higgs boson 
masses of 125 and 165~GeV/$c^2$, respectively.

The SS $\ge$1 Jets search sample focuses solely on signal contributions from {\it VH} 
production, in which like-sign charged leptons result from the decays of the 
associated vector boson and one of two $W$ bosons produced in the Higgs boson decay.
Over 50\% of background events in the sample are predicted to originate from 
$W$+jets production, where the lepton candidate, misidentified from the decay 
products of the jet, is assigned the same charge as the lepton produced in the 
$W$ boson decay.  The 9 kinematic variables used to train the neural network 
used for separating signal and backgrounds are $p_{T}(\ell_1)$, 
$p_{T}(\ell_2)$, $\Met$, $\MetSpec$, $\Delta\phi$($\Met$,$\ell$ or jet), 
$\MetSig$, $\Sigma E_{T}$(jets), $E_{T}(j_1)$, and $N_{{\rm jets}}$.   Distributions 
of the most discriminating among these variables, $\MetSig$ and $N_{{\rm jets}}$, are 
shown in Figs.~\ref{fig::NNInputsDilA4}(a) and~\ref{fig::NNInputsDilA4}(b).  These 
variables are sensitive to the presence of neutrinos and jets associated with 
leptonic and hadronic decays of the multiple vector bosons originating from 
the {\it VH} production process. Examples of neural network output distributions 
for this search sample are shown in Figs.~\ref{fig:TemplatesDilA4}(a) 
and~\ref{fig:TemplatesDilA4}(b) for Higgs boson masses of 125 and 165~GeV/$c^2$, 
respectively.

\clearpage

\subsection{Dilepton search samples with hadronically-decaying tau leptons}
\label{sec:DileptonEventsTaus}

The numbers of expected events from each contributing signal and background 
process are compared in Table~\ref{tbl:YieldsDileptonTaus} with the total 
number of observed events in each of the two dilepton search samples formed 
from one electron or muon candidate and one hadronically-decaying tau lepton
candidate.  Background and signal predictions, referring to potential Higgs 
boson masses of 125 and 165~GeV/$c^2$, are taken from the models described 
in Sec.~\ref{sec:BGmodeling}.

Signal and background kinematic properties of events in these samples are similar 
to those in the other dilepton search samples and the multivariate techniques 
applied to these samples for separating signal and background contributions use a 
subset of the kinematic variables in Table~\ref{tbl:Dilepton Variables} as inputs.  
In addition, identification variables associated with the hadronically-decaying tau 
lepton candidate are strongly discriminating against dominant $W$+jets background 
contributions, in which a particle jet is misidentified as a hadronically-decaying 
tau lepton candidate.  The additional tau lepton identification variables 
used as inputs to the BDT algorithms applied to these samples are listed in 
Table~\ref{tbl:Tau Variables}.

The dominant signal contributions to the OS Hadronic Tau search samples 
originate from $ggH$ production, although contributions from the {\it VH} 
and VBF production mechanisms are also considered.  Over 80\% of events 
in these samples are predicted to originate from $W$+jet production.  A 
BDT algorithm, with a combined set of dilepton kinematic and tau lepton
identification variables as inputs, is used to provide a single output 
variable for distinguishing potential signal events from the large 
background contributions.  The best separation is obtained when the BDT 
algorithm is trained solely to distinguish $ggH$ signal from $W$+jet 
background contributions.  Although the same set of input variables are 
used, independent BDT algorithms are trained for the $e$+$\tau_{\rm{had}}$ 
and $\mu$+$\tau_{\rm{had}}$ search samples to exploit differences in the 
distributions of reconstructed electron and muon candidates.

\begin{table}[t]
   \setlength{\extrarowheight}{3pt}
\begin{ruledtabular} 
\begin{center} 
\caption{\label{tbl:YieldsDileptonTaus}
Summary of predicted and observed event yields for two dilepton 
search samples formed from one electron or muon candidate and one 
hadronically-decaying tau lepton candidate.  Expected signal yields 
are shown for potential SM Higgs boson masses of 125 and 165~GeV/$c^2$.
Normalizations for background event yields are taken directly from 
the modeling of the individual processes.}
%{\scriptsize
\begin{tabular}{lcc}                                                                                                   
\toprule                                                                                                                     
Process                    &  OS Hadronic            &  OS Hadronic              \\
                           &  Tau ($e$+$\tau_{\rm{had}}$) &  Tau ($\mu$+$\tau_{\rm{had}}$) \\   
\hline                                     
$t\bar{t}$                 &     15.6$\pm$2.3        &    11.3$\pm$1.7           \\
{\it WW}, {\it WZ}, and {\it ZZ}       &     25.1$\pm$3.7        &    19.5$\pm$2.9           \\
Multijet and $\gamma+$jet  &        0$^{+34}_{-0}$         &       0$^{+29}_{-0}$            \\
DY ($Z\to\tau\tau$)        &      0.5$\pm$0.2        &     1.2$\pm$0.8           \\
DY ($Z\to ee,\mu\mu$)      &     14.4$\pm$3.6        &      78$\pm$12            \\
$W+$jets                   &      745$\pm$123        &     514$\pm$85            \\ 
$W\gamma$                  &      2.5$\pm$0.4        &     2.3$\pm$0.3           \\ 
\hline                                                                                                                     
Total background           &      803$\pm$126        &     626$\pm$89            \\
\hline                                                                                                                     
\multicolumn{3}{c}{$M_H=$~125~GeV/$c^2$}\\                                                                         
\hline                                             
$ggH$                      &     0.12$\pm$0.02       &     0.09$\pm$0.02         \\  
{\it WH}                       &     0.07$\pm$0.01       &     0.05$\pm$0.01         \\
{\it ZH}                       &     0.04$\pm$0.01       &     0.03$\pm$0.01         \\
$VBF$                      &     0.01$\pm$0.00       &     0.01$\pm$0.00         \\
\hline                                                                                                                    
Total signal               &     0.24$\pm$0.03       &     0.18$\pm$0.02         \\
\hline         
\multicolumn{3}{c}{$M_H=$~165~GeV/$c^2$}\\                                                                         
\hline 
$ggH$                      &     1.07$\pm$0.18       &     0.80$\pm$0.13         \\  
{\it WH}                       &     0.25$\pm$0.03       &     0.17$\pm$0.02         \\
{\it ZH}                       &     0.15$\pm$0.02       &     0.11$\pm$0.02         \\
$VBF$                      &     0.10$\pm$0.02       &     0.08$\pm$0.01         \\
\hline    
Total signal               &     1.56$\pm$0.21       &     1.16$\pm$0.15         \\
\hline                                                                                                                     
Data                       &     792                 &     598                   \\
\bottomrule                                                                                                                     
\end{tabular}
%}
\end{center}
\end{ruledtabular}
\end{table}

\begin{table*}[t]
  \setlength{\extrarowheight}{3pt}
\begin{ruledtabular} 
\begin{center} 
\caption{\label{tbl:Tau Variables}
Summary of identification variables associated with a hadronically-decaying 
tau lepton candidate used as inputs to the multivariate algorithms for 
separating signal and background contributions.}
{\scriptsize
%\begin{tabular}{|p{2.0cm}|p{11.8cm}|p{0.8cm}|p{0.8cm}|}         
\begin{tabular}{llcc}         
\toprule                                                                                                                     
Variable        & Definition            &  OS            &  Trilepton                     \\   
%Name            &                       &  Hadronic      &  {\it WH}                           \\
%                &                       &  Tau           &  $\ell$+$\ell$+$\tau_{\rm{had}}$   \\
                &                       &  Hadronic Tau  &  {\it WH} $\ell$+$\ell$+$\tau_{\rm{had}}$ \\
\hline                                                                 
$p_T^{{\rm seed}}$                  & Transverse momentum of tau candidate seed track                                    &  \checkmark  &  \checkmark  \\
$d_0^{{\rm seed}}$                  & Impact parameter of tau candidate seed track with respect to primary vertex        &  \checkmark  &  \checkmark  \\
$E^{{\rm vis}}_T$                   & Tau candidate visible transverse energy                                            &  \checkmark  &  \checkmark  \\
$M^{{\rm vis}}_T$                   & Tau candidate visible mass                                                         &  \checkmark  &  \checkmark  \\
$I_{\text{track}}$            & Tau candidate track isolation                                                      &  \checkmark  &  \checkmark  \\
$\Sigma P_T(\text{iso cone})$ & Scalar sum of track $p_T$ for all tracks within isolation cone not used in         &  \checkmark  &  \checkmark  \\
                              & reconstruction of tau candidate                                                    &     &     \\
$\Sigma E_T(\text{iso cone})$ & Scalar sum of $\pi^0$ candidate $E_T$ for all candidates within isolation cone     &  \checkmark  &  \checkmark  \\
                              & not used in reconstruction of tau candidate                                        &     &     \\
$p_T^{{\rm closest}}$               & $p_T$ of track closest to direction of tau candidate visible momentum              &  \checkmark  &  \checkmark  \\
$E_T^{{\rm closest}}$               & $E_T$ of $\pi^0$ candidate closest to direction of tau candidate visible momentum  &  \checkmark  &  \checkmark  \\
$\theta_{{\rm track}}^{{\rm closest}}$    & Angle between tau candidate and the closest track                                  &  \checkmark  &  \checkmark  \\
$\theta_{\pi^0}^{{\rm closest}}$    & Angle between tau candidate and the closest $\pi^0$ candidate                      &  \checkmark  &  \checkmark  \\
\bottomrule                                                                                                                     
\end{tabular}
}
\end{center}
\end{ruledtabular}
\end{table*}

\begin{figure*}[]
\begin{center}
\subfigure{
\includegraphics[width=0.45\textwidth]{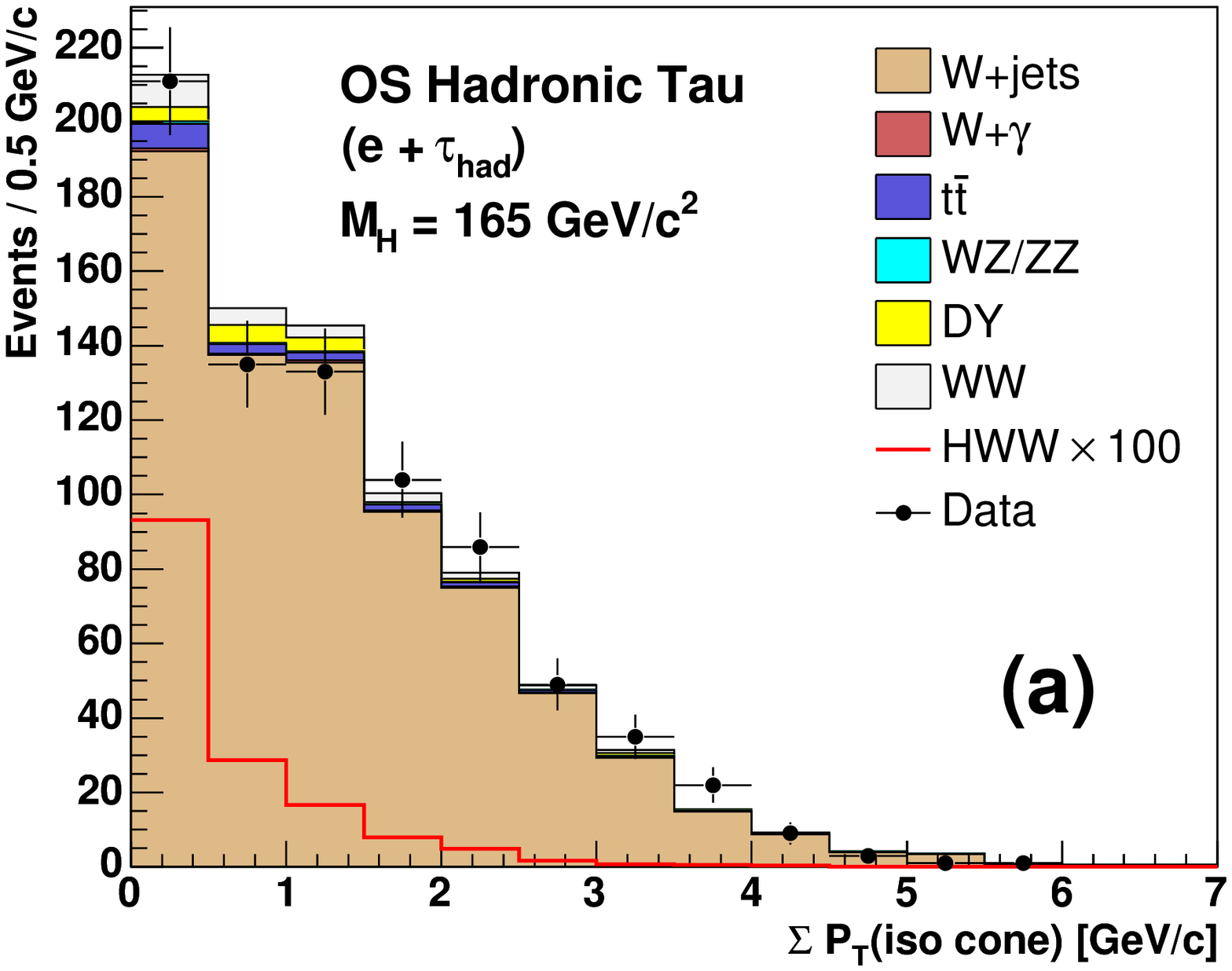}
}
\subfigure{
\includegraphics[width=0.45\textwidth]{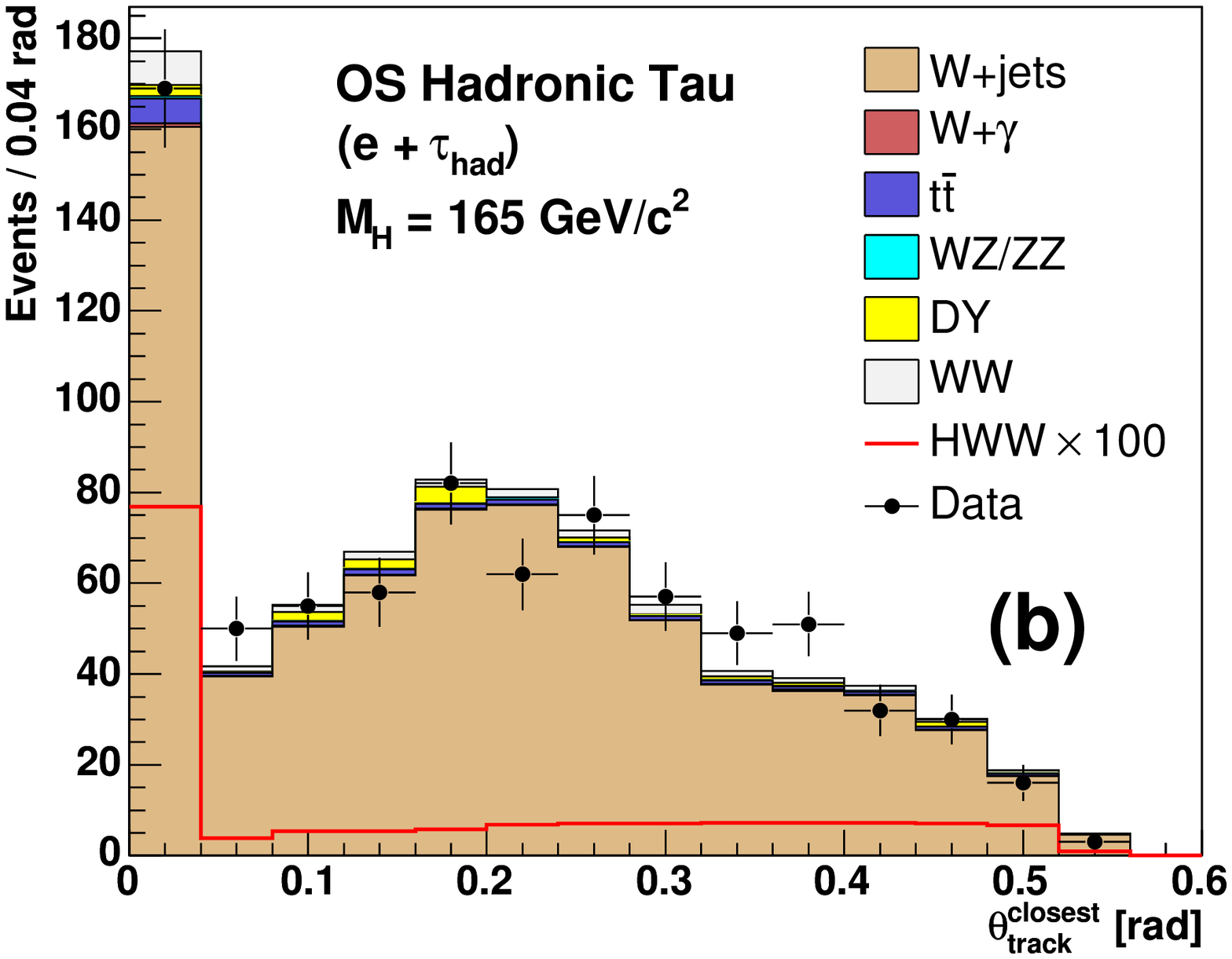}
}
\subfigure{
\includegraphics[width=0.45\textwidth]{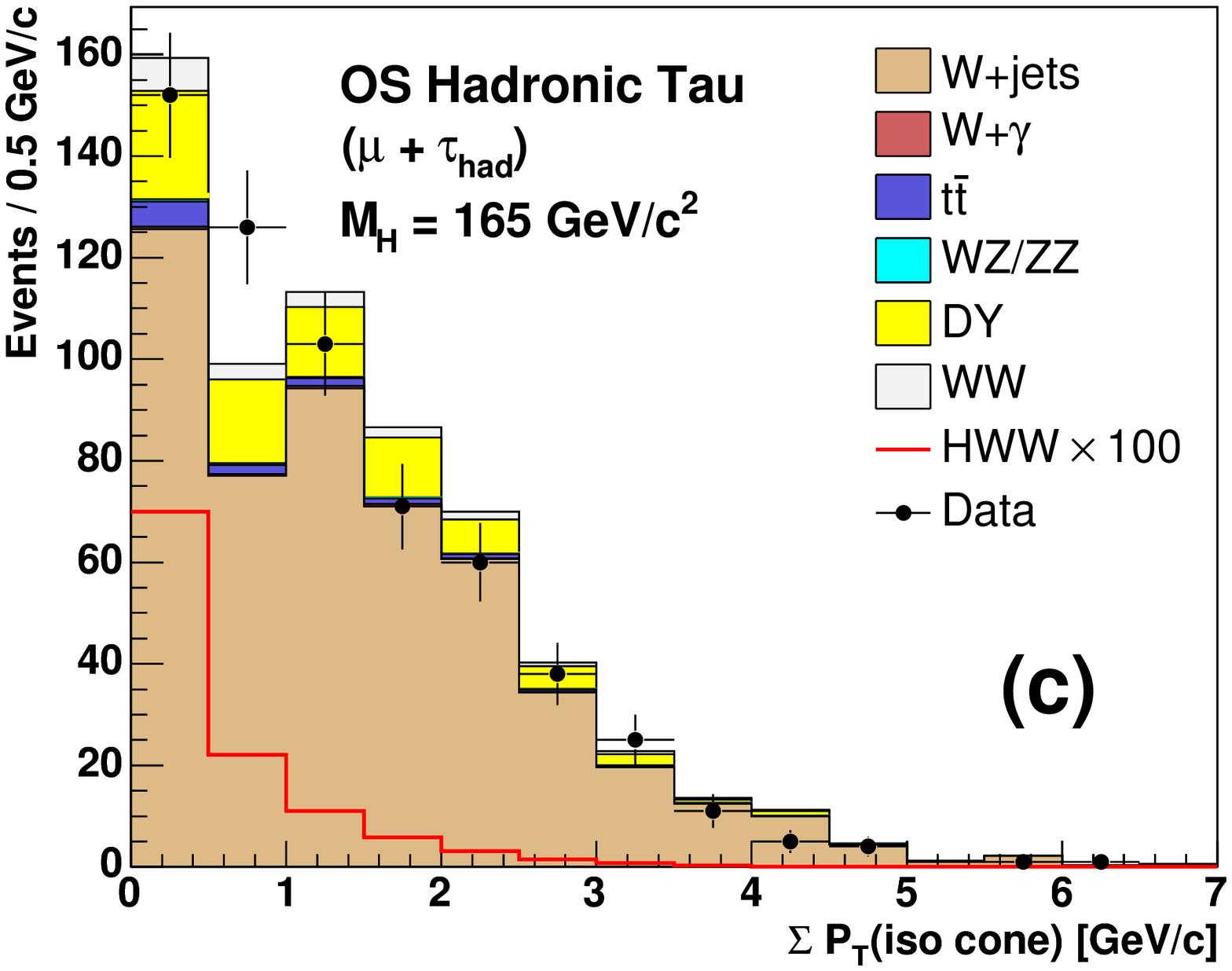}
}
\subfigure{
\includegraphics[width=0.45\textwidth]{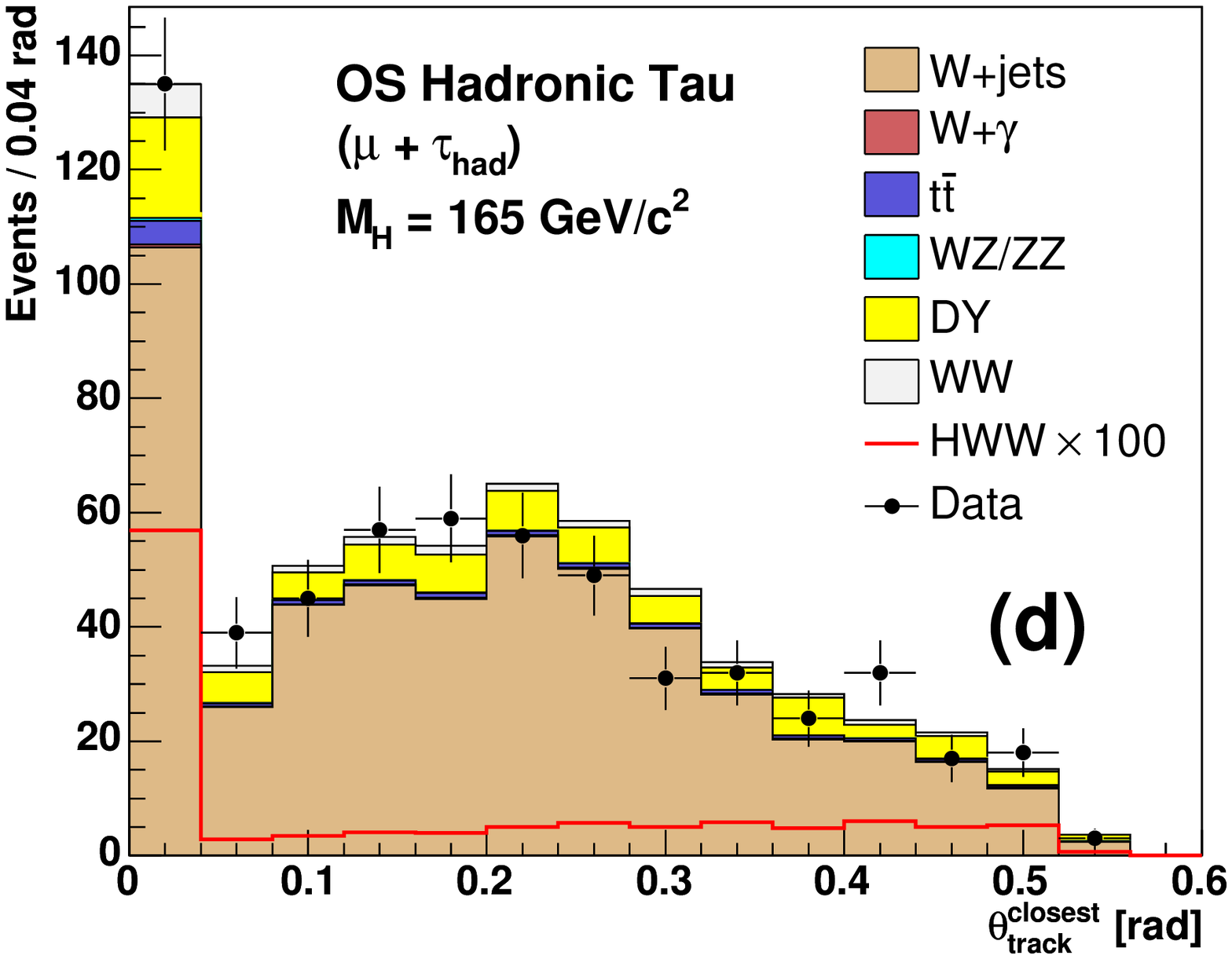}
}
\end{center}
\caption{Predicted and observed distributions of kinematic input variables 
providing the largest separation between potential signal and background 
contributions in the (a,b) OS Hadronic Tau ($e$+$\tau_{\rm{had}}$) and (c,d) 
OS Hadronic Tau ($\mu$+$\tau_{\rm{had}}$) search samples.  The overlaid 
signal predictions correspond to the sum of four production modes ($ggH$, 
{\it WH}, {\it ZH}, and VBF) for a Higgs boson with mass of 165~GeV/$c^2$ 
and are multiplied by a factor of 100 for visibility.  Normalizations for 
background event yields are those obtained from the final fit used to extract 
search limits.}
\label{fig::NNInputsTauDilep}
\end{figure*}

\begin{figure*}[]
\begin{center}
\subfigure{
\includegraphics[width=0.45\textwidth]{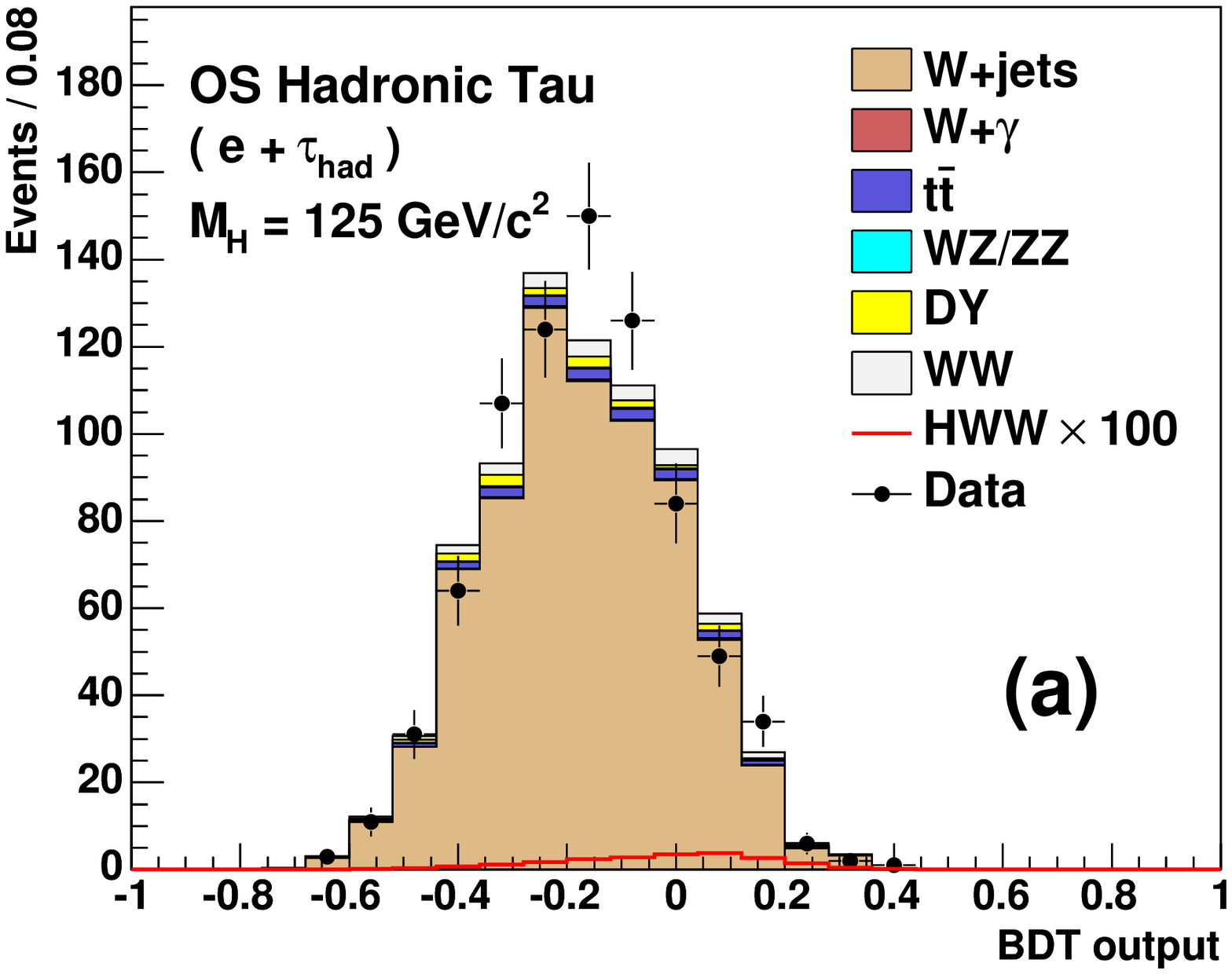}
}
\subfigure{
\includegraphics[width=0.45\textwidth]{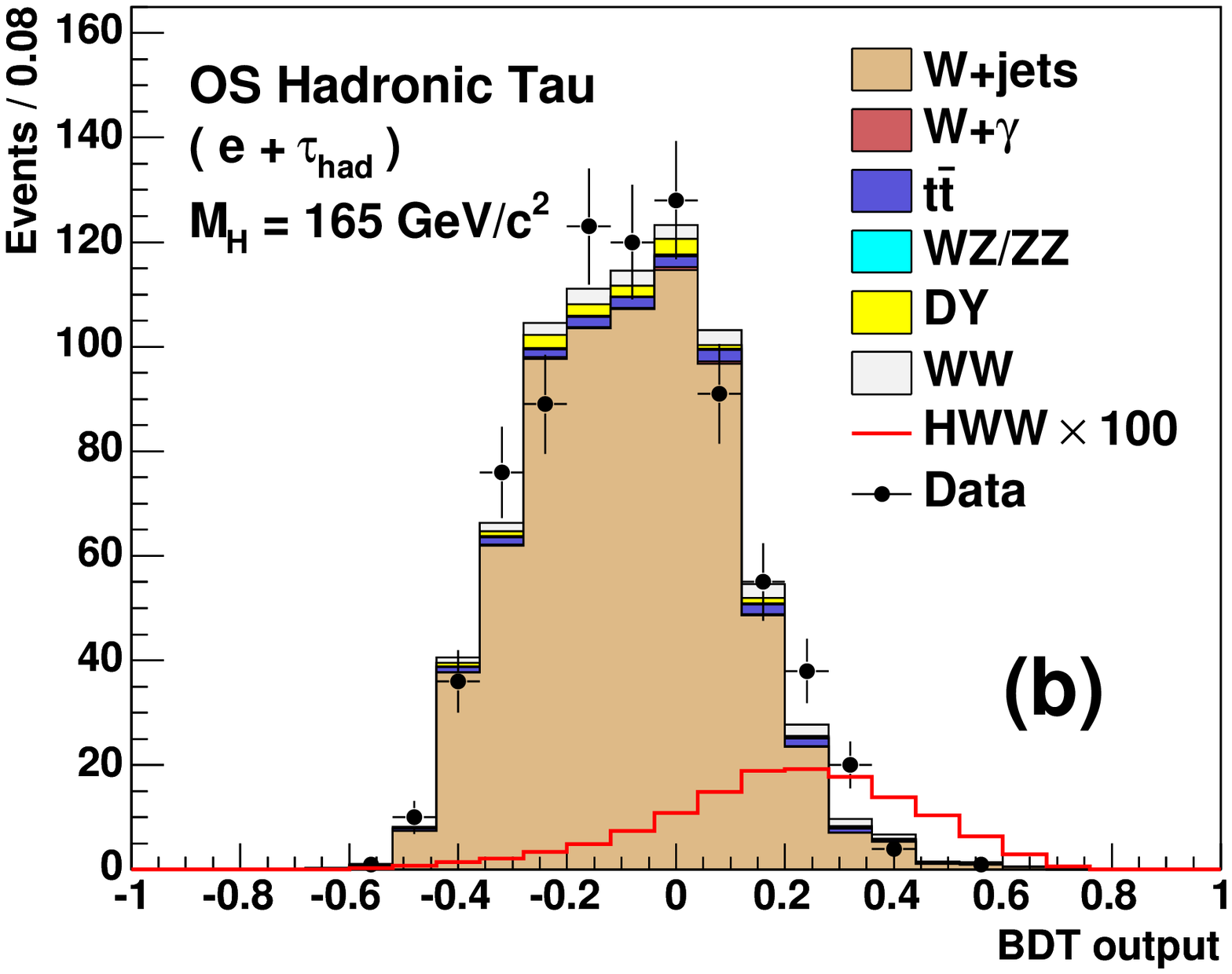}
}
\subfigure{
\includegraphics[width=0.45\textwidth]{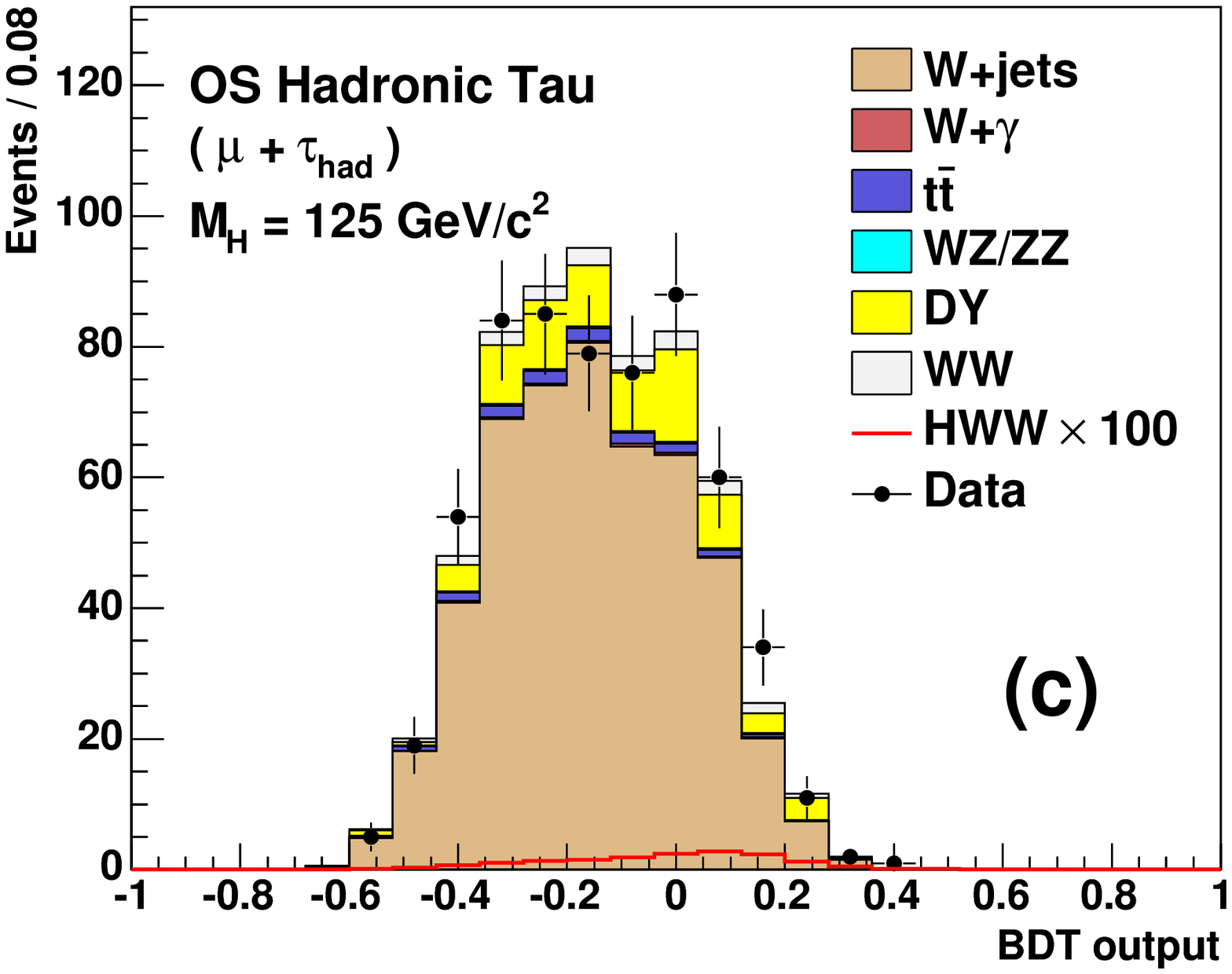}
}
\subfigure{
\includegraphics[width=0.45\textwidth]{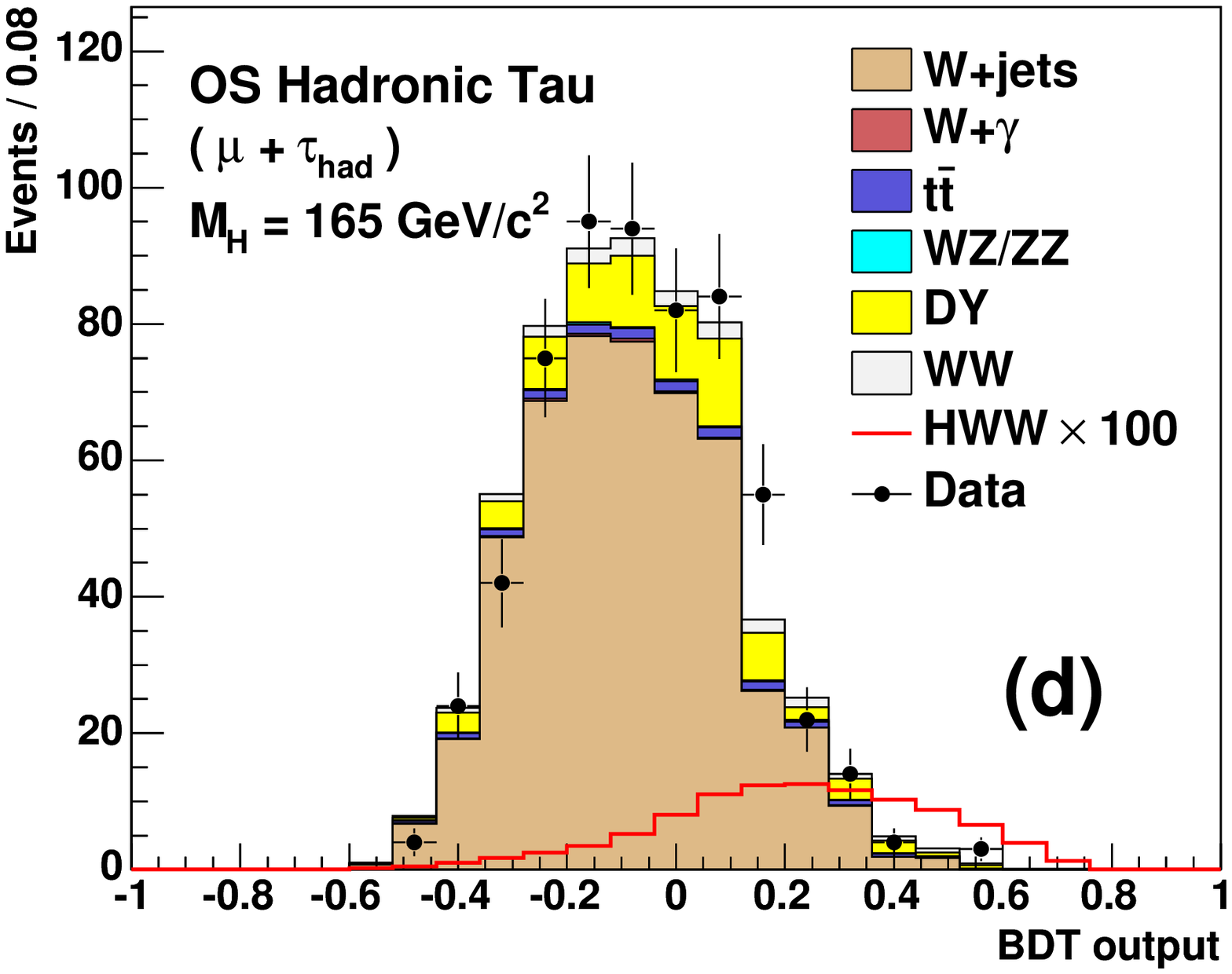}
}
\end{center}
\caption{Predicted and observed distributions of BDT output variables for 
trees trained to separate potential Higgs boson events from background 
contributions in the (a,b) OS Hadronic Tau ($e$+$\tau_{\rm{had}}$) and 
(c,d) OS Hadronic Tau ($\mu$+$\tau_{\rm{had}}$) search samples for Higgs 
boson mass hypotheses of 125~and~165~GeV/$c^2$.  The overlaid signal 
predictions correspond to the sum of four production modes ($ggH$, {\it WH}, 
{\it ZH}, and VBF) and are multiplied by a factor of 100 for visibility.
Normalizations for background event yields are those obtained from the final 
fit used to extract search limits.}
\label{fig:TemplatesTauDilep}
\end{figure*}

The 12 kinematic variables used as inputs to the BDT algorithms are 
$\Delta\phi(\ell\ell)$, $\Delta\eta(\ell\ell)$, $\Delta R(\ell\ell)$,
$M(\ell\ell)$, $p_{T}(\ell_1)$, $\Met$, $\MetSig$, $\Sigma E_{T}$($\ell$,jets), 
$\Delta\phi$($\Met$,$\ell$), $\Delta\phi$($\Met$,$\tau$), $M_T$($\ell$,$\Met$), 
and $M_T$($\tau$,$\Met$).  All 11 tau lepton identification variables 
listed in Table~\ref{tbl:Tau Variables} are also used.  Distributions of 
the most discriminating variables, $\Sigma P_T(\text{iso cone})$ and 
$\theta_{{\rm track}}^{{\rm closest}}$, are shown in Figs.~\ref{fig::NNInputsTauDilep}(a) 
and~\ref{fig::NNInputsTauDilep}(b) for the $e$+$\tau_{\rm{had}}$ sample and in 
Figs.~\ref{fig::NNInputsTauDilep}(c) and~\ref{fig::NNInputsTauDilep}(d) for 
the $\mu$+$\tau_{\rm{had}}$ sample.  These variables primarily separate events 
containing real and misidentified hadronically-decaying tau lepton candidates.
Examples of BDT output distributions for Higgs boson masses of 125 
and 165~GeV/$c^2$ are shown in Figs.~\ref{fig:TemplatesTauDilep}(a) 
and~\ref{fig:TemplatesTauDilep}(b) for the $e$+$\tau_{\rm{had}}$ sample and 
in Figs.~\ref{fig:TemplatesTauDilep}(c) and~\ref{fig:TemplatesTauDilep}(d) 
for the $\mu$+$\tau_{\rm{had}}$ sample.

\clearpage

\subsection{Trilepton search samples}
\label{sec:TrileptonEvents}

The numbers of expected events from each contributing signal and background 
process are compared in Table~\ref{tbl:YieldsTrilepton} with the total number 
of observed events in each of the four trilepton search samples.  Background 
and signal predictions, referring to potential Higgs boson masses of 125 and 
165~GeV/$c^2$, are taken from the models described in Sec.~\ref{sec:BGmodeling}.

A summary of the kinematic variables used as inputs to the multivariate algorithms 
in these four search samples is shown in Table~\ref{tbl:Trilepton Variables}.  
For the Trilepton {\it WH} ($\ell$+$\ell$+$\tau_{\rm{had}}$) sample, identification 
variables associated with the hadronically-decaying tau lepton candidate are also 
important for suppressing the dominant $Z$+jets background contribution and are 
included as inputs to the multivariate algorithm.  These variables are listed in 
Table~\ref{tbl:Tau Variables}.

\begin{table*}[]
  \setlength{\extrarowheight}{3pt}
\begin{ruledtabular} 
\begin{center} 
\caption{\label{tbl:YieldsTrilepton}
Summary of predicted and observed event yields for four trilepton 
search samples.  Expected signal yields are shown for potential SM
Higgs boson masses of 125 and 165~GeV/$c^2$.  Normalizations for 
background event yields are taken directly from the modeling of the 
individual processes.} 
%{\scriptsize
\begin{tabular}{lcccc}                                                                                                   
\toprule                                                                                                                     
Process                 &     Trilepton {\it WH}    &     Trilepton {\it WH}                 &     Trilepton {\it ZH}    &     Trilepton {\it ZH}   \\
                        &                     &     $\ell$+$\ell$+$\tau_{\rm{had}}$   &     1 Jet           &     $\ge$2 Jets        \\
\hline                                     
$t\bar{t}$              &     $0.75\pm0.23$      &    $2.1\pm0.4$       &    $0.12\pm0.05$      &    $0.2\pm0.04$       \\      
{\it WZ} and {\it ZZ}           &     $10.1\pm1.2$       &    $3.7\pm0.7$       &    $19.9\pm2.4$       &    $10.0\pm1.6$       \\
$Z+$jets                &     $4.9\pm1.1$        &    $31.6\pm6.1$      &    $9.9\pm2.3$        &    $7.8\pm1.4$        \\
$Z\gamma$               &     $4.87\pm0.97$      &    $2.6\pm0.4$       &    $7.8\pm1.6$        &    $3.0\pm0.8$        \\  
\hline                                                                                                                     
Total background        &     $20.6\pm2.2$       &    $40.0\pm6.5$      &    $37.7\pm4.6$       &    $20.9\pm3.1$       \\
\hline                                                                                                                     
\multicolumn{5}{c}{$M_H=$~125~GeV/$c^2$}\\                                                                         
\hline                                             
{\it WH}                    &     $0.49\pm0.07$      &    $0.11\pm0.02$     &    $0.02\pm0.01$      &    $0.01\pm0.01$     \\
{\it ZH}                    &     $0.11\pm0.02$      &    $0.05\pm0.01$     &    $0.24\pm0.04$      &    $0.30\pm0.04$      \\
\hline                                                                                                                   
Total signal            &     $0.60\pm0.08$      &    $0.16\pm0.02$     &    $0.26\pm0.04$      &    $0.30\pm0.04$     \\
\hline         
\multicolumn{5}{c}{$M_H=$~165~GeV/$c^2$}\\                                                                         
\hline 
{\it WH}                    &     $1.03\pm0.14$      &    $0.30\pm0.04$     &    $0.04\pm0.01$      &    $0.02\pm0.01$     \\
{\it ZH}                    &     $0.24\pm0.03$      &    $0.11\pm0.02$     &    $0.31\pm0.05$      &    $0.8\pm0.1$        \\
\hline                                                                                                                    
Total signal            &     $1.27\pm0.17$      &    $0.41\pm0.06$     &    $0.35\pm0.05$      &    $0.8\pm0.1$        \\
\hline                                                                                                                     
Data                    &     20                 &      28              &    38                 &    26                  \\
\bottomrule                                                                                                                     
\end{tabular}
%}
\end{center}
\end{ruledtabular}
\end{table*}
 
\begin{table*}[]
  \setlength{\extrarowheight}{3pt}
\begin{ruledtabular} 
\begin{center} 
\caption{\label{tbl:Trilepton Variables}
Summary of kinematic variables used as inputs to the multivariate 
algorithms for separating signal and background contributions in the trilepton 
search samples.}
\vspace{0.2cm}
{\scriptsize
%\begin{tabular}{|p{2.2cm}|p{10.8cm}|p{0.8cm}|p{0.8cm}|p{0.8cm}|}         
\begin{tabular}{llcccc}         
\toprule                                                                                                                     
Variable          & Definition          & Trilepton & Trilepton                  & Trilepton & Trilepton \\
                  &                     &    {\it WH}     & {\it WH}                         & {\it ZH}        & {\it ZH}        \\
                  &                     &           & $\ell$+$\ell$+$\tau_{\rm{had}}$ & 1 Jet     & $\ge$2 Jets   \\
\hline     
$p_{T}(\ell_1)$               & Transverse momentum of leading lepton                                           &     &  \checkmark  &     &     \\        
$p_{T}(\ell_2)$               & Transverse momentum of subleading lepton                                       &  \checkmark  &  \checkmark  &     &     \\        
$p_{T}(\ell_3)$               & Transverse momentum of subsubleading lepton                                   &     &  \checkmark  &     &     \\        
$\Delta R(\ell^{+}\ell^{-})_{{\rm near}}$ & Minimum $\Delta R(\ell\ell)$ among opposite-sign lepton pairs             &  \checkmark  &  \checkmark  &  \checkmark  &     \\ 
$\Delta R(\ell^{+}\ell^{-})_{{\rm far}}$ & Maximum $\Delta R(\ell\ell)$ among opposite-sign lepton pairs              &  \checkmark  &  \checkmark  &  \checkmark  &     \\  
$M_T$($\ell$,$\ell$,$\ell$)   & Transverse mass of the three leptons                                            &  \checkmark  &  \checkmark  &     &     \\
$M(\ell\ell\ell)$             & Invariant mass of the three leptons                                             &     &  \checkmark  &  \checkmark  &     \\
$E_{T}(j_1)$                  & Transverse energy of the leading jet                                            &     &     &  \checkmark  &  \checkmark  \\ 
$E_{T}(j_2)$                  & Transverse energy of the subleading jet                                        &     &     &     &  \checkmark  \\ 
$M(jj)$                       & Invariant mass of the two leading jets                                          &     &     &     &  \checkmark  \\
$N_{{\rm jets}}$                    & Number of jets in event                                                         &  \checkmark  &     &     &     \\
$\Met$                        & Missing transverse energy                                                       &  \checkmark  &  \checkmark  &  \checkmark  &  \checkmark  \\
$\Delta\phi(\ell_2,\Met)$ & Azimuthal angle between the subleading lepton and the \Met                         &  \checkmark  &  \checkmark  &  \checkmark  &     \\ 
$\Delta\phi(\ell_1+\ell_2+\ell_3,\Met)$ & Azimuthal angle between $\vec{p}_{T}(\ell_1)$+$\vec{p}_{T}(\ell_2)$+$\vec{p}_{T}(\ell_3)$ and the \Met
                                                                                                                &     &     &  \checkmark  &     \\
$\MetSig$                     & $\Met$/($\Sigma E_{T}$($\ell$,jets))$^{1/2}$                                    &     &  \checkmark  &     &     \\
$M_T$($\ell_3$,$\Met$)        & Transverse mass of the subsubleading lepton and the \Met                      &  \checkmark  &  \checkmark  &     &     \\
$M_T$($\ell$,$\ell$,$\ell$,$\Met$,jets) & Transverse mass of the three leptons, all jets and the \Met           &  \checkmark  &  \checkmark  &  \checkmark  &     \\
$M$($\ell_3$,$\Met$,jets)     & Invariant mass of the subsubleading lepton, all jets and the \Met             &  \checkmark  &  \checkmark  &  \checkmark  &     \\
$M$($\ell_1$,$\ell_2$,$\Met$) & Invariant mass of the leading and subleading leptons and the \Met              &  \checkmark  &  \checkmark  &     &     \\
$M(\ell^{+}\ell^{-})_{{\rm near}}$  & Invariant mass of opposite-sign lepton pair closest in $\Delta\phi$             &  \checkmark  &  \checkmark  &     &     \\ 
$H_T$                         & Scalar sum of lepton $p_T$, jet $E_T$, and the \Met                             &  \checkmark  &  \checkmark  &  \checkmark  &     \\
$F(\ell\ell\ell)$             & Trilepton flavor combination (3 $\times$ $e$, $\mu$, or unspecified track)      &  \checkmark  &     &  \checkmark  &  \checkmark  \\   
$\Delta\phi(\ell_{{\rm noZ}},\Met)$ & Azimuthal angle between the lepton not associated with the $Z$                  &     &     &  \checkmark  &     \\
                              & and the \Met                                                                    &     &     &     &     \\ 
$\Delta R(\ell_{{\rm noZ}},j_1)$    & $\Delta R$ between lepton not associated with $Z$ and leading jet               &     &     &  \checkmark  &     \\
$\Delta R$($\ell_{{\rm noZ}}$,jet)$_{{\rm near}}$ & $\Delta R$ between lepton not associated with $Z$ and closest jet       &     &     &     &  \checkmark  \\
$M_T$($\ell_{{\rm noZ}}$,$\Met$)    & Transverse mass of lepton not associated with $Z$ and the \Met                  &     &     &  \checkmark  &  \checkmark  \\
$M_T$($\ell_{{\rm noZ}}$,$\Met$,jets) & Transverse mass of lepton not associated with $Z$, all jets and               &     &     &  \checkmark  &  \checkmark  \\
                                & the \Met                                                                      &     &     &     &     \\
$M(\ell_{{\rm noZ}},\Met)$          & Invariant mass of lepton not associated with $Z$ and the \Met                   &     &     &  \checkmark  &  \checkmark  \\
$\Delta R(WW)$                & $\Delta R$ between hadronically and leptonically decaying $W$ bosons            &     &     &     &  \checkmark  \\
\bottomrule                                                                                                                     
\end{tabular}
}
\end{center}
\end{ruledtabular}
\end{table*}

\begin{figure*}
\begin{center}
\subfigure{
\includegraphics[width=0.45\linewidth]{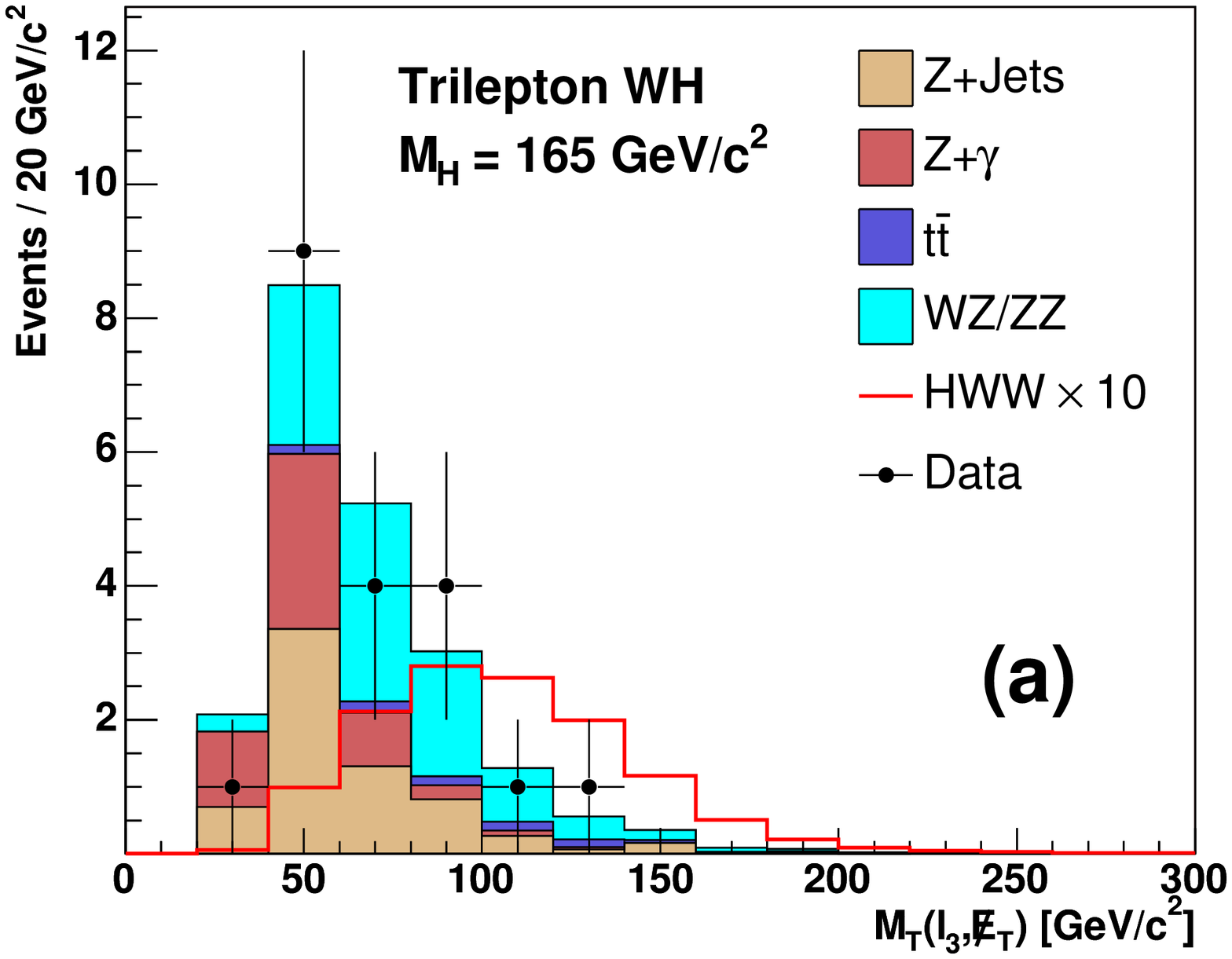}
}
\subfigure{
\includegraphics[width=0.45\linewidth]{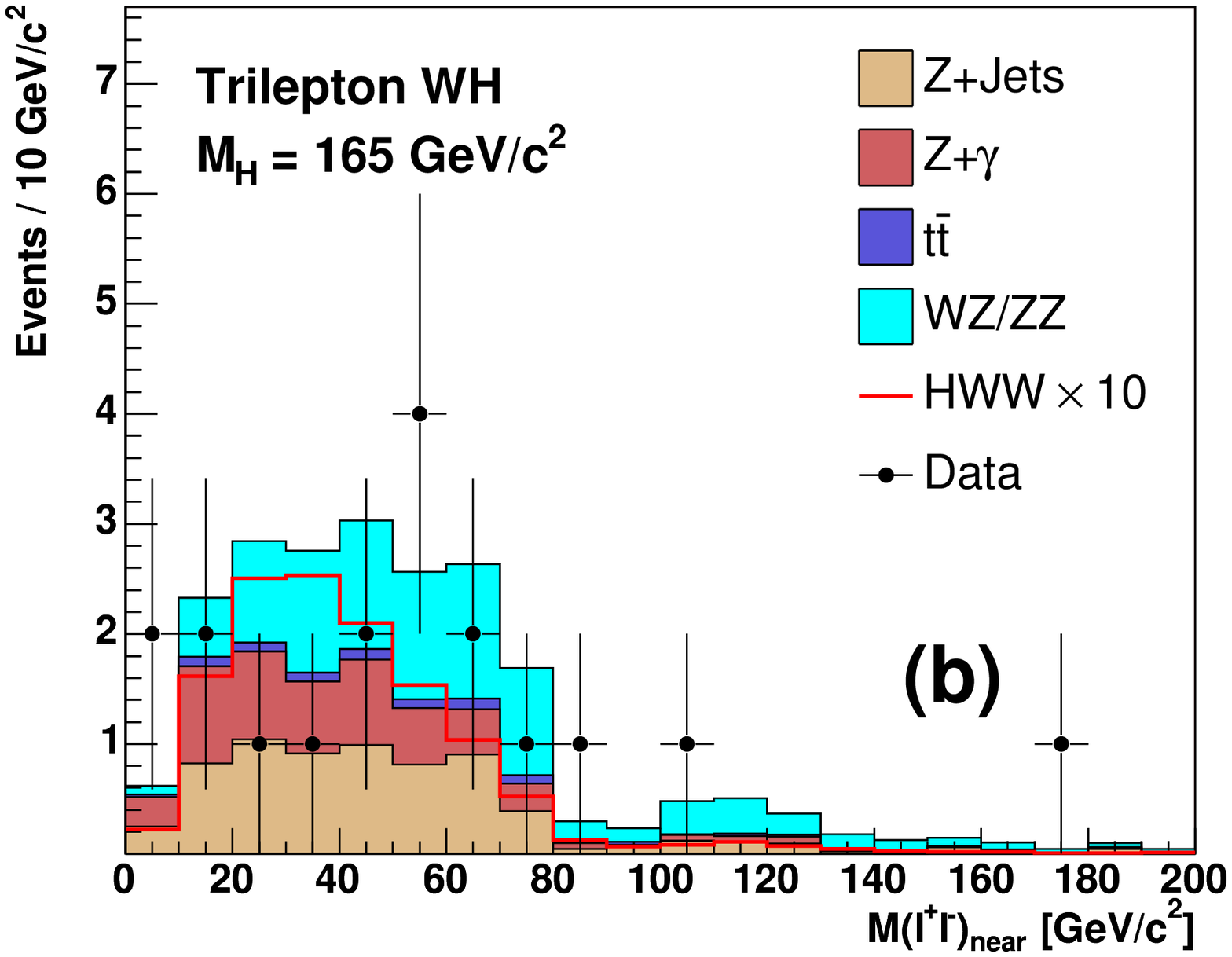}
}
\subfigure{
\includegraphics[width=0.45\linewidth]{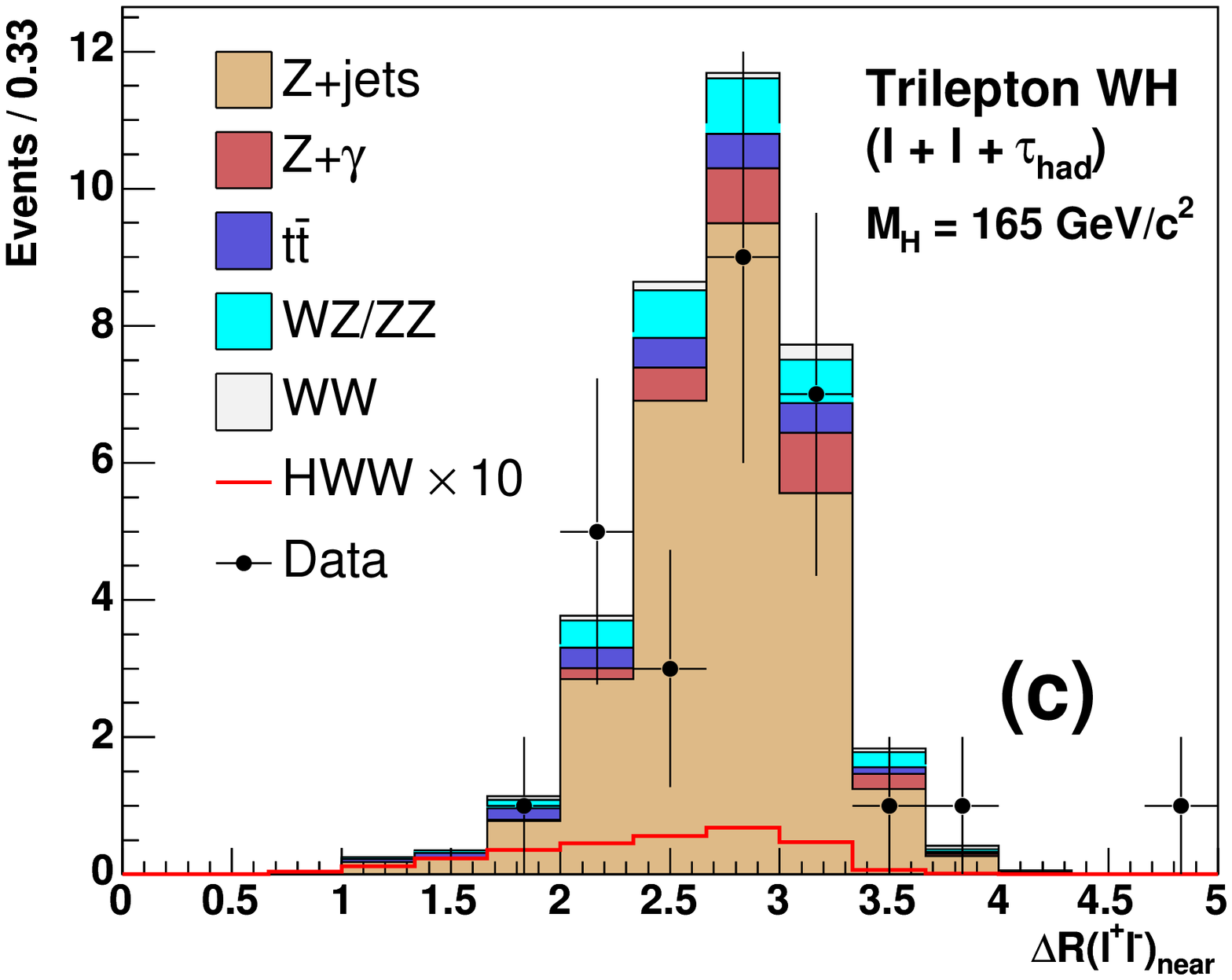}
}
\subfigure{
\includegraphics[width=0.45\linewidth]{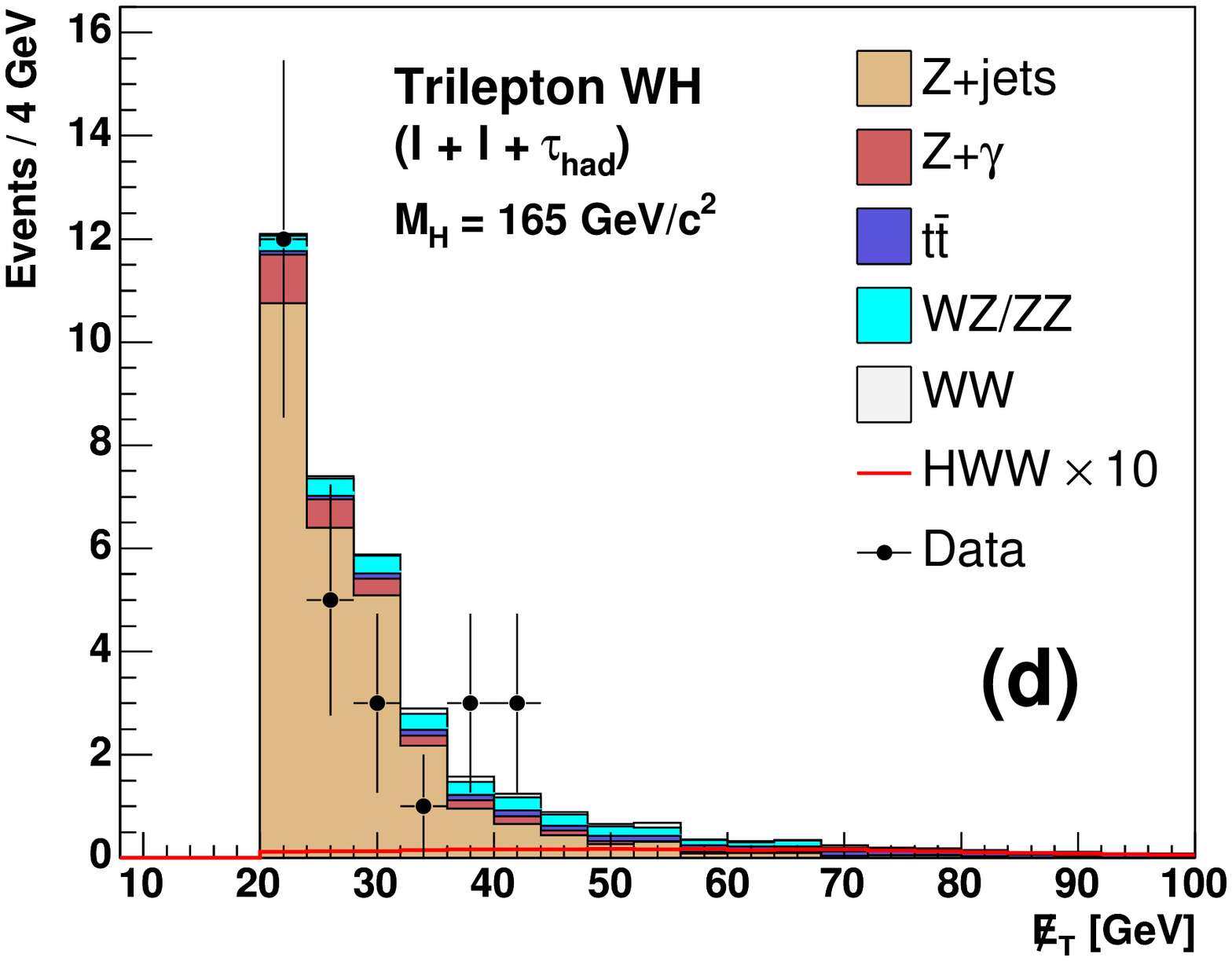}
}
\end{center}
\caption{Predicted and observed distributions of kinematic input variables 
providing the largest separation between potential signal and background 
contributions in the (a,b) Trilepton {\it WH} and (c,d) Trilepton {\it WH} 
($\ell$+$\ell$+$\tau_{\rm{had}}$) search samples.  The overlaid signal 
predictions correspond to the sum of two production modes ({\it WH} and 
{\it ZH}) for a Higgs boson with mass of 165~GeV/$c^2$ and are multiplied 
by a factor of 10 for visibility.  Normalizations for background event 
yields are those obtained from the final fit used to extract search limits.}
\label{fig:TriDiscvarA1}
\end{figure*}

\begin{figure*}
\begin{center}
\subfigure{
\includegraphics[width=0.45\linewidth]{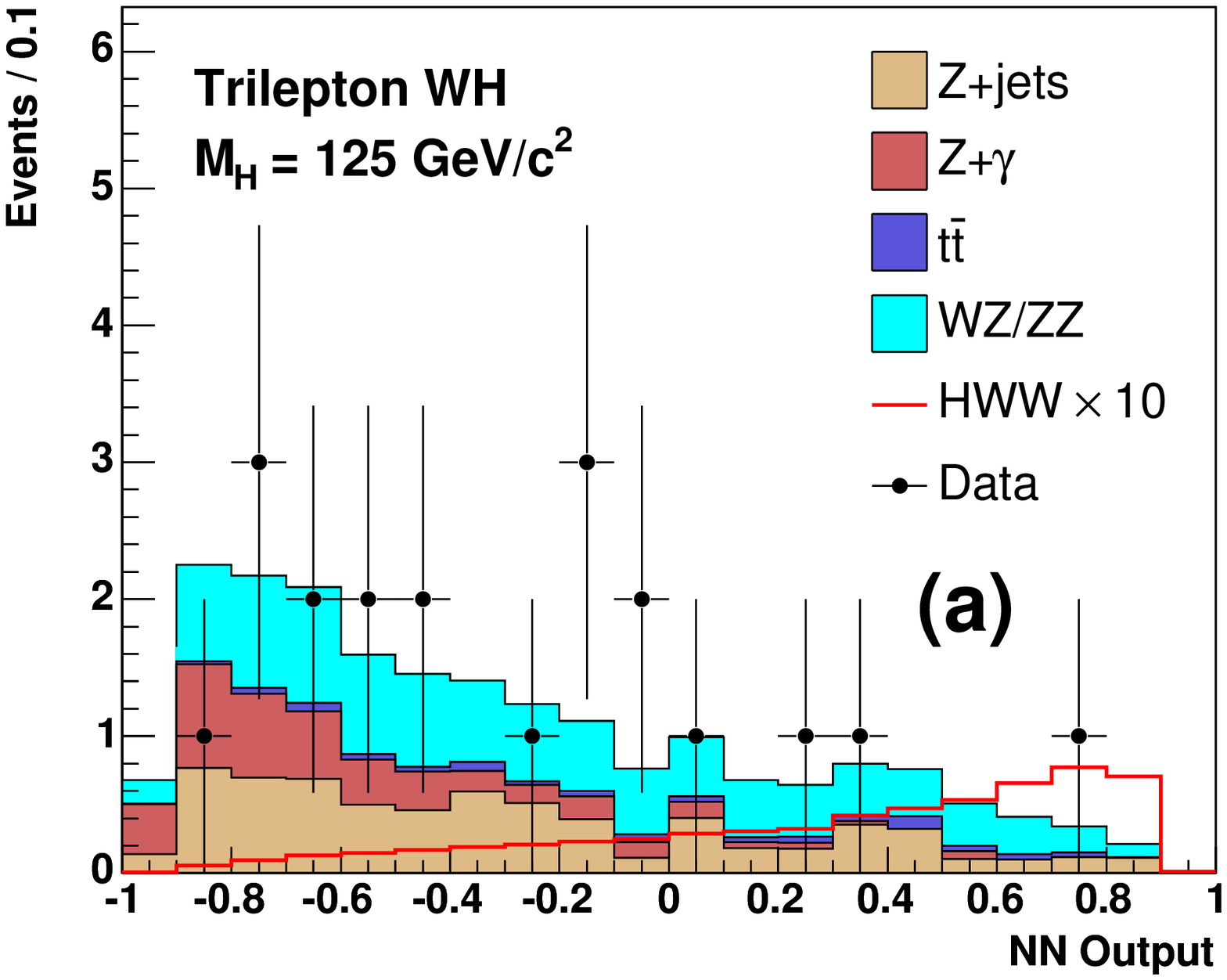}
}
\subfigure{
\includegraphics[width=0.45\linewidth]{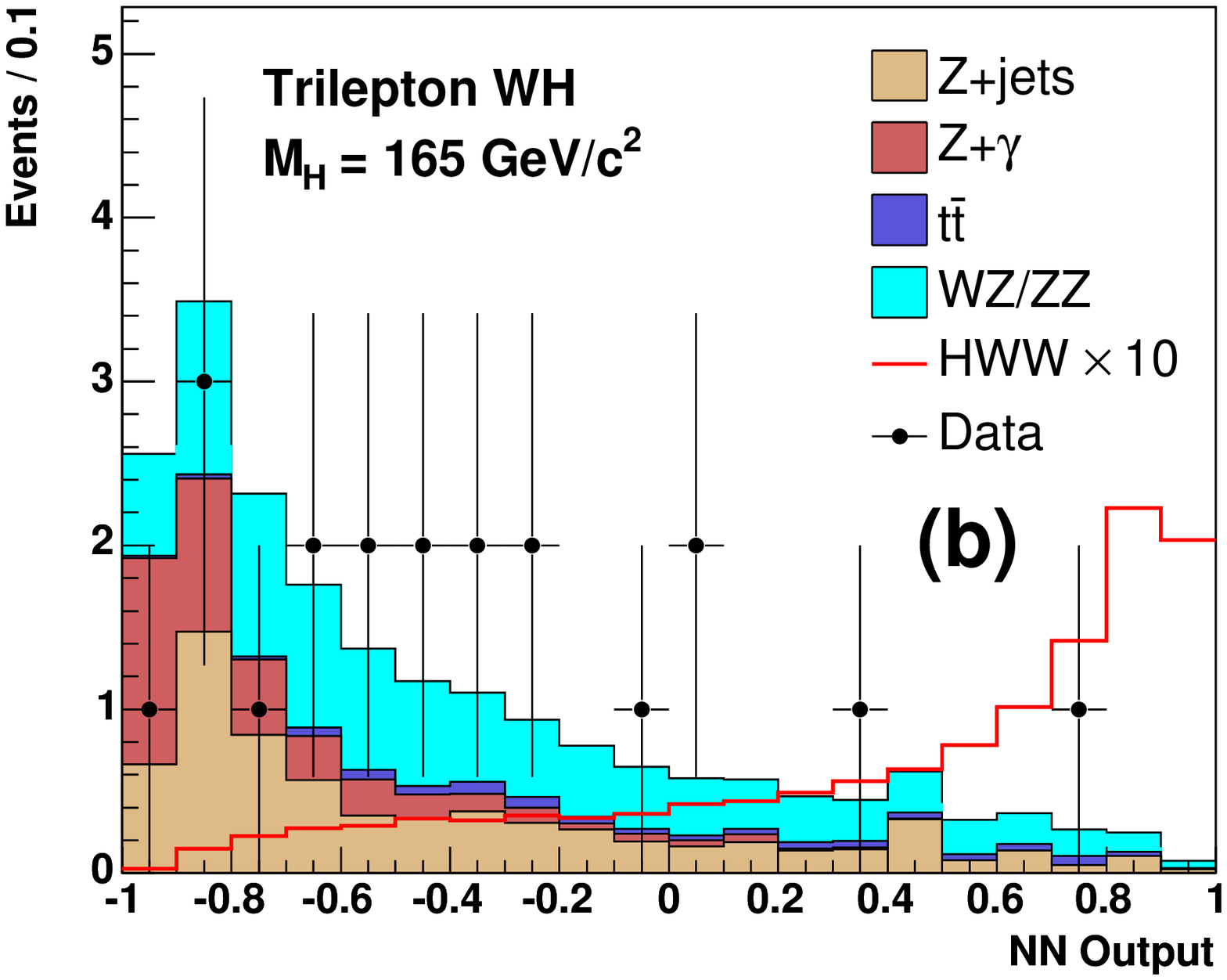}
}
\subfigure{
\includegraphics[width=0.45\linewidth]{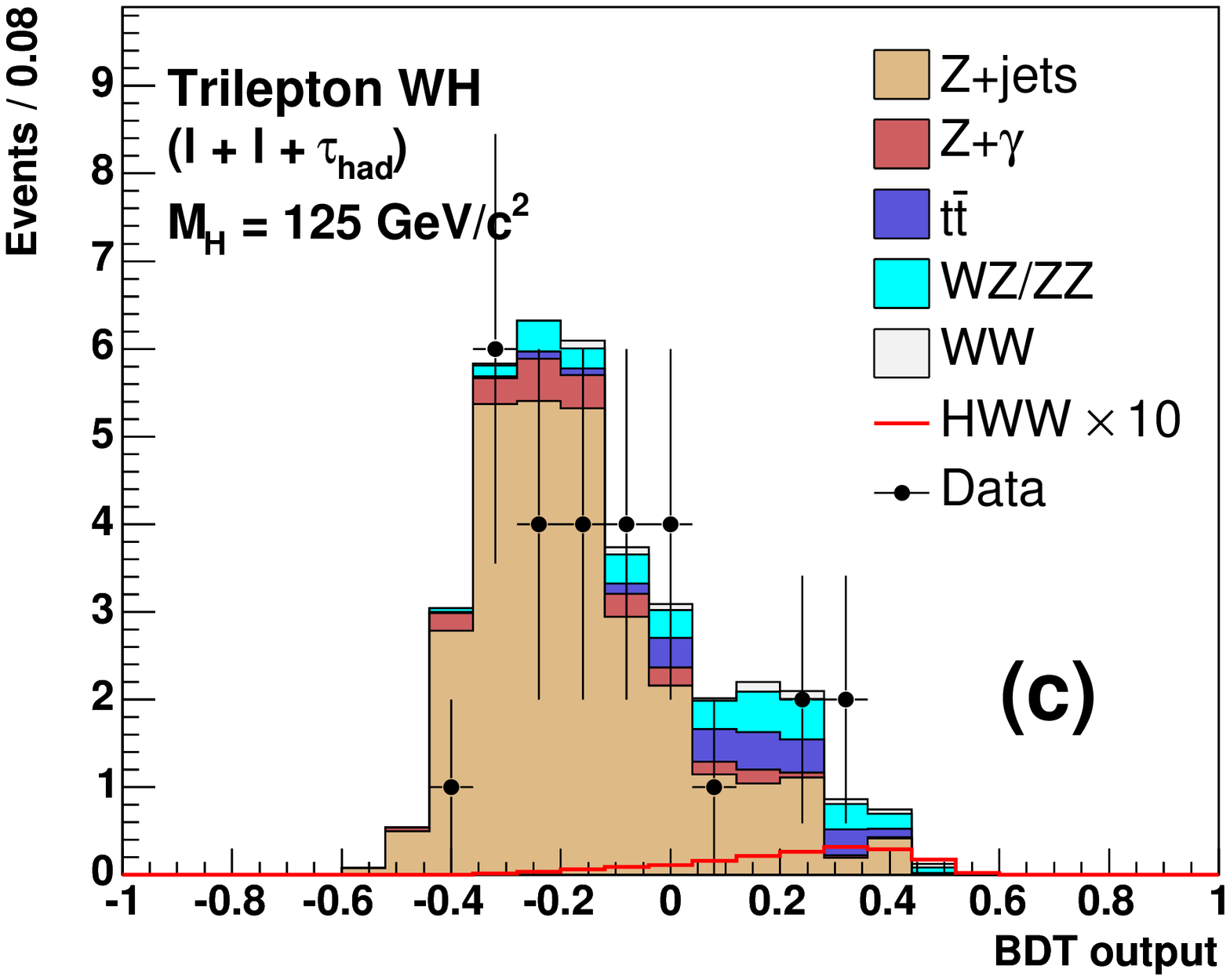}
}
\subfigure{
\includegraphics[width=0.45\linewidth]{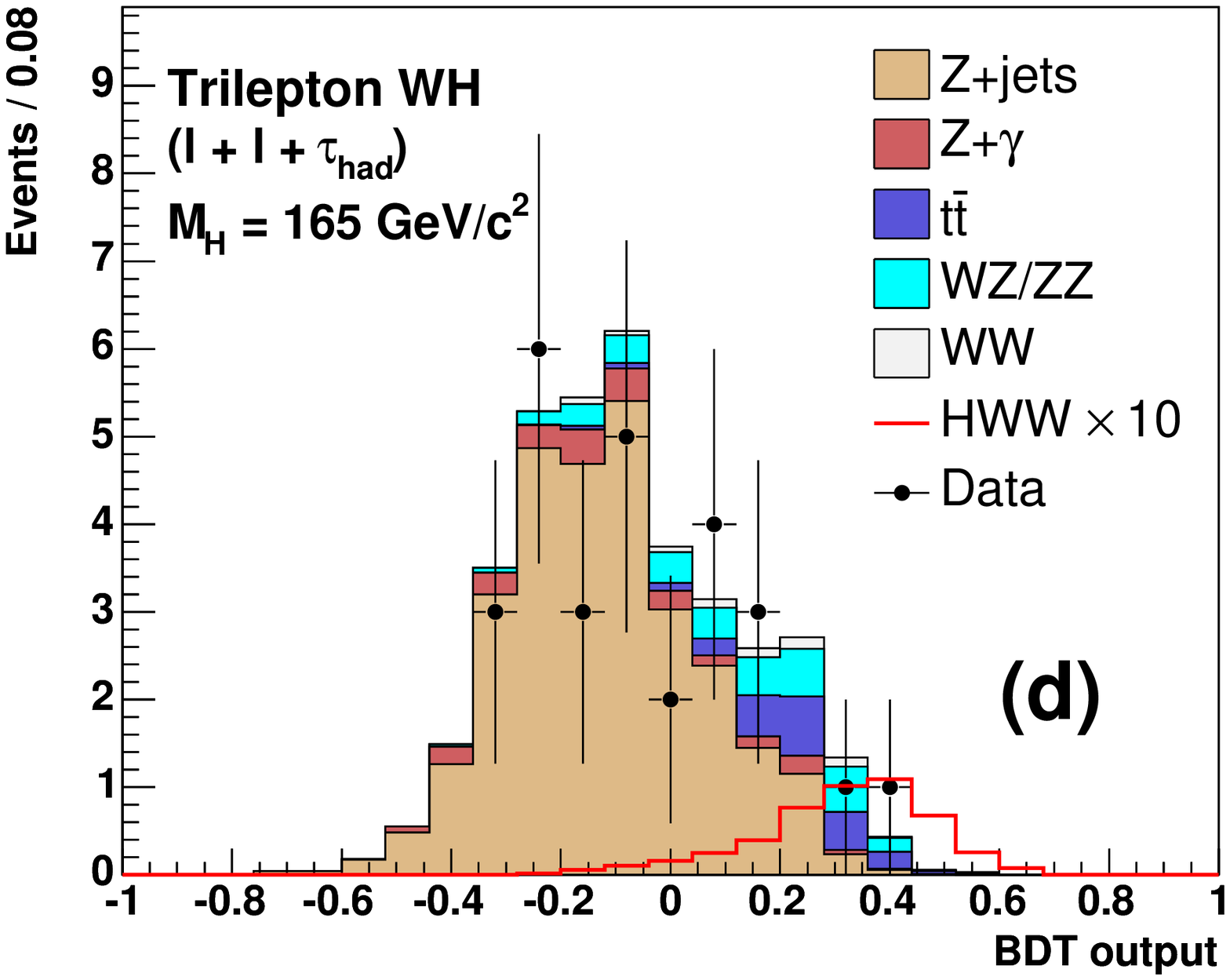}
}
\end{center}
\caption{Predicted and observed distributions of output variables from multivariate 
algorithms trained to separate potential Higgs boson events from background 
contributions in the (a,b) Trilepton {\it WH} and (c,d) Trilepton {\it WH} 
($\ell$+$\ell$+$\tau_{\rm{had}}$) search samples for Higgs boson mass hypotheses 
of 125~and~165~GeV/$c^2$.  The overlaid signal predictions correspond to the sum 
of two production modes ({\it WH} and {\it ZH}) and are multiplied by a factor of 
10 for visibility.  Normalizations for background event yields are those obtained 
from the final fit used to extract search limits.}
\label{fig:TriNNtemplatesA1}
\end{figure*}

In all Trilepton search samples, signal contributions from $ggH$ and 
VBF production are negligible, and we consider potential event yields 
from {\it VH} production only.  For the Trilepton {\it WH} search sample, 
approximately 50\% of background events originate from direct {\it WZ} 
production.  The neural network trained for this sample uses the 
following 14 kinematic variables as inputs: $p_{T}(\ell_2)$,
$\Delta R(\ell^{+}\ell^{-})_{{\rm near}}$, 
$\Delta R(\ell^{+}\ell^{-})_{{\rm far}}$, $M_T$($\ell$,$\ell$,$\ell$), 
$N_{{\rm jets}}$, $\Met$, $\Delta\phi(\ell_2,\Met)$, $M_T$($\ell_3$,$\Met$), 
$M_T$($\ell$,$\ell$,$\ell$,$\Met$,jets), $M$($\ell_3$,$\Met$,jets), 
$M$($\ell_1$,$\ell_2$,$\Met$), $M(\ell^{+}\ell^{-})_{{\rm near}}$, 
$H_T$, and $F(\ell\ell\ell)$.  Distributions of the most discriminating 
among these variables, $\Delta R(\ell^{+}\ell^{-})_{{\rm near}}$ and $\Met$, 
are shown in Figs.~\ref{fig:TriDiscvarA1}(a) and~\ref{fig:TriDiscvarA1}(b).  
The purpose of these variables is to isolate the collinear leptons 
originating from the spin correlations between the two $W$ bosons 
produced in the decay of the spin-0 Higgs boson and the large 
missing transverse energy associated with the neutrinos produced 
in the leptonic decays of three $W$ bosons.  Examples of neural 
network output distributions for this search sample are shown in 
Figs.~\ref{fig:TriNNtemplatesA1}(a) and~\ref{fig:TriNNtemplatesA1}(b) 
for Higgs boson masses of 125 and 165~GeV/$c^2$, respectively.     
       
For the Trilepton {\it WH} ($\ell$+$\ell$+$\tau_{\rm{had}}$) search sample, 
$\approx$80\% of background events originate from $Z$+jets production.  
A BDT is used to combine both kinematic and tau lepton identification 
variables as inputs to the multivariate algorithm.  The 16 kinematic 
variables used as inputs to the BDT algorithm are $p_{T}(\ell_1)$, 
$p_{T}(\ell_2)$, $p_{T}(\ell_3)$, $\Delta R(\ell^{+}\ell^{-})_{{\rm near}}$, 
$\Delta R(\ell^{+}\ell^{-})_{{\rm far}}$, $M_T$($\ell$,$\ell$,$\ell$), 
$M(\ell\ell\ell)$, $\Met$, $\Delta\phi(\ell_2,\Met)$, $\MetSig$, 
$M_T$($\ell_3$,$\Met$), $M_T$($\ell$,$\ell$,$\ell$,$\Met$,jets),  
$M$($\ell_3$,$\Met$,jets), $M$($\ell_1$,$\ell_2$,$\Met$), 
$M(\ell^{+}\ell^{-})_{{\rm near}}$, and $H_T$.  The 11 tau lepton 
identification variables listed in Table~\ref{tbl:Tau Variables} are 
also used.  Distributions of the most discriminating among these 
variables, $\Delta R(\ell^{+}\ell^{-})_{{\rm far}}$ and $\Met$, are shown in 
Figs.~\ref{fig:TriDiscvarA1}(c) and~\ref{fig:TriDiscvarA1}(d).  Examples of 
BDT output distributions are shown in Figs.~\ref{fig:TriNNtemplatesA1}(c) 
and~\ref{fig:TriNNtemplatesA1}(d) for Higgs boson masses of 125 and 
165~GeV/$c^2$, respectively.   

For the Trilepton {\it ZH} search samples, the presence of an opposite-sign dilepton 
pair with a mass consistent with the $Z$ boson mass ensures that potential 
signal contributions originate almost exclusively from {\it ZH} production.  Likewise, 
most background event contributions originate from processes containing a real 
$Z$ boson ($\approx$~50\% from direct {\it WZ} and {\it ZZ} production).  Neural networks 
are trained to separate these background contributions from signal.  Typically, 
one of the $W$ bosons decays hadronically, yielding potentially multiple 
reconstructed jets within each event.  Hence, potential signal contributions in 
the (1 Jet) search sample are smaller than those in the ($\ge$2 Jets) sample.  In 
addition, the possibility of reconstructing all Higgs boson decay products in 
the ($\ge$2 Jets) sample events allows for the full reconstruction of a Higgs boson 
mass, which provides an additional highly-discriminating variable to enhance 
signal-to-background separation. 

\begin{figure*}
\begin{center}
\subfigure{
\includegraphics[width=0.45\linewidth]{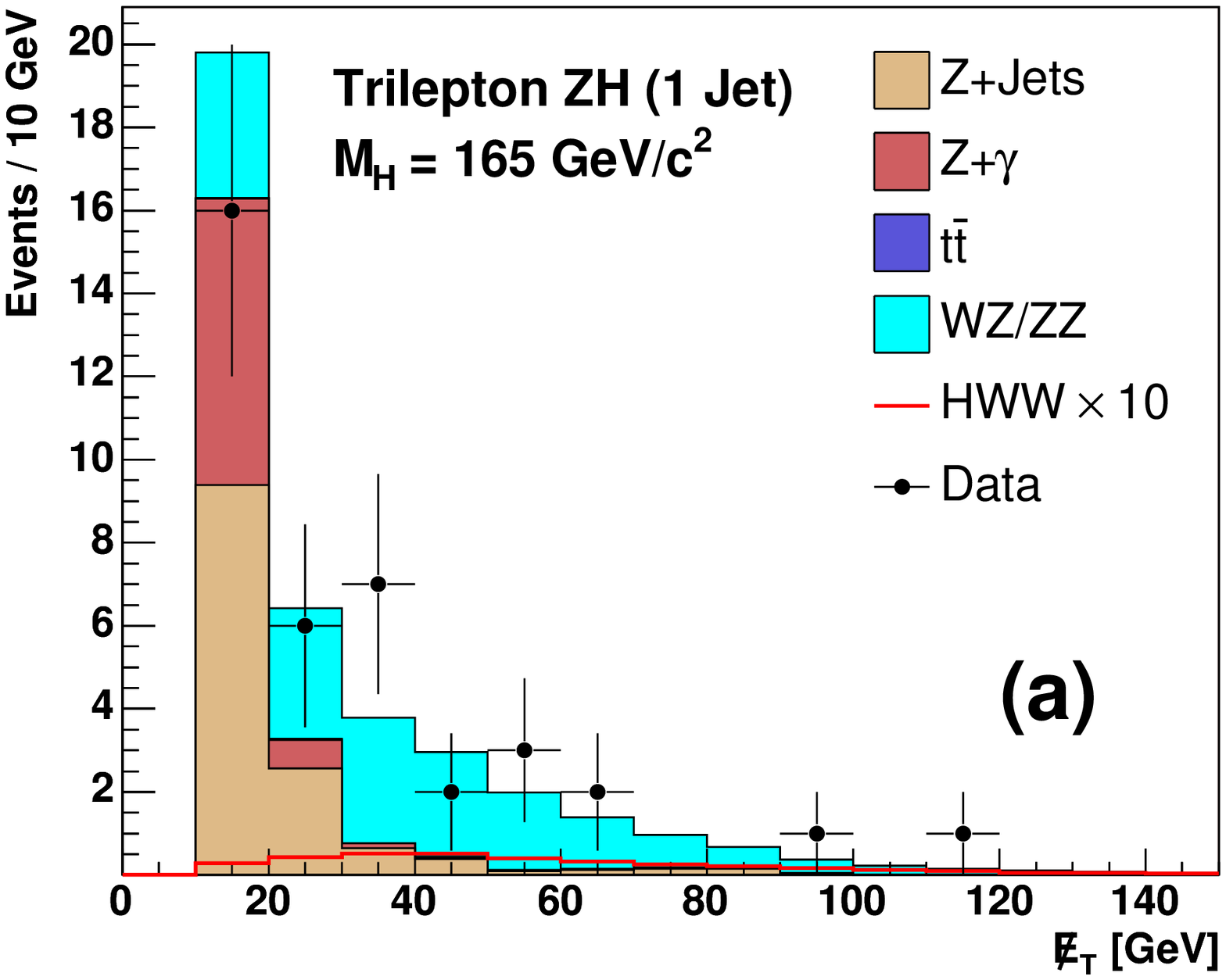}
}
\subfigure{
\includegraphics[width=0.45\linewidth]{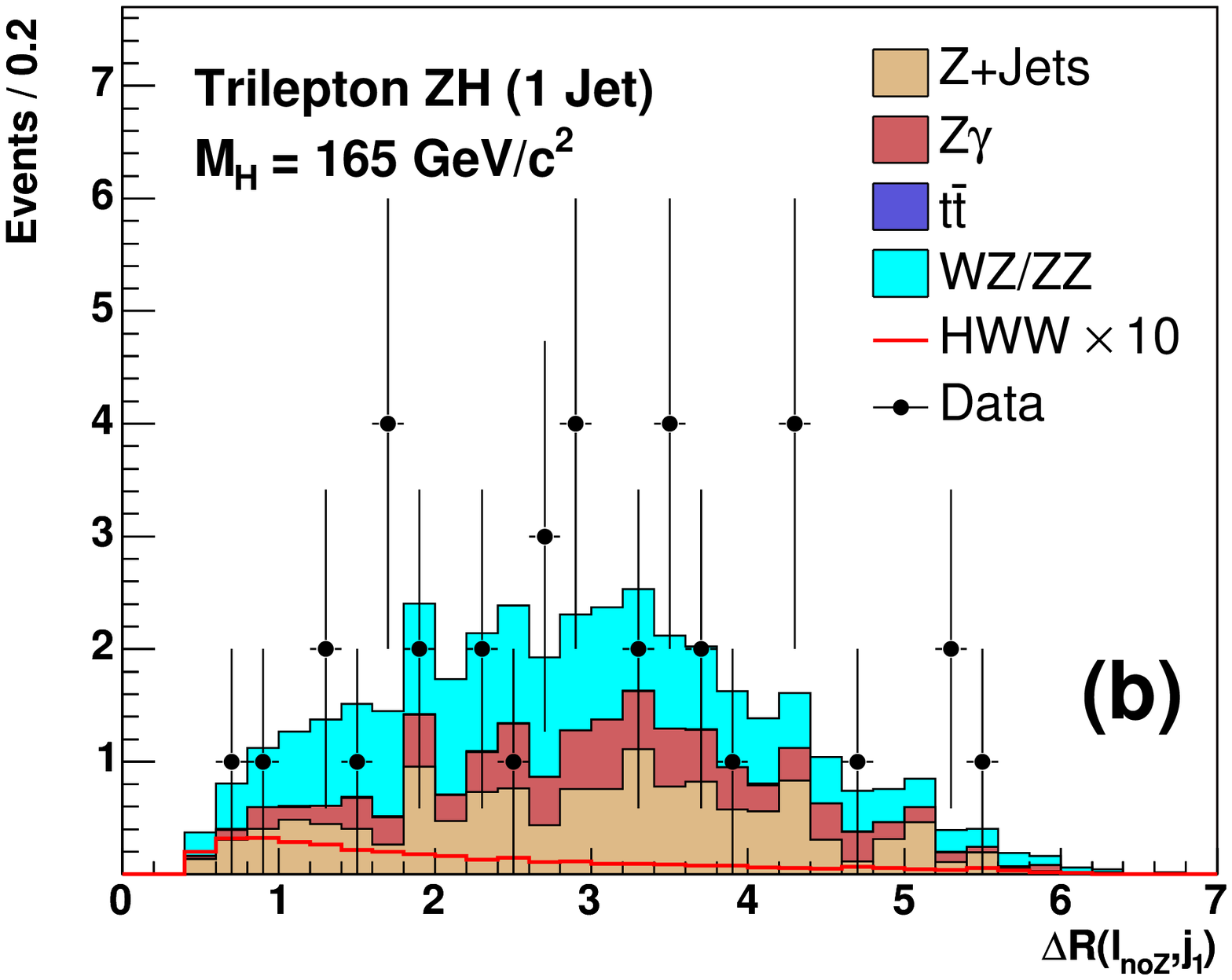}
}
\subfigure{
\includegraphics[width=0.45\linewidth]{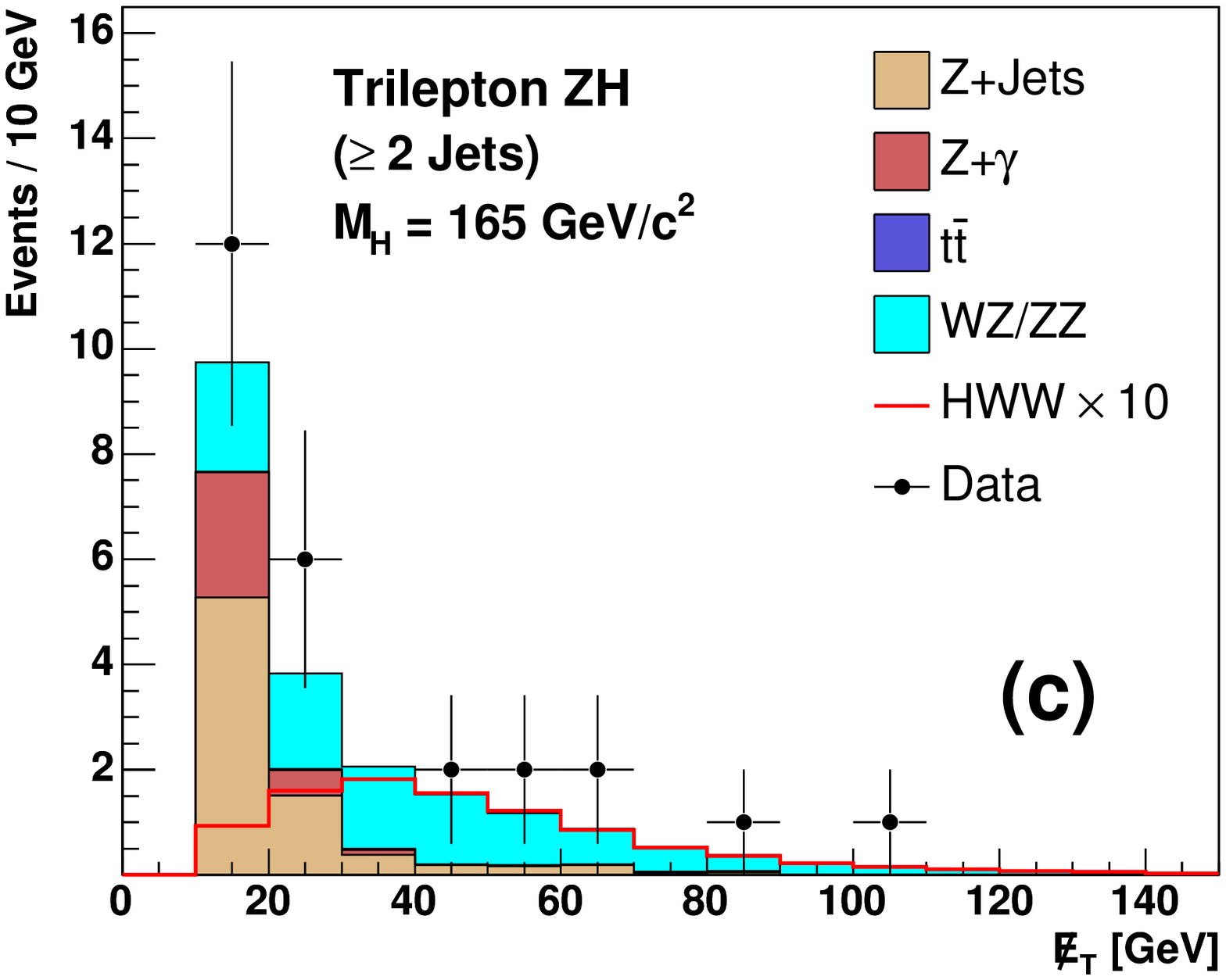}
}
\subfigure{
\includegraphics[width=0.45\linewidth]{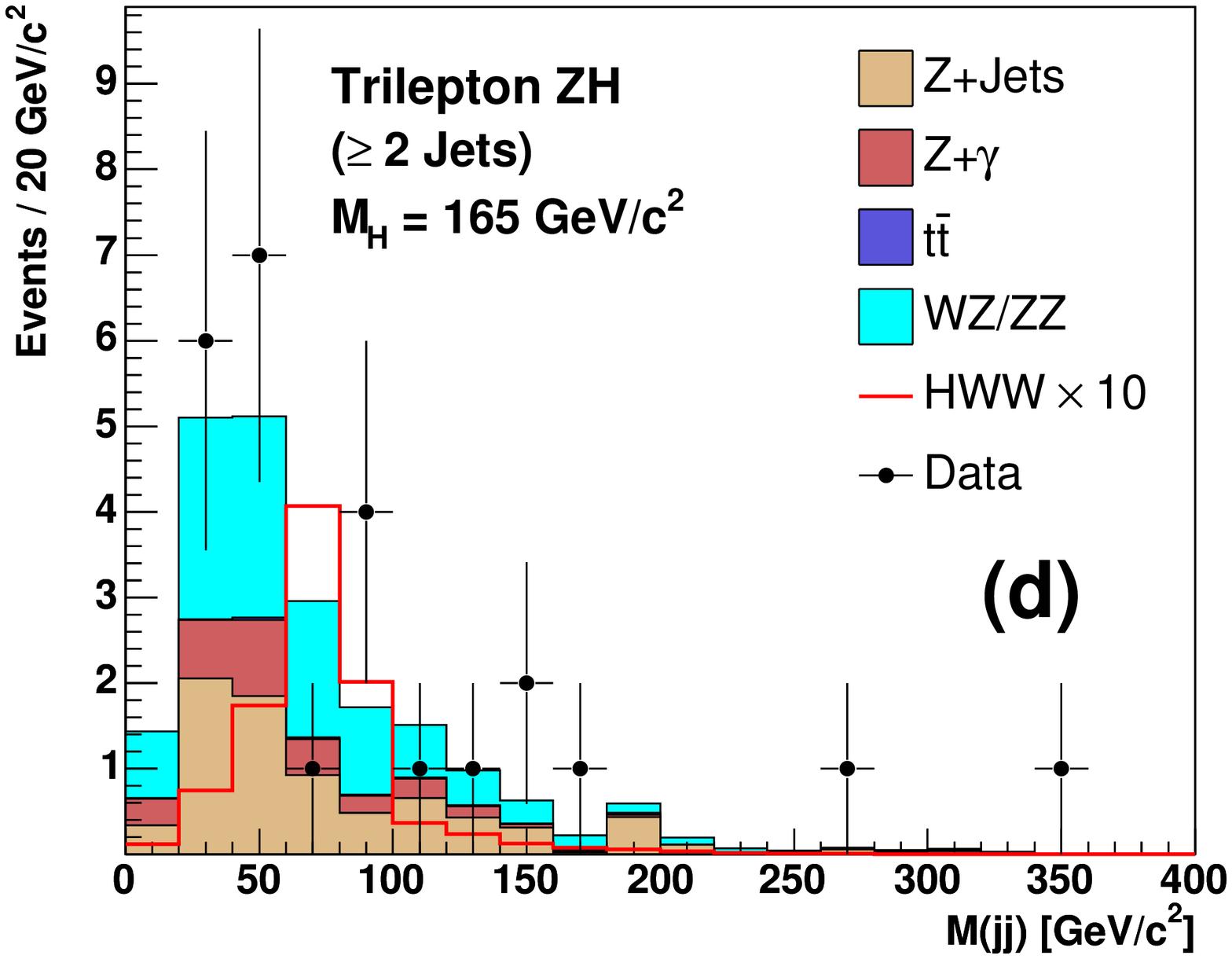}
}
\end{center}
\caption{Predicted and observed distributions of kinematic input variables 
providing the largest separation between potential signal and background 
contributions in the (a,b) Trilepton {\it ZH} (1 Jet) and (c,d) Trilepton 
{\it ZH} ($\ge$2 Jets) search samples.  The overlaid signal predictions 
correspond to the sum of two production modes ({\it WH} and {\it ZH}) for 
a Higgs boson with mass of 165~GeV/$c^2$ and are multiplied by a factor of 
10 for visibility.  Normalizations for background event yields are those 
obtained from the final fit used to extract search limits.}
\label{fig:TriDiscvarA2}
\end{figure*}

\begin{figure*}
\begin{center}
\subfigure{
\includegraphics[width=0.45\linewidth]{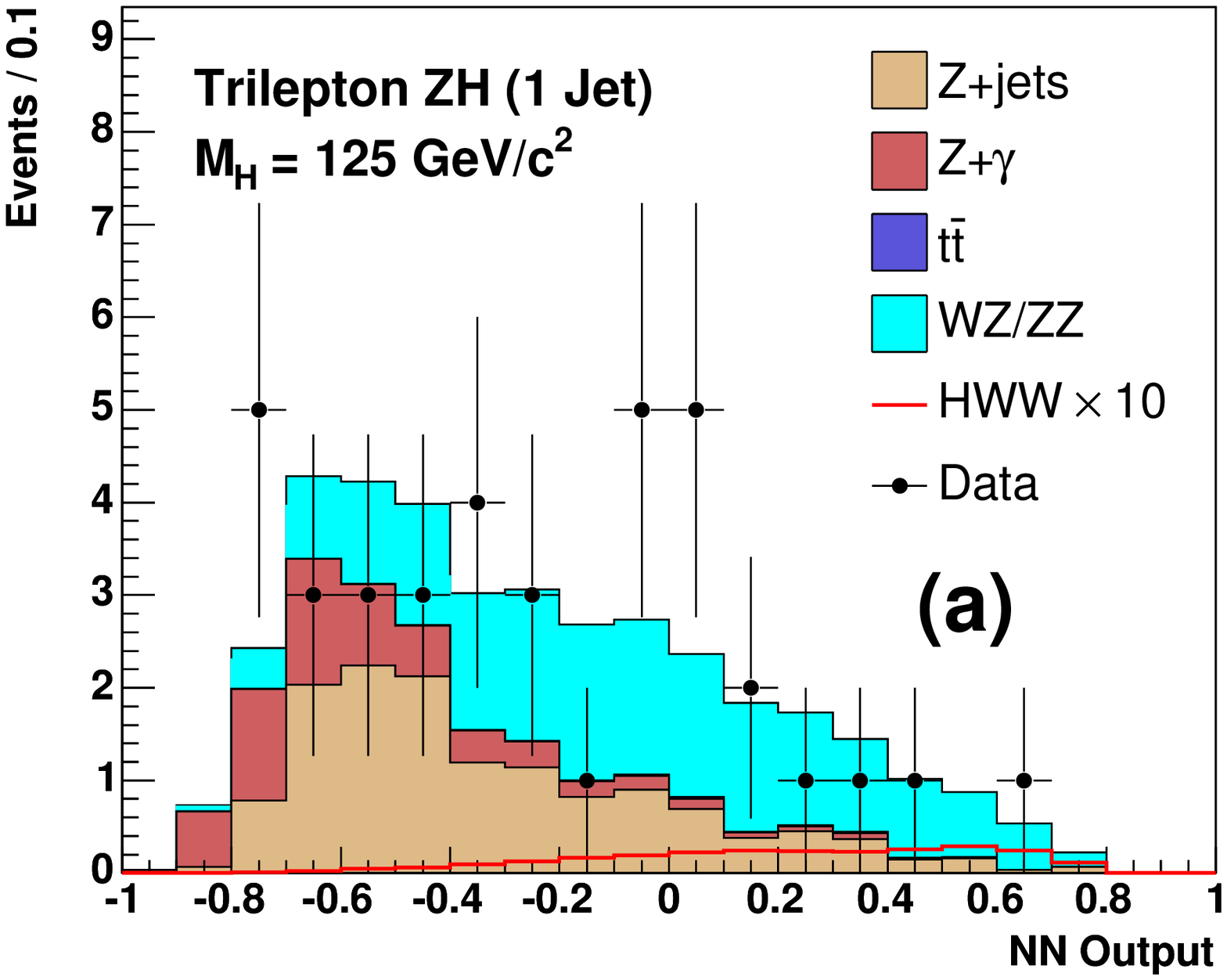}
}
\subfigure{
\includegraphics[width=0.45\linewidth]{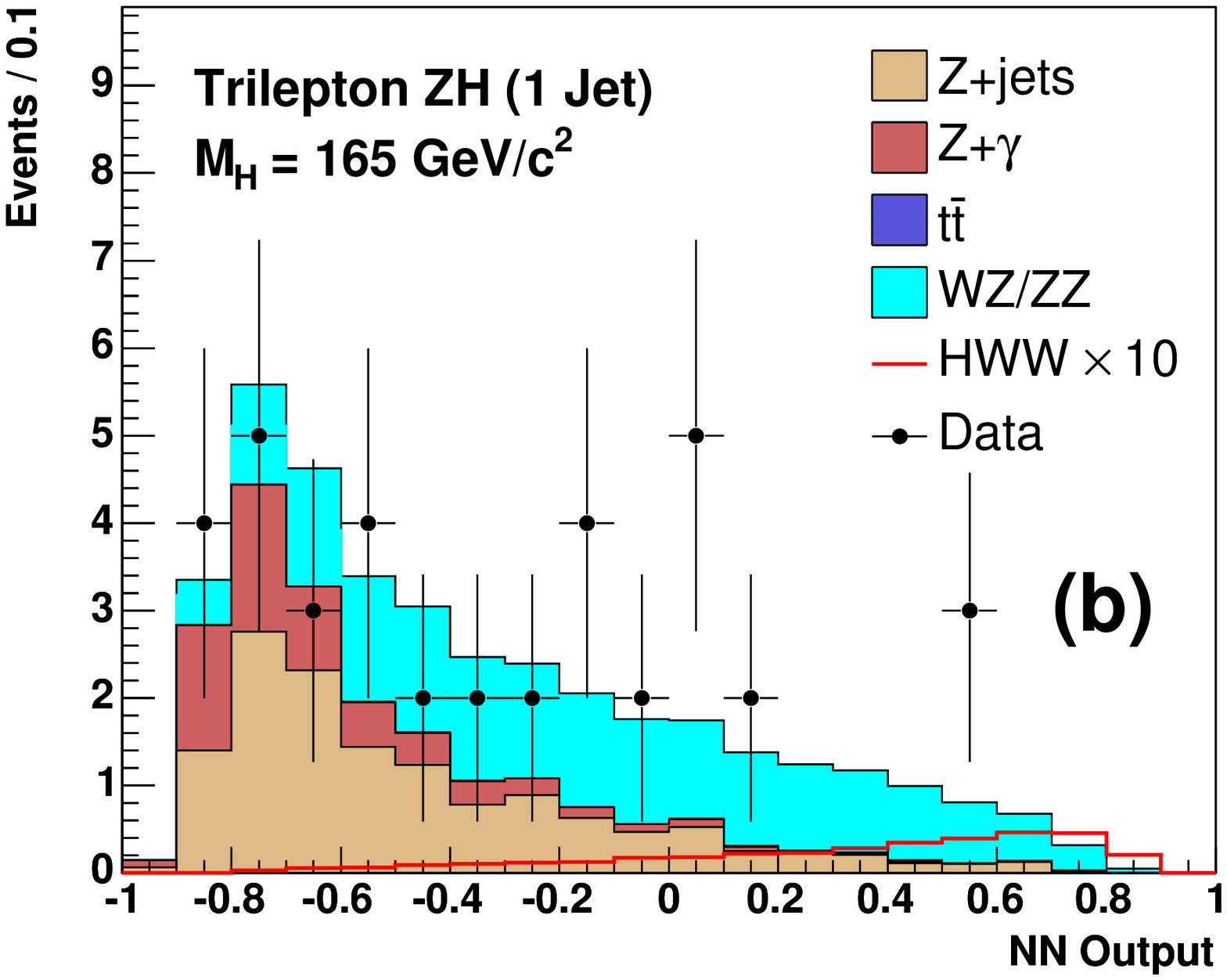}
}
\subfigure{
\includegraphics[width=0.45\linewidth]{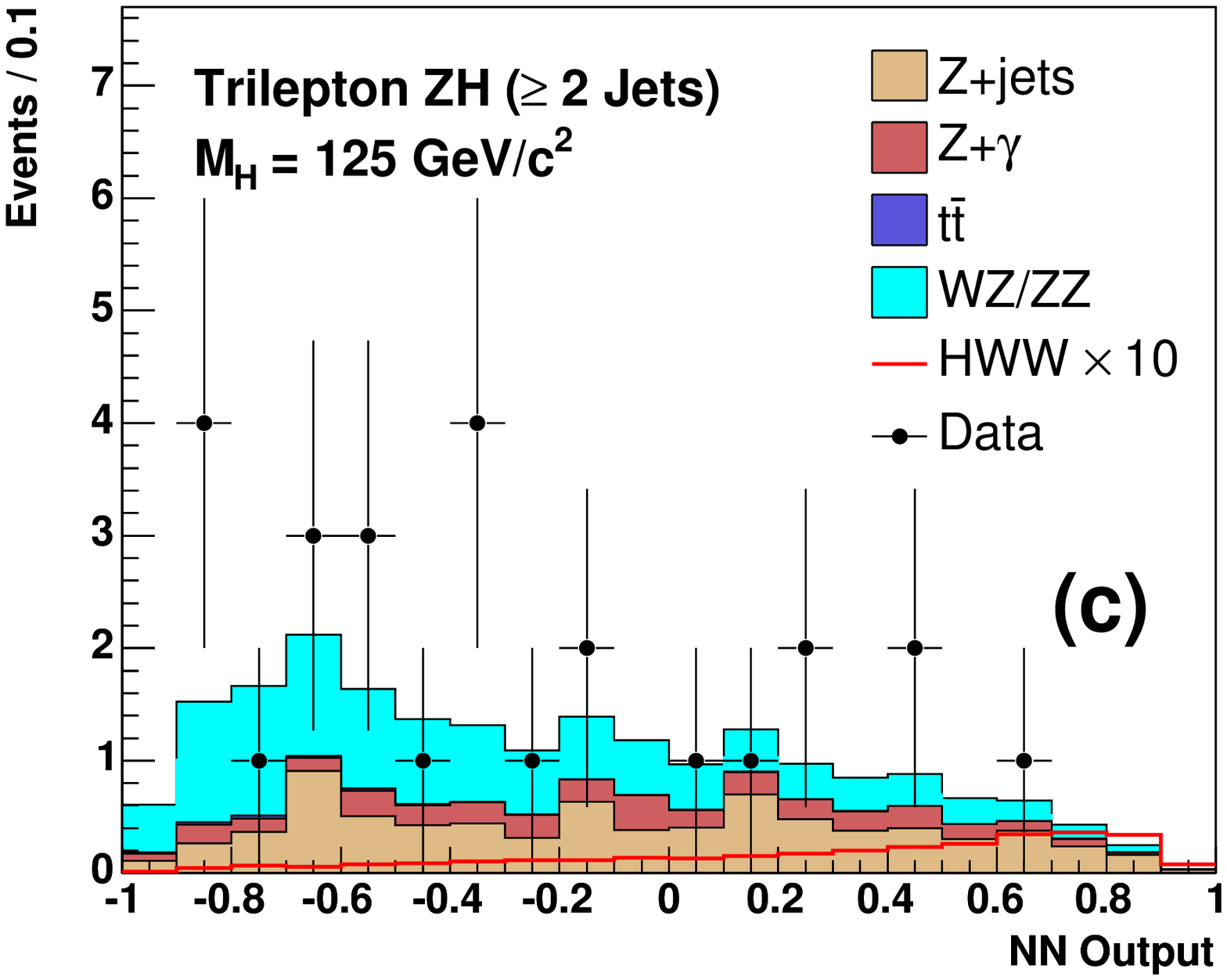}
}
\subfigure{
\includegraphics[width=0.45\linewidth]{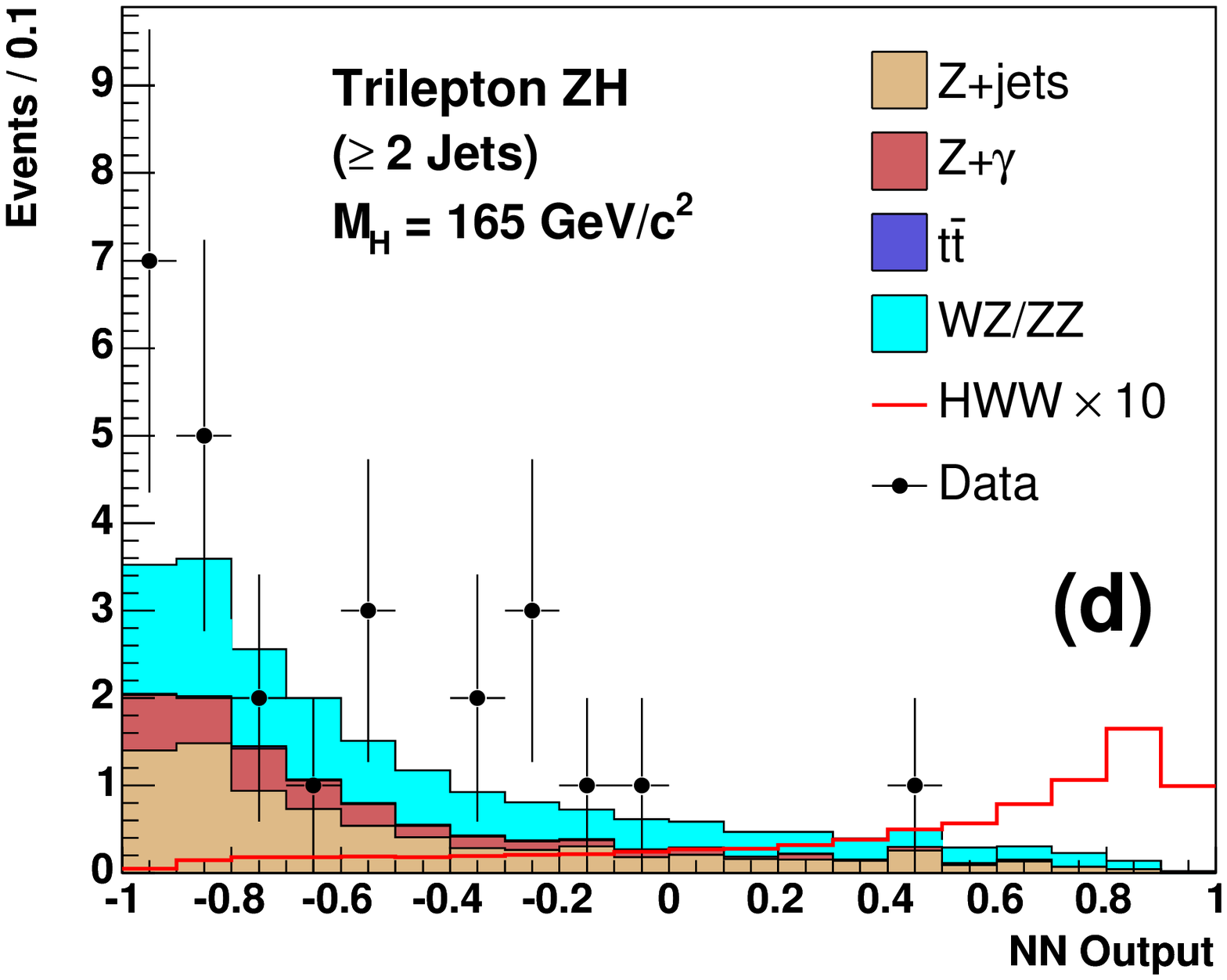}
}
\end{center}
\caption{Predicted and observed distributions of neural network output variables 
for networks trained to separate potential Higgs boson events from background 
contributions in the (a,b) Trilepton {\it ZH} (1 Jet) and (c,d) Trilepton {\it ZH}
($\ge$2 Jets) search samples for Higgs boson mass hypotheses of 125~and~165~GeV/$c^2$.  
The overlaid signal predictions correspond to the sum of two production modes 
({\it WH} and {\it ZH}) and are multiplied by a factor of 10 for visibility.
Normalizations for background event yields are those obtained from the final fit 
used to extract search limits.}
\label{fig:TriNNtemplatesA2}
\end{figure*}

For the Trilepton {\it ZH} (1 Jet) sample, a large number of kinematic variables 
are used as inputs to the neural network to maximally constrain the missing kinematic 
information associated to the unreconstructed jet.  The 16 kinematic input variables 
to the neural network are $\Delta R(\ell^{+}\ell^{-})_{{\rm near}}$, 
$\Delta R(\ell^{+}\ell^{-})_{{\rm far}}$, $M(\ell\ell\ell)$, $E_{T}(j_1)$, 
$\Met$, $\Delta\phi(\ell_1+\ell_2+\ell_3,\Met)$, $\Delta\phi(\ell_2,\Met)$, 
$M_T$($\ell$,$\ell$,$\ell$,$\Met$,jets), $M$($\ell_3$,$\Met$,jets), $H_T$, 
$F(\ell\ell\ell)$, $\Delta\phi(\ell_{{\rm noZ}},\Met)$, $\Delta R(\ell_{{\rm noZ}},j_1)$, 
$M_T$($\ell_{{\rm noZ}}$,$\Met$), $M_T$($\ell_{{\rm noZ}}$,$\Met$,jets), and 
$M(\ell_{{\rm noZ}},\Met)$.  Distributions of the most discriminating among these 
variables, $\Delta R(\ell_{{\rm noZ}},j_1)$ and $\Met$, are shown in 
Figs.~\ref{fig:TriDiscvarA2}(a) and~\ref{fig:TriDiscvarA2}(b).  Examples 
of neural network output distributions for this search sample are shown in 
Figs.~\ref{fig:TriNNtemplatesA2}(a) and~\ref{fig:TriNNtemplatesA2}(b) for 
Higgs boson masses of 125 and 165~GeV/$c^2$, respectively.

Fewer kinematic input variables are required for the neural network used in the 
Trilepton {\it ZH} ($\ge$2 Jets) sample due to the additional discrimination contributed
by variables related to the reconstructed Higgs boson mass.  The following 
10 kinematic variables are used as input to the network: $E_{T}(j_1)$, $E_{T}(j_2)$, 
$M(jj)$, $\Met$, $F(\ell\ell\ell)$, $\Delta R$($\ell_{{\rm noZ}}$,jet)$_{{\rm near}}$, 
$M_T$($\ell_{{\rm noZ}}$,$\Met$), $M_T$($\ell_{{\rm noZ}}$,$\Met$,jets), $M(\ell_{{\rm noZ}},\Met)$, 
and $\Delta R(WW)$.   Distributions of the most discriminating among these 
variables, $\Delta R$($\ell_{{\rm noZ}}$,jet)$_{{\rm near}}$ and $\Met$, are shown in 
Figs.~\ref{fig:TriDiscvarA2}(c) and~\ref{fig:TriDiscvarA2}(d).  Examples of 
neural network output distributions for this search sample are shown in 
Figs.~\ref{fig:TriNNtemplatesA2}(c) and~\ref{fig:TriNNtemplatesA2}(d) for 
Higgs boson masses of 125 and 165~GeV/$c^2$, respectively.

\clearpage

\section{Systematic uncertainties}
\label{sec:syst}

The discriminant output distributions in each of the 13 search samples are 
combined in a single simultaneous fit to determine the Higgs boson signal 
rate.  Inputs to the fit include both {\it rate} uncertainties on expected 
event yields from each specific background and signal process and also 
{\it shape} uncertainties on the expected distribution of events within the 
discriminant outputs for each process.  The treatment of these systematic 
uncertainties in the fit is described in Sec.~\ref{sec:finalresults}.  
The fit procedure does account for correlations between uncertainties 
across the different search samples and the different background and 
signal processes.  Rate and shape uncertainties associated with a common 
source are also treated as correlated. 

\begin{table*}[t]
  \setlength{\extrarowheight}{3pt}
\begin{ruledtabular} 
\begin{center} 
\caption{\label{tbl:backrate1}
Uncertainties on background process event yields for seven dilepton search 
samples formed from electron and muon candidates.  The symbol $^{\ast}$ indicates 
uncertainty sources applied only in the SS ($\ge$1 Jets) search sample.  The symbol
$^{\dagger}$ indicates uncertainty sources applied only in the OS Base ($\ge$2 Jets) 
search sample.  The DY$^{a}$ column corresponds to uncertainties on the untuned 
Monte Carlo models of DY background contributions to the OS Inverse 
$M_{\ell\ell}$, SS ($\ge$1 Jets), and OS Base ($\ge$2 Jets) search samples.  The DY$^{b}$ 
column corresponds to uncertainties on the tuned Monte Carlo model of DY 
background contributions to the OS Base (0 Jet) and OS Base (1 Jet) search samples.} 
%{\scriptsize
\begin{tabular}{l*{8}{c}}
\toprule
Uncertainty source                                  & {\it WW}      & {\it WZ}     & {\it ZZ}     & $t\bar{t}$ & DY$^{a}$ & DY$^{b}$ & $W\gamma$ & $W$+Jets \\
\hline
Theoretical cross section                           & 6\%       & 6\%      & 6\%      & 7\%        & 5\%      &          &           &          \\
Luminosity                                          & 5.9\%     & 5.9\%    & 5.9\%    & 5.9\%      & 5.9\%    &          &           &          \\
Lepton ($e$ or $\mu$) identification efficiency     & 3.8\%     & 3.8\%    & 3.8\%    & 3.8\%      & 3.8\%    &          &           &          \\
Trigger efficiency                                  & 2.0\%     & 2.0\%    & 2.0\%    & 2.0\%      & 2.0\%    &          &           &          \\
Higher-order amplitudes                             & 2.3--17\%  & 10\%     & 10\%     & 10\%       & 10\%     &          & 0--10\%    &          \\
Jet energy scale                                    & 1.2--21\%  & 1.1--13\% & 2.0--13\% & 0.3--28\%   & 4.9--33\% & 6.5--18\% & 1.2--22\%  &          \\
Lepton charge mismeasurement$^{\ast}$               & 25\%      &          &          &            & 25\%     &          &           &          \\
$b$-quark jet veto modeling$^{\dagger}$             &           &          &          & 3.6\%      &          &          &           &          \\
$\Met$ modeling                                     &           &          &          &            &          & 19--21\%  &           &          \\
Photon conversion modeling                          &           &          &          &            &          &          & 6.8--8.4\% &          \\
Jet to lepton ($e$ or $\mu$) misreconstruction rate &           &          &          &            &          &          &           & 14--38\%  \\
\bottomrule
\end{tabular}
%}
\end{center}
\end{ruledtabular}
\end{table*}

Rate uncertainties on the contributing background processes are summarized in
Table~\ref{tbl:backrate1} for the seven dilepton search samples formed from 
electron and muon candidates, Table~\ref{tbl:backrate2} for the additional 
two dilepton search samples formed from one electron or muon candidate and 
one hadronically-decaying tau lepton candidate, and Table~\ref{tbl:backrate3} 
for the four trilepton search samples.  Ranges are used to indicate cases 
where the effect of a specific uncertainty source on the estimated event 
yield for a contributing background process varies across the different 
search samples grouped within the individual tables.  

All estimated event yields obtained directly from the Monte Carlo simulation 
are assigned uncertainties from the theoretical cross section calculation, 
the data luminosity measurement, and the lepton identification and trigger 
efficiency measurements used to normalize the simulated event samples.  
In the case of other simulated background samples, whose normalization is 
obtained from data control samples, these uncertainties are not applicable.

\begin{table*}[t]
  \setlength{\extrarowheight}{3pt}
\begin{ruledtabular} 
\begin{center} 
\caption{\label{tbl:backrate2}
Uncertainties on background process event yields for two dilepton search 
samples formed from one electron or muon candidate and one hadronically-decaying 
tau lepton candidate.}
%{\scriptsize
\begin{tabular}{l*{7}{c}}
\toprule
Uncertainty source                                   & {\it WW}      & {\it WZ}      & {\it ZZ}      & $t\bar{t}$ & DY        & $W\gamma$ & $W$+jets \\
\hline
Theoretical cross section                            & 6\%       & 6\%       & 6\%       & 7\%        & 5\%       &           &          \\
Luminosity                                           & 5.9\%     & 5.9\%     & 5.9\%     & 5.9\%      & 5.9\%     &           &          \\
Lepton ($e$ or $\mu$) identification efficiency      & 2.8\%     & 2.8\%     & 2.8\%     & 2.8\%      & 2.8\%     &           &          \\
Lepton ($\tau$) identification efficiency            & 1.3\%     & 1.3\%     & 1.3\%     & 2.0\%      & 3.3--3.5\% &           &          \\
Trigger efficiency                                   & 2.0\%     & 2.0\%     & 2.0\%     & 2.0\%      & 2.0\%     &           &          \\
Higher-order amplitudes                              & 10\%      & 10\%      & 10\%      & 10\%       & 10\%      & 10\%      &          \\
Lepton ($e$ or $\mu$) to lepton ($\tau$) misreconstruction rate  & 0.1--0.2\% & 0.1--0.2\% & 0.1--0.2\% & 0.1--0.2\%  & 2.1--2.3\% & 1.2--2.1\% &          \\
Photon conversion modeling                           &           &           &           &            &           & 6.8\%     &          \\
$V$+jets control region normalization                &           &           &           &            &           &           & 12.1\%   \\
Jet to lepton ($\tau$) misreconstruction rate        & 5.8\%     & 5.8\%     & 5.8\%     & 4.4--5.1\%  & 0.1--0.2\% &           & 8.8\%    \\
\bottomrule
\end{tabular}
%}
\end{center}
\end{ruledtabular}
\end{table*}

Theoretical diboson production cross sections are taken from {\sc mcfm}~\cite{Campbell:1999ah} 
with a renormalization scale of $\mu_{0} = M^{2}_{V} + p^{2}_{T}(V)$, where $M_V$ is 
the boson mass, and the MSTW2008~\cite{Martin:2009iq} PDF set.  Calculations of 
{\it WZ} and {\it ZZ} production rates necessarily include contributions from $\gamma^{\ast} 
\rightarrow \ell^{+} \ell^{-}$ processes, where the invariant dilepton mass from the 
neutral current exchange is restricted to the range 75~$< m_{\ell^{+}\ell^{-}} 
<$~105~GeV/$c^2$.  The calculated cross sections are 11.34~pb for {\it WW} production, 
3.22~pb for {\it WZ} production, and 1.20~pb for {\it ZZ} production.  We assign a 
6\% uncertainty based on the effects of different scale choices and the application 
of MSTW2008 PDF uncertainties on the calculations.  For $t\bar{t}$ production we 
assign a cross section of 7.04~pb~\cite{ref::ttbar_xsec}, based on a top-quark mass of 
173.1~$\pm$~1.2~GeV/$c^2$ and the MSTW2008NNLO PDF set, yielding an uncertainty of 7\%.  
Similarly, for DY production we rely on a NLO cross section calculation~\cite{dy_xsec},
yielding a central value of 251.3~pb with 5\% uncertainty.  In the case of $Z\gamma$ 
production, simulated samples are generated using specific requirements on the minimum 
$p_T$ of the photon and the minimum separation between the photon and the leptons 
originating from the decay of the $Z$ boson.  Because the production cross section 
depends significantly on these requirements, we use the cross section determined by 
the LO generator to normalize the event sample and assign a larger 10\% uncertainty.          
      
\begin{table*}[t]
  \setlength{\extrarowheight}{3pt}
\begin{ruledtabular} 
\begin{center} 
\caption{\label{tbl:backrate3}
Uncertainties on background process event yields for four trilepton search 
samples.  The symbol $^{\ddagger}$ indicates uncertainty sources applied only in 
the Trilepton {\it WH} ($\ell$+$\ell$+$\tau_{\rm{had}}$) search sample.  The $Z$+jets$^{c}$ 
column corresponds to uncertainties on the tuned Monte Carlo model of $Z$+jets 
background contributions to the Trilepton {\it WH} ($\ell$+$\ell$+$\tau_{\rm{had}}$) search 
sample.  The $Z$+jets$^{d}$ column corresponds to uncertainties on the data-driven
model of $Z$+jets background contributions to the remaining three trilepton search 
samples.}
%{\scriptsize
\begin{tabular}{l*{6}{c}}
\toprule
Uncertainty source                                               & {\it WZ}      & {\it ZZ}      & $t\bar{t}$ & $Z\gamma$ & $Z$+jets$^{c}$ & $Z$+jets$^{d}$ \\
\hline
Theoretical cross section                                        & 6\%       & 6\%       & 7\%        & 10\%      &                &                \\
Luminosity                                                       & 5.9\%     & 5.9\%     & 5.9\%      & 5.9\%     &                &                \\
Lepton ($e$ or $\mu$) identification efficiency                  & 3.8--5.0\% & 3.8--5.0\% & 3.8--5.0\%  & 3.8--5.0\% &                &                \\
Lepton ($\tau$) identification efficiency$^{\ddagger}$           & 2.1\%     & 2.1\%     & 1.1\%      & 1.4\%     &                &                \\
Trigger efficiency                                               & 2.0\%     & 2.0\%     & 2.0\%      & 2.0\%     &                &                \\
Higher-order amplitudes                                          & 10\%      & 10\%      & 10\%       & 15\%      & 10\%           &                \\
Jet energy scale                                                 & 0--18\%    & 0--15\%    & 0--2.3\%    & 2.7--17\%  &                &                \\
Modeling of leptons from $b$-quark jets                          &           &           & 22--42\%    &           &                &                \\
Lepton ($e$ or $\mu$) to lepton ($\tau$) misreconstruction rate$^{\ddagger}$ & 0.5\%     & 0.5\%     & 0.2\%      & 0.7\%     & 0.3\%          &                \\
$V$+jets control region normalization                            &           &           &            &           & 12.1\%         &                \\
Jet to lepton ($e$ or $\mu$) misreconstruction rate              &           &           &            &           &                & 18--24\%        \\
Jet to lepton ($\tau$) misreconstruction rate$^{\ddagger}$       & 4.5\%     & 4.5\%     & 5.1\%      & 0.1\%     & 6.5\%          &                \\
\bottomrule
\end{tabular}
%}
\end{center}
\end{ruledtabular}
\end{table*}

The uncertainty in the measured luminosity is $\pm$~5.9\%, of which 4.4\% comes 
from detector acceptance and operation of the luminosity monitor and 4.0\% 
comes from uncertainty on the inelastic $p\bar{p}$ cross section~\cite{lumi_xs}.
Electron and muon identification efficiencies are measured from trigger-unbiased
final state leptons reconstructed in $Z\rightarrow\ell^{+}\ell^{-}$ decays 
collected with single lepton triggers, and associated uncertainties originate 
from the limited statistical power of these samples.  The lepton-identification 
uncertainty applied to specific search samples depends on the required number 
of reconstructed leptons in each event.  Tau lepton identification efficiencies 
are measured from the OS Hadronic Tau ($\mu + \tau_{\rm{had}}$, low $\Met$, 
low $\Delta\phi(\vec{p}_T(\ell),\VecMet)$) control sample with associated 
uncertainty due to the limited sample size and subtraction of non-DY background 
contributions.  Single-lepton trigger efficiencies are also measured from the 
trigger-unbiased final state lepton in $Z\rightarrow\ell^{+}\ell^{-}$ decays 
collected with single lepton triggers, and uncertainties originate from the 
limited sample size.

Acceptance uncertainties originate from approximations employed within the signal
and background process generators and mismodeling in the detector simulation.  
To account for the potential acceptance effects of higher-order amplitudes
not incorporated in event generators, additional rate uncertainties are included 
on the predicted event yields.  For samples generated with {\sc pythia},
we assign an uncertainty of 10\%, which is the observed acceptance difference 
obtained from {\it WW} event samples generated at LO with {\sc pythia} and at NLO 
using the {\sc mc@nlo}~\cite{Frixione:2002ik} program.  In the specific case 
of {\it WW} production, we use {\sc pythia} to model observed differences in the 
{\it WW} $p_T$ spectrum, when applying harder and softer fragmentation scales in 
the parton shower algorithms used for modeling higher-order effects.  Events 
from the simulated {\sc mc@nlo} {\it WW} event sample are reweighted as a function 
of {\it WW} $p_T$ to match the changes in the spectra obtained from increasing or
decreasing the size of the fragmentation scales, and uncertainties are assigned 
based on changes in acceptance resulting from these reweightings.  Normalization of 
the simulated $W\gamma$ event samples is obtained from a control sample containing 
SS dileptons with invariant mass $M_{\ell\ell} <$~16~GeV/$c^2$.  Because modeling 
of higher-order amplitudes can affect the extrapolation of this normalization to 
predicted $W\gamma$ event yields for the search samples containing dileptons 
with $M_{\ell\ell} >$~16~GeV/$c^2$, the 10\% rate uncertainty is retained for 
these cases.  Because the simulated $Z\gamma$ event sample is generated with 
an incomplete luminosity profile, we assign a slightly higher 15\% uncertainty.

Event yields obtained from simulated event samples also have uncertainties
associated with mismodelings in the detector simulation.  We vary the energy 
scale of reconstructed jets in simulated events within an uncertainty range 
determined from $p_T$ balancing studies performed on $\gamma^{\ast}/Z$ plus 
one-jet events in data and simulation.  The resulting differences in predicted 
event yields are taken as additional rate uncertainties.  Since search samples 
are typically defined by the number of reconstructed jets within each event, 
changes to the jet energy scale can result in simulated events moving from 
one search sample to another.  Hence, correlations and anti-correlations are   
included in the jet energy scale uncertainties applied across the different 
search samples.  Modeling of lepton charge mismeasurement rates has a 
significant impact on predicted background event yields only in the SS 
($\ge$1 Jets) search sample.  Uncertainties are obtained from a comparison of 
the predicted and observed numbers of SS candidate events contained in an 
inclusive DY control sample.

Other uncertainties related to the detector simulation include modeling 
of the $b$-quark jet tagging algorithm used for vetoing events in the 
OS Base ($\ge$2 Jets) search sample and modeling of isolated lepton candidates 
from $b$-quark decays in the Trilepton search samples.  These rate 
uncertainties apply only to background predictions for $t\bar{t}$ production, 
for which resulting events necessarily contain two $b$-quark jets.  As 
discussed in Sec.~\ref{sec:BGmodeling}, scale factors are applied to 
simulated events with jets identified as originating from bottom quarks 
to account for differences in tagging algorithm performance between data 
and Monte Carlo and the small subset of data events, for which silicon 
tracking detector information is not available.  Uncertainties associated 
with these scale factors come primarily from the limited size of the data 
samples used to estimate them.

For simulated samples normalized to the observed event rate in a specific 
data control sample, we assign rate uncertainties based on the limited 
control-sample size and subtraction of residual background contributions.
The scale factors applied to $W\gamma$ simulated event samples to account 
for uncertainties in photon-conversion modeling is obtained from the 
SS Inverse $M_{\ell\ell}$ control sample.  The normalization applied to
simulated $W$+jet and $Z$+jet event samples, which are used for modeling 
contributions of these processes to the OS Hadronic Tau and Trilepton 
WH ($\ell$+$\ell$+$\tau_{\rm{had}}$) search samples, is obtained from the OS
Hadronic Tau (high $\Delta\phi(\vec{p}_T(\tau),\vec{p}_T(\ell))$) control
sample.  The construction of the {\sc pythia} sample tuned to model DY 
contributions in the OS Base (0 Jet) and OS Base (1 Jet) search samples 
is described in Sec.~\ref{sec:BGmodeling}.  The $\Met$ in each simulated 
event is shifted to account for effects of multiple interactions and the 
resulting sample is normalized to event counts in data obtained from the 
OS Base (Intermediate $\MetSpec$) control sample.  Uncertainties from 
$\Met$ modeling applied to the corresponding event yield predictions are 
obtained through additional $\pm2$~GeV shifts with respect to the nominal 
$\Met$ correction and renormalization of the retuned event samples.

The data-driven procedure for modeling $W$+jet and $Z$+jet contributions 
to search samples that do not incorporate hadronically-decaying tau lepton
candidates is also described in Sec.~\ref{sec:BGmodeling}.  Jet-to-lepton 
misidentification rates are measured in inclusive jet samples collected 
using single jet triggers and applied as weights to events containing 
both reconstructed leptons and jets.  Differences in the measured jet 
misidentification rate from event samples collected with varied $E_T$ 
thresholds are observed due to changes in the relative contributions 
of quark and gluon jets in these samples.  Rate uncertainties on the 
predicted event yields are obtained by propagating these differences 
through the modeling procedure.  For the search samples that incorporate 
hadronically-decaying tau lepton candidates, lepton-to-tau and jet-to-tau 
misidentification rates are modeled within the event simulation and 
validated using data control samples.  Assigned uncertainties are based 
on differences between predicted and observed event yields for these 
control samples.    

\begin{table*}[t]
  \setlength{\extrarowheight}{3pt}
\begin{ruledtabular} 
\begin{center} 
\caption{\label{tbl:sigrate}
Uncertainties on signal process event yields for all search samples.  
The symbol $^{\perp}$ indicates uncertainty sources applied only in 
the two OS Hadronic Tau and Trilepton {\it WH} ($\ell$+$\ell$+$\tau_{\rm{had}}$) 
search samples.}
%{\scriptsize
\begin{tabular}{l*{4}{c}}
\toprule
Uncertainty source                                            & $ggH$             & {\it WH}      & {\it ZH}      & VBF       \\
\hline
Theoretical cross section                                     & 14--44\%           & 5\%       & 5\%       & 10\%      \\
Luminosity                                                    & 5.9\%             & 5.9\%     & 5.9\%     & 5.9\%     \\
Lepton ($e$ or $\mu$) identification efficiency               & 2.8--3.8\%         & 2.8--5.0\% & 2.8--5.0\% & 2.8--3.8\% \\
Lepton ($\tau$) identification efficiency$^{\perp}$           & 4.1\%             & 1.4--2.1\% & 1.6--2.2\% & 4.0\%     \\
Trigger efficiency                                            & 2.0\%             & 2.0\%     & 2.0\%     & 2.0\%     \\
Higher-order amplitudes                                       & 2.3--13\%          & 10\%      & 10\%      & 10\%      \\
Jet energy scale                                              & 0--15\%            & 0--20\%    & 0--7.8\%   & 0--13\%    \\
Lepton ($e$ or $\mu$) to lepton ($\tau$) misreconstruction rate$^{\perp}$ & 0.1\%             & 0.1\%     & 0.1\%     & 0.1\%     \\
Jet to lepton ($\tau$) misreconstruction rate$^{\perp}$       &                   & 3.5--4.5\% & 2.9--4.2\% & 0--0.4\%   \\
\bottomrule
\end{tabular}
%}
\end{center}
\end{ruledtabular}
\end{table*}

In the context of a combined search, assumptions are needed on the relative 
sizes of the expected contributions originating from each production process.  
We incorporate full rate uncertainties on estimated event yields within the 
final fit.  Rate uncertainties applied to estimated signal contributions 
from each production mode are summarized in Table~\ref{tbl:sigrate}.  Here,
uncertainty ranges cover variations across all 13 search samples, which 
depend on the same set of simulated samples for modeling potential signal.
Contributions from $ggH$ and VBF production are not considered in the SS 
($\ge$1 Jets) and Trilepton search samples.

Theoretical cross section calculations used to normalize simulated signal event 
samples and associated uncertainties are described in Sec.~\ref{sec:theory}.
Uncertainties on $ggH$ production are much larger for higher jet multiplicity 
search samples, and an algorithm is used to assign correlated rate uncertainties 
to each search sample.  The inputs to this algorithm are the theoretical 
uncertainties associated with calculations of the inclusive, exclusive 
one-or-more parton, and exclusive two-or-more parton $ggH$ production cross 
sections.  The $ggH$ theoretical cross section uncertainty range reported in 
Table~\ref{tbl:sigrate} is obtained from the quadrature sum of all contributions 
as applied within each of the 13 search samples.  The other rate uncertainties 
applied to estimated signal event yields correspond directly to those applied 
to background predictions and are obtained following the same methodology.  

\begin{figure*}
\begin{center}
\subfigure{
\includegraphics[width=0.42\linewidth]{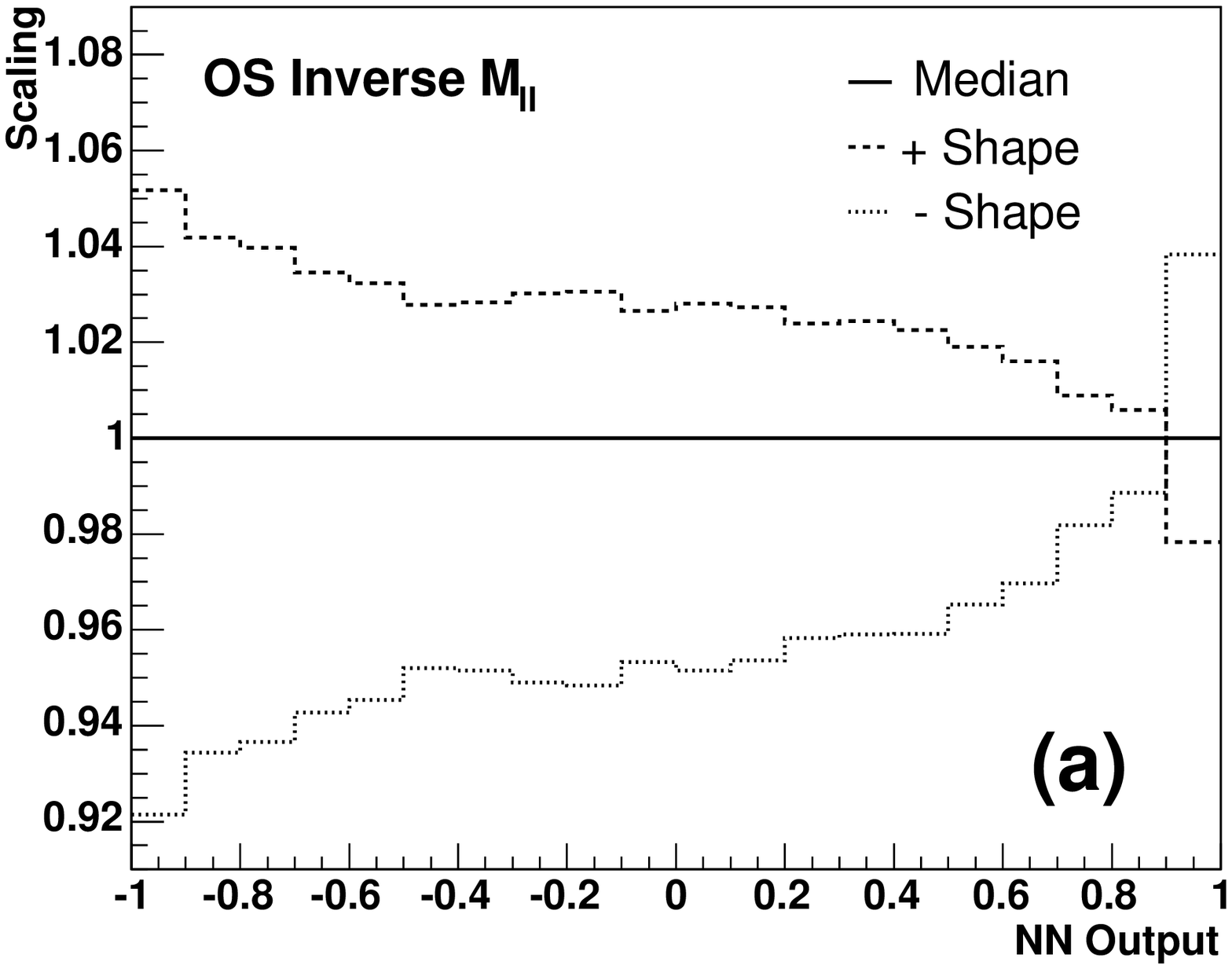}
}
\subfigure{
\includegraphics[width=0.42\linewidth]{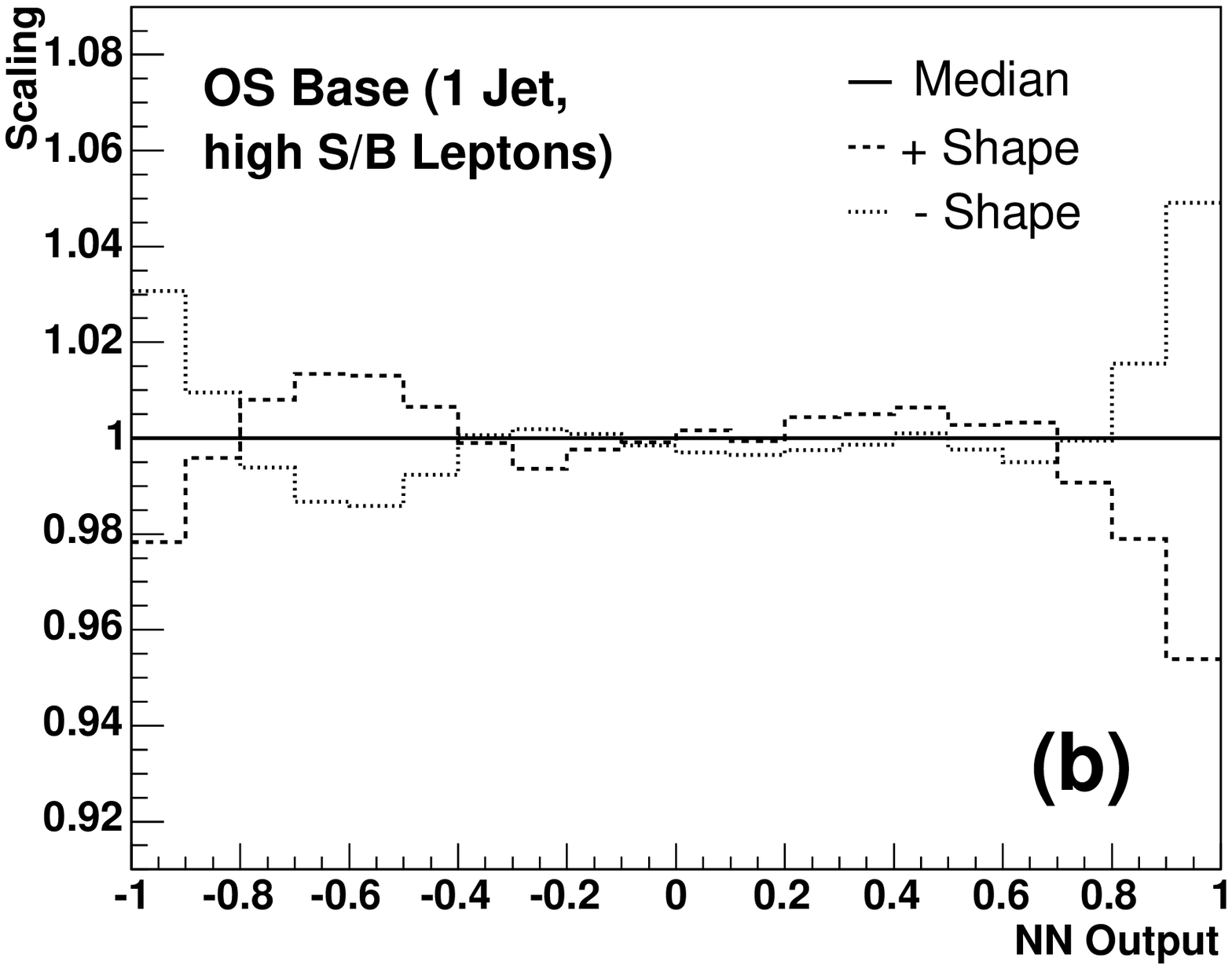}
}
\subfigure{
\includegraphics[width=0.42\linewidth]{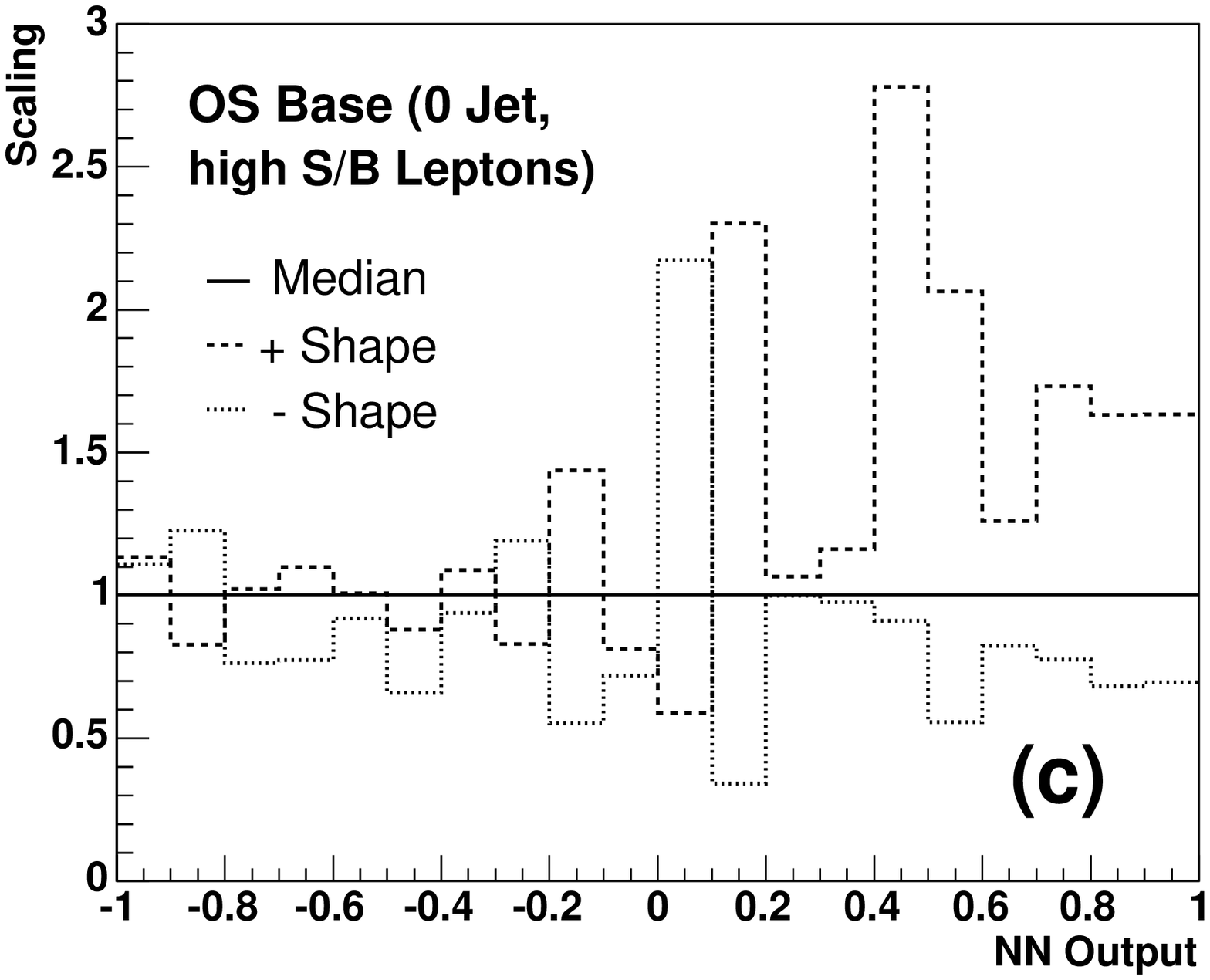}
}
\subfigure{
\includegraphics[width=0.42\linewidth]{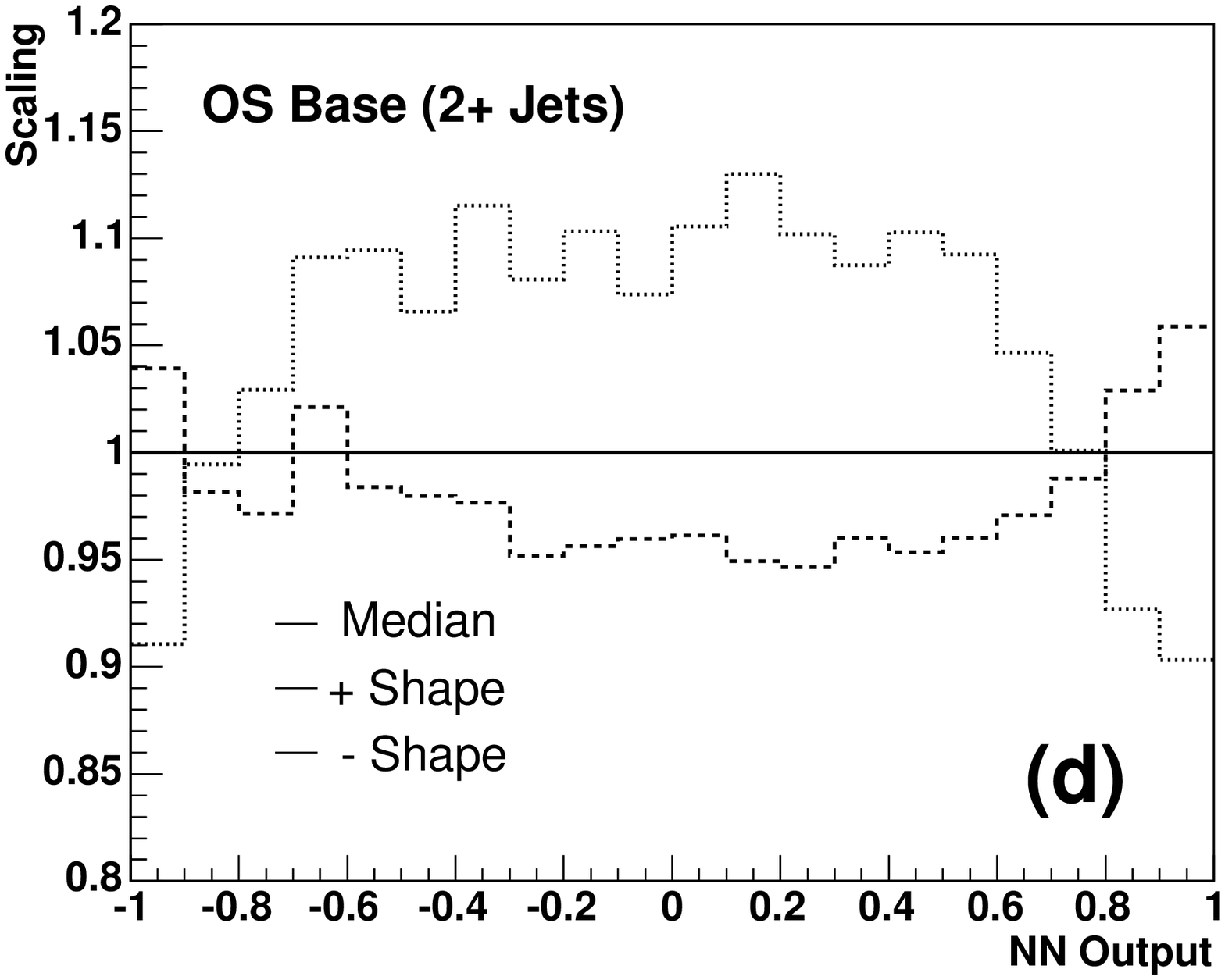}
}
\subfigure{
\includegraphics[width=0.42\linewidth]{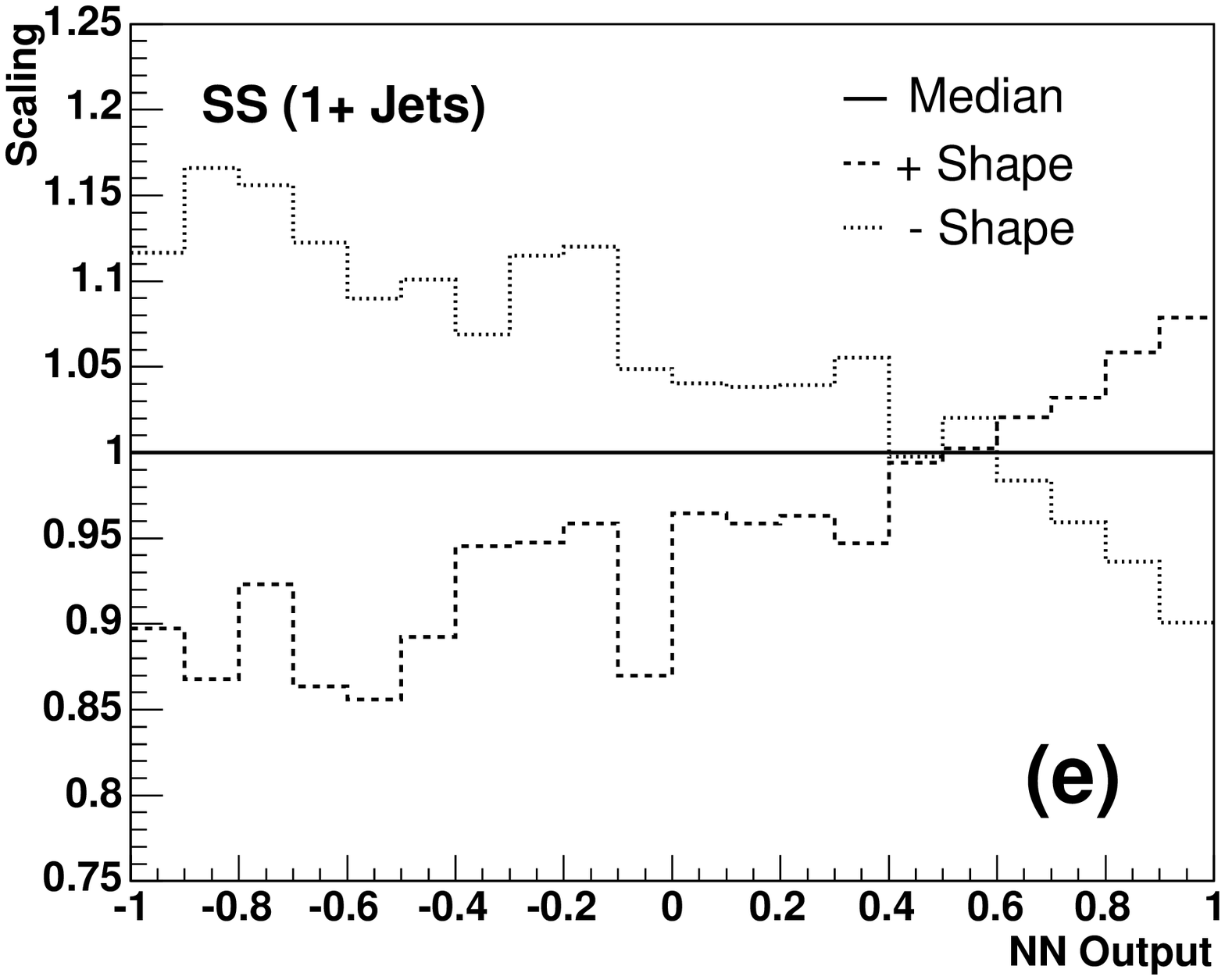}
}
\subfigure{
\includegraphics[width=0.42\linewidth]{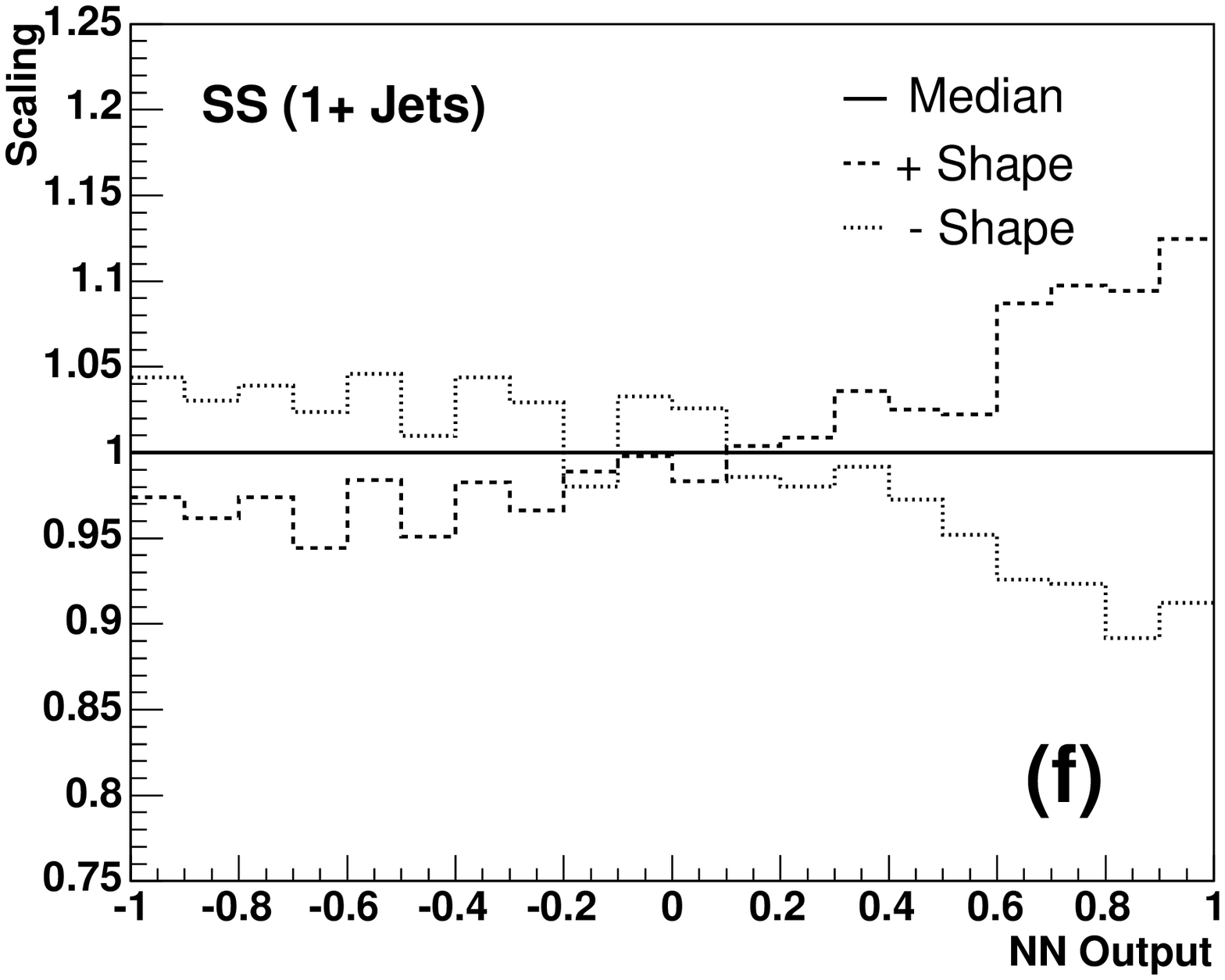}
}
\end{center}
\caption{Example bin-by-bin scalings used to obtain alternative neural network 
discriminant outputs associated with (a) higher-order diagrams uncertainty 
on the $ggH$ contribution in the OS Inverse $M_{\ell\ell}$ search sample, (b) 
higher-order diagrams uncertainty on the {\it WW} background contribution in the 
OS Base (1 Jet, high $s/b$ Leptons) search sample, (c) $\Met$ modeling uncertainty 
on the DY background contribution in the OS Base (0 Jet, high $s/b$ Leptons) search 
sample, (d) jet energy scale uncertainty on the DY contribution in the OS Base 
($\ge$2 Jets) search sample, (e) jet energy scale uncertainty on the {\it WH} contribution 
in the SS ($\ge$1 Jets) search sample, and (f) jet energy scale uncertainty on the 
{\it WZ} contribution in the SS ($\ge$1 Jets) search sample.}        
\label{fig::NNShapeSyst}
\end{figure*}

Each source contributing to the rate uncertainties assigned to background 
and signal predictions can also affect the shapes of discriminant outputs 
associated with the corresponding processes.  The effects of all uncertainty 
sources on discriminant distributions are studied and found to be mostly 
negligible.  In the remaining cases shape uncertainties, which correspond 
to correlated but nonuniform bin-by-bin rate uncertainties applied across 
a single discriminant distribution, are incorporated.  In particular, we 
account for the uncertainty originating from missing higher-order amplitudes 
to the modeled Higgs boson $p_T$ spectrum on the shapes of the $ggH$ 
discriminant outputs for each of the six OS dilepton search samples.  
Similarly, the effects of uncertainties from missing higher-order amplitudes 
to the modeled {\it WW} $p_T$ spectrum on the shapes of {\it WW} discriminant outputs 
are also included.  Figures~\ref{fig::NNShapeSyst}(a) and~\ref{fig::NNShapeSyst}(b) 
show resulting examples of the bin-by-bin scalings applied in individual 
search samples to generate alternative $ggH$ and {\it WW} discriminant shapes.

The shapes of DY discriminant outputs are also found to be significantly altered 
by uncertainties associated with $\Met$ modeling in the four OS Base (0 or 1 Jet) 
search samples and by uncertainties associated with jet energy scale modeling 
in the OS Base ($\ge$2 Jets) search sample.  Figures~\ref{fig::NNShapeSyst}(c) 
and~\ref{fig::NNShapeSyst}(d) show examples of the bin-by-bin scalings applied 
in these search samples to generate the alternative DY discriminant shapes.  
For the SS ($\ge$1 Jets) search sample, uncertainties associated with jet energy 
scale modeling are determined to significantly affect the shapes of discriminant 
outputs associated with both signal ({\it WH} and {\it ZH}) and background 
({\it WW}, {\it WZ}, and DY) contributions.  Figures~\ref{fig::NNShapeSyst}(e) 
and~\ref{fig::NNShapeSyst}(f) show resulting examples of the bin-by-bin scalings 
used to generate alternative signal and background discriminant shapes.

\section{Results}
\label{sec:finalresults}

The primary goal is to test for the presence of signal events 
originating from Higgs boson production and decay.  We adopt 
a Bayesian approach to estimate or bound the signal strength
most consistent with the observed data.  If the SM prediction 
of the signal strength for a specific value of $m_H$ is larger 
than the observed 95\% C.L. upper limit, that mass value is 
excluded at the 95\% C.L.  We quantify the search sensitivity 
using the median of the expected upper limit distribution as 
obtained in an ensemble of experiments simulated without signal. 

The extraction of results is complicated by the presence of multiple 
signal production processes, each potentially contributing signal 
events with differing kinematic signatures.  Combination of results 
from multiple search samples is pursued to optimize the search 
sensitivity.  The results are binned in their respective discriminant 
variables, and the data are assumed to be Poisson distributed in 
each bin.  Predictions of expected signal and background rates within 
each bin of the discriminant distributions associated with the different 
search samples are affected by systematic uncertainties.  Many of these 
systematic uncertainties are correlated across discriminant bins, 
between signal and background components, and between search samples.  
Uncertainty sources that result in events migrating between search 
samples need to be treated as anti-correlated with respect to those 
samples.  To address these issues correctly, we use the methodology 
described in Ref.~\cite{Aaltonen:2010sy} as summarized below.

The contents of low signal-to-background ($s/b$) bins serve to constrain 
the values of nuisance parameters, corresponding to each of the 
individual sources of systematic uncertainty on signal and background 
modeling.  The same sources of systematic uncertainty affect predictions 
for signal and background yields in the high-$s/b$ bins, which are more 
sensitive to the presence of a Higgs boson signal and its production 
rate.

We group the systematic uncertainties in three classes, according to 
their impact on the interpretation of results.  The first class includes
systematic effects affecting the event rates, which uniformly scale the
predicted yields in each bin of the modeled discriminants.  A second 
category corresponds to uncertainties affecting the shapes of the 
discriminants' distributions, which are also parametrized using common 
nuisance parameters and therefore applied as correlated across all bins 
within a modeled discriminant.  In this case, bin-to-bin scalings are 
not required to be uniform, allowing for distortions in the shape of 
the distribution of the discriminant.  A final category is for bin-by-bin 
independent uncertainties, which arise from the limited size of simulated 
and experimental data samples.  Uncertainties associated with the last 
two categories reduce the constraining power of low-$s/b$ bins on nuisance 
parameters. 

The likelihood function, $L({\rm{data}}|s,b,\vec\nu)$, is the same as that 
used in Ref.~\cite{Aaltonen:2010sy}, with $\vec\nu$ representing the nuisance 
parameters.  Shape uncertainties are applied first in an additive fashion, 
interpolating and extrapolating the contents in each bin according to the 
value of the nuisance parameter governing the shape distortion and the 
difference between the central and alternative shapes of the modeled 
discriminant.  The prior probability densities assumed for the systematic 
uncertainties are Gaussian, and bin contents are constrained to be 
positive in this procedure.  Bin-by-bin uncertainties are then applied 
to the signal and background predictions as Gaussians that are again 
truncated to prevent negative values of predictions.  Finally, rate 
uncertainties are applied multiplicatively, scaling all discriminant bins 
by the same factor.  Gaussian prior densities are also used for rate 
uncertainties with constraints to avoid negative scale factors.  Asymmetric 
rate and shape uncertainties are parametrized as in Ref.~\cite{Aaltonen:2010sy}.  
Correlations in the predictions for different signal and background 
processes are accounted for by applying effects of shared uncertainty 
sources consistently across the modeled discriminants for each search 
sample.  Because of the requirement for combining the results from 
several different search samples, a single parameter $R$ is used to scale 
all signal contributions.

We integrate the likelihood function multiplied by the product of the prior 
densities for the nuisance parameters, over the nuisance parameters
\begin{equation}
L^\prime({\rm{data}}|Rs,b) = \int L({\rm{data}}|Rs,b,\vec\nu)\pi(\vec\nu)d{\vec{\nu}},
\end{equation}
where $\pi(\vec\nu)d{\vec{\nu}}$ is the joint prior probability density for 
all of the nuisance parameters as described in Ref.~\cite{Nakamura:2010zzi}.
In this case the joint prior density is the product of individual prior
densities as systematic uncertainty sources are treated as uncorrelated.  

As described in Ref.~\cite{Nakamura:2010zzi}, a limit on $R$ is obtained 
from
\begin{equation}
0.95 = \frac{\int_0^{R_{\rm{limit}}}L^\prime({\rm{data}}|Rs,b)\pi(R)}{\int_0^\infty L^\prime({\rm{data}}|Rs,b)\pi(R)},
\end{equation}
where $\pi(R)$ is a uniform prior density over all positive values 
of $R$.  The value of $R$ that maximizes $L^\prime({\rm{data}}|Rs,b)$ 
is defined as the best-fit value.  The interval for quoting one 
standard deviation uncertainties is given by the shortest interval 
$[R_{\rm{low}},R_{\rm{high}}]$ satisfying
\begin{equation}
0.68 = \frac{\int_{R_{\rm{low}}}^{R_{\rm{high}}}L^\prime({\rm{data}}|Rs,b)\pi(R)}{\int_0^\infty L^\prime({\rm{data}}|Rs,b)\pi(R)}.
\end{equation}

Search sensitivity is estimated by generating multiple simulated test 
experiments according to background-only predictions and determining 
the observed limits for each trial.  Values of nuisance parameters are 
separately varied according to their prior densities for each simulated 
experiment.  The median observed limit, $R^{\rm{med}}_{\rm{limit}}$, is 
used as a gauge of analysis sensitivity.  The distribution of possible 
limits, quantified as those values of $R$ for which 2.3\%, 16\%, 50\%, 
84\%, and 97.7\% of background-only simulated experiments fall on one 
side of those requirements, are used to illustrate the dispersion of 
possible outcomes associated with a single experiment.

\subsection{Diboson cross section measurements}

Measurements of diboson production cross sections using the same tools 
and techniques applied within the Higgs boson search provide an important 
validation of the analysis framework.  A measurement of the $p\bar{p} 
\rightarrow W^+W^-$ cross section based on the $\ell^+\bar{\nu}\ell^-\nu$ 
decay mode was obtained from the OS Base (0 Jet, high $s/b$) search sample 
using 3.6 fb$^{-1}$ of integrated luminosity~\cite{cdfwwpub}.  A value 
of $\sigma(p\bar{p}\rightarrow W^+W^- + X) = 12.1 \pm1.8$~pb, which is 
in good agreement with the NLO prediction, was obtained using the same 
matrix-element based discriminants employed within the Higgs boson search.  
Similarly, a measurement of the $p\bar{p} \rightarrow ZZ$ cross section 
based on the $\ell^+\ell^-\nu\bar{\nu}$ decay mode was obtained from the 
OS Base (0 and 1 Jet) search samples using 6.0 fb$^{-1}$ of integrated 
luminosity~\cite{cdfzzpub}.  Neural network based discriminants were used to 
extract a value of $\sigma(p\bar{p}\rightarrow ZZ + X) = 1.34 \pm0.56$~pb, 
which is in good agreement with the NLO zero-width calculation, to which the 
result was normalized.  Finally, a measurement of the $p\bar{p} \rightarrow 
W^{\pm}Z$ cross section based on the $\ell^{\pm}\nu\ell^+\ell^-$ decay mode 
was obtained from the Trilepton {\it WH} search sample using 7.1 fb$^{-1}$ of 
integrated luminosity~\cite{cdfwzpub}.  Neural network discriminants were 
again used to extract a value of $\sigma(p\bar{p}\rightarrow W^{\pm}Z + X) 
= 3.93 \pm0.84$~pb, in good agreement with the NLO prediction. 

\subsection{SM Higgs boson interpretation}

We determine limits on SM Higgs boson production for the combination of 
all search samples and for groups of samples with analogous final states.  
The limit calculations are performed separately for each of the 19 Higgs 
boson mass hypotheses considered.  Because we account for potential 
contributions from all four Higgs boson production modes, the resulting 
limits are determined as ratios with respect to SM expectations.  
Based on the (N)NLO Higgs boson production cross sections and decay 
branching ratios for $H \to W^+W^-$ presented in Sec.~\ref{sec:theory}, 
the largest potential signal contributions would originate from a 
Higgs boson with a mass of 165~GeV/$c^2$, and the best combined search
sensitivity is indeed obtained for this mass hypothesis.  The actual 
sensitivity of an individual search sample under a specific mass hypothesis 
depends both on the signal-to-background ratio of events in the sample and 
the ability of the neural network to separate background contributions 
from the potential signal contributions associated with the hypothesized 
Higgs boson mass. 

The OS Base (0 Jet) search samples have the highest sensitivity to SM 
Higgs boson production.  The dominant signal contributions originate 
from $ggH$ production.  Similar sensitivity is obtained from the OS 
Base (1 Jet) and OS Base ($\ge$2 Jets) samples, where additional signal 
contributions from {\it VH} and VBF production have a more significant 
impact.  The OS Inverse $M_{\ell\ell}$ search sample, with dominant 
signal contributions from $ggH$ production, is approximately 50\% 
less sensitive than the OS Base samples for the $m_H =$165~GeV/$c^2$ 
hypothesis.  But for the $m_H =$125~GeV/$c^2$ hypothesis the sensitivities 
are comparable since a higher fraction of potential signal events satisfy 
the kinematic criteria of this sample.  The SS ($\ge$1 Jets), Trilepton WH, 
and Trilepton {\it ZH} search samples, which focus exclusively on {\it VH} 
production, contribute sensitivities of typically 20-50\% of the best 
OS Base samples.  However, the inclusion of these samples has a nonnegligible 
impact on the combined search sensitivity, and important information on the 
potential couplings of heavy vector bosons to a potential Higgs boson can 
be extracted directly from these samples.  Because they contain much larger 
background contributions, the OS Hadronic Tau search samples contribute 
significantly less to the combined search sensitivity.  Since the neural 
networks are unable to separate background and signal contributions in 
these samples for low Higgs boson masses, these samples are incorporated 
into combined limits only for mass hypotheses of 130~GeV/$c^2$ and above.

\begin{table}[t]
   \setlength{\extrarowheight}{3pt}
\begin{ruledtabular} 
\begin{center} 
\caption{\label{tab:smlimits125}
Median expected 95\% C.L. upper limits assuming the background-only 
hypothesis, and corresponding observed limits on Higgs boson production 
relative to SM expectations for the $m_H=$125 and 165~GeV/$c^2$ mass hypotheses 
obtained from combinations of search samples with analogous final states and 
the combination of all search samples.}   
%{\scriptsize
\begin{tabular}{l*{4}{c}}
\toprule
%\cline{2-5}
\multicolumn{1}{l}{} & \multicolumn{2}{c}{$m_H=$125~GeV/$c^2$} & \multicolumn{2}{c}{$m_H=$165~GeV/$c^2$} \\ 
\hline
Search sample(s)          & Obs/$\sigma_{{\rm SM}}$ & Exp/$\sigma_{{\rm SM}}$  & Obs/$\sigma_{{\rm SM}}$  & Exp/$\sigma_{{\rm SM}}$ \\ 
\hline 
OS Base (0 Jet)           & 4.76 & 7.30 & 1.36  & 1.41  \\
OS Base (1 Jet)           & 9.86 & 9.76 & 1.45  & 1.85  \\
OS Base ($\ge$2 Jets)         & 18.1 & 7.34 & 2.83  & 1.95  \\
OS Inverse $M_{\ell\ell}$ & 11.9 & 11.0 & 1.71  & 2.76  \\
SS ($\ge$1 Jets)              & 13.9 & 11.7 & 4.20  & 3.95  \\
Trilepton {\it WH}              & 12.1 & 12.2 & 4.79  & 4.36  \\
Trileptons {\it ZH}             & 19.9 & 23.2 & 4.94  & 6.59  \\ 
OS Hadronic Tau           &      &      & 15.7  & 11.7  \\
All samples               & 3.26 & 3.25 & 0.493 & 0.701 \\ 
\bottomrule
\end{tabular}
%}
\end{center}
\end{ruledtabular}
\end{table}

\begin{table*}[t]
   \setlength{\extrarowheight}{3pt}
\begin{ruledtabular} 
\begin{center} 
\caption{\label{tbl:FinalCombLimits}
Median expected 95\% C.L. upper limits assuming the 
background-only hypothesis, and corresponding observed limits 
on Higgs boson production relative to SM expectations from the 
combination of all search samples for 19 mass hypotheses within 
the range 110~$< m_H <$~200~GeV/$c^2$.  The boundaries of the 
one and two standard deviations assuming the background-only 
hypothesis are also provided.} 
%{\scriptsize
\begin{tabular}{l*{19}{c}}
\toprule
$m_H$                   &       110 &       115 &       120 &       125 &       130 &       135 &       140 &       145 &       150 &       155 &       160 &       165 &       170 &       175 &       180 &       185 &       190 &       195 &       200 \\
\hline
$-2\sigma/\sigma_{{\rm SM}}$  &      7.11 &      3.78 &      2.47 &      1.67 &      1.25 &      1.04 &      0.84 &      0.73 &      0.64 &      0.54 &      0.41 &      0.37 &      0.43 &      0.53 &      0.66 &      0.80 &      0.98 &      1.21 &      1.23 \\
$-1\sigma/\sigma_{{\rm SM}}$  &      9.60 &      5.25 &      3.25 &      2.32 &      1.70 &      1.37 &      1.13 &      0.98 &      0.85 &      0.71 &      0.53 &      0.50 &      0.58 &      0.71 &      0.87 &      1.08 &      1.34 &      1.54 &      1.74 \\
Exp./$\sigma_{{\rm SM}}$      &      13.4 &      7.41 &      4.51 &      3.25 &      2.33 &      1.89 &      1.60 &      1.37 &      1.16 &      0.98 &      0.74 &      0.70 &      0.83 &      1.00 &      1.18 &      1.51 &      1.88 &      2.12 &      2.48 \\
$+1\sigma/\sigma_{{\rm SM}}$  &      18.8 &      10.4 &      6.36 &      4.52 &      3.20 &      2.62 &      2.28 &      1.91 &      1.60 &      1.38 &      1.04 &      0.99 &      1.18 &      1.39 &      1.63 &      2.15 &      2.63 &      3.01 &      3.49 \\
$+2\sigma/\sigma_{{\rm SM}}$  &      26.0 &      14.3 &      8.90 &      6.19 &      4.34 &      3.60 &      3.22 &      2.62 &      2.19 &      1.94 &      1.45 &      1.37 &      1.66 &      1.92 &      2.23 &      3.01 &      3.62 &      4.26 &      4.81 \\
\hline
Obs./$\sigma_{{\rm SM}}$      &      14.1 &      9.49 &      5.26 &      3.26 &      2.66 &      2.01 &      2.02 &      1.25 &      0.95 &      0.74 &      0.60 &      0.49 &      0.84 &      1.28 &      1.50 &      2.53 &      3.47 &      4.64 &      5.65 \\
\bottomrule
\end{tabular}
%}
\end{center}
\end{ruledtabular}
\end{table*}

\begin{figure*}[]
\begin{center}
\subfigure{
\includegraphics[width=0.45\textwidth]{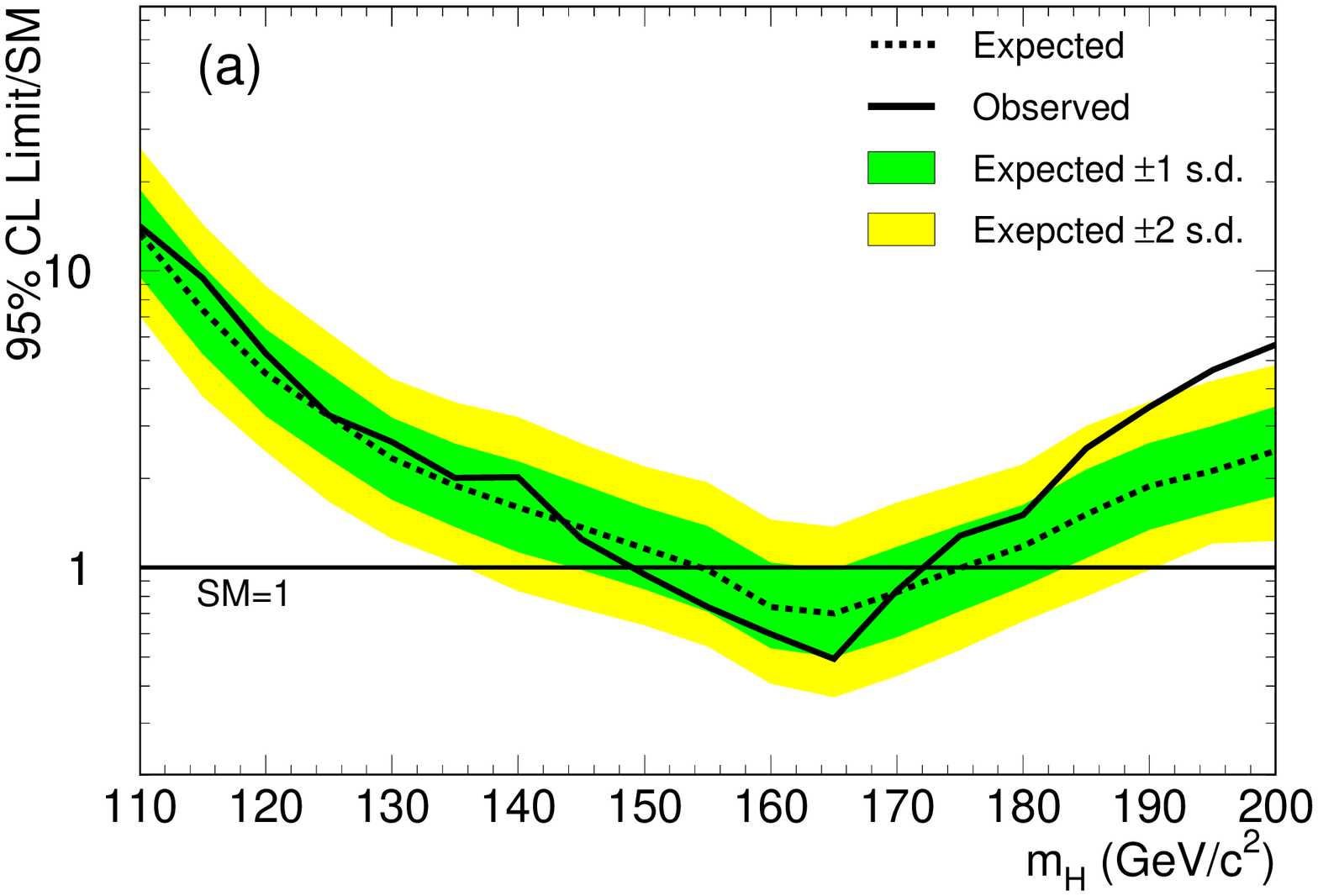}
}
\subfigure{
\includegraphics[width=0.45\textwidth]{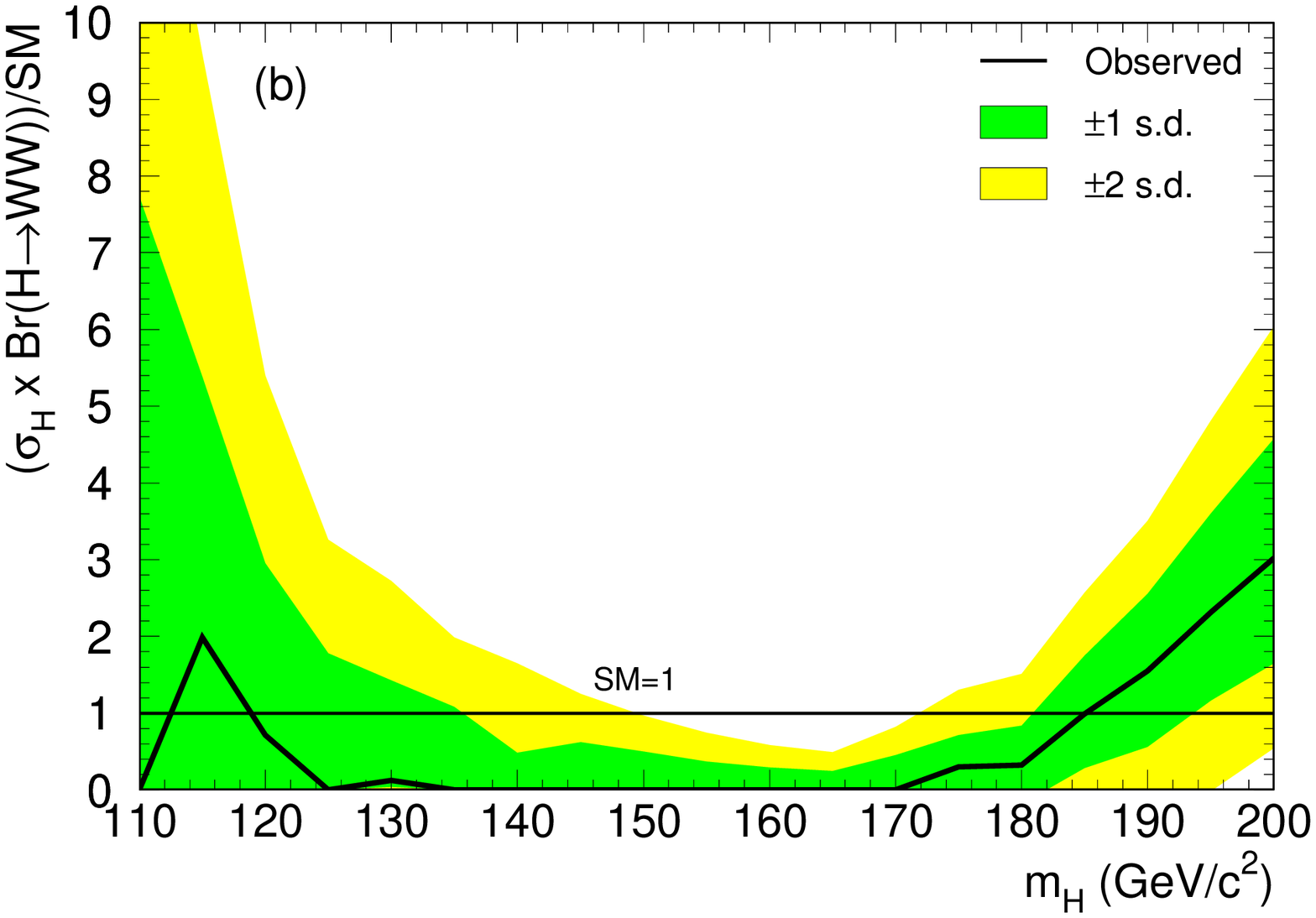}
}
\end{center}
\caption{(a) Median expected, assuming the background-only hypothesis, 
(dashed line) and observed (solid line) 95\% C.L. upper limits on Higgs 
boson production relative to SM expectations from the combination of 
all search samples as a function of the Higgs boson mass.  The dark and 
light shaded bands show the one and two standard deviations assuming 
the background-only hypothesis.  (b) Best-fit cross section for inclusive 
Higgs boson production, normalized to the SM expectation, for the 
combination of all search samples as a function of the Higgs boson mass.
The solid line indicates the fitted cross section, and the associated 
dark and light shaded regions show the 68\% and 95\% credibility intervals.} 
\label{fig:LimitsPlotFinalComb}
\end{figure*}
             
Table~\ref{tab:smlimits125} presents limits on Higgs boson production 
obtained from combinations of search samples with analogous final 
states and from the combination of all search samples.  Median expected 
95\% C.L. upper limits assuming the background-only hypothesis, and 
corresponding observed limits on Higgs boson production relative to 
SM expectations are shown for the 125 and 165~GeV/$c^2$ mass hypotheses.  
Limits obtained from the combination of all search samples for 19 Higgs 
boson mass hypotheses within the range 110~$< m_H <$~200~GeV/$c^2$ 
are presented in Table~\ref{tbl:FinalCombLimits} along with boundaries 
on the one and two standard deviations assuming the background-only 
hypothesis.  These limits are also presented graphically in 
Fig.~\ref{fig:LimitsPlotFinalComb}(a).  SM Higgs boson mass values are 
excluded at the 95\% C.L. in the range over which the observed limits 
lie below one (the expected SM production rate).  The data excludes
Higgs boson masses in the range 149~$< m_H <$~172~GeV/$c^2$, where the 
median expected exclusion range, assuming the background-only hypothesis, 
is 155~$< m_H <$~175~GeV/$c^2$.

We also fit for the Higgs boson production rate most compatible 
with the observed data.  Best-fit cross sections normalized to SM 
expectations are displayed as a function of the Higgs boson mass 
in Fig.~\ref{fig:LimitsPlotFinalComb}(b).  In the cross section 
fit, the SM ratios for the relative rates of the four contributing 
production mechanisms are assumed.  Over a significant fraction 
of the tested mass range, the fit to the data indicates little 
or no contribution associated with Higgs boson production.  For 
the $m_H=$~125~GeV/$c^2$ mass hypothesis, the fitted Higgs boson 
production rate relative to the SM expectation is $0.00^{+1.78}_{-0.00}$, 
which is compatible at the level of one standard deviation with 
both the SM Higgs boson and the background-only expectations.                   

\subsection{Limits on $ggH$ production and Higgs boson constraints in SM4}
\label{sec:sm4results}

Because Higgs boson $ggH$ production proceeds at lowest order via a 
virtual loop containing strongly-interacting particles, the production 
rate from this mechanism is sensitive to the existence of particles 
that may be too massive for direct observation.  The presence of a 
fourth generation of heavy fermions beyond the three families described 
in the SM enhances the $ggH$ production cross section by a factor 
between seven and nine in the range of $m_H$ accessible at the Tevatron.  
The presence of a fourth fermion generation affects $ggH$ production 
only, and neither enhances nor suppresses {\it WH}, {\it ZH}, and VBF 
production.  

In order to interpret the search in terms of the SM4 and other extensions 
to the SM that would affect the $ggH$ production rate, we first extract 
upper bounds on the $ggH$ production cross section times decay branching  
ratio $H\rightarrow W^+W^-$ assuming negligible contributions from {\it WH}, 
{\it ZH}, and VBF production.  This assumption ensures that resulting 
limits are the most conservative with respect to possible enhancements or 
suppressions of the other production mechanisms within the context of a 
particular new physics model.  Because we are focusing on enhancements 
in the production, which could lie significantly above SM expectations, 
we extend on the search mass range to 300~GeV/$c^2$.  

Since we are in this case setting limits on the rate of a specific Higgs 
boson production and decay mode, no theoretical rate uncertainties are 
incorporated.  However, because we analyze opposite-sign dilepton events 
with zero, one, and two or more reconstructed jets in different search 
samples, the uncertainties on the relative fractions of Higgs boson 
signal events within these samples are retained.  Median expected 
95\% C.L. upper limits assuming the background-only hypothesis, and 
corresponding observed limits on $\sigma(ggH)\times \mathcal{B}(H\rightarrow W^+W^-)$ 
are listed in Table~\ref{tab:xseclimits} along with the boundaries of 
one and two standard deviations assuming the background-only hypothesis.  

\begin{table*}[t]
   \setlength{\extrarowheight}{3pt}
\begin{ruledtabular} 
\begin{center} 
\caption{\label{tab:xseclimits}
Median expected 95\% C.L. upper limits assuming the background-only 
hypothesis, and corresponding observed limits on $\sigma(ggH) 
\times \mathcal{B}(H\rightarrow W^+W^-)$ in picobarns (pb) from 
the combination of all search samples for 29 mass hypotheses within 
the range 110~$< m_H <$~300~GeV/$c^2$.  The boundaries of one and 
two standard deviations assuming the background-only hypothesis are 
also provided.  The {\it WH}, {\it ZH}, and VBF Higgs boson 
production mechanisms are assumed to contribute no events to the 
search samples.} 
%{\scriptsize
\begin{tabular}{l*{15}{c}}
\toprule
$m_H$       &       110 &       115 &       120 &       125 &       130 &       135 &       140 &       145 &       150 &       155 &       160 &       165 &       170 &       175 &       180 \\
\hline
$-2\sigma$  &      0.70 &      0.53 &      0.52 &      0.46 &      0.45 &      0.43 &      0.39 &      0.37 &      0.32 &      0.28 &      0.21 &      0.18 &      0.21 &      0.21 &      0.23 \\
$-1\sigma$  &      0.95 &      0.77 &      0.70 &      0.65 &      0.61 &      0.58 &      0.52 &      0.51 &      0.44 &      0.38 &      0.28 &      0.24 &      0.27 &      0.29 &      0.31 \\
Exp.        &      1.32 &      1.09 &      0.97 &      0.92 &      0.85 &      0.81 &      0.74 &      0.71 &      0.62 &      0.52 &      0.38 &      0.33 &      0.37 &      0.40 &      0.44 \\
$+1\sigma$  &      1.84 &      1.47 &      1.36 &      1.29 &      1.19 &      1.14 &      1.05 &      1.00 &      0.86 &      0.72 &      0.52 &      0.46 &      0.51 &      0.56 &      0.61 \\
$+2\sigma$  &      2.54 &      1.94 &      1.86 &      1.78 &      1.64 &      1.59 &      1.48 &      1.38 &      1.17 &      0.99 &      0.72 &      0.64 &      0.71 &      0.77 &      0.85 \\
\hline
Obs.        &      1.42 &      1.18 &      1.04 &      0.97 &      0.82 &      0.69 &      0.71 &      0.67 &      0.41 &      0.33 &      0.26 &      0.28 &      0.33 &      0.46 &      0.54 \\
\hline
$m_H$       &       180 &       185 &       190 &       195 &       200 &       210 &       220 &       230 &       240 &       250 &       260 &       270 &       280 &       290 &       300 \\
\hline
$-2\sigma$  &      0.23 &      0.25 &      0.27 &      0.28 &      0.32 &      0.32 &      0.32 &      0.33 &      0.35 &      0.30 &      0.29 &      0.27 &      0.25 &      0.25 &      0.22 \\
$-1\sigma$  &      0.31 &      0.34 &      0.36 &      0.39 &      0.42 &      0.42 &      0.44 &      0.45 &      0.47 &      0.41 &      0.39 &      0.37 &      0.33 &      0.33 &      0.29 \\
 Exp.       &      0.44 &      0.47 &      0.50 &      0.55 &      0.58 &      0.59 &      0.60 &      0.62 &      0.65 &      0.56 &      0.53 &      0.51 &      0.46 &      0.46 &      0.40 \\
$+1\sigma$  &      0.61 &      0.65 &      0.69 &      0.77 &      0.80 &      0.84 &      0.83 &      0.86 &      0.91 &      0.79 &      0.74 &      0.70 &      0.64 &      0.65 &      0.57 \\
$+2\sigma$  &      0.85 &      0.90 &      0.94 &      1.05 &      1.10 &      1.19 &      1.12 &      1.19 &      1.25 &      1.10 &      1.01 &      0.95 &      0.89 &      0.91 &      0.80 \\
\hline
Obs.        &      0.54 &      0.66 &      0.81 &      1.01 &      1.01 &      1.38 &      1.10 &      1.14 &      1.34 &      1.19 &      0.97 &      0.95 &      0.92 &      1.01 &      0.81 \\
\bottomrule
\end{tabular}
%}
\end{center}
\end{ruledtabular}
\end{table*}

\begin{figure*}[]
\begin{center}
\subfigure{
\includegraphics[width=0.45\linewidth]{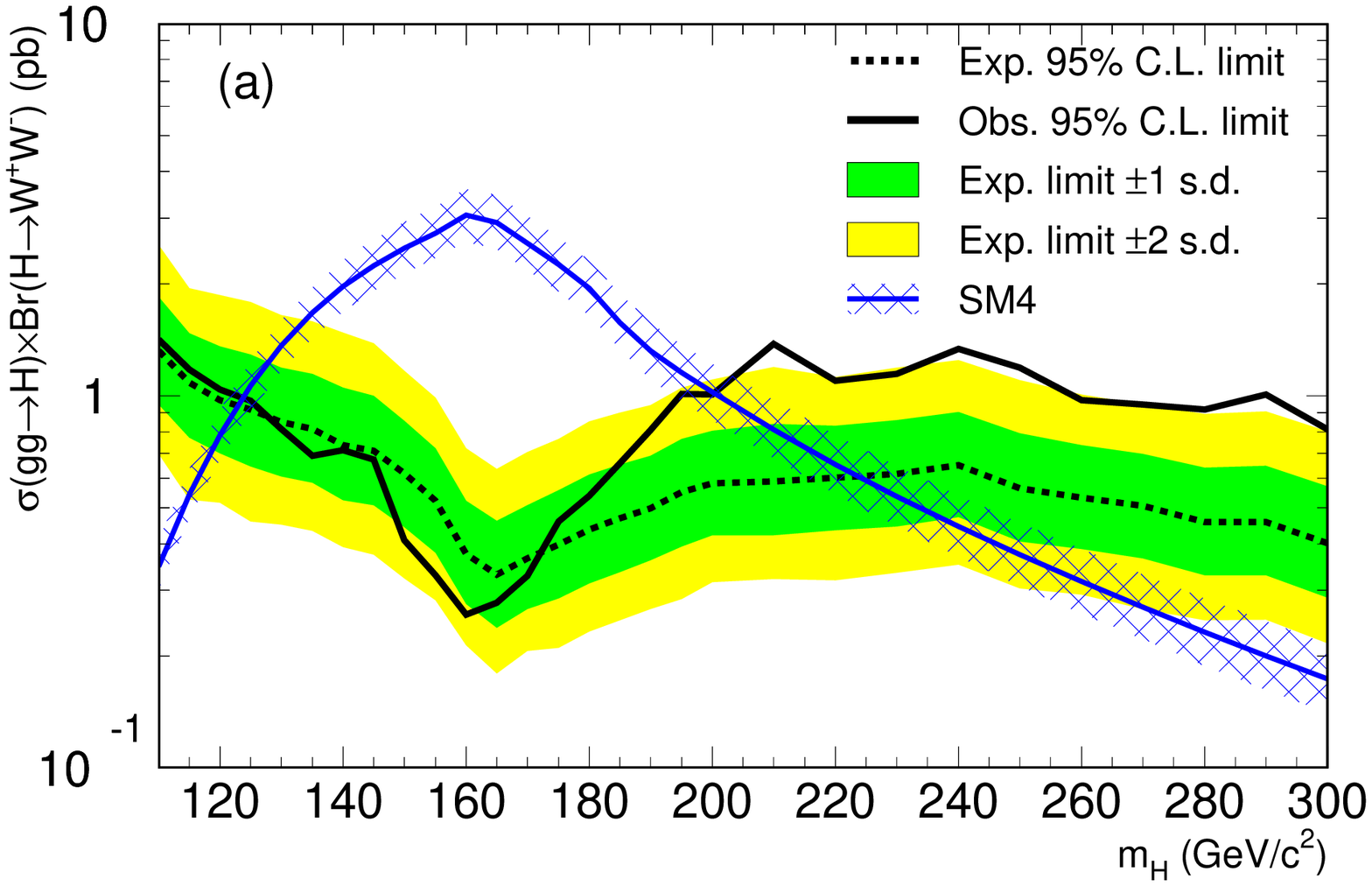}
}
\subfigure{
\includegraphics[width=0.45\linewidth]{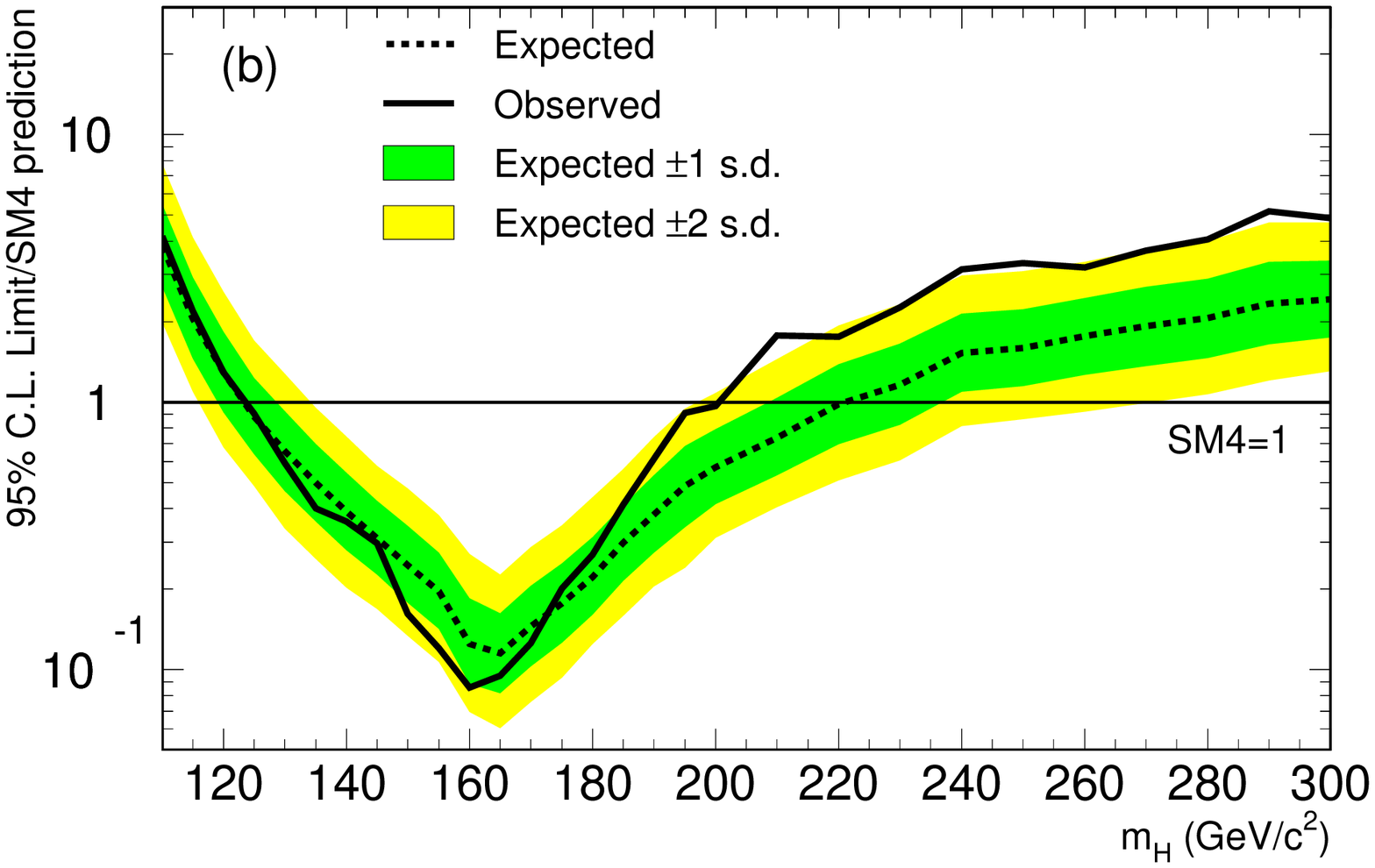}
}
\end{center}
\caption{Median expected 95\% C.L. upper limits assuming the background-only 
hypothesis (dashed line), and corresponding observed limits (solid line) on 
(a) $\sigma(ggH)\times \mathcal{B}(H\rightarrow W^+W^-)$ in picobarns (pb) 
and (b) Higgs boson production relative to SM4 expectations from the 
combination of all search samples as a function of the Higgs boson mass.  
The dark and light shaded bands show the one and two standard deviations 
assuming the background-only hypothesis.  In the (a) panel, the lighter 
colored line indicates the SM4 expectation and the hatched region 
encompasses the associated theoretical uncertainties.}  
\label{fig:xseclimits}
\end{figure*}

A comparison between observed upper limits on $\sigma(ggH)\times 
\mathcal{B}(H\rightarrow W^+W^-)$ and SM4 expectations based on 
the production cross sections and decay branching ratios listed 
in Table~\ref{table:xsec} as a function of $m_H$ is shown in 
Fig.~\ref{fig:xseclimits}(a).  To extract SM4 model constraints, 
rate uncertainties associated with the theoretical cross sections 
and branching ratios are included within the limit calculation.  
The resulting median expected 95\% C.L. upper limits assuming 
the background-only hypothesis, and corresponding observed limits 
on Higgs boson production relative to SM4 expectations are shown 
in Table~\ref{tab:sm4limits}.  The same limits are shown graphically 
in Fig.~\ref{fig:xseclimits}(b).  Within the SM4 model we exclude 
Higgs boson masses in the range 124~$< m_H <$~200~GeV/$c^2$, 
to be compared against a median expected exclusion range of 
124~$< m_H <$~221~GeV/$c^2$.

\begin{table*}[t]
   \setlength{\extrarowheight}{3pt}
\begin{ruledtabular} 
\begin{center} 
\caption{\label{tab:sm4limits}
Median expected 95\% C.L. upper limits assuming the background-only 
hypothesis, and corresponding observed limits on Higgs boson production 
relative to SM4 expectations from the combination of all search samples 
for 29 mass hypotheses within the range 110~$< m_H <$~300~GeV/$c^2$.  The 
boundaries of one and two standard deviations assuming the background-only 
hypothesis are also provided.  The {\it WH}, {\it ZH}, and VBF Higgs boson production 
mechanisms are assumed to contribute no events to the search samples.} 
%{\scriptsize
\begin{tabular}{l*{15}{c}}
\toprule
$m_H$                    &       110 &       115 &       120 &       125 &       130 &       135 &       140 &       145 &       150 &       155 &       160 &       165 &       170 &       175 &       180 \\
\hline
$-2\sigma/\sigma_{\mathrm{SM4}}$  &      1.98 &      1.09 &      0.66 &      0.45 &      0.33 &      0.25 &      0.20 &      0.17 &      0.13 &      0.10 &      0.07 &      0.06 &      0.08 &      0.09 &      0.12 \\
$-1\sigma/\sigma_{\mathrm{SM4}}$  &      2.69 &      1.46 &      0.91 &      0.62 &      0.45 &      0.34 &      0.27 &      0.22 &      0.18 &      0.14 &      0.09 &      0.08 &      0.10 &      0.13 &      0.16 \\
Exp./$\sigma_{\mathrm{SM4}}$      &      3.83 &      2.05 &      1.29 &      0.87 &      0.62 &      0.47 &      0.38 &      0.31 &      0.25 &      0.19 &      0.12 &      0.11 &      0.14 &      0.18 &      0.22 \\
$+1\sigma/\sigma_{\mathrm{SM4}}$  &      5.49 &      2.92 &      1.80 &      1.22 &      0.87 &      0.67 &      0.53 &      0.43 &      0.34 &      0.27 &      0.17 &      0.16 &      0.20 &      0.25 &      0.31 \\
$+2\sigma/\sigma_{\mathrm{SM4}}$  &      7.77 &      4.13 &      2.49 &      1.69 &      1.20 &      0.92 &      0.73 &      0.59 &      0.46 &      0.37 &      0.24 &      0.23 &      0.28 &      0.35 &      0.43 \\
\hline
Obs./$\sigma_{\mathrm{SM4}}$      &      4.17 &      2.19 &      1.29 &      0.91 &      0.59 &      0.39 &      0.35 &      0.28 &      0.16 &      0.12 &      0.09 &      0.10 &      0.13 &      0.20 &      0.27 \\
\hline
$m_H$                    &       180 &       185 &       190 &       195 &       200 &       210 &       220 &       230 &       240 &       250 &       260 &       270 &       280 &       290 &       300 \\
\hline
$-2\sigma/\sigma_{\mathrm{SM4}}$  &      0.12 &      0.16 &      0.19 &      0.23 &      0.29 &      0.36 &      0.44 &      0.53 &      0.64 &      0.62 &      0.71 &      0.74 &      0.78 &      0.83 &      0.88 \\
$-1\sigma/\sigma_{\mathrm{SM4}}$  &      0.16 &      0.21 &      0.26 &      0.32 &      0.39 &      0.49 &      0.59 &      0.70 &      0.86 &      0.86 &      0.93 &      0.99 &      1.03 &      1.11 &      1.15 \\
Exp./$\sigma_{\mathrm{SM4}}$      &      0.22 &      0.29 &      0.36 &      0.45 &      0.53 &      0.69 &      0.82 &      0.97 &      1.20 &      1.22 &      1.27 &      1.37 &      1.43 &      1.54 &      1.61 \\
$+1\sigma/\sigma_{\mathrm{SM4}}$  &      0.31 &      0.41 &      0.51 &      0.64 &      0.75 &      0.97 &      1.15 &      1.38 &      1.66 &      1.73 &      1.75 &      1.92 &      2.03 &      2.17 &      2.27 \\
$+2\sigma/\sigma_{\mathrm{SM4}}$  &      0.43 &      0.58 &      0.70 &      0.90 &      1.03 &      1.34 &      1.60 &      1.94 &      2.28 &      2.42 &      2.41 &      2.64 &      2.87 &      3.02 &      3.17 \\
\hline
Obs./$\sigma_{\mathrm{SM4}}$      &      0.27 &      0.38 &      0.56 &      0.81 &      0.78 &      1.43 &      1.29 &      1.58 &      2.10 &      2.07 &      1.83 &      2.06 &      2.18 &      2.67 &      2.37 \\
\bottomrule
\end{tabular}
%}
\end{center}
\end{ruledtabular}
\end{table*}

\subsection{Higgs boson constraints in fermiophobic (FHM) model}
\label{sec:fpresults}

Within the FHM model described in Sec.~\ref{sec:theory}, the allowed 
fermiophobic Higgs boson, $H_f$, production mechanisms are {\it WH}$_f$, 
{\it ZH}$_f$, and VBF.  Contributions from the dominant SM gluon 
fusion production mechanism, $ggH_f$, are negligibly small.  Despite 
a smaller overall production rate, potential signal contributions of 
a fermiophobic Higgs boson are actually larger for lower Higgs boson 
masses due to increases in the branching ratio, 
$\mathcal{B}(H_f\rightarrow W^+W^-)$, relative to the SM.

\begin{figure}[]
\begin{center}
\includegraphics[width=\linewidth]{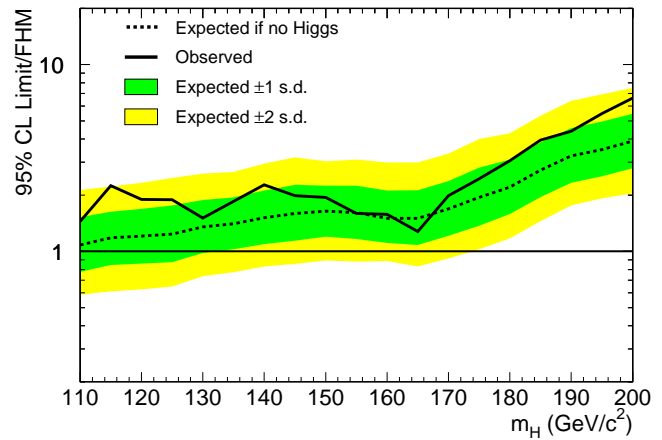}
\end{center}
\caption{Median expected 95\% C.L. upper limits assuming the 
background-only hypothesis (dashed line), and corresponding 
observed limits (solid line) on fermiophobic Higgs boson production 
relative to FHM model expectations from the combination of all 
relevant search samples as a function of the Higgs boson mass.  
The dark and light shaded bands correspond to one and two standard 
deviations assuming the background-only hypothesis.}  
\label{fig:fplimits}
\end{figure}

We extract FHM model constraints from the SS ($\ge$1 Jets) and Trilepton 
search samples, for which the potential signal contributions originate 
solely from {\it WH}$_f$ and {\it ZH}$_f$ production.  Potential 
{\it WH}$_f$, {\it ZH}$_f$, and VBF signal contributions to the 
OS Base search samples are also incorporated.  In the specific case 
of the OS Base ($\ge$2 Jets) sample, the discriminant output used is 
that from the neural network trained to distinguish signal events 
originating from the production mechanisms relevant to the FHM model.  
From the combination of these search samples, we determine 95\% C.L. 
upper bounds on the fermiophobic Higgs boson production rate normalized 
to FHM model expectations using the SM theoretical cross section 
predictions for {\it WH}, {\it ZH}, and VBF production and branching 
ratios as predicted by the FHM model for $H_f\rightarrow W^+W^-$ listed 
in Table~\ref{table:xsec}.  Median expected 95\% C.L. upper limits 
assuming the background-only hypothesis, and corresponding observed 
limits on fermiophobic Higgs boson production relative to FHM model 
expectations are listed in Table~\ref{tab:fplimits} and presented 
graphically in Fig.~\ref{fig:fplimits}.  

\begin{table*}[t]
   \setlength{\extrarowheight}{3pt}
\begin{ruledtabular} 
\begin{center} 
\caption{\label{tab:fplimits}
Median expected 95\% C.L. upper limits assuming the 
background-only hypothesis, and corresponding observed limits
on fermiophobic Higgs boson production relative to FHM model 
expectations from the combination of all relevant search samples 
for 19 mass hypotheses within the range 110~$< m_H <$~200~GeV/$c^2$.  
The boundaries of the one and two standard deviations bands 
assuming the background-only hypothesis are also provided.} 
%{\scriptsize
\begin{tabular}{l*{19}{c}}
\toprule
$m_H$                   &       110 &       115 &       120 &       125 &       130 &       135 &       140 &       145 &       150 &       155 &       160 &       165 &       170 &       175 &       180 &       185 &       190 &       195 &       200 \\
\hline
$-2\sigma/\sigma_{\mathrm{FHM}}$  &      0.59 &      0.61 &      0.63 &      0.65 &      0.74 &      0.77 &      0.83 &      0.86 &      0.89 &      0.88 &      0.89 &      0.83 &      0.92 &      1.03 &      1.17 &      1.45 &      1.76 &      1.92 &      2.04 \\
$-1\sigma/\sigma_{\mathrm{FHM}}$  &      0.78 &      0.85 &      0.86 &      0.88 &      0.98 &      1.03 &      1.09 &      1.14 &      1.20 &      1.17 &      1.11 &      1.08 &      1.21 &      1.37 &      1.58 &      1.95 &      2.33 &      2.53 &      2.79 \\
Exp./$\sigma_{\mathrm{FHM}}$      &      1.08 &      1.18 &      1.21 &      1.24 &      1.35 &      1.41 &      1.51 &      1.60 &      1.64 &      1.61 &      1.50 &      1.50 &      1.69 &      1.95 &      2.21 &      2.72 &      3.24 &      3.51 &      3.90 \\
$+1\sigma/\sigma_{\mathrm{FHM}}$  &      1.53 &      1.63 &      1.69 &      1.76 &      1.89 &      1.95 &      2.12 &      2.27 &      2.25 &      2.25 &      2.12 &      2.13 &      2.38 &      2.80 &      3.10 &      3.83 &      4.57 &      4.96 &      5.45 \\
$+2\sigma/\sigma_{\mathrm{FHM}}$  &      2.13 &      2.22 &      2.33 &      2.47 &      2.61 &      2.67 &      2.95 &      3.19 &      3.04 &      3.10 &      2.99 &      2.99 &      3.33 &      4.00 &      4.30 &      5.32 &      6.39 &      6.95 &      7.51 \\
\hline
Obs./$\sigma_{\mathrm{FHM}}$      &      1.45 &      2.25 &      1.90 &      1.89 &      1.51 &      1.85 &      2.28 &      1.98 &      1.95 &      1.60 &      1.58 &      1.28 &      1.99 &      2.45 &      3.05 &      3.94 &      4.40 &      5.48 &      6.63 \\
\bottomrule
\end{tabular}
%}
\end{center}
\end{ruledtabular}
\end{table*}

\section{Conclusion}
\label{sec:summary}

We present the results of CDF searches for the Higgs boson 
focusing on the $H\rightarrow W^+W^-$ decay mode.  The 
searches are based on the final CDF~II data set corresponding 
to an integrated luminosity of 9.7~fb$^{-1}$.  In the context 
of the SM, we exclude at the 95\% C.L. Higgs bosons with masses 
in the range $149<m_H<172$~GeV/$c^2$.  The expected exclusion 
range, in the absence of a signal, is $155<m_H<175$~GeV/$c^2$. 
In the case of a SM-like Higgs boson in the presence of a fourth 
generation of fermions with the lowest lepton and neutrino masses
allowed by current experimental constraints, we exclude the range 
$124<m_H<200$~GeV/$c^2$ at the 95\% C.L., where the expected 
exclusion region is $124<m_H<221$~GeV/$c^2$.  Upper limits on 
fermiophobic Higgs boson production are also presented.   

\section{Acknowledgments}

We thank the Fermilab staff and the technical staffs of the 
participating institutions for their vital contributions. 
This work was supported by the U.S. Department of Energy and 
National Science Foundation; the Italian Istituto Nazionale di 
Fisica Nucleare; the Ministry of Education, Culture, Sports, 
Science and Technology of Japan; the Natural Sciences and
Engineering Research Council of Canada; the National Science 
Council of the Republic of China; the Swiss National Science 
Foundation; the A.P. Sloan Foundation; the Bundesministerium 
f\"ur Bildung und Forschung, Germany; the Korean World Class 
University Program, the National Research Foundation of Korea; 
the Science and Technology Facilities Council and the Royal 
Society, UK; the Russian Foundation for Basic Research; 
the Ministerio de Ciencia e Innovaci\'{o}n, and Programa
Consolider-Ingenio 2010, Spain; the Slovak R\&D Agency; 
the Academy of Finland; the Australian Research Council (ARC); 
and the EU community Marie Curie Fellowship contract 302103. 

% BIBLIOGRAPHY
%\bibliography{HWW_PRD_submit}

\end{document}

%% file: November2012_Authors-1.tex
\affiliation{Institute of Physics, Academia Sinica, Taipei, Taiwan 11529, Republic of China}
\affiliation{Argonne National Laboratory, Argonne, Illinois 60439, USA}
\affiliation{University of Athens, 157 71 Athens, Greece}
\affiliation{Institut de Fisica d'Altes Energies, ICREA, Universitat Autonoma de Barcelona, E-08193, Bellaterra (Barcelona), Spain}
\affiliation{Baylor University, Waco, Texas 76798, USA}
\affiliation{Istituto Nazionale di Fisica Nucleare Bologna, $^{ee}$University of Bologna, I-40127 Bologna, Italy}
\affiliation{University of California, Davis, Davis, California 95616, USA}
\affiliation{University of California, Los Angeles, Los Angeles, California 90024, USA}
\affiliation{University of California, San Diego, La Jolla, California  92093} 
\affiliation{Instituto de Fisica de Cantabria, CSIC-University of Cantabria, 39005 Santander, Spain}
\affiliation{Carnegie Mellon University, Pittsburgh, Pennsylvania 15213, USA}
\affiliation{Enrico Fermi Institute, University of Chicago, Chicago, Illinois 60637, USA}
\affiliation{Comenius University, 842 48 Bratislava, Slovakia; Institute of Experimental Physics, 040 01 Kosice, Slovakia}
\affiliation{Joint Institute for Nuclear Research, RU-141980 Dubna, Russia}
\affiliation{Duke University, Durham, North Carolina 27708, USA}
\affiliation{Fermi National Accelerator Laboratory, Batavia, Illinois 60510, USA}
\affiliation{University of Florida, Gainesville, Florida 32611, USA}
\affiliation{Laboratori Nazionali di Frascati, Istituto Nazionale di Fisica Nucleare, I-00044 Frascati, Italy}
\affiliation{University of Geneva, CH-1211 Geneva 4, Switzerland}
\affiliation{Glasgow University, Glasgow G12 8QQ, United Kingdom}
\affiliation{Harvard University, Cambridge, Massachusetts 02138, USA}
\affiliation{Division of High Energy Physics, Department of Physics, University of Helsinki and Helsinki Institute of Physics, FIN-00014, Helsinki, Finland}
\affiliation{University of Illinois, Urbana, Illinois 61801, USA}
\affiliation{The Johns Hopkins University, Baltimore, Maryland 21218, USA}
\affiliation{Institut f\"{u}r Experimentelle Kernphysik, Karlsruhe Institute of Technology, D-76131 Karlsruhe, Germany}
\affiliation{Center for High Energy Physics: Kyungpook National University, Daegu 702-701, Korea; Seoul National University, Seoul 151-742, Korea; Sungkyunkwan University, Suwon 440-746, Korea; Korea Institute of Science and Technology Information, Daejeon 305-806, Korea; Chonnam National University, Gwangju 500-757, Korea; Chonbuk National University, Jeonju 561-756, Korea; Ewha Womans University, Seoul, 120-750, Korea}
\affiliation{Ernest Orlando Lawrence Berkeley National Laboratory, Berkeley, California 94720, USA}
\affiliation{University of Liverpool, Liverpool L69 7ZE, United Kingdom}
\affiliation{University College London, London WC1E 6BT, United Kingdom}
\affiliation{Centro de Investigaciones Energeticas Medioambientales y Tecnologicas, E-28040 Madrid, Spain}
\affiliation{Massachusetts Institute of Technology, Cambridge, Massachusetts 02139, USA}
\affiliation{Institute of Particle Physics: McGill University, Montr\'{e}al, Qu\'{e}bec H3A~2T8, Canada; Simon Fraser University, Burnaby, British Columbia V5A~1S6, Canada; University of Toronto, Toronto, Ontario M5S~1A7, Canada; and TRIUMF, Vancouver, British Columbia V6T~2A3, Canada}
\affiliation{University of Michigan, Ann Arbor, Michigan 48109, USA}
\affiliation{Michigan State University, East Lansing, Michigan 48824, USA}
\affiliation{Institution for Theoretical and Experimental Physics, ITEP, Moscow 117259, Russia}
\affiliation{University of New Mexico, Albuquerque, New Mexico 87131, USA}
\affiliation{The Ohio State University, Columbus, Ohio 43210, USA}
\affiliation{Okayama University, Okayama 700-8530, Japan}
\affiliation{Osaka City University, Osaka 588, Japan}
\affiliation{University of Oxford, Oxford OX1 3RH, United Kingdom}
\affiliation{Istituto Nazionale di Fisica Nucleare, Sezione di Padova-Trento, $^{ff}$University of Padova, I-35131 Padova, Italy}
\affiliation{University of Pennsylvania, Philadelphia, Pennsylvania 19104, USA}
\affiliation{Istituto Nazionale di Fisica Nucleare Pisa, $^{gg}$University of Pisa, $^{hh}$University of Siena and $^{ii}$Scuola Normale Superiore, I-56127 Pisa, Italy, $^{mm}$INFN Pavia and University of Pavia, I-27100 Pavia, Italy}
\affiliation{University of Pittsburgh, Pittsburgh, Pennsylvania 15260, USA}
\affiliation{Purdue University, West Lafayette, Indiana 47907, USA}
\affiliation{University of Rochester, Rochester, New York 14627, USA}
\affiliation{The Rockefeller University, New York, New York 10065, USA}
\affiliation{Istituto Nazionale di Fisica Nucleare, Sezione di Roma 1, $^{jj}$Sapienza Universit\`{a} di Roma, I-00185 Roma, Italy}
\affiliation{Rutgers University, Piscataway, New Jersey 08855} 
\affiliation{Mitchell Institute for Fundamental Physics and Astronomy, Texas A\&M University, College Station, Texas 77843, USA}
\affiliation{Istituto Nazionale di Fisica Nucleare Trieste/Udine; $^{nn}$University of Trieste, I-34127 Trieste, Italy; $^{kk}$University of Udine, I-33100 Udine, Italy}
\affiliation{University of Tsukuba, Tsukuba, Ibaraki 305, Japan}
\affiliation{Tufts University, Medford, Massachusetts 02155, USA}
\affiliation{University of Virginia, Charlottesville, Virginia 22906, USA}
\affiliation{Waseda University, Tokyo 169, Japan}
\affiliation{Wayne State University, Detroit, Michigan 48201, USA}
\affiliation{University of Wisconsin, Madison, Wisconsin 53706, USA}
\affiliation{Yale University, New Haven, Connecticut 06520, USA}

\author{T.~Aaltonen}
\affiliation{Division of High Energy Physics, Department of Physics, University of Helsinki and Helsinki Institute of Physics, FIN-00014, Helsinki, Finland}
\author{S.~Amerio}
\affiliation{Istituto Nazionale di Fisica Nucleare, Sezione di Padova-Trento, $^{ff}$University of Padova, I-35131 Padova, Italy}
\author{D.~Amidei}
\affiliation{University of Michigan, Ann Arbor, Michigan 48109, USA}
\author{A.~Anastassov$^x$}
\affiliation{Fermi National Accelerator Laboratory, Batavia, Illinois 60510, USA}
\author{A.~Annovi}
\affiliation{Laboratori Nazionali di Frascati, Istituto Nazionale di Fisica Nucleare, I-00044 Frascati, Italy}
\author{J.~Antos}
\affiliation{Comenius University, 842 48 Bratislava, Slovakia; Institute of Experimental Physics, 040 01 Kosice, Slovakia}
\author{G.~Apollinari}
\affiliation{Fermi National Accelerator Laboratory, Batavia, Illinois 60510, USA}
\author{J.A.~Appel}
\affiliation{Fermi National Accelerator Laboratory, Batavia, Illinois 60510, USA}
\author{T.~Arisawa}
\affiliation{Waseda University, Tokyo 169, Japan}
\author{A.~Artikov}
\affiliation{Joint Institute for Nuclear Research, RU-141980 Dubna, Russia}
% ...

\collaboration{CDF Collaboration\footnote{With visitors from
$^a$University of British Columbia, Vancouver, BC V6T 1Z1, Canada,
$^b$Istituto Nazionale di Fisica Nucleare, Sezione di Cagliari, 09042 Monserrato (Cagliari), Italy,
$^c$University of California Irvine, Irvine, CA 92697, USA,
$^e$Institute of Physics, Academy of Sciences of the Czech Republic, 182~21, Czech Republic,
$^f$CERN, CH-1211 Geneva, Switzerland,
$^g$Cornell University, Ithaca, NY 14853, USA,
$^{dd}$The University of Jordan, Amman 11942, Jordan,
$^h$University of Cyprus, Nicosia CY-1678, Cyprus,
$^i$Office of Science, U.S. Department of Energy, Washington, DC 20585, USA,
$^j$University College Dublin, Dublin 4, Ireland,
$^k$ETH, 8092 Z\"{u}rich, Switzerland,
$^l$University of Fukui, Fukui City, Fukui Prefecture, Japan 910-0017,
$^m$Universidad Iberoamericana, Lomas de Santa Fe, M\'{e}xico, C.P. 01219, Distrito Federal,
$^n$University of Iowa, Iowa City, IA 52242, USA,
$^o$Kinki University, Higashi-Osaka City, Japan 577-8502,
$^p$Kansas State University, Manhattan, KS 66506, USA,
$^q$Brookhaven National Laboratory, Upton, NY 11973, USA,
$^r$University of Manchester, Manchester M13 9PL, United Kingdom,
$^s$Queen Mary, University of London, London, E1 4NS, United Kingdom,
$^t$University of Melbourne, Victoria 3010, Australia,
$^u$Muons, Inc., Batavia, IL 60510, USA,
$^v$Nagasaki Institute of Applied Science, Nagasaki 851-0193, Japan,
$^w$National Research Nuclear University, Moscow 115409, Russia,
$^x$Northwestern University, Evanston, IL 60208, USA,
$^y$University of Notre Dame, Notre Dame, IN 46556, USA,
$^z$Universidad de Oviedo, E-33007 Oviedo, Spain,
$^{aa}$CNRS-IN2P3, Paris, F-75205 France,
$^{cc}$Universidad Tecnica Federico Santa Maria, 110v Valparaiso, Chile,
$^{ll}$Universite catholique de Louvain, 1348 Louvain-La-Neuve, Belgium,
$^{oo}$University of Z\"{u}rich, 8006 Z\"{u}rich, Switzerland,
$^{pp}$Massachusetts General Hospital and Harvard Medical School, Boston, MA 02114 USA,
$^{qq}$Hampton University, Hampton, VA 23668, USA,
$^{rr}$Los Alamos National Laboratory, Los Alamos, NM 87544, USA
}}
\noaffiliation

%% file: HWW_PRD_submit.bbl
\begin{thebibliography}{123}
\bibitem[{coo()}]{coordsys}
\bibinfo{note}{Positions and angles are expressed in a cylindrical coordinate
  system, with the $z$ axis directed along the proton beam. The origin of the
  coordinate system lies at the center of the detector. The azimuthal angle
  $\phi$ around the beam axis is defined with respect to a horizontal vector
  running outwards from the center of the Tevatron, and radii are measured with
  respect to the beam axis. The polar angle $\theta$ is defined with respect to
  the proton beam direction, and the pseudorapidity $\eta$ is defined to be
  $\eta = -\ln[\tan(\theta/2)]$. The transverse energy and momentum of a
  particle are defined as $E_T = E \sin \theta$ and $p_T = p \sin \theta$,
  respectively.}
\end{thebibliography}
